\documentclass[12pt,preprint]{aastex}
\usepackage{apjfonts}

\usepackage{lscape}

\usepackage{graphicx,subfigure}
\usepackage{natbib}
\newcommand{\etal}{et al.}
\newcommand{\hbeta}{H{$\beta$}}
\newcommand{\halpha}{H{$\alpha$}}

\newcommand{\OIII}{[O{\sevenrm\,III}]}
\newcommand{\OIIIa}{[O{\sevenrm\,III}]\,$\lambda$4959}
\newcommand{\OIIIb}{[O{\sevenrm\,III}]\,$\lambda$5007}
\newcommand{\NII}{[N{\sevenrm\,II}]}
\newcommand{\NIIb}{[N{\sevenrm\,II}]\,$\lambda$6584}

 \font\sevenrm=cmr7 scaled 1000

\begin{document}

\title{Type 2 Active Galactic Nuclei with Double-Peaked \OIII\ Lines. II. Single AGNs with Complex Narrow-Line Region Kinematics are More
Common than Binary AGNs}


\shorttitle{DOUBLE-PEAKED NARROW LINE AGNS}


\shortauthors{SHEN ET AL.}
\author{Yue Shen\altaffilmark{1}, Xin Liu\altaffilmark{1,2,4},
Jenny E. Greene\altaffilmark{3,4}, Michael A. Strauss\altaffilmark{4}}
\altaffiltext{1}{Harvard-Smithsonian Center for Astrophysics, 60
Garden Street, MS-51, Cambridge, MA 02138, USA.}
\altaffiltext{2}{Einstein Fellow.}
\altaffiltext{3}{Department of Astronomy, UT Austin, Austin, TX 78712, USA.}
\altaffiltext{4}{Princeton University Observatory, Princeton, NJ 08544, USA.}

\begin{abstract}
Approximately $1\%$ of low redshift ($z\la 0.3$)
optically-selected type 2 Active Galactic Nuclei (AGNs) show a
double-peaked \OIII\ narrow emission line profile in their
spatially-integrated spectra. Such features are usually
interpreted as due either to kinematics, such as biconical
outflows and/or disk rotation of the narrow line region (NLR)
around single black holes, or to the relative motion of two
distinct NLRs in a merging pair of AGNs. Here we report
follow-up near infrared (NIR) imaging and optical slit
spectroscopy of $31$ double-peaked \OIII\ type 2 AGNs drawn
from the Sloan Digital Sky Survey (SDSS) parent sample
presented in \citet{Liu_etal_2010a}. The NIR imaging traces the
old stellar population in each galaxy, while the optical slit
spectroscopy traces the NLR gas. These data reveal a mixture of
origins for the double-peaked feature. Roughly $10\%$ of our
objects are best explained by binary AGNs at (projected)
kpc-scale separations, where two stellar components with
spatially coincident NLRs are seen. $\sim 50\%$ of our objects
have \OIII\ emission offset by a few kpc, corresponding to the
two velocity components seen in the SDSS spectra, but there are
no corresponding double stellar components seen in the NIR
imaging. For those objects with sufficiently high quality slit
spectra, we see velocity and/or velocity dispersion gradients
in \OIII\ emission, suggestive of the kinematic signatures of a
single NLR. The remaining $\sim 40\%$ of our objects are
ambiguous, and will need higher spatial resolution observations
to distinguish between the two scenarios. Our observations
therefore favor the kinematics scenario with a single AGN for
the majority of these double-peaked \OIII\ type 2 AGNs. We
emphasize the importance of combining imaging and slit
spectroscopy in identifying kpc-scale binary AGNs, i.e., in no
cases does one of these alone allow an unambiguous
identification. We estimate that $\sim 0.5-2.5\%$ of the $z\la
0.3$ type 2 AGNs are kpc-scale binary AGNs of comparable
luminosities, with a relative orbital velocity $\ga 150\,{\rm
km\,s^{-1}}$.\\

Based in part on observations obtained with the 6.5 m Magellan
telescopes located at Las Campanas Observatory, Chile, and with
the Apache Point Observatory 3.5 m telescope, which is owned
and operated by the Astrophysical Research Consortium.

\end{abstract}

\section{Introduction}\label{sec:intro}

Binary supermassive black holes (SMBHs) have long been
proposed theoretically \citep[e.g.,][]{Begelman_etal_1980}.
Galaxy mergers frequently occur and it is now widely
appreciated that most bulge-dominated merging galaxies harbor
central SMBHs. Thus the formation of binary SMBHs from galaxy
mergers seems inevitable. Major mergers between two gas-rich
galaxies of comparable mass also provide a viable means to
channel a large amount of gas into the central region to
trigger starbursts and quasar activity
\citep[e.g.,][]{Hernquist_1989}. On the other hand,
observationally confirmed binary SMBHs are surprisingly scarce.
About $0.1\%$ of quasars are in physical pairs with projected
separations of tens to hundreds of kpc
\citep[][]{Hennawi_etal_2006,Hennawi_etal_2010,Myers_etal_2008},
and only a handful of confirmed binaries with projected
separations on kpc scales are known in the literature
\citep[e.g.,][]{Komossa_etal_2003,Bianchi_etal_2008}. On pc
scales, there is only one confirmed binary SMBH separated by
$\sim 7\,$pc in projection \citep[][]{Rodriguez_etal_2006},
detected with the Very Long Baseline Array (VLBA). There are
essentially no confirmed sub-pc binaries, although several
candidates have been proposed, such as OJ287
\citep[e.g.,][]{Valtonen_etal_2008} and SDSSJ1536+0441
\citep[][]{Boroson_Lauer_2009}, for both of which alternative,
{\em non-binary} interpretations exist. The frequency of
observed SMBH binaries is thus far below the expectations from
the observed (and simulation-predicted) galaxy merger rate, and
is in tension with the hypothesis that quasar activity is
triggered by major mergers.

If mergers do produce binary SMBHs, then the most natural
explanation for the apparent deficit of SMBH binaries is that
only a tiny fraction of them are detectable; both BHs must be
active for a detection and either their spatial separations
must be resolvable, or orbital motion can be unambiguously
inferred. So far the searches for binary AGNs/quasars with
separations below several kpc are highly incomplete due to the
stringent spatial resolution requirement. If quasar activity
only occurs at the late stage of a merger (as simulations
suggest), then most binaries will be difficult to detect. A
further complication is that at relatively low AGN luminosity
(i.e., $L_{\rm bol}\la 10^{45}\, {\rm erg\,s^{-1}}$), mergers
may not be necessary to trigger BH activity, and alternative
fueling routes (secular processes) may suffice. Simulations
predict that the effects of mergers on BH accretion start to
become important at late stages when a pair of BHs is separated
by a few kpc \citep[e.g.,][]{Hopkins_etal_2005a}; this is a
scale of which one can still resolve the two nuclei with
current instruments. It is therefore important to quantify the
fraction of kpc-scale binary AGNs. This fraction will shed
light on the AGN triggering mechanism, and individual binaries
will provide ideal test beds to understand the interplay
between AGNs and hosts in these merging systems.


The most efficient and systematic way to find these kpc-scale
binary AGNs is an imaging survey conducted in the optical/NIR
with sub-arcsec resolution (to find kpc separation pairs) and
the ability to identify AGNs. The seeing in imaging data from
the SDSS \citep[][]{SDSS} is only $\sim 1.4$\arcsec, but SDSS
does recover some very low-redshift kpc-scale binary AGNs (Liu
et al. 2010c, in preparation), where the two AGNs are separated
by more than a few arcsec and were observed spectroscopically
with separate fibers. At higher redshift, such kpc-scale binary
AGNs will fall within the 3\arcsec\ fiber and spatial
information within the fiber is completely lost. A different
approach is to select candidates with spectroscopic features
that may hint at a binary and follow them up to confirm their
binary AGN nature. \citet{Gerke_etal_2007} and
\citet{Comerford_etal_2009a} revived the idea of selecting
candidate kpc-scale binary AGNs\footnote{They dubbed these
objects as ``dual AGN''.} based on double-peaked \OIII\ lines
\citep[e.g.,][]{Heckman_etal_1981,Zhou_etal_2004}, where the
hope is that the two velocity components of the \OIII\ line
originate from two distinct narrow line regions (NLRs), each
associated with its own BH. These authors found that in two
objects at $z\sim 0.7$ the two \OIII\ components were spatially
offset by a few kpc in their slit spectra, and claimed them to
be kpc-scale binary AGNs. A third candidate was reported by
\citet{Comerford_etal_2009b} based on spatially offset \OIII\
emission in a slit spectrum and a pair of resolved double
nuclei in {\em HST} imaging \citep[but see][ for an alternative
interperation for this system]{Civano_etal_2010}.

Several independent searches for AGNs with double-peaked \OIII\
lines have since been conducted using the SDSS spectroscopic
database: \citet[][hereafter Paper I]{Liu_etal_2010a} and
\citet{Wang_etal_2009a} selected such objects from type 2 AGNs,
while \citet{Smith_etal_2010} used a sample of type 1 AGNs
(mixed with some type 2 objects due to SDSS pipeline
misclassification). More than 200 double-peaked \OIII\ AGNs
were identified (based on visual inspection of SDSS spectra)
with a median redshift $\bar{z}\sim 0.15$, amounting to about
$1\%$ of the total AGN population. With very few exceptions,
these objects do not appear to have double nuclei in their
optical SDSS images \citep[e.g., Paper I;][]{Smith_etal_2010};
and of course the SDSS fiber spectra do not indicate if the two
velocity components are spatially offset by a few kpc. The bulk
properties of the double-peaked AGNs are not dramatically
different from those of the normal AGN population (Paper I).

Not all of these double-peaked \OIII\ objects are kpc-scale
binary AGNs. Spatially resolved studies of nearby Seyferts
frequently show a quite dynamic picture of the NLR, involving
outflows, inflows, rotation, and interactions with a radio jet.
Indeed, double-peaked \OIII\ features were already seen in
early NLR studies
\citep[e.g.,][]{Sargent_1972,Heckman_etal_1981,Veilleux_1991}.
One famous example is Mrk 78, which has a bipolar outflowing
NLR structure, broadly aligned with the linear radio jet
\citep[e.g.,][]{Whittle_Wilson_2004,Whittle_etal_2005,Fischer_etal_2010};
its spatially integrated spectrum shows a clear double-peaked
\OIII\ line profile \citep[see fig. 1 of][]{Heckman_etal_1981}.
Spatially resolved studies with {\em HST} have revealed more
such cases \citep[such as NGC 1068 and NGC 4151,
e.g.,][]{Crenshaw_Kraemer_2000,Crenshaw_etal_2000a,Veilleux_etal_2001,Das_etal_2005},
where the NLR is dominated by outflows and/or rotation. For
these dynamic NLRs, double-peaked features will naturally arise
in spatially-integrated spectra when viewed at proper angles.
Moreover, these outflows or rotating gaseous disks can extend
as far as a few kpc. Thus detecting two \OIII\ peaks offset by
a few kpc in slit spectra alone is {\em not sufficient} to
conclude that the object is a kpc-scale binary AGN. All these
complications simply reflect the diverse nature of NLR gas
dynamics, and follow-up observations are needed to identify
bona fide kpc-scale binary AGNs from the double-peaked \OIII\
sample.

To that end, we have been conducting follow-up observations of
our double-peaked \OIII\ sample (Paper I) with ground-based NIR
imaging and slit spectroscopy. The NIR imaging offers spatial
resolution $\sim 0.6$\arcsec\ (much better than the typical
$\sim 1.4$\arcsec\ seeing for optical SDSS imaging), and probes
the old stellar populations (mostly stars in the bulge). Unlike
the NLR gas, the stellar populations are unlikely to be
involved in outflows, thus are a much less ambiguous indicator
of the binary nature than using the NLR gas as a tracer. Slit
spectroscopy probes the spatial distribution of NLR gas
emission, giving the spatial information unavailable in the
SDSS fiber spectra. Once a pair of nuclei are resolved in the
NIR imaging, slit spectroscopy is required to associate the two
\OIII\ peaks with each of the two nuclei. If, as is the case
for Mrk 78, the NIR image shows a smooth stellar distribution
while the optical spectrum shows two spatially offset \OIII\
peaks without NIR counterparts, then the double-peaked feature
seen in the spatially integrated spectrum is more likely due to
the kinematics of the NLR of a single AGN. Our imaging and
spectroscopy follow-up observations have led to the discovery
of several candidate binaries \citep[][]{Liu_etal_2010b}. In
these objects, we see spatially resolved double nuclei in the
NIR images, whose locations are coincident with the two \OIII\
components in the slit spectra. The two \OIII\ components have
a relative velocity offset of a few hundred ${\rm kms^{-1}}$,
which leads to the double-peaked profile in the SDSS spectrum.
The two components in each case have emission line flux ratios
that place them in the AGN region of the BPT diagram
\citep[][]{BPT}. These cases strongly suggest that they are
binary AGNs at $\sim $ kpc (projected) separations, which are
co-rotating along with their stellar bulges and are ionizing
their individual NLRs in a merging pair of galaxies.

Recently, the interest of these double-peaked \OIII\ objects
has led to two NIR imaging studies with ground-based adaptive
optics (AO) systems \citep[][]{Fu_etal_2010,Rosario_etal_2011}.
With diffraction-limited resolution ($\sim 0.1$\arcsec), these
studies were able to resolve even closer double nucleus in the
NIR than our imaging data under natural seeing. However,
optical spectroscopy is still needed to register the resolved
double nucleus with the \OIII\ emission in order to confirm the
binary AGN nature, which becomes challenging on these smallest
scales from the ground.

In this paper we present a summary of our imaging and
spectroscopy follow-up for 31 objects in our parent sample of
double-peaked \OIII\ type 2 AGNs presented in Paper I. The
details of the parent sample (167 objects) can be found in that
paper. Here we briefly describe the construction of this
sample. We started from a sample of $\sim 15,000$ type 2 AGNs
mostly from the MPA-JHU SDSS DR7 galaxy
sample\footnote{http://www.mpa-garching.mpg.de/SDSS/} with the
following criteria: (1) the rest-frame wavelength ranges [4700,
5100]\AA\ and [4982, 5035]\AA\ centered on the \OIIIb\ line
have median signal-to-noise ratio (S/N) $> 5$ pixel$^{-1}$ and
bad pixel fraction $<30\%$; (2) the \OIIIb\ line is detected at
$>5\sigma$ and has a rest-frame equivalent width (EW) $>4$\AA;
3) the line flux ratio \OIIIb/\hbeta $>3$ if $z > 0.33$, or the
diagnostic line ratios \OIIIb/\hbeta and \NIIb/\halpha\ lie
above the theoretical upper limits for star-formation
excitation from Kewley et al.\ (2001) on the BPT diagram
(Baldwin et al.\ 1981) if $z < 0.33$. All AGNs were then
visually inspected and those with well-detected double peaks in
both \OIIIa\ and \OIIIb\ with similar profiles were included in
our final sample. We did not include those with complex line
profiles such as lumpy, winged, or multi-component features.

The structure of the paper is as follows. We describe our
observations and data reduction in \S\ref{sec:data}, followed
by discussions of individual objects in \S\ref{sec:inter},
where we categorize objects into binaries (\S\ref{sec:binary}),
NLR kinematics around single AGNs (\S\ref{sec:kinematics}), and
ambiguous cases (\S\ref{sec:amb}). We discuss the frequency of
kpc-scale binary AGNs in \S\ref{sec:dist}, and conclude in
\S\ref{sec:con}. Throughout this paper we adopt a flat
$\Lambda$CDM cosmology with $\Omega_0=0.3$,
$\Omega_\Lambda=0.7$ and $H_0=70\,{\rm kms^{-1}Mpc^{-1}}$.


\section{Observations and Data Reduction}\label{sec:data}


\subsection{NIR Imaging}\label{subsec:panic}

We obtained $K_s$-band (or $J$-band if $K_s$ is unavailable) images for $61$ objects in our double-peaked sample during six nights in two observing
runs using the Persson's Auxiliary Nasmyth Infrared Camera \citep[PANIC;][]{martini04} on the 6.5 m Magellan I (Baade) telescope. Here we
report\footnote{For SDSS J1146+5110 we only have 2MASS images because this object is unobservable from the Magellan site.} 31 of these objects for
which we also have slit spectroscopy, and will publish the remaining imaging data when slit spectroscopy data become available. The first observing
run was on the nights of 2009 December 29 through 2010 January 2 UT and the second was on the night of 2010 May 30 UT.  The typical observing
procedure consisted of two sequences each dithered at nine positions with 25\arcsec\ offsets.  We obtained four 15-second exposures at each position
so that the typical total exposure time per target was 18 min.  We observed standard stars \citep{persson98} at the beginning, middle, and end of
each
night.  The observing conditions were clear but not photometric, with seeing ranging between 0\arcsec.5 and 0\arcsec.8 in the optical (through an
RG610 filter) during the first run and 0\arcsec.7 and 1\arcsec.5 during the second.  Table \ref{table:log} lists total exposure times and seeing
measured from field stars whenever available or from adjacent observations otherwise.

PANIC has a $2'\times2'$ field of view (FOV) and 0\arcsec.125 pixels. We reduced PANIC data using the Carnegie Supernova Project pipeline
\citep{hamuy06} following standard procedures, including spatial-distortion correction, dark subtraction, bad-pixel masking, flat fielding (using
twilight flats), sky subtraction, and aligning and stacking of the dithered frames.
We determined $K_s$ and $J$ photometric zero points using 2MASS \citep[][]{2MASS} magnitudes of field stars
when available, or the standard stars we observed during each night.

\subsection{Optical Slit Spectroscopy}\label{subsec:ldss3}

We conducted optical slit spectroscopy using the Low-Dispersion Survey Spectrograph (LDSS3) on the 6.5 m Magellan II (Clay) telescope and the Dual
Imaging Spectrograph (DIS) on the Apache Point Observatory 3.5 m telescope. To date we have observed 31 objects with NIR imaging data. All objects
which appear to have double nuclei in their NIR images were observed with slit spectroscopy.

Our LDSS3 runs were on the nights of 2010 January 12 through 14
UT. The observing conditions were clear but not photometric,
with seeing ranging between 0\arcsec.7 and 1\arcsec.1.  LDSS3
has a 8$'$.3 diameter FOV and 0\arcsec.188 pixels.  We employed
a 1$''\times$4$'$ long-slit with the VPH-Blue grism to cover
\hbeta\ and \OIII\ in the blue ($\sim$ 4260--7020\AA), and the VPH-Red grism (with
the OG590 filter) to cover \halpha\ and \NII\ in the red ($\sim$ 5800--9800\AA). The spectral resolution was 3.1 (6.3) \AA\ FWHM
in the blue (red). For
most of our LDSS3 targets we only obtained spectra in the blue
due to observing time constraints.  For objects with
double stellar components identified either from our NIR imaging or
from SDSS images, we oriented the slit to go through the two
stellar components; for the others we oriented the slit along the
major axis of the galaxy. For a few targets whose \OIII\ emission
along the primary slit position was either not spatially
resolved or was particularly complex, we obtained spectra at a
second slit position usually perpendicular to the primary slit
position. We list slit positions,
total exposure times, and seeing measured from field stars in
acquisition images in Table \ref{table:log}. We took wavelength calibration
spectra and flat fields after observing each object. During the
course of each night, we observed two white dwarfs at different
airmasses for spectrophotometric calibration and several K and
M giants for velocity calibration.


Our DIS observations were carried out during eight nights
between 2009 July 15 and 2010 July 14 UT. The observing
conditions were partly cloudy or clear but not photometric on
the nights of 2009 July 15, September 23, December 12 and 19,
and 2010 July 07 and 14, and photometric on the nights of 2010
February 15 and June 10 UT. During most nights the seeing was
poor, ranging between 1\arcsec.5 and 2\arcsec.7 (except the
nights of 2009 December 12 and 19 with seeing around
1\arcsec.0).  DIS has a 4$'\times$6$'$ FOV and 0\arcsec.414
pixels. The spectral resolution was 1.8 (1.3) \AA\ FWHM
in the blue (red) channel.  We adopted a 1\arcsec.5$\times$6$'$
slit and the B1200+R1200 gratings centered at 5000 and 7000
\AA\ (or 5200 and 7050 or 5500 and 7450 or 5550 and 7550
\AA, depending on target redshifts).  The slit
orientation was determined in the same way as in our LDSS3
runs.  

We reduced the LDSS3 and DIS data following standard
IRAF\footnote{IRAF is distributed by the National Optical
Astronomy Observatory, which is operated by the Association of
Universities for Research in Astronomy (AURA) under cooperative
agreement with the National Science Foundation.} procedures
\citep{tody86} and with the COSMOS reduction
pipeline\footnote{http://www.ociw.edu/Code/cosmos}. The 2d data
reduction included bias subtraction, flat fielding, cosmic ray
removal, wavelength calibration and spatial rectification, flux
calibration and extinction correction, sky subtraction, and
aligning and stacking of individual frames. We applied telluric
correction over the extracted 1d spectra using standard stars;
the O$_2$ A-band (7580--7740\AA) is the major feature that
affects the region of interest in the red. The quality of the
LDSS3 data is substantially better than that of the DIS data
due to the much better seeing and larger telescope, hence we
will always present LDSS3 data whenever available.

\begin{deluxetable}{llcccccccc}
\tabletypesize{\scriptsize}  
\tablecaption{Observing Summary
\label{table:log}} \tablehead{ & \multicolumn{4}{c}{NIR Imaging} & \multicolumn{5}{c}{Slit Spectroscopy}\\
\hline
SDSS Name & Instrument & Seeing & Obs Date & Exp Time & Instrument & Seeing  & Obs Date & PA  & Exp Time \\
          &             & (\arcsec)          &  UT      & (min)           &              & (\arcsec)        &    UT      & ($^\circ$) & (sec)
}  \startdata
0002$+$0045 &    PANIC    $K_s$ &    0.48 &    091230 &    9  &    DIS    & 1.1  &    091212 &    71 &    1800 \\
0009$-$0036 &    PANIC    $K_s$ &    0.63 &    091231 &    18 &    LDSS3  & 0.94 &    100113 &    278 &    900 \\
0116$-$1025 &    PANIC    $K_s$ &    0.52 &    091229 &    18 &    LDSS3  & 0.83 &    100112 &    95 &    1800  \\
0135$-$0058 &    PANIC    $K_s$ &    0.58 &    091229 &    18 &    LDSS3  & 0.86 &    100112 &    101 &    1800  \\
0135$+$1435 &    PANIC    $J$   &    0.55 &    100101 &    18 &    LDSS3  & 0.92 &    100113 &    200 &    900  \\
0156$-$0007 &    PANIC    $K_s$ &    0.64 &    091229 &    18 &    LDSS3  & 1.0  &    100113 &    299 &    1800  \\
0400$-$0652 &    PANIC    $K_s$ &    0.48 &    091229 &    18 &    LDSS3  & 0.58/0.57 &    100112/100114 &  95/173  &    2700/1800 \\
0837$+$1500 &    PANIC    $K_s$ &    0.66 &    091229 &    27 &    LDSS3  & 1.1 &    100113 &    322 &    1800 \\
0851$+$1327 &    PANIC    $K_s$ &    0.49 &    091230 &    18 &    LDSS3  & 0.91 &    100114 &   201 &    900 \\
0942$+$1254 &    PANIC    $K_s$ &    0.50 &    091229/100102 &    36 &    LDSS3  &  0.88/0.95 &    100112/100113 &  213/348  &    2700/1800 \\
0958$-$0051 &    PANIC    $K_s$ &    0.45 &    091230 &    18 &    LDSS3  &  0.87 &    100113 &    200 &    900  \\
1009$+$0133 &    PANIC    $K_s$ &    0.44 &    091230 &    18 &    LDSS3  &  0.84 &    100114 &    217 &    900  \\
1019$+$0134 &    PANIC    $K_s$ &    0.70 &    091229 &    18 &    LDSS3  &  0.94 &    100114 &    303 &    900  \\
1038$+$0255 &    PANIC    $K_s$ &    0.45 &    091231 &    18 &    LDSS3  &  0.75 &    100114 &    270 &    900  \\
1108$+$0659 &    PANIC    $K_s$ &    0.54 &    091229 &    18 &    LDSS3  &  0.86/0.95 &    100112/100113 &    320 &    4800  \\
1131$-$0204 &    PANIC    $K_s$ &    0.47 &    091231/100102 &    36 &    LDSS3  &  0.94/0.71 &    100112/100113 &    267 &    2700/3600  \\
1146$-$0226 &    PANIC    $K_s$ &    0.64 &    100102 &    18 &    LDSS3  &  0.98 &    100114 &    233$^\dag$ &    900 \\
1146$+$5110 &    2MASS    $K$   &    ...  &     ...   &   ... &    DIS  &  1.7 &    100215 &    51 &    4800 \\
1322$+$2631 &    PANIC    $K_s$ &    0.73 &    100530 &    18 &    DIS   & 2.0/1.8 &    100707/100714 &    79 &    1800/3600 \\
1332$+$0606 &    PANIC    $J$   &    0.52 &    100101 &    18 &    LDSS3   & 1.2/0.86 &    100112/100114 &    196 &    2400/2100 \\
1341$+$2219 &    PANIC    $K_s$ &    0.74 &    100530 &    18 &    DIS   & 1.4 &    100707 &    175 &    3600 \\
1356$+$1026 &    PANIC    $K_s$ &    0.79 &    100530 &    18 &    LDSS3$^*$ & ... & ... & ... & ... \\
1450$+$0838 &    PANIC    $K_s$ &    0.85 &    100530 &    18 &    DIS   & 2.4/2.5 &    100707/100714 &    55 &    3600/1800 \\
1552$+$0433 &    PANIC    $K_s$ &    0.71 &    100530 &    18 &    DIS   & 1.3 &    100610 &    303 &    3600 \\
1556$+$0948 &    PANIC    $K_s$ &    0.58 &    100530 &    18 &    DIS   & 1.5 &    100610 &    158 &    3000 \\
1630$+$1649 &    PANIC    $K_s$ &    0.86 &    100530 &    18 &    DIS   & 1.6 &    100610 &    39 &    3000  \\
2252$+$0029 &    PANIC    $K_s$ &    0.83 &    100530 &    27 &    DIS   & 1.5 &    091219 &    162 &    2400 \\
2255$-$0812 &    PANIC    $K_s$ &    0.86 &    100530 &    18 &    DIS   & 1.5 &    091219 &    180 &    2400 \\
2304$-$0933 &    PANIC    $K_s$ &    0.80 &    100530 &    18 &    DIS   & 2.0 &    090923 &    61 &    1800 \\
2310$-$0900 &    PANIC    $K_s$ &    0.62 &    091230 &    18 &    DIS   & 2.0 &    090923 &    73 &    2700 \\
2333$+$0049 &    PANIC    $K_s$ &    0.76 &    091229 &    18 &    DIS   & 1.5 &    091219 &    117 &    3600 \\
\enddata
\tablecomments{Summary of our NIR imaging and optical slit
spectroscopy observations. The full SDSS designations are given
in Table 2. For some objects we have two observations, which
are listed separately. $^\dag$Parallactic angle. $^*$The
spectroscopic observation was reported in
\citet{Greene_etal_2011}.}
\end{deluxetable}

\begin{deluxetable}{lcccccc}
\tabletypesize{\scriptsize}  
\tablecaption{Object Properties
\label{table:prop}} \tablehead{
SDSS Name  & redshift & PANIC offset \arcsec\ (kpc) &  Spec offset \arcsec\ (kpc) & Category
 }  \startdata
 000249.07$+$004504.8  &  0.0868        &                & 0.8 (1.3) & NLR kinematics \\
 000911.58$-$003654.7  &  0.0733        &                & $<0.2$ ($<0.3$) & ambiguous\\
 011659.59$-$102539.1  &  0.1503        &                & 0.9 (2.3) & NLR kinematics \\
 013546.93$-$005858.5  &  0.1595        &                & 0.2 (0.6) & NLR kinematics \\
 013555.82$+$143529.7  &  0.0719        &                & 0.8 (1.1) & NLR kinematics \\
 015605.14$-$000721.7  &  0.0806        &                & 0.6 (0.9) & NLR kinematics \\
 040001.59$-$065254.1  &  0.1707        &                & 1.3/$<0.2$ (3.8/$<0.6$) & NLR kinematics \\
 083713.49$+$150037.2  &  0.1408        &                & 1.3 (3.2) & NLR kinematics \\
 085121.94$+$132702.2  &  0.0931        &                & 0.6 (1.0) & NLR kinematics \\
 094205.83$+$125433.7  &  0.1543        &                & 0.2/$<0.2$ (0.5/$<0.5$) & ambiguous \\
 095833.20$-$005118.6  &  0.0860        &                & 0.8 (1.3) & NLR kinematics \\
 100921.26$+$013334.6  &  0.1437        &                & 0.2 (0.5) & ambiguous \\
 101927.56$+$013422.5  &  0.0730        &                & 0.3 (0.4) & ambiguous \\
 103850.13$+$025555.1  &  0.0762        &                & 0.9 (1.3) & NLR kinematics \\
 110851.04$+$065901.4  &  0.1816        &    0.5 (1.5)   & 0.9 (2.7) & binary AGN$^a$ \\
 113126.08$-$020459.2  &  0.1463        &    0.6 (1.5)   & 0.6 (1.5) & binary AGN$^a$ \\
 114610.04$-$022619.2  &  0.1225        &                & 1.3 (2.9) & NLR kinematics \\
 114642.47$+$511029.6  &  0.1300        &    2.7 (6.2)   & 2.5 (5.8) & binary AGN$^a$ \\
 132231.86$+$263159.1  &  0.1441        &                & 2.1 (5.3) & ambiguous \\      
 133226.34$+$060627.4  &  0.2070        &    1.5 (5.1)   & 1.5 (5.1) & binary AGN$^a$ \\
 134114.87$+$221957.8  &  0.1152        &                & 1.7 (3.5) & NLR kinematics \\
 135646.11$+$102609.1  &  0.1231        &    1.3 (2.9)   & 1.3 (2.9)$^b$ & binary AGN$^b$ \\
 145050.60$+$083832.6  &  0.1168        &                & $<0.4$ ($<0.8$) & ambiguous \\
 155205.93$+$043317.5  &  0.0803        &                & 1.2 (1.8) & NLR kinematics \\
 155619.30$+$094855.6  &  0.0678        &                & 0.4 (0.5) & ambiguous \\
 163056.75$+$164957.2  &  0.0341        &                & 0.8 (0.5) & NLR kinematics \\
 225252.94$+$002928.4  &  0.1525        &                & $<0.4$ ($<1.1$) & ambiguous \\
 225510.12$-$081234.4  &  0.1494        &                & 0.4 (1.0) & ambiguous \\
 230442.82$-$093345.3  &  0.0319        &                & 0.8 (0.5) & NLR kinematics \\
 231051.95$-$090011.9  &  0.0944        &                & $<0.4$ ($<0.7$) & ambiguous\\
 233313.17$+$004911.8  &  0.1699        &                & $<0.4$ ($<1.2$) & ambiguous \\
\enddata
\tablecomments{Properties of the 31 objects in our sample. In
those cases which showed two nuclei in NIR imaging, column 3
shows the spatial offset (in units of \arcsec\ and kpc) between
the two nuclei, which is measured from the emission peaks of
the two nuclei. Column 4 shows the spatial offset (in units of
\arcsec\ and kpc) between the two velocity components of the
narrow line emission, measured from the emission peaks of the
two velocity components in the slit spectrum. The smallest
spatial offset of the two velocity components we can measure is
the pixel scale, i.e., $\sim 0.2$\arcsec\ for LDSS3 and $\sim
0.4$\arcsec\ for DIS. For J0400-0652 and J0942+1254 we have
measurements for two slit position angles. The last column
shows our classification of these objects (see
\S\ref{sec:inter}). $^a$Published in
\citet[][]{Liu_etal_2010b}; $^b$Slit spectrum reported in
\citet{Greene_etal_2011}.}
\end{deluxetable}

\section{Interpretation}\label{sec:inter}

Our combined NIR imaging and slit spectroscopy data reveal that
double-peaked \OIII\ emission arises from a diverse set of
circumstances. We classify them by examining the NIR images and
the two-dimensional spectra simultaneously. Given the typical
seeing conditions in our NIR imaging, two stellar components
with $K_s$ luminosities that are within an order of magnitude
of each other can be identified at separations $\ga
0.6$\arcsec.  The seeing is poorer in our optical slit
spectroscopy. However, the two \OIII\ components are spectrally
resolved, which makes it easier to deblend them spatially in
the 2d spectrum. We measure the spatial offset between the peak
emission of the two velocity components in the 2d spectra
(Table \ref{table:prop}), which can be measured down to one
pixel scale\footnote{Measuring sub-pixel offsets by fitting
model line profiles to the data requires both high
signal-to-noise ratio and symmetric spatial line profiles.
These criteria are generally not satisfied by our spectroscopic
data.} ($0.188$\arcsec\ for LDSS3 and 0.414\arcsec\ for DIS).

Based on the NIR imaging and slit spectroscopy data, we classify the 31 objects in three categories (Table \ref{table:prop}):
\begin{enumerate}

\item {\em Kpc-scale binary AGNs:} There are two spatially
    resolved (or marginally resolved) stellar nuclei in NIR
    imaging. The two velocity components of the narrow line
    emission are spatially coincident with the pair of
    nuclei seen in the NIR. These are the best candidates
    for kpc-scale binary AGNs in our sample. Five objects
    are included in this category, four of which were
    reported in \citet{Liu_etal_2010b}.

\item {\em NLR kinematics in single AGNs.} The NIR imaging
    shows a single smooth stellar component, but the two
    velocity components of the narrow lines are spatially
    offset by $\ga 0.6$\arcsec. We performed tests with
    simulated images of two stellar components with
    different luminosity contrasts, structural parameters
    and seeing conditions, and concluded that two stellar
    components with comparable ($0.1\la L_1/L_2\la 10$)
    luminosities under the actual seeing would have been
    resolved in NIR imaging at the separation indicated by
    the two \OIII\ components. There could still be two
    stellar components hidden in the system if, e.g., the
    luminosity contrast is larger than an order of
    magnitude, the separation of the two stellar components
    is smaller than that inferred from the narrow line
    emission, or one of the stellar components has an
    unusual surface brightness profile. But given the
    similar appearances of these objects in NIR imaging and
    slit spectroscopy to the classic example of Mrk 78 (see
    \S\ref{sec:intro}), the simplest explanation is that
    the double peaks emerge from multiple kinematic
    components in a single NLR for the majority of these
    sources. Fifteen objects are included in this category.
    We refer to these objects below as having NLR
    kinematics origin.

\item {\em Ambiguous cases.} For the remaining objects, a
    single nucleus is seen in the NIR imaging, and the two
    velocity components of the narrow lines are spatially
    offset by $\la 0.4$\arcsec, smaller than the resolution
    of our NIR imaging. These objects could be single AGNs
    with NLR structure on smaller scales, or binary AGNs at
    smaller separations. We need better spatial-resolution
    observations and/or different slit positions to test
    these scenarios. Eleven objects are included in this
    category.

\end{enumerate}

These three categories represent our best effort to interpret
these objects based on current data; they are by no means
exact. In particular, the binary AGN and NLR kinematics
classifications both have caveats that may cause us to
misinterpret their nature. Keeping this in mind, we adopt this
classification scheme in our following discussion. Below we
summarize the objects in each category, and discuss the
caveats.

\subsection{Kpc-Scale Binary AGNs}\label{sec:binary}

Our primary interest is to identify bona fide kpc-scale binary
AGNs, which motivated our original search for these
double-peaked objects in SDSS spectroscopic database. Most of
our NIR imaging preceded the slit spectroscopy, and we obtained
spectra of all our targets with resolved double nuclei in the
NIR. Five out of $\sim 60$, or $\sim 10\%$ of our objects show
resolved double stellar nuclei in their NIR images. Their
two-dimensional (2d) spectra show that the two \OIII\ velocity
components seen in the spatially-integrated spectra are
spatially coincident with the two stellar continuum peaks.

\begin{figure*}
  \centering
    \includegraphics[width=0.3\textwidth]{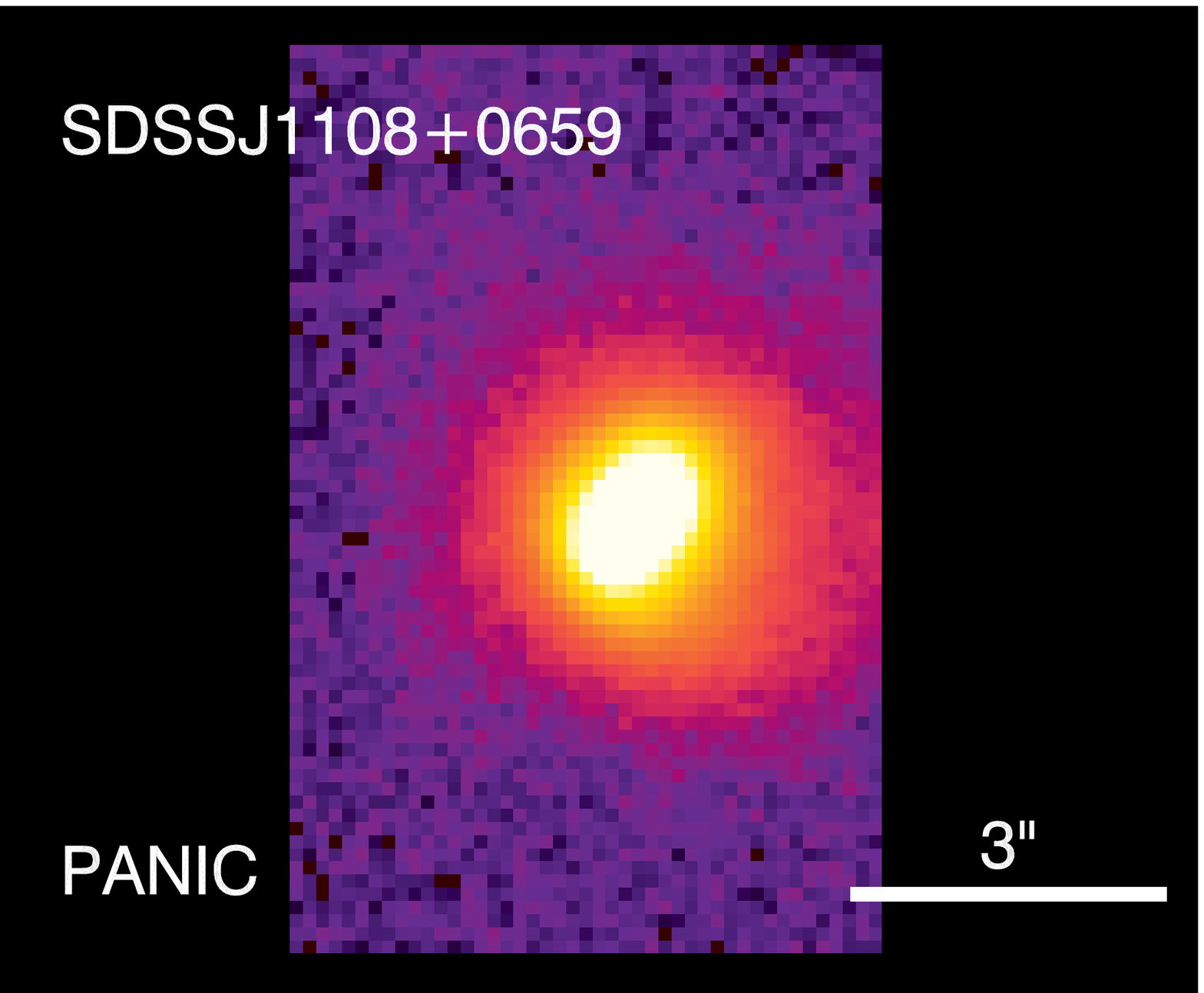}
    \includegraphics[width=0.3\textwidth]{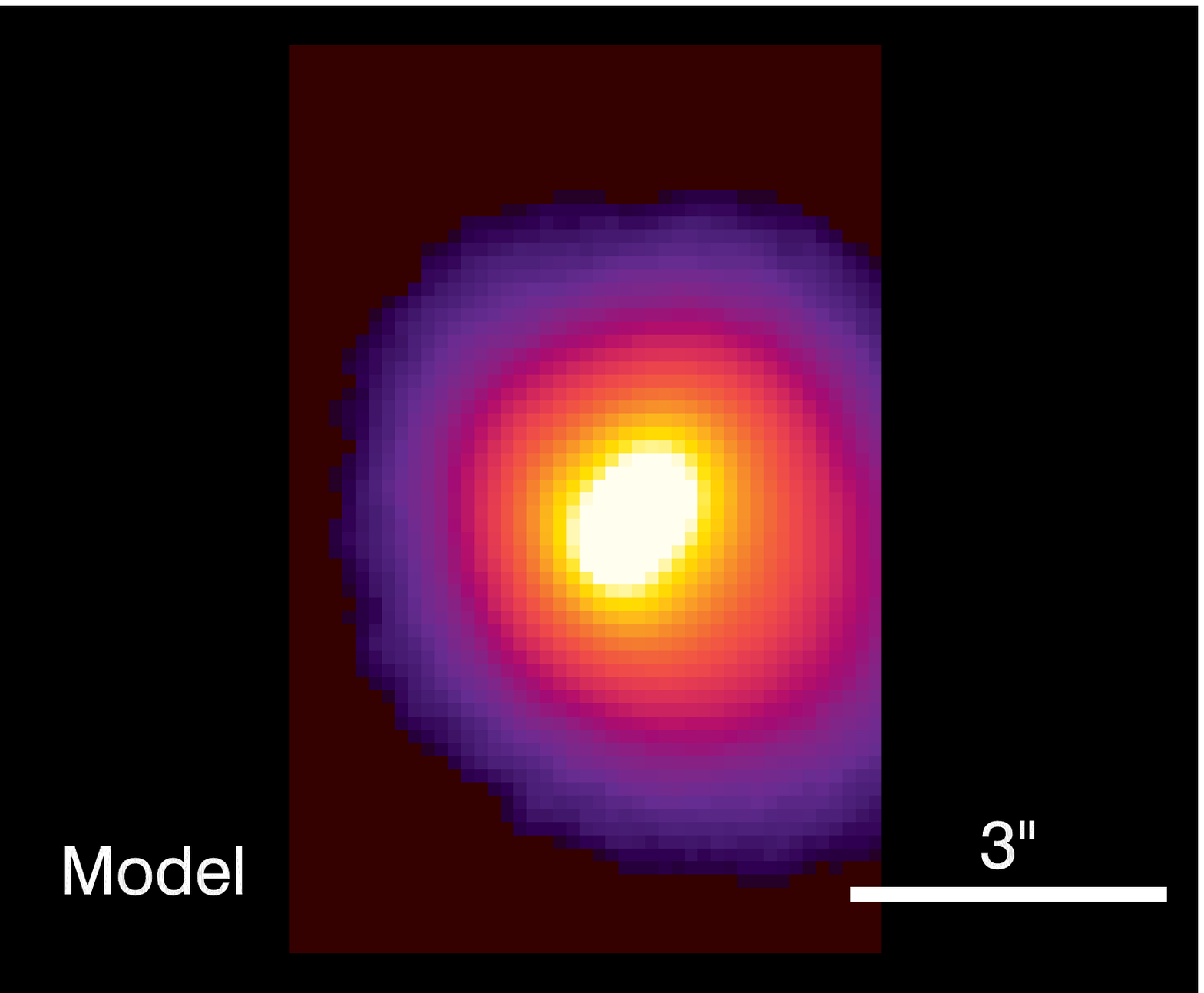}
    \includegraphics[width=0.3\textwidth]{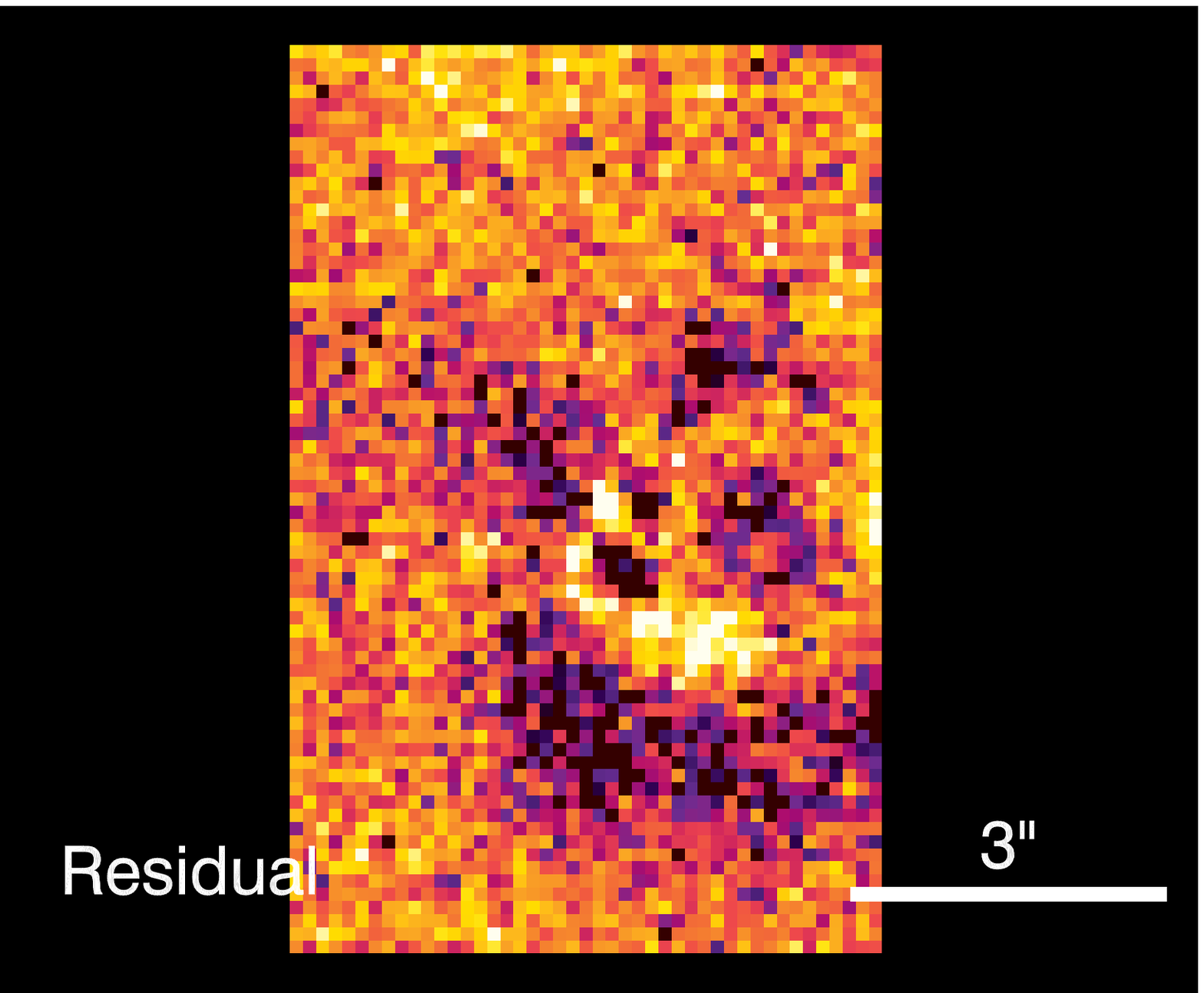}\vspace{3pt}
    \includegraphics[width=0.3\textwidth]{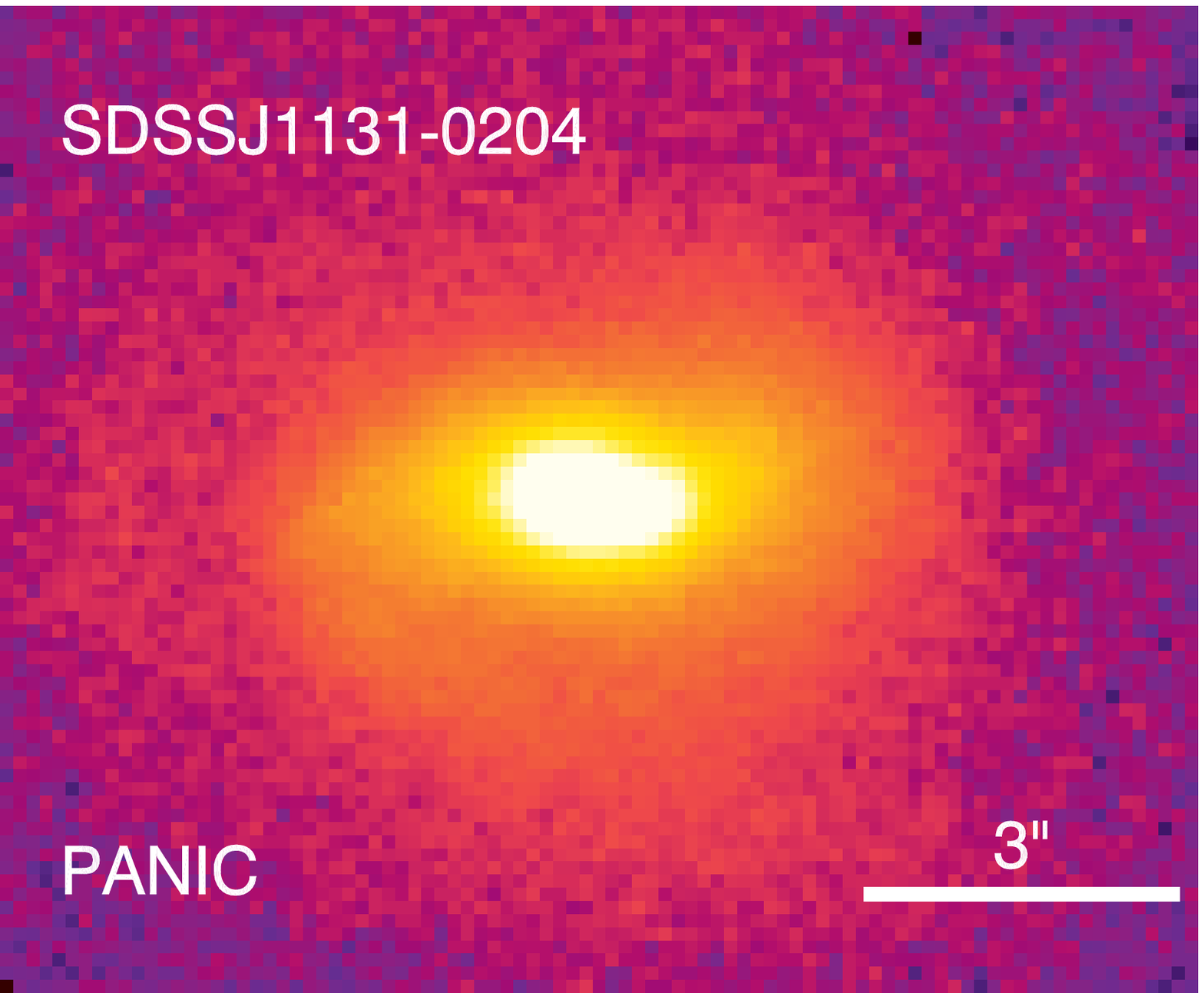}
    \includegraphics[width=0.3\textwidth]{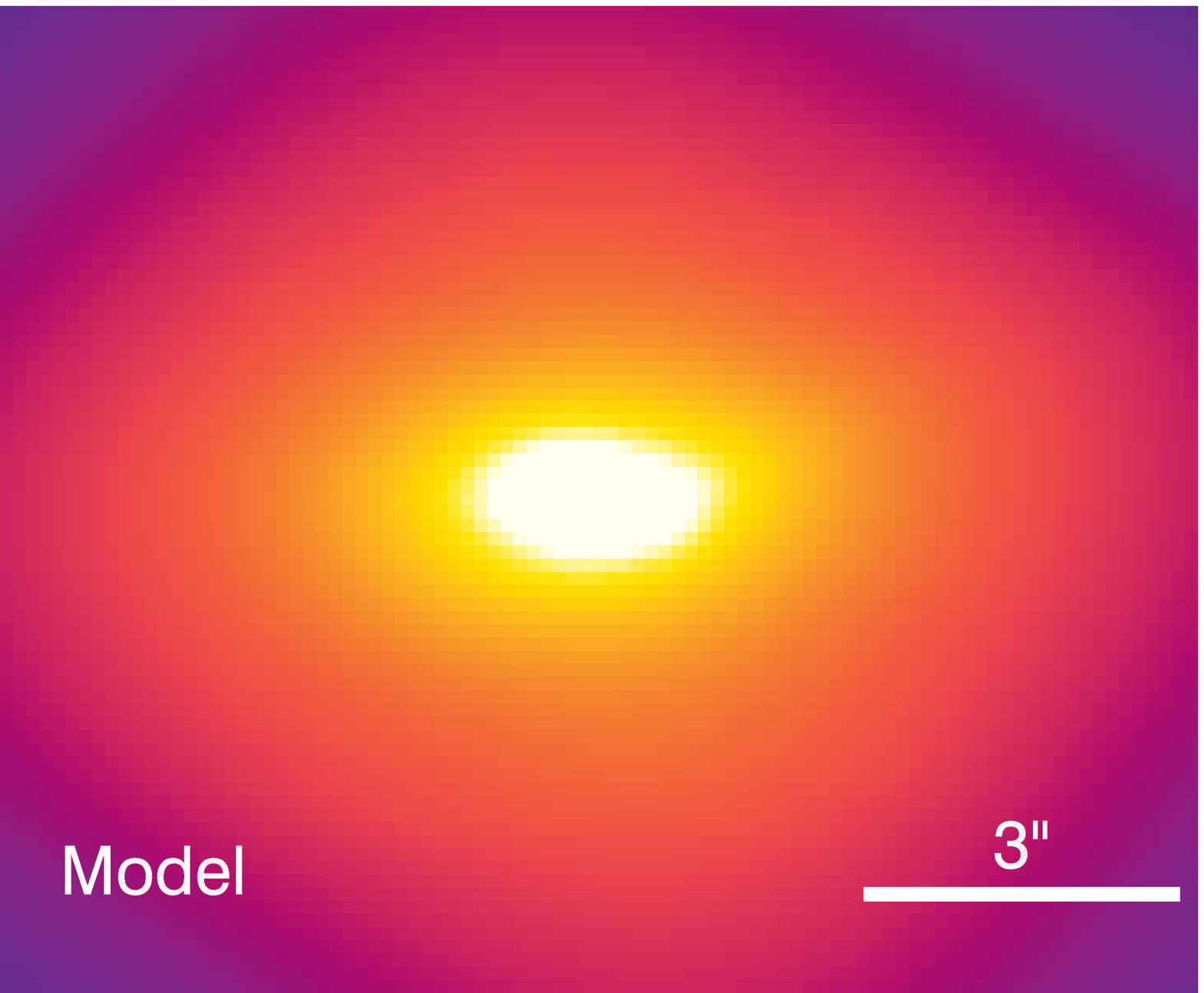}
    \includegraphics[width=0.3\textwidth]{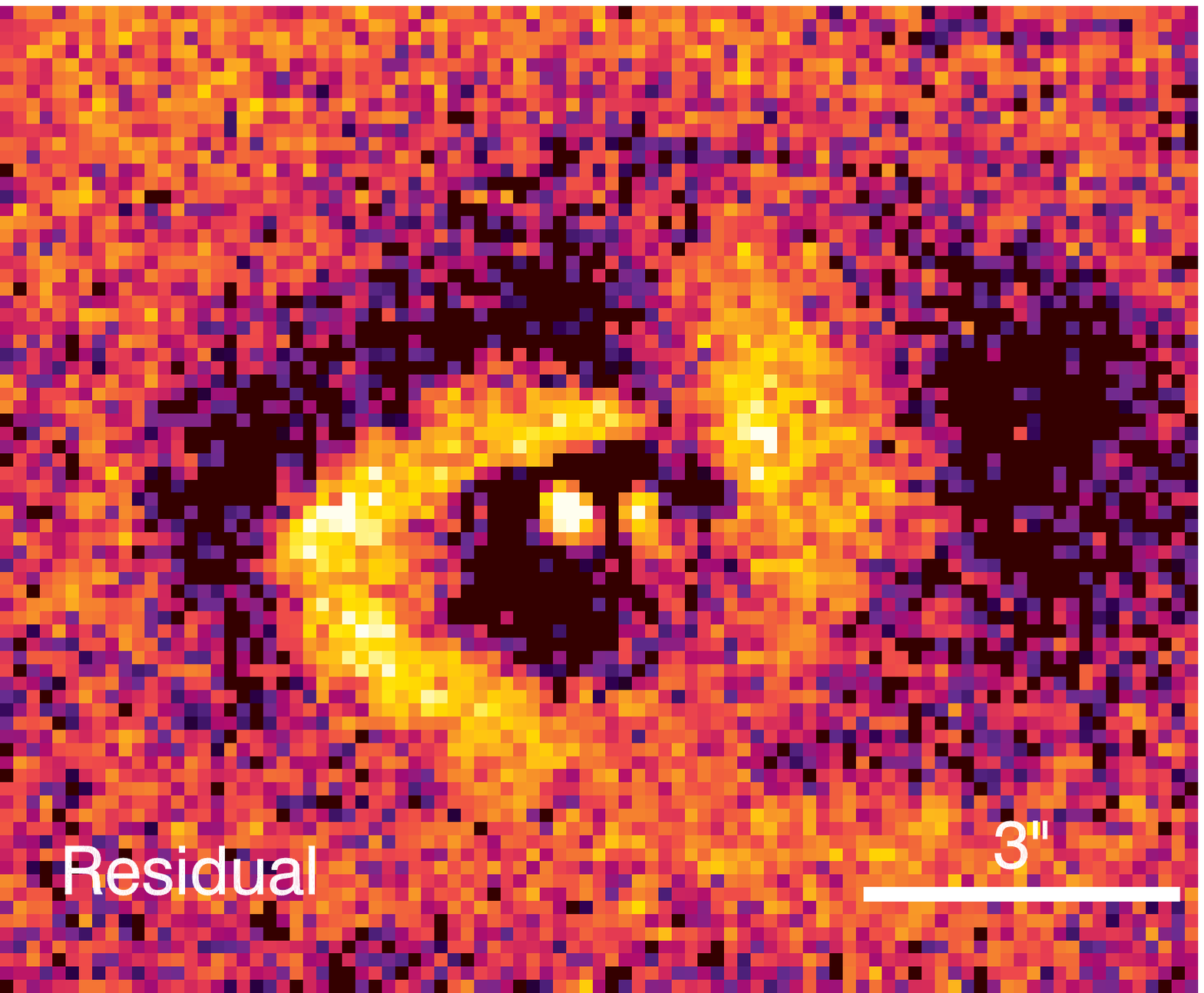}\vspace{3pt}
    \includegraphics[width=0.3\textwidth]{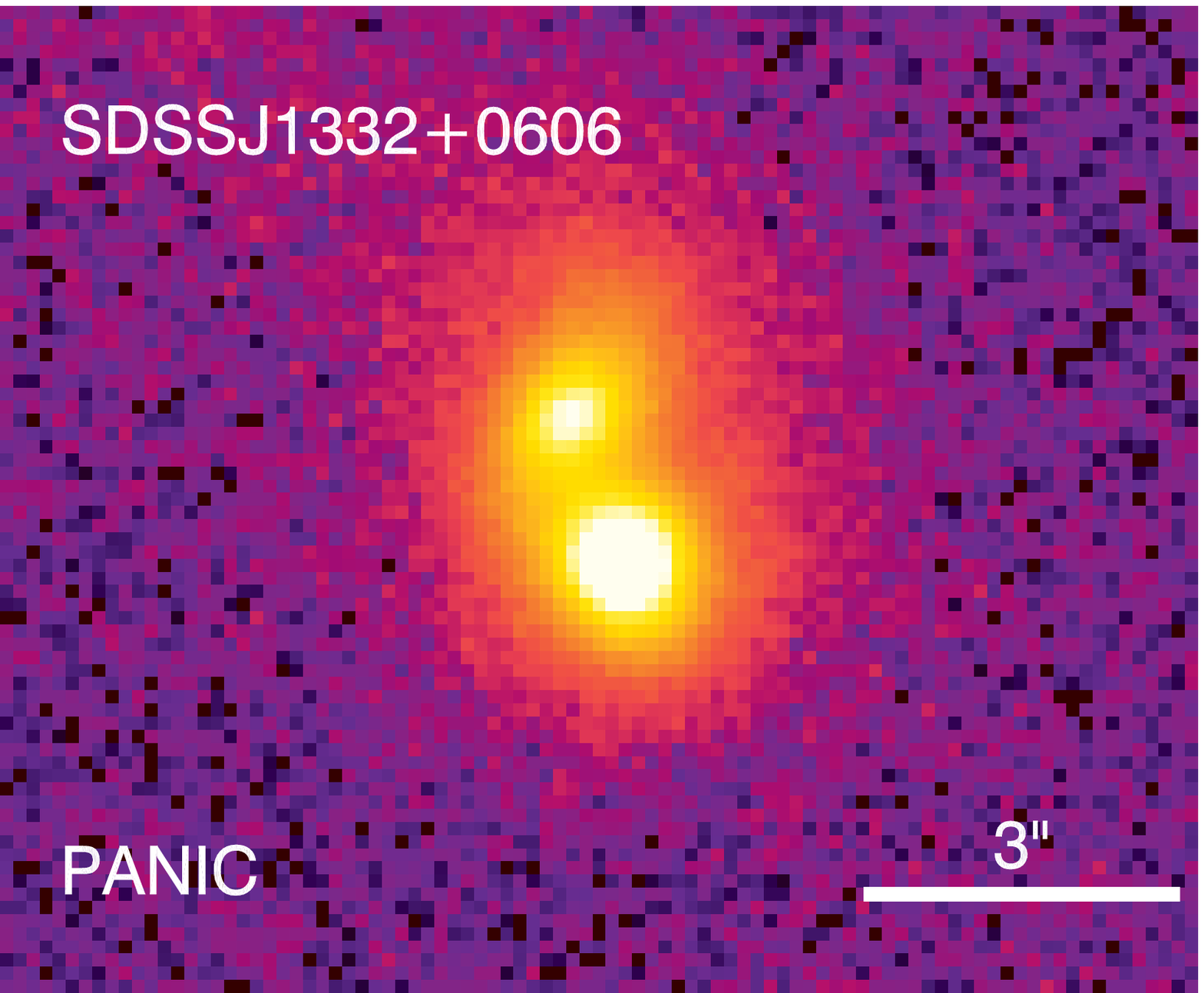}
    \includegraphics[width=0.3\textwidth]{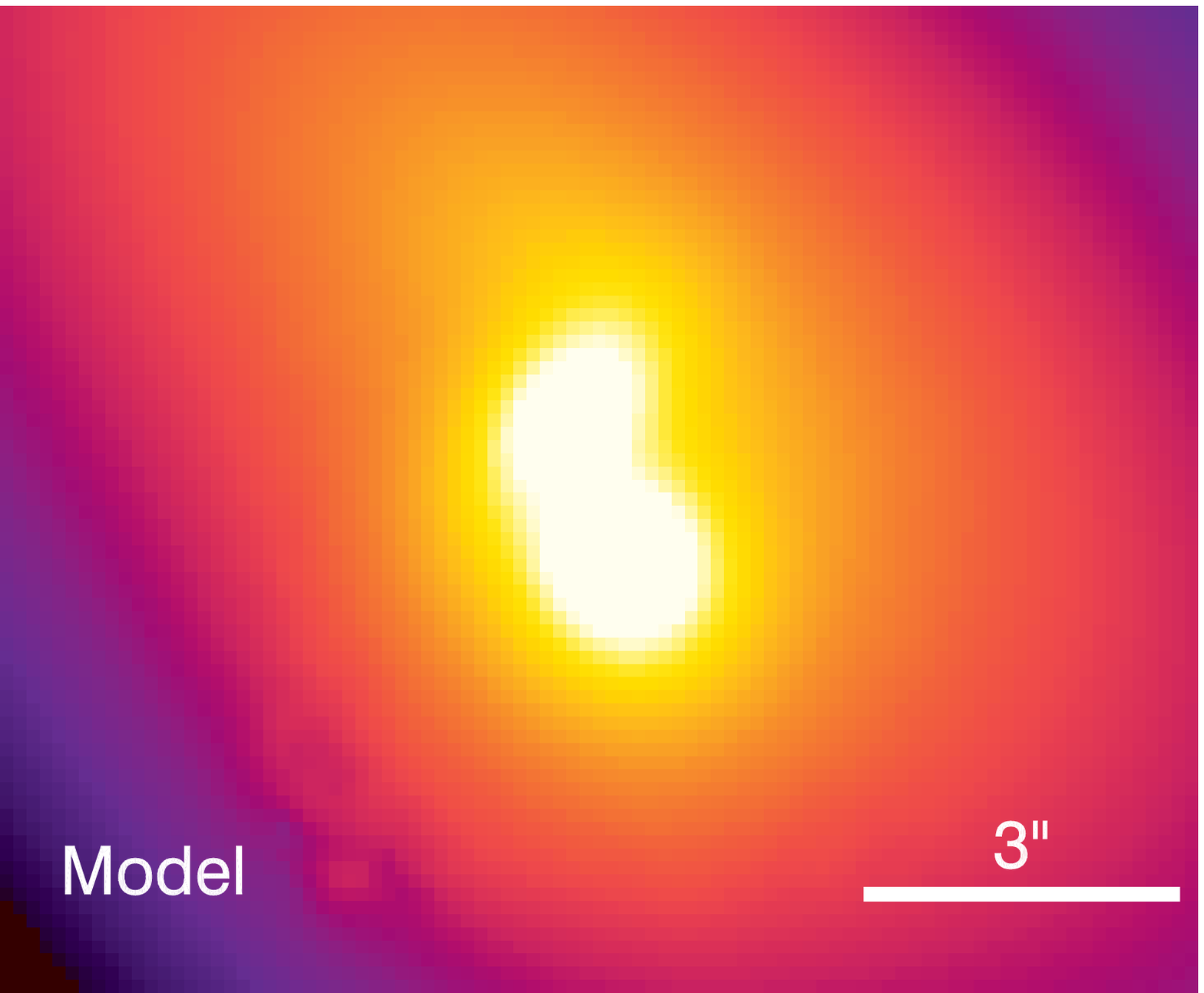}
    \includegraphics[width=0.3\textwidth]{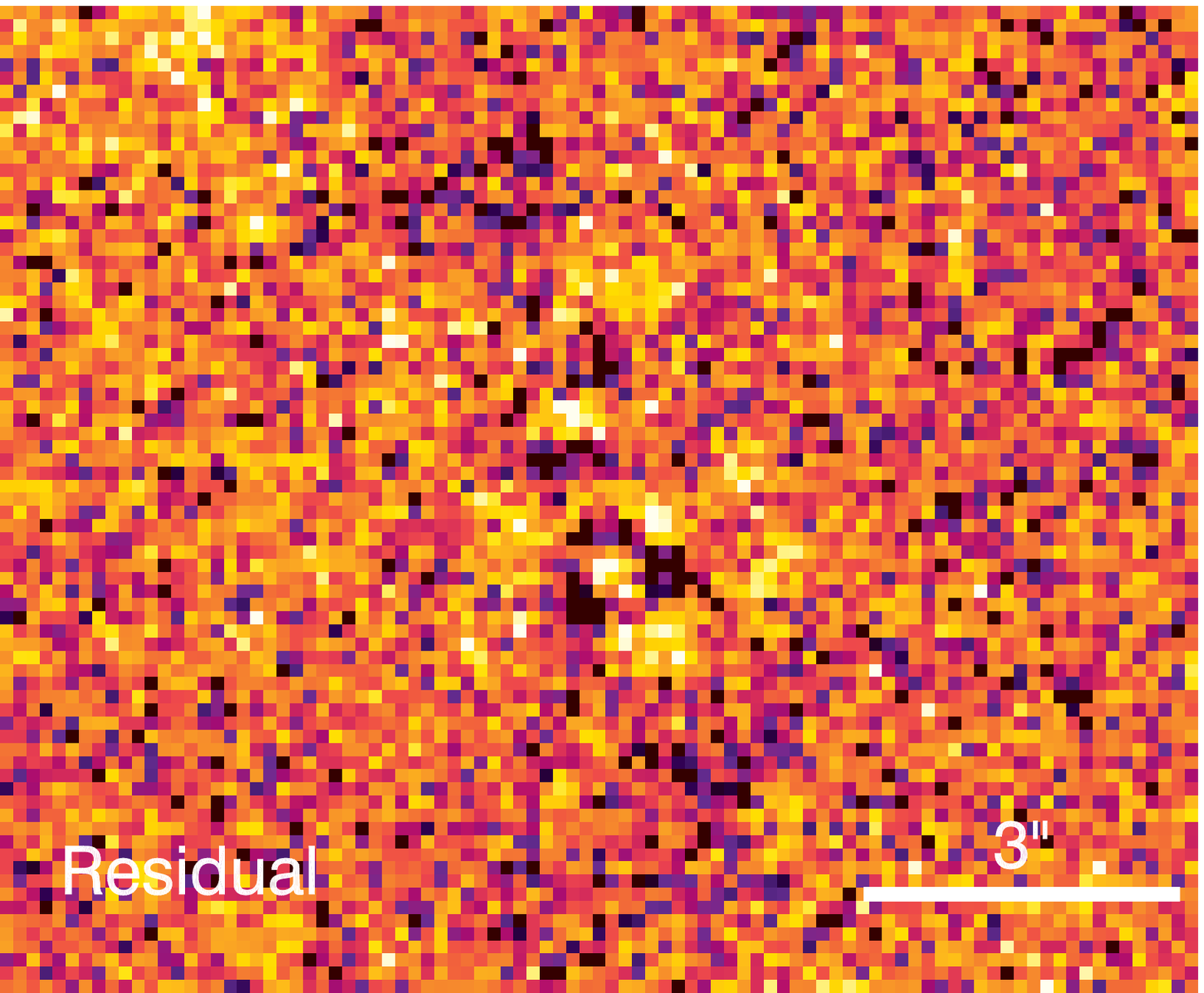}\vspace{3pt}
    \includegraphics[width=0.3\textwidth]{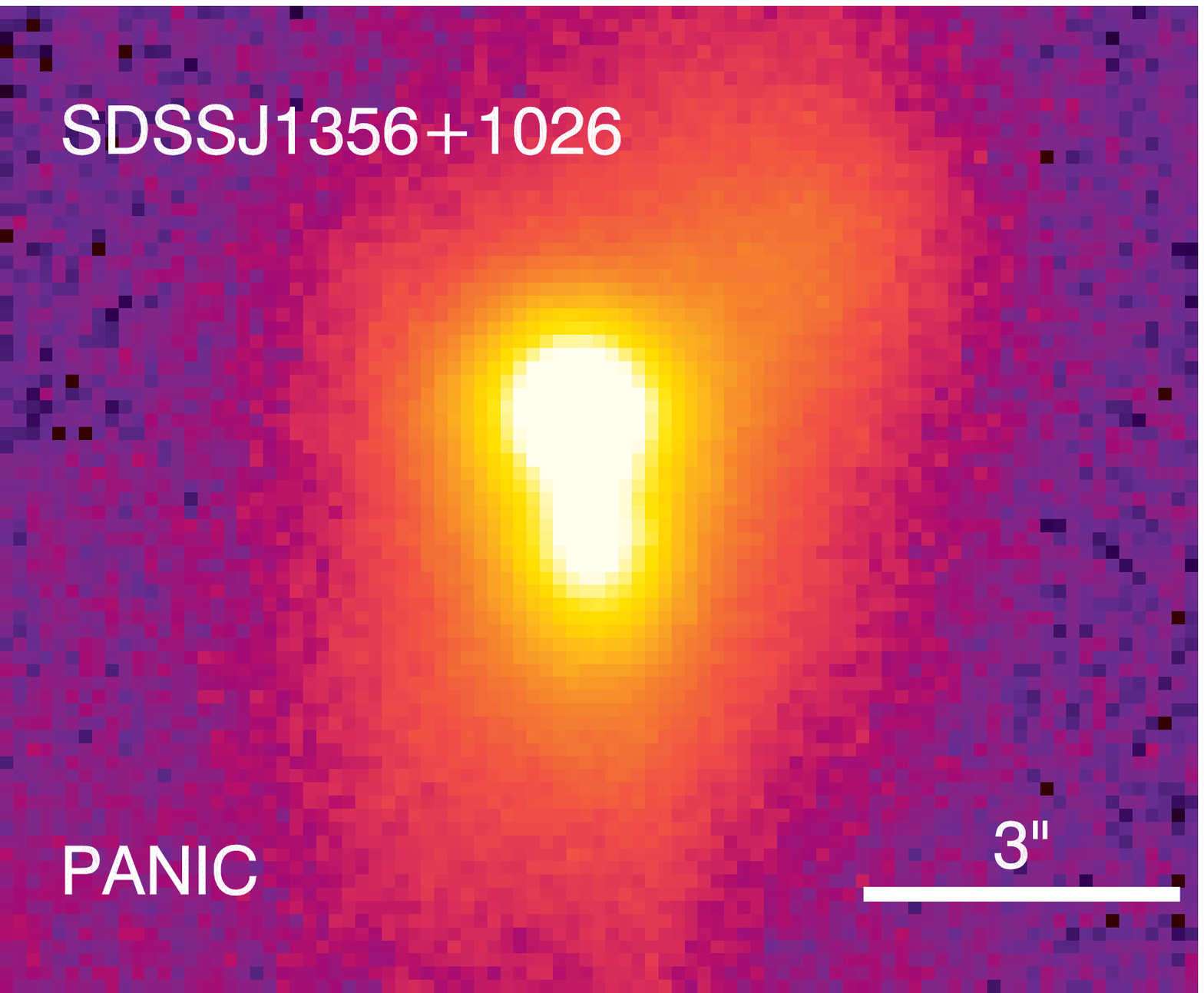}
    \includegraphics[width=0.3\textwidth]{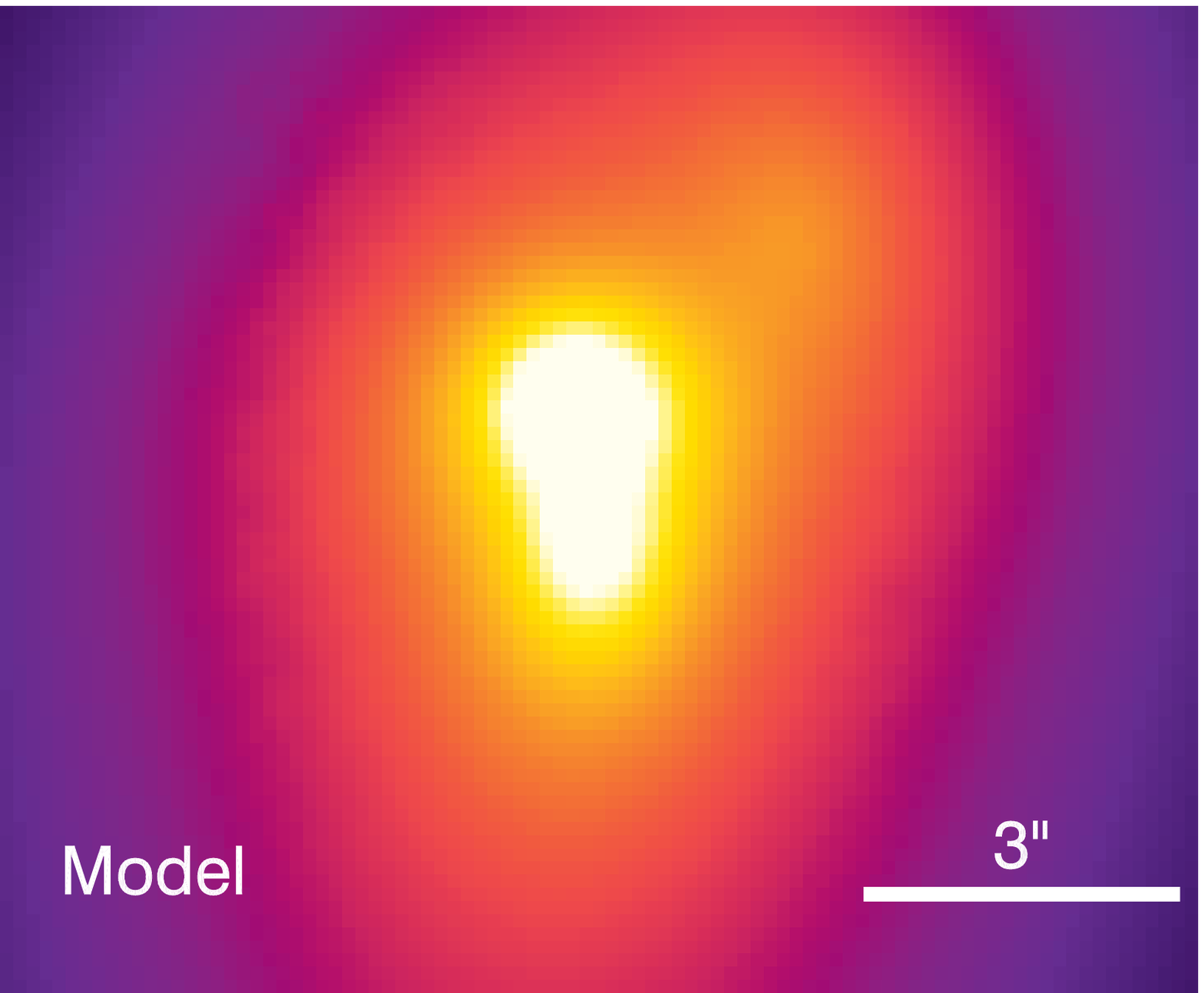}
    \includegraphics[width=0.3\textwidth]{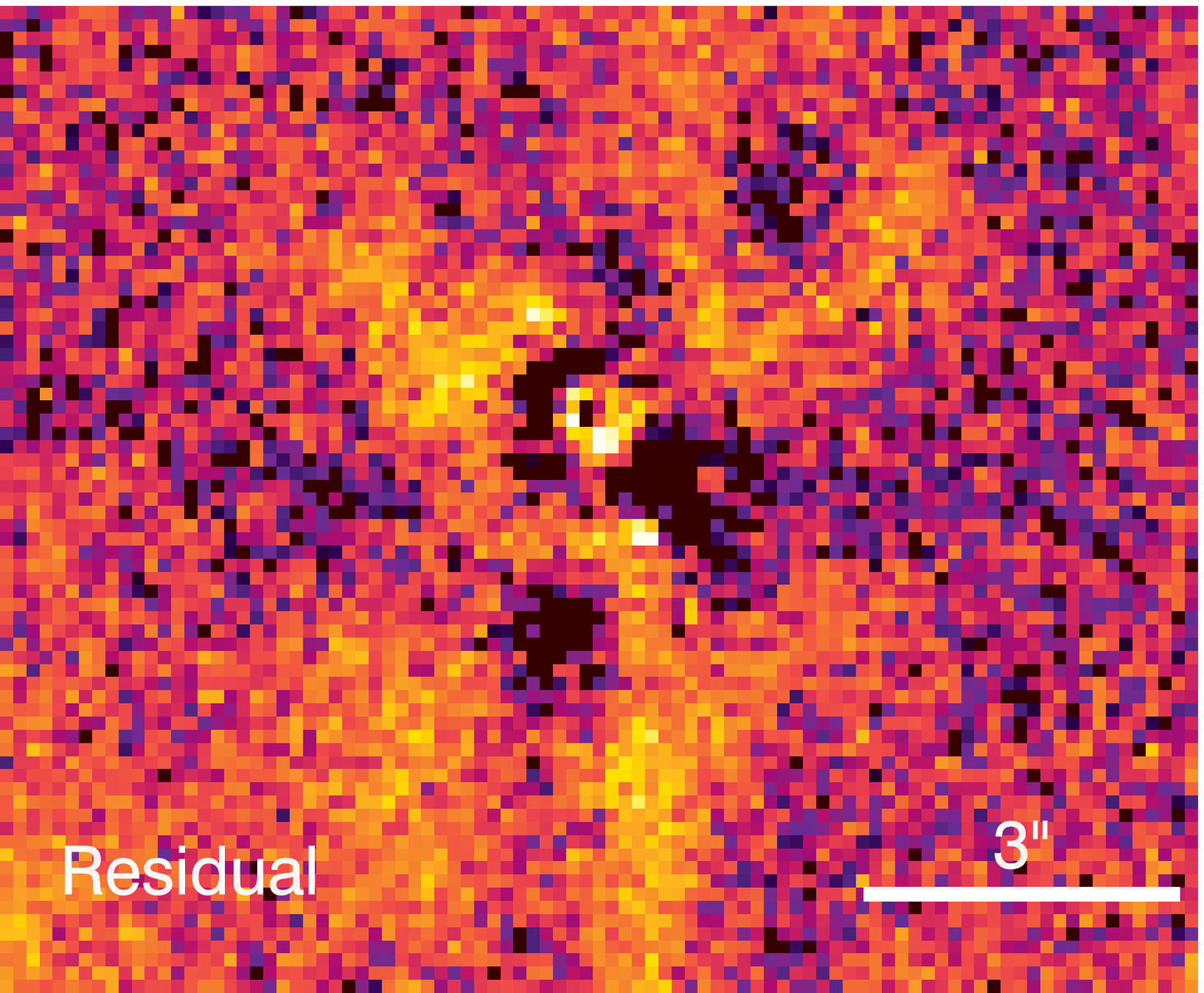} 
    \caption{Model fits of the NIR surface brightness for four objects classified as binary AGNs (\S\ref{sec:binary}) using {\sevenrm GALFIT}. North
    is up and East is left. The first column shows the data, the second column shows the models and the last column shows the residuals. See the text
    in \S\ref{sec:binary} for details regarding the model fits. Note that the residual maps have been re-stretched to enhance the contrast and the
    residuals are not important compared to the observed fluxes (the apparent excess in some of the residual maps has values less than $5\%$ of the
    observed fluxes at the corresponding locations). }
    \label{fig:galfit}
\end{figure*}

For the sky-subtracted NIR imaging data, we use {\sevenrm
GALFIT} \citep[][]{Peng_etal_2010} to model the light
distribution with multiple components. PSFs were taken from
stars within the same image. Due to the complexity of merging
systems, and the fact that the results depend on the quality of
the data, we only use the exponential disk (expdisk), de
Vaucouleurs (devauc) and S\'{e}rsic (sersic) profiles in each
fits, and we urge caution on the interpretation of the best-fit
model for some of these objects. During the fits we did not fix
any of the model parameters, and for each fit we tried
different combinations of the above three profiles until it
reaches the minimum reduced $\chi^2$. Fig.\ \ref{fig:galfit}
shows the best-fit model (and the residuals) along with data
for the four objects with PANIC data. The best-fit models are
summarized in Table \ref{table:galfit}, where the best-fit
parameters have been rounded using the 1$\sigma$ statistical
uncertainties from {\sevenrm GALFIT}. Due to the nature of
nonlinear multi-component models and possible systematic
effects involved in the data (such as bad PSF or sky
subtraction), the statistical uncertainties reported by
{\sevenrm GALFIT} are an approximation of the actual
uncertainties at best. We use these model fits to estimate the
luminosity ratios of multiple stellar components in these
merging systems, and to determine the galaxy type of each
component.

For the slit spectra, we extract one-dimensional (1d) spectra
at different spatial locations along the slit. We model the 1d
spectra in different spatial bins with double-Gaussian (or
double-Lorentzian if a better fit can be achieved) profiles for
the double-peaked lines, plus a power-law model for the local
continuum. Fig.\ \ref{fig:1108+0659_diag} shows an example of
the results of our modeling. Our goal is to see if there are
observable trends in the velocity, line width, and line flux
ratio of the two narrow line components. Unfortunately, given
the typical seeing in our slit spectroscopy, these trends are
only apparent in a few cases. There are several objects for
which the spectral quality is too poor to perform such
analysis. Below we briefly comment on individual objects.

\begin{figure*}
  \centering
    \includegraphics[width=0.6\textwidth]{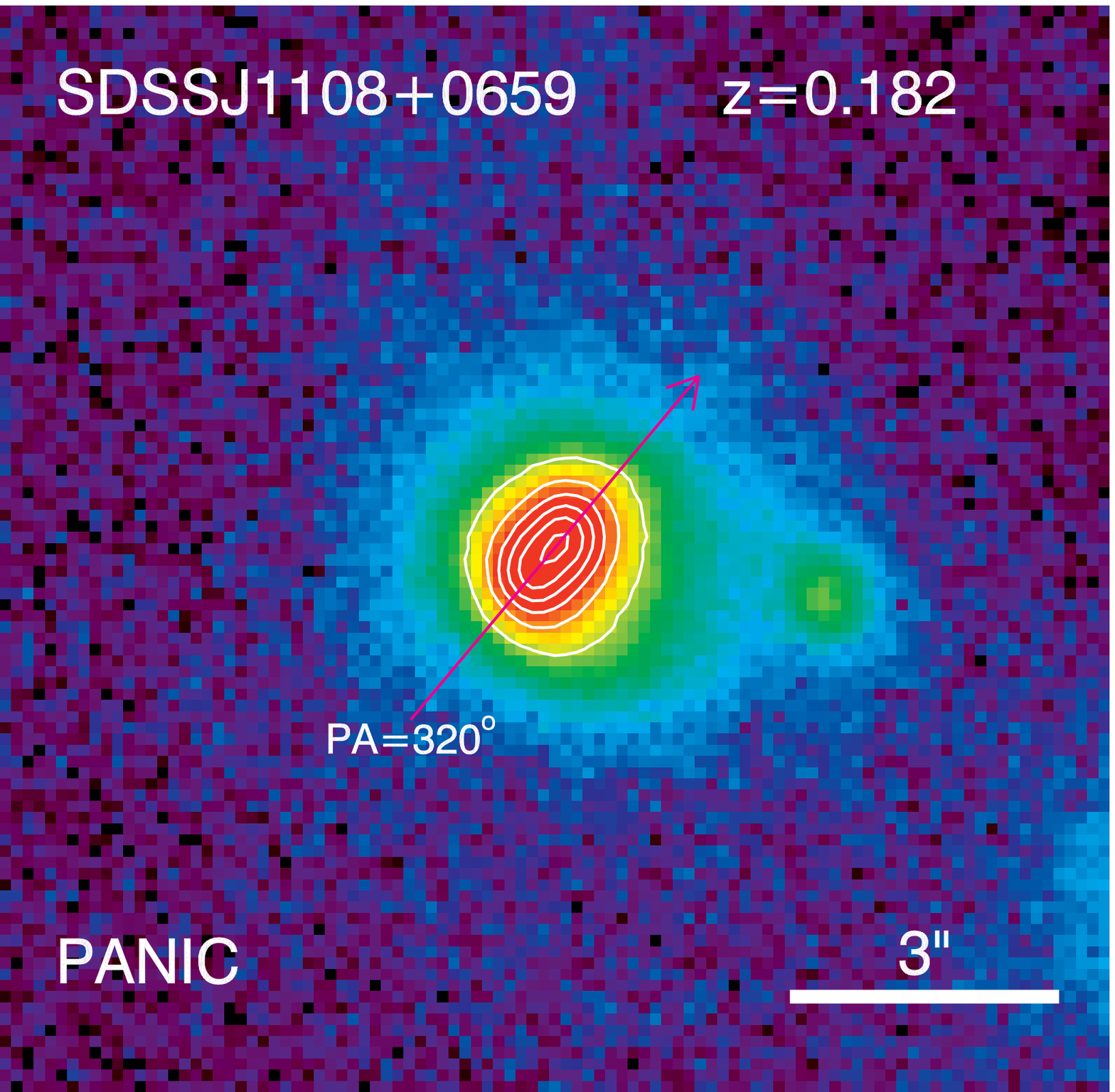}\vspace{9pt}
    \includegraphics[width=0.9\textwidth]{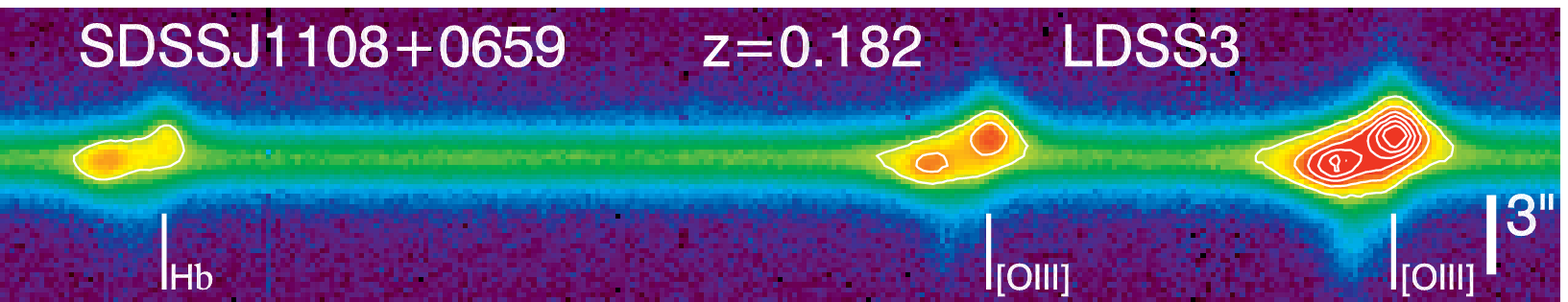}
    \caption{SDSS J1108+0659 (binary). {\em Top:} PANIC NIR image in $K_s$. North is up and East is left. The magenta line shows the direction of the
    slit used in the spectroscopy, with the arrow indicating the upper direction in the extracted 2d spectrum. This object has two (marginally)
    resolved stellar nuclei separated by $\sim 0.5$\arcsec\ in the cental galaxy. There is also a faint companion to the right which is not covered
    by
    the slit spectrum. {\em Bottom:} LDSS3 2d spectrum for the \hbeta-\OIII\ region with corresponding lines marked (note that the locations of these
    line marks are approximate). The two velocity components are spatially offset by 0.9\arcsec. The stellar continua are not separated in the 2d
    spectrum due to the limited seeing and the proximity of the two stellar nuclei. }
    \label{fig:1108+0659}
\end{figure*}

\begin{figure*}
  \centering
    \includegraphics[width=0.9\textwidth]{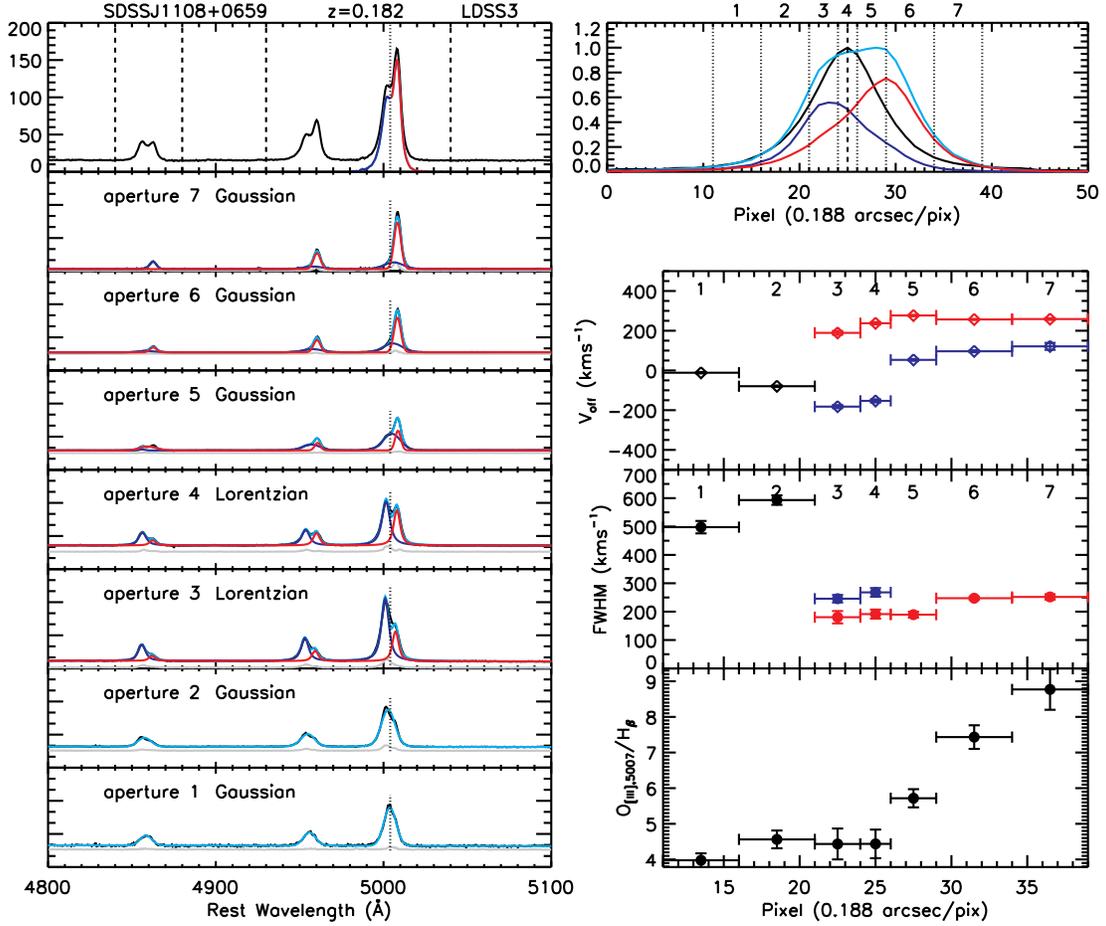}
    \caption{\footnotesize Diagnosis of the 2d spectrum of J1108+0659. The left panel shows the 1d spectra summed over the entire slit (uppermost) and
    for slit
    slices at 7 different
    spatial locations (flux scale is arbitrary). The dotted vertical lines divide the \OIIIb\ line into the blueshifted and redshifted components.
    The
    dashed vertical lines separate the emission line regions from the continuum regions. The upper-right panel shows the spatial profiles of the
    continuum (black) and the \OIIIb\ emission line region (cyan), while the blue and red lines show the spatial profile of the blueshifted and
    redshifted \OIIIb\ emission. We slice the 2d spectra in seven spatial bins (marked by the numbers), centered on the spatial peak of the continuum
    emission, and the 1d spectra for these spatial bins are shown in the left panel. For the 1d spectrum in each spatial bin, we try to deblend the
    two velocity components in the emission lines by fitting the spectrum with double Gaussian/Lorentzian functions and a local continuum model. We
    consider the deblending successful if the two components have a flux ratio greater than 0.3 and both components were detected at $>3\sigma$. If
    the deblending is successful we over-plot the two velocity components in blue and red respectively in the aperture spectra, otherwise we
    over-plot
    the whole model in cyan. We plot the velocity offset (relative to the systemic redshift) and FWHM of the two components, and the \OIIIb/\hbeta\
    flux ratio measured from the model in the bottom-right panels. In a few cases the apparent ``successful'' deblending in a spatial bin can still
    be
    spurious as judged by eye; we nevertheless plot results for these bins, but such bins should be ignored in assessing the spatial gradients of
    these quantities. }
    \label{fig:1108+0659_diag}
\end{figure*}

\textbf{J1108+0659}. This object was reported in
\citet{Liu_etal_2010b}. It shows two nuclei in its NIR image,
separated by $\sim 0.5$\arcsec. The two \OIII\ velocity
components are spatially coincident with the two NIR nuclei.
Its $K_s$-band image and 2d spectrum for the \OIII-\hbeta\
region are shown in Fig.\ \ref{fig:1108+0659}. The pair of
nuclei are well resolved in a recent NIR adaptive-optics (AO)
imaging observation \citep[][]{Fu_etal_2010}. The best-fit
$K_s$-band model consists of two de Vaucouleurs bulges embedded
in an exponential disk. The luminosity contrast of the two
bulges is $\sim 0.15$ magnitude (0.06 dex).

\begin{figure*}
  \centering
    \includegraphics[width=0.6\textwidth]{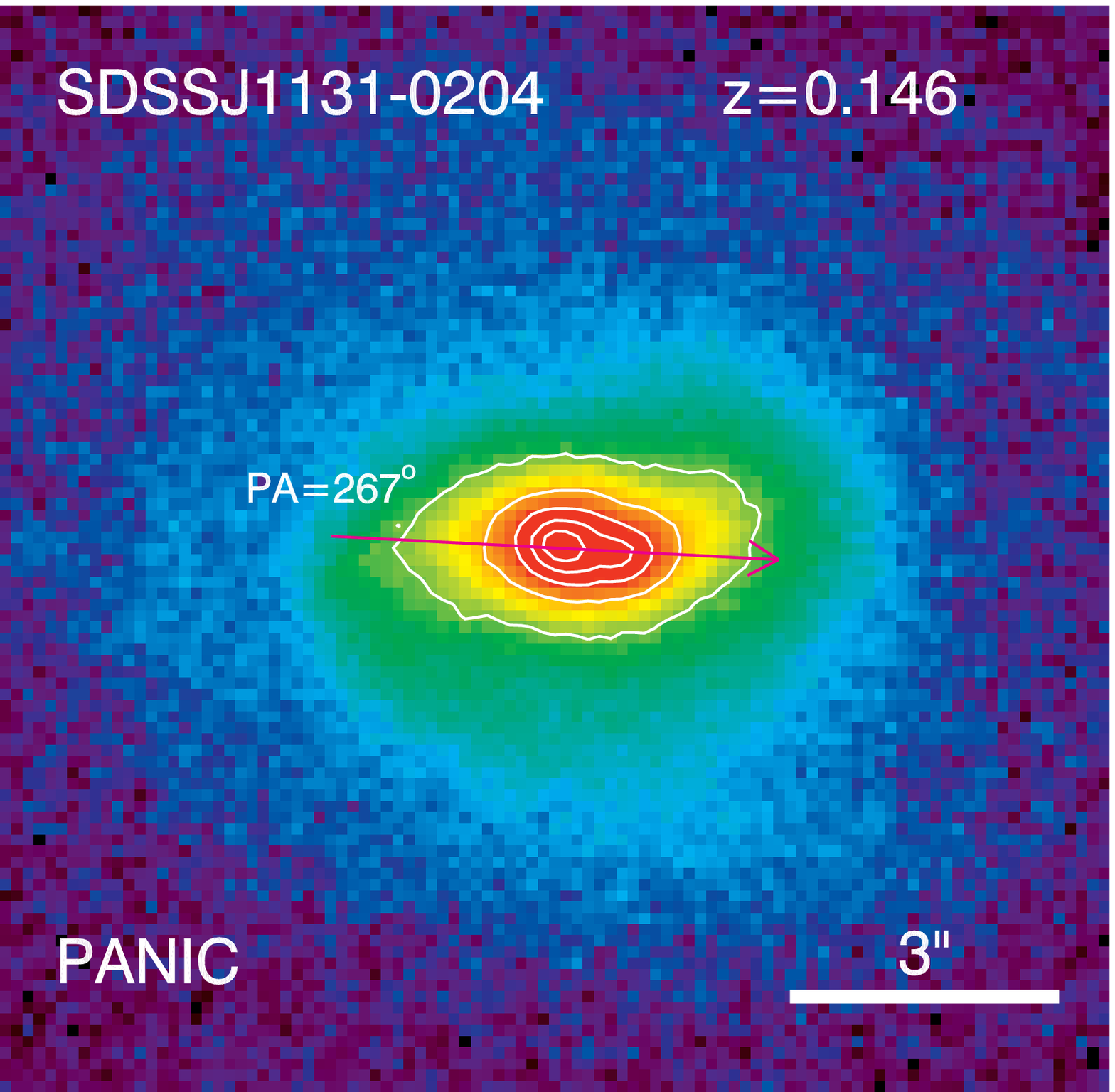}\vspace{9pt}
    \includegraphics[width=0.9\textwidth]{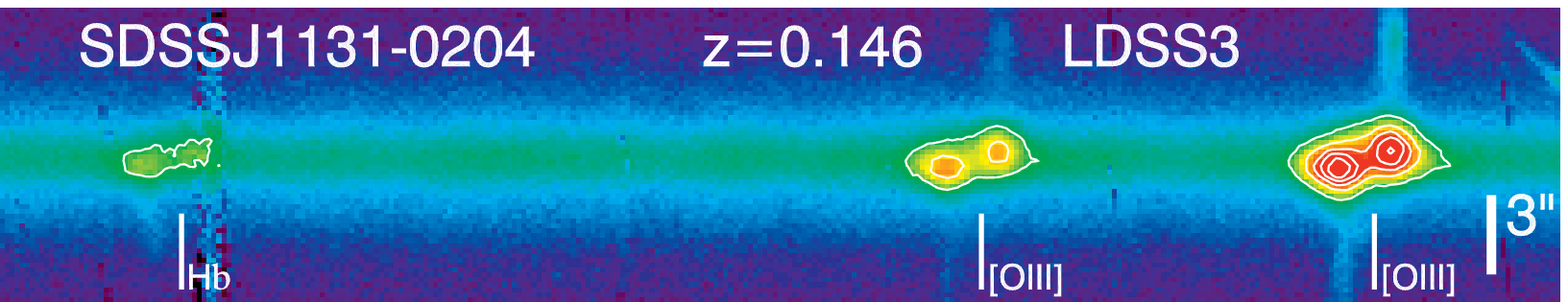}
    \caption{SDSS J1131$-$0204 (binary). {\em Upper:} PANIC NIR image in $K_s$. This object has two stellar nuclei separated by $\sim 0.6$\arcsec.
    {\em Bottom:} LDSS3 2d spectrum for the \hbeta-\OIII\ region with corresponding lines marked (note that the locations of these line marks are
    approximate). The two velocity components are spatially offset by $\sim 0.6$\arcsec. Notation is the same as Fig.\ \ref{fig:1108+0659}. For this
    object, the two stellar nuclei are embedded in a disk, and the long ``spur'' features seen in the 2d spectrum are from ionized gas emission in
    the
    disk. }
    \label{fig:1131-0204}
\end{figure*}

\textbf{J1131$-$0204}. This object was reported in
\citet{Liu_etal_2010b}. It shows two resolved stellar nuclei
separated by $\sim 0.6$\arcsec. The slit spectroscopy confirmed
the coincidence of the two \OIII\ velocity components with the
two NIR nuclei. Its $K_s$-band image and 2d spectrum for the
\OIII-\hbeta\ region are shown in Fig.\ \ref{fig:1131-0204}.
The two stellar nuclei are embedded within a galactic disk, and
the long ``spur'' features seen in the 2d spectrum to $>15$ kpc
are ionized gas emission from the disk which traces the
galactic rotation curve. The best-fit $K_s$-band model consists
of one S\'{e}rsic bulge ($n\approx 3$) and one de Vaucouleurs
bulge embedded in an exponential disk. The luminosity contrast
of the two bulges is $\sim 2$ magnitude (0.8 dex). The model
fit is imperfect and there is some residual spiral structure
indicative of interactions.

\textbf{J1146+5110}. This object was reported in
\citet{Liu_etal_2010b}. The double stellar nuclei were
marginally resolved in the 2MASS $K_s$ image, and our slit
spectroscopy subsequently confirmed the coincidence of the two
\OIII\ velocity components with the two NIR nuclei. This is
also one of the few cases that show resolved optical double
nuclei in the SDSS images. The southern nucleus itself seems to
have a complex NLR geometry and two NLR velocity components.
The double nucleus was well resolved in the NIR AO imaging in
\citet{Fu_etal_2010}, with a luminosity contract of 0.7
magnitude (0.28 dex). We do not have a PANIC image for this
object.

\textbf{J1332+0606}. This object was reported in
\citet{Liu_etal_2010b}. The two stellar nuclei were clearly
resolved in optical (SDSS) and $J$-band images, and are
spatially coincident with the two \OIII\ velocity components
seen in our slit spectroscopy. Its $J$-band image and 2d
spectrum for the \OIII-\hbeta\ region are shown in Fig.\
\ref{fig:1332+0606}. The best-fit $J$-band model consists of
three components: a S\'{e}rsic bulge ($n\approx 2$) for the
southern nucleus, an exponential disk for the northern nucleus,
and an exponential disk with a scale-length of $r_s\sim
1.6$\arcsec. The luminosity ratio of the three components is
$\sim 2:1:1$. The less luminous northern stellar component
corresponds to the stronger \OIII\ emission component seen in
the slit spectrum, as in J1322+2631.

\textbf{J1356+1026}. This object was studied in
\citet{Greene_etal_2011}. It shows two continuum sources in the
optical separated by $\sim 3$ kpc, corresponding to the north
and south knots of \OIII\ emission seen in the slit spectrum in
\citet{Greene_etal_2011}. The two knots of \OIII\ emission have
a relative velocity offset of $\sim 200\ {\rm km\,s^{-1}}$. The
slit spectrum in \citet{Greene_etal_2011} shows a rich \OIII\
emission structure, including a giant \OIII\ bubble to the
South of the southern continuum. Its $K_s$-band image is shown
in Fig.\ \ref{fig:1356+1026}. This is a galaxy with a highly
disturbed morphology and its {\it IRAS} fluxes indicate that it
is a ULIRG. The best-fit $K_s$-band model consists of three
components: a S\'{e}rsic bulge ($n\approx 5$) for the southern
nucleus, a S\'{e}rsic bulge ($n\approx 3$) for the northern
nucleus, and an exponential disk component ($r_s\sim
0.8$\arcsec) towards the north-western corner. The luminosity
ratio of the three components is $\sim 15:5:1$. However, due to
the highly disturbed morphology of this system, we urge caution
on the best-fit model. This object was also observed with NIR
AO imaging in \citet{Fu_etal_2010}.

\begin{deluxetable}{lcccccc}
\tabletypesize{\footnotesize}  
\tablewidth{0.8\textwidth}
\tablecaption{Model fits of NIR images
\label{table:galfit}} \tablehead{
Object & $n_s$ & $R_e/r_s$ & $(x_c,y_c)$ &  $q$ & PA  & Mag \\
          &       &  (pixel/kpc)    & (pixel)   &      & ($^\circ$) & (Vega)
 }  \startdata
J1108+0659 ($K_s$) &     &     & $\chi^2_{\nu}=0.19$   &      &     &  \\
devauc (S)         &  4  & 3.2/1.2 & $(94.1,99.8)$  & 0.60 & 46  & 15.46\\ 
devauc (N)         &  4  & 3.2/1.2 & $(97.3,103.6)$ & 0.61 & $-$42 & 15.32\\ 
expdisk            &  1  & 8.5/3.2 & $(101.4,99.8)$ & 0.80 & 46  & 15.70\\ 
\hline\\
J1131$-$0204 ($K_s$) &     &     & $\chi^2_{\nu}=1.00$   &      &     & \\
sersic (E)         & 3.2 & 25.2/8.1  & $(98.1,101.6)$ & 0.46 & $-$88 & 15.12\\ 
devauc (W)         &  4  & 14.8/4.7  & $(104.2,100.3)$& 0.28 & $-$69 & 17.24\\ 
expdisk            &  1  & 22.2/7.1  & $(101.5,97.9)$ & 0.71 & 13  & 15.67\\ 
\hline\\
J1332+0606 ($J$)   &     &     & $\chi^2_{\nu}=1.08$   &      &     & \\
expdisk (N)        &  1  & 3.0/1.3 & $(98.2,107.6)$ & 0.53 & $-$33 & 17.92\\ 
sersic  (S)        & 2.0 & 4.4/1.9 & $(102.2,96.4)$ & 0.85 & 27  & 17.06\\ 
expdisk            &  1  & 13.1/5.5  & $(100.1,115.9)$& 0.55 & 46  & 17.79\\ 
\hline\\
J1356+1026$^*$ ($K_s$) &     &     & $\chi^2_{\nu}=0.91$    &      &     &  \\
sersic (N)         & 3.2 & 4.6/1.3 & $(99.0,108.0)$ & 0.79 & $-$80 & 15.55\\ 
sersic (S)         & 5.6 & 63.1/17.4    & $(100.2,97.4)$ & 0.53 & $-$8  & 14.34\\ 
expdisk            & 1   & 6.7/1.9     & $(115.6,121.2)$& 0.67 & $-$1  & 17.32\\ 
\enddata
\tablecomments{Best-fit surface brightness models for four objects that show multiple stellar components in the NIR. Only the S\'{e}rsic
(including de Vaucouleurs) and the exponential disk profiles were used in these fits. We report best-fit parameters for the
S\'{e}rsic index ($n$), the effective radius ($R_e$) or scale-length ($r_s$), the centroid of each component ($x_c,y_c$),
the aspect ratio ($q$), the position angle of the major axis (PA), and the integrated magnitude for each component
(normalized using the total 2MASS flux). These parameters are rounded using the statistical errors from the fits. Due to the complexity
of these systems, we caution that a ``successful'' model may not be the unique model, and the actual errors of these parameters are expected
to be substantially larger. This is especially a concern for J1356 (marked with a ``*''), whose morphology is quite disturbed. We report scales
in units of pixels, where 1 pixel corresponds to 0.125\arcsec.}
\end{deluxetable}

All of these objects appear to be major mergers (the $K_s$-band luminosity ratio between the two main stellar components is less than 10), which is
partly due to our selection based on comparable \OIII\ luminosities. But we also note that the more massive stellar component does not necessarily
have stronger \OIII\ emission.

We classify these objects as kpc-scale binary AGNs based on
spatially coincident double stellar nuclei and \OIII\ emission.
While the binary scenario seems to be the most natural
explanation, it is possible that only one SMBH is active and
ionizing the gas clouds in both nuclei. In this single-AGN
scenario, the \OIII\ gas clouds in the non-AGN host are further
away from the ionizing source than those in the AGN host. This
difference will lead to different ionization states in the
\OIII\ clouds in the two hosts. If we assume that the electron
density is similar in both hosts, we expect very different
\OIII/\hbeta\ flux ratios of the two narrow line components,
contrary to what we observe \citep[][]{Liu_etal_2010b}.
However, the electron density may well be quite different in
the two hosts, allowing a single AGN within one host to be
responsible for the \OIII\ emission in both hosts. This is
particularly relevant for the ULIRG J1356+1026, where the
interstellar medium (ISM) may be quite clumpy. Our current data
are insufficient to completely rule out the single AGN
possibility. We are currently acquiring images with {\em HST}
and {\em Chandra}, as well as higher S/N slit spectroscopy for
these objects, and further investigations of these objects will
be presented in future work.

\subsection{NLR Kinematics in Single AGNs}\label{sec:kinematics}
The remaining ($\sim 90\%$) objects with NIR imaging data do
not show resolved (or marginally resolved) double nuclei at the
limit of our resolution ($\sim 0.6$\arcsec). These unresolved
cases fall into one of the following categories: a) they are
binaries at smaller projected separations; b) they are two
accreting BHs each with its own NLR, co-rotating within a
single merged stellar bulge, or c) they are single AGNs with
complex NLR kinematics.

About $60\%$ of the objects that appear single in the NIR
imaging show spatially resolved \OIII\ emission (typically $\ga
0.6$\arcsec) in the 2d spectra, which correspond to the two
velocity components seen in the spatially-integrated spectra.
Double stellar nuclei separated on these scales would have been
identified in our NIR imaging. Thus the lack of spatially
coincident stellar nuclei favors either scenario (b) or
scenario (c). However, the NLR dynamics are presumably affected
by the bulge potential more than by the BH, thus it is
difficult to maintain two distinct NLRs in scenario (b). On the
other hand, NLR kinematics involving rotation or outflows on
sub-kpc to kpc scales are quite common for local Seyferts (see
\S\ref{sec:intro}). In fact, in some of the objects with good
spatial quality we can see velocity/velocity dispersion
gradients along the slit direction in the 2d spectrum (see
below), which strongly supports the kinematics scenario.
Therefore we believe that the double velocity peaks in the vast
majority of these objects arise from complex kinematics in a
single NLR . We now comment on each of these objects in detail.


\textbf{J0002+0045}. For this object the two velocity
components of \OIII\ are spatially offset by $\sim 0.8$\arcsec,
but they do not have a corresponding pair of nuclei seen in the
NIR image. Fig.\ \ref{fig:0002+0045} shows the $K_s$ image and
the 2d spectrum. Note that this object has a north-east
companion $\sim 5$\arcsec\ away, which does not have \OIII\
emission; this companion galaxy is at the same redshift as
J0002+0045 measured from stellar absorption features in the slit spectrum.

\textbf{J0116$-$1025}. This object has two spatially offset \OIII\ emission peaks, which correspond to the blue- and red-shifted components in the
double-peaked line profile respectively. The spatial offset of the two \OIII\ components is $\sim 1.1$\arcsec, and there is no corresponding double
nucleus seen in the NIR image. Fig.\ \ref{fig:0116-1025} shows the $K_s$ image and the 2d spectrum. The host galaxy clearly has a disk component.
There is also a small galaxy about $\sim 4$\arcsec\ away from the center of J0116-1025 to the West, which was not covered by our slit observation.
Fig.\ \ref{fig:0116-1025_diag} shows the 1d slices of the 2d spectrum at different distances from the peak of the continuum emission. The
\OIIIb/\hbeta\ flux ratio is almost independent of position, but the \OIII\ line width increases towards the center of the continuum emission. No
obvious velocity gradient is seen for either of the two \OIII\ components. Combining the NIR imaging and slit spectroscopic data, this object is best
explained by a rotational \OIII\ disk co-planar with the stellar disk.

\textbf{J0135$-$0058}. Fig.\ \ref{fig:0135-0058} shows the $K_s$ image and the 2d spectrum for this object. The long slit was placed to cover the
tidal feature to the south-east of the galaxy seen in the NIR image, and hence was misaligned with the major axis of the disk. Fig.\
\ref{fig:0135-0058_diag} shows the 1d slices of the 2d spectrum at different distances from the peak of the continuum emission. The \OIIIb/\hbeta\
flux ratio is almost constant until the outermost apertures (6 and 7), where the \OIIIb/\hbeta\ decreases as it is now tracing the faint tidal
feature
seen from the 2d spectrum shown in Fig.\ \ref{fig:0135-0058}. There is a slight velocity gradient for the blue- and red-shifted components. Although
the spatial offset between the two \OIII\ components is only $\sim 0.2$\arcsec\ (possibly due to the misaligned slit position angle), we believe a
rotational \OIII\ disk is the best explanation given the velocity gradient seen in the 2d spectrum as well as the apparent disk morphology seen in
the
NIR.

\textbf{J0135+1435}. Fig.\ \ref{fig:0135+1435} shows the $K_s$
image and the 2d spectrum for this object. The two velocity
components of \OIII\ are spatially offset by $\sim 0.8$\arcsec,
while no corresponding pair of nuclei was seen in the NIR.
Fig.\ \ref{fig:0135+1435_diag} shows the 1d slices of the 2d
spectrum at different distances from the peak of the continuum
emission. The \OIIIb/\hbeta\ flux ratio increases towards the
center of the continuum emission. Velocity gradients for both
\OIII\ components are clearly seen in Fig.\
\ref{fig:0135+1435_diag}, and there is some indication of
increasing line width toward the center of the continuum
emission, although decomposition into two components is not
always successful at each aperture. This object is best
explained by a rotational \OIII\ disk, with an asymptotic flat
rotation velocity $V_{c}\sin i\sim 200\,{\rm kms^{-1}}$.

\textbf{J0156$-$0007}. Fig.\ \ref{fig:0156-0007} shows the $K_s$ image and the 2d spectrum for this object. A disk morphology is apparent in the NIR
image. The two velocity components of \OIII\ are spatially offset by $\sim 0.7$\arcsec, with no corresponding pair of nuclei seen in the NIR. Fig.\
\ref{fig:0156-0007_diag} shows the 1d slices of the 2d spectrum at different distances from the peak of the continuum emission. The \OIIIb/\hbeta\
flux ratio is almost constant at each aperture location. Weak velocity gradients can been seen for both \OIII\ components. At most locations the line
width is narrow, which suggests that coherent rotation is the dominant motion. This object is best explained by a rotational \OIII\ disk, with an
asymptotic flat rotation velocity $V_{c}\sin i\sim 150\,{\rm kms^{-1}}$.

\textbf{J0400$-$0652}. This object shows a smooth single
profile in the $K_s$ image, while its slit spectra show
spatially resolved \OIII\ emission extending to $\sim
3$\arcsec. A subsequent NIR AO image obtained by
\citet{Fu_etal_2010} did not reveal a double nucleus at $\sim
0.1$\arcsec\ resolution. In Fig.\ \ref{fig:0400-0652} we show
its $K_s$ image and 2d spectra at two position angles, and in
Fig.\ \ref{fig:0400-0652_diag} we show the 1d spectral
diagnostics. The 2d spectrum along PA$=95^\circ$ shows rather
complicated \OIII\ emission region kinematics. It shows a high
velocity dispersion near the center, and some velocity gradient
along the slit direction, which is indicative of disk rotation
and/or outflows. Clearly there are more than two \OIII\
components. On the other hand, the 2d spectrum along
PA$=173^\circ$ shows much less structure, which is presumably
caused by the different spatial coverage. This object is best
explained by NLR kinematics, possibly involving both outflows
and rotation. A more detailed follow-up of this object (i.e.,
with IFU or multiple-slit spectroscopy) is highly desirable to
resolve the \OIII\ kinematics map.

\textbf{J0837+1500}. Fig.\ \ref{fig:0837+1500} shows the $K_s$ image and the 2d spectrum for this object. The two velocity components of \OIII\ are
spatially offset by $\sim 1.5$\arcsec, with no corresponding pair of nuclei seen in the NIR. Fig.\ \ref{fig:0837+1500_diag} shows the 1d slices of
the
2d spectrum at different distances from the peak of the continuum emission.


\textbf{J0851+1327}. Fig.\ \ref{fig:0851+1327} shows the $K_s$ image and the 2d spectrum for this object. The two velocity components of \OIII\ are
spatially offset by $\sim 0.6$\arcsec, with no corresponding pair of nuclei seen in the NIR. This galaxy has an edge-on disk component seen in the
NIR, which is also seen in its SDSS optical image and the 2d spectrum shown in Fig.\ \ref{fig:0851+1327_diag}. This object is best explained as a
rotational \OIII\ disk, given the disk morphology seen in the NIR and in the optical.

\textbf{J0958$-$0051}. Fig.\ \ref{fig:0958-0051} shows the $K_s$ image and the 2d spectrum for this object. A disk component is seen in the NIR
image.
The two velocity components of \OIII\ are spatially offset by $\sim 0.8$\arcsec, with no corresponding pair of nuclei seen in the NIR. Fig.\
\ref{fig:0958-0051_diag} shows the 1d slices of the 2d spectrum at different distances from the peak of the continuum emission. Velocity gradients
are
clearly seen for both \OIII\ components. This object is best explained by a rotational \OIII\ disk.

\textbf{J1038+0255}. Fig.\ \ref{fig:1038+0255} shows the $K_s$ image and the 2d spectrum for this object. The two velocity components of \OIII\ are
spatially offset by $\sim 1$\arcsec, with no corresponding pair of nuclei seen in the NIR. This object is best explained by either a rotational
\OIII\
disk or biconical outflows.

\textbf{J1146$-$0226}. Fig.\ \ref{fig:1146-0226} shows the $K_s$ image and the 2d spectrum for this object. The $K_s$ image shows a smooth single
stellar bulge, while the 2d spectrum shows a clear rotation-curve like velocity gradient across the slit. Under poor spatial resolution or if the
object were observed at higher redshifts, the rotation-curve would be unresolvable, and the \OIII\ emission would appear as two distinct velocity
components spatially offset by $\sim$ a few kpc. This object is best explained by a rotational \OIII\ disk.

\textbf{J1341+2219}. Fig.\ \ref{fig:1341+2219} shows the $K_s$ image and the 2d spectrum for this object. A disk component is seen in the NIR image.
The \OIII-\hbeta\ part of our 2d spectrum has low quality so we show the \halpha\ region instead. The two NEL velocity components are spatially
offset
by $\sim 1.7$\arcsec, with no corresponding pair of nuclei seen in the NIR. This object is best explained by a rotational \OIII\ disk.

\textbf{J1552+0433}. For this object the two \OIII\ components
are spatially offset by $\sim 1.2$\arcsec\ in their peak
emission, as seen from its 2d spectrum shown in Fig.\
\ref{fig:1552+0433}. Its $K_s$ image shows a disk component,
which seems to be warped at the north-west edge. Nevertheless,
two stellar bulge components separated by $\sim 1.2$\arcsec\
would have been seen in the NIR image if this object were a
kpc-scale binary AGN. Thus we classify this object in the NLR
kinematics category.

\textbf{J1630+1649}. Fig.\ \ref{fig:1630+1649} shows the $K_s$ image and the 2d spectrum for this object. The two \OIII\ velocity components are
spatially offset by $\sim 0.8$\arcsec, with no corresponding pair of nuclei seen in the NIR. A recent NIR AO image of this object at $\sim
0.1$\arcsec\ resolution did not show a double nucleus either \citep{Fu_etal_2010}. This object is best explained by a rotational \OIII\ disk.

\textbf{J2304$-$0933}. For this object the NIR image shows a clear disk component. The seeing was quite poor for the slit spectroscopy observation,
but the two NEL components seem to be spatially offset by $\sim 0.8$\arcsec\ in their peak emission. A recent NIR AO image of this object reveals no
double nucleus at $\sim 0.1$\arcsec\ resolution \citep{Fu_etal_2010}. It is thus most likely a case of NLR kinematics.

While many of these NLR kinematics cases show a clear disk
morphology in imaging data, a few objects, such as J0837+1500
and J1146-0226, show an early-type morphology. The ionized gas
emission in these early-type hosts is extended on kpc scales.
Emission line kinematics studies of local early-type galaxies
often show ionized gas emission to kpc scales, and the gas
kinematics is usually decoupled from the stellar kinematics
\citep[e.g.,][]{Sarzi_etal_2006}, which is consistent with our
findings here. However, the emission line strength is much
stronger in our AGNs, which may indicate a larger gas reservoir
in these early-type hosts than their local counterparts.


\subsection{Ambiguous Cases}\label{sec:amb}
About $40\%$ of objects that appear single in NIR imaging were classified as ambiguous cases, i.e., the combined NIR imaging and slit spectroscopic
data are insufficient to distinguish between the binary scenario and the kinematics scenario. In essentially all of these cases, the two velocity
components of the \OIII\ emission have undetectable or small spatial offset that is below the resolution limit of the NIR imaging (i.e., $\la
0.4$\arcsec). In the binary scenario, the two BHs and their associated stellar components would have smaller projected separations, which would need
higher-resolution observations such as ground AO imaging \citep[e.g.,][]{Fu_etal_2010} or space-based observations to resolve the pair of nuclei.
Nevertheless the real separation between the two BHs is unlikely to be much less than $\sim$ kpc because NLRs have intrinsic sizes of hundreds of pc
or above. In the case of NLR kinematics (rotation or outflows), either the relevant scale is below one or two kpc, or the slit position happens to be
perpendicular to the outflow axis or the rotational disk (cf., see Fig.\ \ref{fig:0400-0652} for J0400$-$0652). Fig.\ \ref{fig:amb1} shows the NIR
images and 2d spectra for these objects. We now briefly comment on individual objects in this category.

\textbf{J0009$-$0036}. In this object the two \OIII\ components are spatially offset by $<0.2$\arcsec\ in the 2d spectrum, and the NIR image shows a
single nucleus. Its NIR image shows a minor structure $\sim 1$\arcsec\ south-east from the center of the peak NIR emission, which is not contributing
to the \OIII\ emission. A better slit position and higher resolution imaging and spectroscopy are needed to rule out the existence of a companion at
$<0.2$\arcsec\ separation.

\textbf{J0942+1254}. This object was observed with two slit positions. Both of the 2d spectra show no spatial offset ($<0.2$\arcsec) between the two
\OIII\ velocity components. No double nucleus was seen in the NIR image. However, the optical and NIR images of this object show spectacular tidal
features, indicative of recent mergers. This object deserves a higher-resolution observation, and we suspect it is likely to be a small projected
separation binary AGN.

\textbf{J1009+0133}. This object shows a minor companion $\sim 1$\arcsec\ north-west of the peak emission in the NIR image. Unfortunately due to our
slit position, we couldn't confirm that these two stellar nuclei correspond to the two \OIII\ components seen in the 2d spectrum. Although the NIR
image alone is suggestive of a binary, a new slit position is needed to test this hypothesis.

\textbf{J1019+0134}. The NIR image shows a disturbed morphology in the outskirts, indicative of a recent merger event. The two \OIII\ components are
spatially offset by $\sim 0.3$\arcsec\ as shown in the 2d spectrum. Seeing in the NIR is $\sim 0.7$\arcsec. This could either be a binary or NLR
kinematics. New slit spectroscopy with a different position angle and/or better spatial resolution are needed to distinguish the two scenarios.

\textbf{J1322+2631}. For this object we only have a slit spectrum covering the \halpha\ region. The two narrow line components are spatially offset
by
$\sim 2.1$\arcsec\ as shown in the 2d spectrum. In the $K_s$-band image we see two main stellar nuclei separated by $\sim 2.3$\arcsec, and there is
also a third (minor) companion to the southwest, about $2.5$\arcsec\ from the central brightest stellar component. A close examination of the 2d
spectrum suggests that the two \OIII\ components are probably not spatially coincident with the two main stellar nuclei, even though their separation
happens to be consistent with that of the two stellar nuclei. If this were true, it would imply a kinematic origin for the double-peaked \OIII\
emission in this object. However, the seeing in our 2d optical spectrum is poor and the locations of the stellar continua are not well determined.
Hence it is still possible that the two spatially offset \OIII\ components are coincident with two of the three stellar components seen in the NIR
image. The poor seeing in the spectrum also prohibited measurements of the absorption redshifts of the northeast and southwest companions, whose
continua are blended with the continuum of the brighter central source. A new slit spectrum with better spatial resolution is needed to draw firm
conclusions on this object.


\textbf{J1450+0838}. The 2d spectra shows that the two \OIII\ components are consistent with no spatial offset ($<0.2$\arcsec). The NIR image shows a
single nucleus at $\sim 0.8$\arcsec\ resolution.

\textbf{J1556+0948}. The two \OIII\ components are offset by
$\sim 0.4$\arcsec\ in the 2d spectrum, and the NIR image shows
a single nucleus at $\sim 0.6$\arcsec\ resolution. The
double-peaked emission lines in this source are likely to be
due to NLR kinematics, but higher-resolution NIR imaging data
is needed to rule out the binary scenario.

\textbf{J2252+0029}. The two NEL components are consistent with no spatial offset in the 2d spectrum. Higher-resolution imaging observation and/or
different slit positions are required to distinguish the binary and the NLR kinematics scenarios.

\textbf{J2255$-$0812}. The NIR image shows a disturbed disk component, and the two NEL components show a small spatial offset, $\la 0.5$\arcsec, in
the 2d spectrum. Extended line emission is seen on opposite sides of the central continuum, which is caused by ionized gas in the rotating galactic
disk. This object is likely a case of NLR kinematics, but a higher resolution imaging observation and/or different slit positions are needed to rule
out the binary scenario.

\textbf{J2310$-$0900}. This object has a companion $\sim
4$\arcsec\ away, which does not contribute to the \OIII\
emission seen in the SDSS fiber spectrum. The companion is too faint to measure its redshift in our slit spectrum.
The seeing was quite poor for the slit spectroscopy observation, and the two NEL
components are consistent with no spatial offset. A better
quality slit spectrum and/or different slit positions are
needed to distinguish the binary and the NLR kinematics
scenarios.

\textbf{J2333+0049}. This object has a minor companion about $6$\arcsec\ to the southeast, which is too faint to measure its redshift in our slit
spectrum. The two NEL components are consistent with
no spatial offset in the 2d spectrum. It appears single in our
NIR imaging and in the recent NIR AO imaging
\citep{Fu_etal_2010} at $\sim 0.1$\arcsec\ resolution.

\section{Discussion}\label{sec:dist}

\subsection{The Bulk Properties of Kpc-scale Binary Type 2 AGNs}\label{sec:dis0}

Bearing in mind the small-number statistics of our sample, in
Fig.\ \ref{fig:bulk_prop} we compare the bulk properties of
kpc-scale binary AGNs (circles) and those that originate from
NLR kinematics (crosses), measured from spatially-integrated
SDSS fiber spectra (Paper I). The left panel shows the FWHM of
each \OIII\ component as a function of the fiber-integrated
stellar velocity dispersion. The red and blue symbols are for
the redshifted component and the blueshifted component,
respectively; the gray symbols are for all the objects in our
parent sample (Paper I). The five binary AGNs all reside in the
right half of the plot, with larger stellar velocity
dispersions on average. This is consistent with the existence
of multiple stellar systems in these binary AGNs. The middle
panel shows the velocity offset of the two \OIII\ components as
a function of the stellar velocity dispersion. The right panel
shows the total \OIIIb\ luminosity versus the \OIII\ velocity
offset. The binary AGNs seem to occupy a different luminosity
regime from those classified as NLR kinematics. Fig.\
\ref{fig:bulk_prop} suggests that the bulk properties of
kpc-binary AGNs might be statistically different from those
with an origin in NLR kinematics, which, if true, can be used
to refine the selection of binary candidates from
spatially-integrated spectroscopy. However, our current sample
of kpc-scale binary AGNs is still small, and a larger
statistical sample of binary AGNs is needed to confirm these
trends.


In these binary systems, the smaller stellar component does not
necessarily correspond to the more luminous \OIII\ component.
If BH mass is proportional to bulge mass/luminosity and \OIII\
luminosity is proportional to the intrinsic AGN luminosity,
then the smaller BH must be accreting at a substantially higher
Eddington ratio than the larger BH in J1131$-$0204 and
J1332+0606. It is likely that the smaller component in these
merging systems is more gas-rich and/or easier for the nuclear
region to be disturbed during the merger, and hence the BH
accretion is more efficient.


\subsection{The Frequency of Kpc-scale Binary Type 2 AGNs}\label{sec:dis1}
Although we have only followed up a small fraction of the
double peaked \OIII\ type 2 AGNs in our sample (Paper I), the
$\sim 30$ objects with both NIR imaging and slit spectroscopic
data presented in this paper allow us to estimate the fraction
of kpc-scale AGNs among these double-peaked \OIII\ AGNs. For
most of our targets we carried out the imaging first, and
promising targets seen in the imaging data were preferentially
observed spectroscopically. We have obtained slit spectra for
only about half of the imaged objects. Therefore we must
estimate the fraction of binary AGNs and NLR kinematics taking
into account the spectroscopic completeness. Of the $61$
objects that we have observed in the NIR, five were classified
as kpc-scale binary AGNs with resolved double nuclei that are
within the $3$\arcsec\ diameter of SDSS
fibers\footnote{Multiple stellar components separated by more
than 3\arcsec\ is unlikely to be covered within the SDSS fiber,
hence do not correspond to the two \OIII\ velocity components
seen in SDSS spectra.}, and subsequently confirmed with slit
spectroscopy. Hence the frequency of kpc-scale binary AGNs
among these double-peaked \OIII\ objects is $\sim 10\%$. Of the
remaining $\sim 90\%$ of the objects, $\sim 50\%$ are best
explained by NLR kinematics (\S\ref{sec:kinematics}) and $\sim
40\%$ are ambiguous cases. At least some of the latter case are
of NLR kinematics origin, and the remainder are probably binary
AGNs with somewhat smaller separations, and could be identified
with better spatial resolution observations
\citep[e.g.,][]{Fu_etal_2010}. If we take the results of
\citet{Fu_etal_2010} and assume all the objects with resolved
double nuclei in their NIR AO imaging are binary AGNs (but see
\S\ref{sec:dis2}), then the binary AGN fraction among the
double-peaked \OIII\ objects increases to $\sim 20\%$. It is
conceivable (although unlikely) that kpc-scale binary AGNs with
even smaller separations would not have been revealed with NIR
AO imaging, and if we assume this entire ambiguous sample is
binaries, we get a hard upper limit of the kpc-scale binary AGN
fraction of $\sim 50\%$. Therefore, a conservative estimate of
the genuine kpc-binary AGN fraction of the double-peaked \OIII\
type 2 AGNs is $10\%\la f_1< 50\%$ and the upper bound should
be considered as a hard limit. Our observations thus indicate
that the majority of these double-peaked \OIII\ objects reflect
NLR kinematics and are not kpc-scale binary AGNs. However, in
some of the systems that we classified as kinematic cases, we
see disturbed morphology or close companions, which may
indicate that the (single) AGN activity was triggered by a
recent merger event.

Only $\sim 1\%$ of low-redshift ($z<0.3$) type 2 AGNs have
double-peaked \OIII\ profiles in their spatially-integrated
spectra (Paper I). If the orbit inclination is random for
binary AGNs, then some of them will not show a double-peaked
\OIII\ profile since the line itself has a finite width (due to
the intrinsic line width and instrument broadening), and thus
will not enter our sample. The real situation is more
complicated. The selection completeness depends on the spectral
resolution of SDSS, and the intrinsic distributions of binary
properties, such as the relative velocity of the two BHs at
$\sim$ kpc separations, and the FWHM and strength of individual
narrow line components. The SDSS spectral resolution will
exclude double-peaked \OIII\ objects with a relative velocity
offset $\la 150\,{\rm kms^{-1}}$, and our sample in Paper I is
strongly biased against weak \OIII\ objects.
In Paper I we performed simple Monte Carlo simulations of mock
spectra to estimate the selection completeness of kpc binaries
with double-peaked \OIII\ with assumed distributions of
kpc-scale binary AGN properties, combined with the SDSS
spectral resolution and random line-of-sight (LOS). We found
that the selection completeness rapidly decreases with
decreasing the LOS velocity offset, decreasing the \OIII\
equivalent width, and increasing the line FWHM; and all these
effects are well coupled.

Here we perform similar Monte Carlo simulations, with
reasonable properties that are appropriate for our sample. We
assume a physical relative velocity offset of $500\,{\rm
km\,s^{-1}}$ between the two velocity components (i.e., both
NLRs are on circular orbits with circular velocities of
$250\,{\rm km\,s^{-1}}$), a fixed FWHM$=250\, {\rm km\,s^{-1}}$
for both components, and assign an \OIIIb\ rest equivalent
width EW$_{\rm [OIII]5007}\equiv 3$EW$_{\rm [OIII]4959}$
uniformly distributed within $15-45\,$\AA\ for each component.
The assumed circular velocity of the binary is reasonable given
that our objects are massive galaxies with stellar mass
$M_*\sim 10^{11}\,M_\odot$, and the adopted FWHM and EW values
are also typical of our sample objects (Paper I). We further
assume random orientations of the binary orbits, and generate
mock spectra with the assumed \OIII\ properties, a continuum
S/N$\sim 5$\,pixel$^{-1}$, and the SDSS spectral resolution. We
visually inspect these mock spectra and flag double-peaked
objects. We found \textbf{$\sim 20\%$} of the simulated objects
are identified as double-peaked \OIII\ objects. 
As discussed earlier, 10-50\% of double-peaked objects have
compelling evidence for two active nuclei ($f_1$). Now assuming
that we can detect only $20\%$ of true double-peaked objects,
we expect that $5\%$ of the $z\la 0.3$ type 2 AGN population is
double, and thus that a fraction $0.5\%<f_2=5\%\times
f_1<2.5\%$ of the total population is composed of kpc-scale
binaries. It is very likely that there are populations of
kpc-scale binary AGNs with a relative (unreduced) orbital
velocity $\la 150\, {\rm kms^{-1}}$ or with more discrepant
\OIII\ luminosity ratios that are missed in our SDSS sample.

At face value, this low fraction of kpc-scale binary AGNs is at
odds with the hypothesis that AGNs are triggered by major
mergers. It either implies that the fraction of time that a
binary AGN spends during this kpc phase is very small compared
to the total AGN lifetime, or only a tiny fraction of kpc-scale
binary AGNs have two distinct NLRs with comparable luminosity
and are lit up simultaneously. Alternatively, it could be that
major mergers are not the dominant mechanism for triggering
low-luminosity AGN activity at low redshift, since the fueling
rate is not as stringent for AGNs as for luminous quasars, and
alternative fueling routes (i.e., secular processes and minor
mergers) may suffice. Refining the binary AGN fraction at
various separations and combining with numerical simulation
results are needed to sort out these possibilities, and our
current study represents the first step towards quantifying the
demographics of binary AGNs on kpc scales.

\subsection{The Different Behaviors of Gas and Stars}\label{sec:dis2}

Our NIR imaging and optical slit spectroscopy demonstrate the
importance of combining imaging and spectroscopy data in
confirming the binary AGN nature of double-peaked objects.
Objects that show spatially offset \OIII\ emission in slit
spectra are not necessarily kpc-scale binary AGNs, as seen in
most of the cases ascribed to NLR kinematics discussed in
\S\ref{sec:kinematics}. Likewise, objects that show apparent
double nuclei in the NIR imaging are not necessarily kpc-scale
binary AGNs, as the companion may not have \OIII\ emission at
all. Objects such as J0002$+$0045 presented in this paper
clearly demonstrate the latter case. For the same reason, not
all of the objects with resolved multiple nuclei in
\citet{Fu_etal_2010} are binary AGNs, even if these multiple
nuclei are covered in a single SDSS fiber. For example,
J1157$+$0816 shows three components in both optical and NIR
imaging \citep[][]{Smith_etal_2010,Fu_etal_2010}. However,
neither the upper or lower companions in the slit spectrum
(Fig.\ \ref{fig:1157+0816}) show \OIII\ emission (although both
are at the redshift of the central object), and the
double-peaked \OIII\ emission is likely caused by the NLR
kinematics of the central source.

\section{Conclusions}\label{sec:con}

We have presented NIR imaging and optical slit spectroscopic
data for 31 double-peaked \OIII\ type 2 AGNs drawn from the
sample of Paper I. The combination of imaging and spectroscopy
allows us to identify the origin of the double-peaked \OIII\
emission in these peculiar objects. Our main conclusions are:

\begin{itemize}

\item $\sim 10\%$ of these objects show spatially coincident double stellar nuclei and NLR emission. The two spatially offset \OIII\ emission
    components are also separated in velocity, and lead to the double-peaked profile seen in the spatially-integrated spectra. These objects are
    best explained by a pair of merging AGNs separated on kpc scales.

\item $\sim 50\%$ of these objects show a single smooth stellar component at $\sim 0.6$\arcsec\ resolution, while the two \OIII\ velocity
    components are spatially offset by $\ga 0.6$\arcsec. For objects with sufficient spectral quality we usually see velocity (or velocity
    dispersion) gradients indicative of a kinematic signature. These objects are best explained by NLR kinematics in single AGNs, involving
    rotation and/or outflows. A recent study of a single double-peaked \OIII\ AGN (SDSS J1517+3353) also favors the kinematics explanation
    \citep{Rosario_etal_2010}.

\item The remaining $\sim 40\%$ of these objects show a small ($\la 0.4$\arcsec) spatial offset between the two \OIII\ velocity components and a
    single smooth stellar component. Some of these objects are likely binaries at smaller separations \citep[e.g.,][]{Fu_etal_2010}, with the
    remainder likely due to NLR kinematics at smaller scales. Follow-up observation with better spatial resolution is needed to distinguish
    between the two scenarios.

\item Our observations demonstrate the necessity of combining imaging and slit spectroscopy data to identify bona fide kpc-scale binary AGNs. In
    particular, objects that show spatially offset \OIII\ components are not necessarily binary AGNs, nor are those objects that show spatially
    resolved stellar components.

\item We estimate that $\sim 0.5-2.5\%$ of the local
    ($z<0.3$) type 2 AGNs are kpc-scale binary AGNs in a
    major merger, where both components are obscured type 2
    AGN with comparable \OIII\ luminosities.

\end{itemize}

Our follow-up observations of the double-peaked \OIII\ sample
demonstrate the feasibility of selecting kpc-scale binary AGN
candidates using this particular spectroscopic feature, with a
$\ga 10\%$ success rate. So far we have only followed up about
one third of our parent sample and already doubled the number
of known kpc-scale binary AGNs. Our imaging and spectroscopy
follow-up is still ongoing, accompanied by detailed
investigations on individual systems in other wavelengths
and/or with better spatial resolution. A natural step forward
would be AO imaging and IFU spectroscopy of merging systems and
systems with outflows to obtain detailed spatial and kinematic
maps. By the end of our survey, we expect to increase the
number of known kpc-scale binary AGNs by an order of magnitude.
The increased statistics of the kpc-scale binary AGN sample
will improve the estimate of the binary fraction and the study
of bulk properties of kpc-scale binary AGNs, and will shed
light on the effects of mergers on the host and AGN activity in
these systems. In the meantime, there is strong need to explore
the parameter space in merger simulations to confront the
observed statistics (e.g., Blecha \etal\ 2010, in preparation).

\acknowledgements We thank the anonymous referee for useful comments, and Laura Blecha for useful discussions
and comments on the draft. Y.S. acknowledges support from a
Clay Postdoctoral Fellowship through the Smithsonian
Astrophysical Observatory. Support for the work of X.L. was
provided by NASA through Einstein Postdoctoral Fellowship grant
number PF0-110076 awarded by the Chandra X-ray Center, which is
operated by the Smithsonian Astrophysical Observatory for NASA
under contract NAS8-03060. X.L. and M.A.S. acknowledge the
support of NSF grant AST-0707266. Support for this work was provided by the National Aeronautics and Space Administration through Chandra Award Number GO1-12127X issued by the Chandra X-ray Observatory Center, which is operated by the Smithsonian Astrophysical Observatory for and on behalf of the National Aeronautics Space Administration under contract NAS8-03060.

Funding for the SDSS and SDSS-II has been provided by the
Alfred P. Sloan Foundation, the Participating Institutions, the
National Science Foundation, the U.S. Department of Energy, the
National Aeronautics and Space Administration, the Japanese
Monbukagakusho, the Max Planck Society, and the Higher
Education Funding Council for England. The SDSS Web Site is
http://www.sdss.org/.

Facilities: Sloan, Magellan: Baade (PANIC), Magellan: Clay
(LDSS3), ARC 3.5m (DIS)

\appendix
\section{Model Fits for the NIR Images of Kinematic and Ambiguous Cases}\label{sec:app}

For completeness we here provide the {\sevenrm GALFIT} results
for objects that we classified as NLR kinematics around single
AGNs or ambiguous cases. Even though these AGNs do not appear
to have substantial multiple components within the host galaxy,
reproducing the exact surface brightness profile is still
challenging for some cases. We restrict ourselves to a basic
model of a de Vaucouleurs bulge $+$ an exponential disk for the
whole galaxy. We also tried a S\'{e}rsic bulge plus an
exponential disk to improve the fits if possible. In a few
objects there are one or two companions overlapping with the
light from the galaxy, and we fit additional components to
these companions. The fitting results are shown in Figs.\
\ref{fig:NLR_galfit} (for the NLR kinematics cases) and
\ref{fig:amb_galfit} (for the ambiguous cases). In the NLR
kinematics cases, no substantial residuals are seen at the
locations indicated by the offset \OIII\ emission seen in the
2d spectra.

\newpage

\newpage
\begin{figure*}
  \centering
    \includegraphics[width=0.9\textwidth]{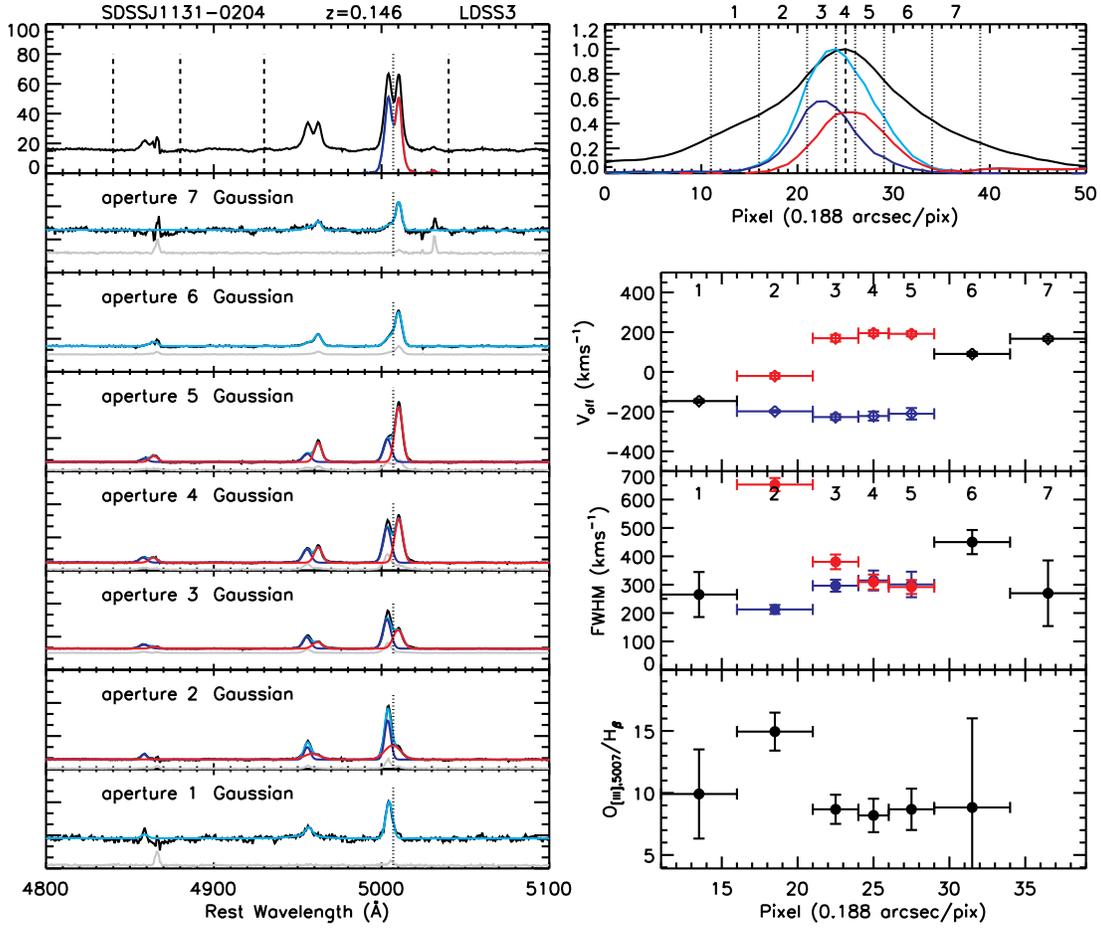}
    \caption{Diagnosis of the 2d spectrum of J1131-0204. Notation is the same as Fig.\ \ref{fig:1108+0659_diag}. The spatial distribution of the
    continuum shown in the upper-right panel has extended wings, which arise from the prominent disk as seen in the optical and NIR images. There are
    no coherent trends seen in the velocity offsets, line widths and \OIII/\hbeta\ flux ratio, as functions of location. }
    \label{fig:1131-0204_diag}
\end{figure*}

\begin{figure*}
  \centering
    \includegraphics[width=0.9\textwidth]{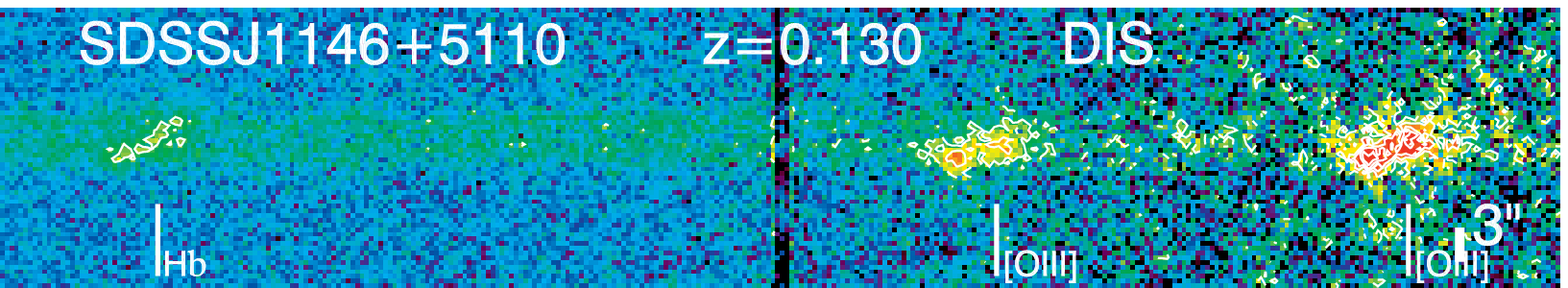}\vspace{9pt}
    \includegraphics[width=0.9\textwidth]{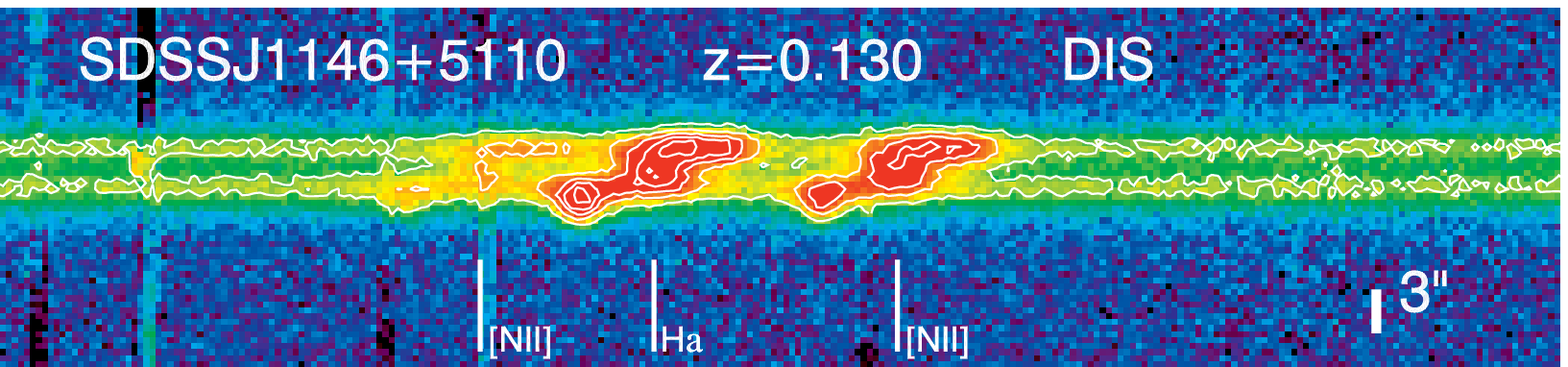}
    \caption{SDSS J1146+5110 (binary). {\em Upper:} DIS 2d spectrum for the \hbeta-\OIII\ region. {\em Bottom:} DIS 2d spectrum for the \halpha\
    region with corresponding lines marked (note that the locations of these line marks are approximate). We do not have a PANIC image for this
    object, but its 2MASS image shows two nuclei separated by $\sim 2.7$\arcsec. The two continua of this system are apparent in the bottom spectrum
    due to the relatively large separation of the two nuclei. The line emission in the southern nucleus itself has two velocity components. The line
    emission in the northern nucleus is spatially offset by $\sim 2.5$\arcsec\ from the continuum of the southern nucleus, and is very weak in the
    \hbeta-\OIII\ region due to poor spectral quality. }
    \label{fig:1146+5110}
\end{figure*}

\begin{figure*}
  \centering
    \includegraphics[width=0.6\textwidth]{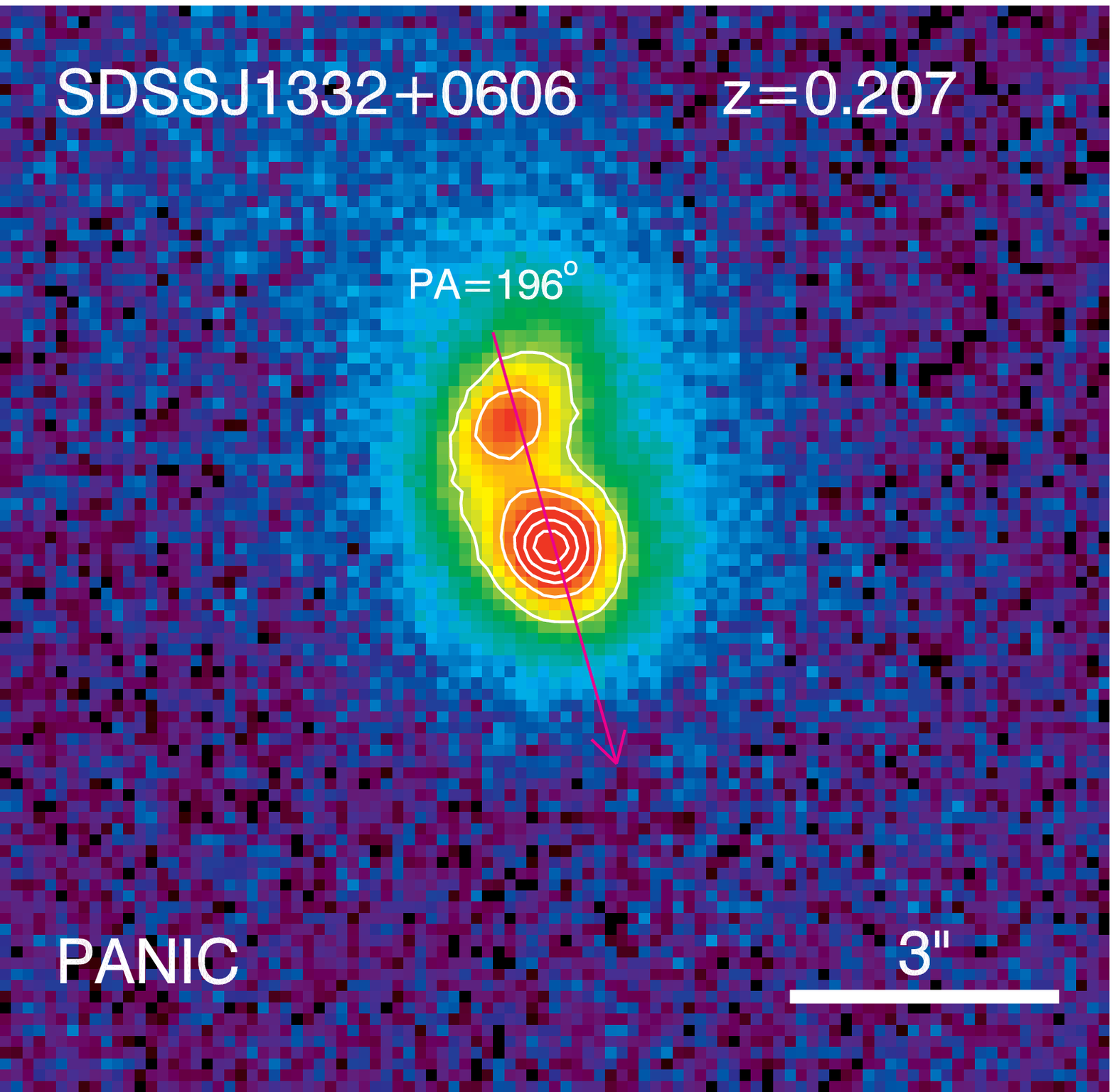}\vspace{9pt}
    \includegraphics[width=0.9\textwidth]{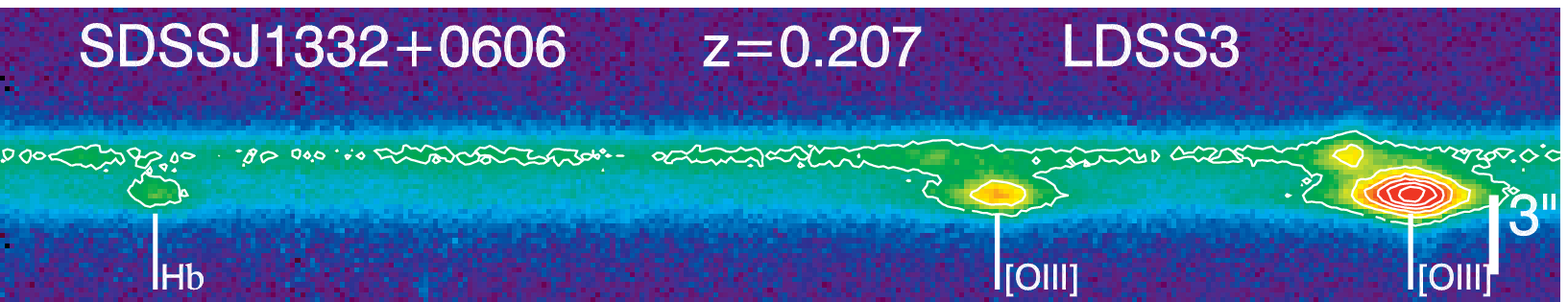}
    \caption{SDSS J1332+0606 (binary). {\em Upper:} PANIC NIR image in $J$. This object has two stellar nuclei separated by $\sim 1.5$\arcsec. {\em
    Bottom:} LDSS3 2d spectrum for the \hbeta-\OIII\ region with corresponding lines marked (note that the locations of these line marks are
    approximate). The two velocity components are spatially offset by $\sim 1.5$\arcsec. Notation is the same as Fig.\ \ref{fig:1108+0659}. The two
    continua are also separated in the 2d spectrum. The northern nucleus with the weaker continuum has stronger \OIII\ emission.}
    \label{fig:1332+0606}
\end{figure*}

\begin{figure*}
  \centering
    \includegraphics[width=0.9\textwidth]{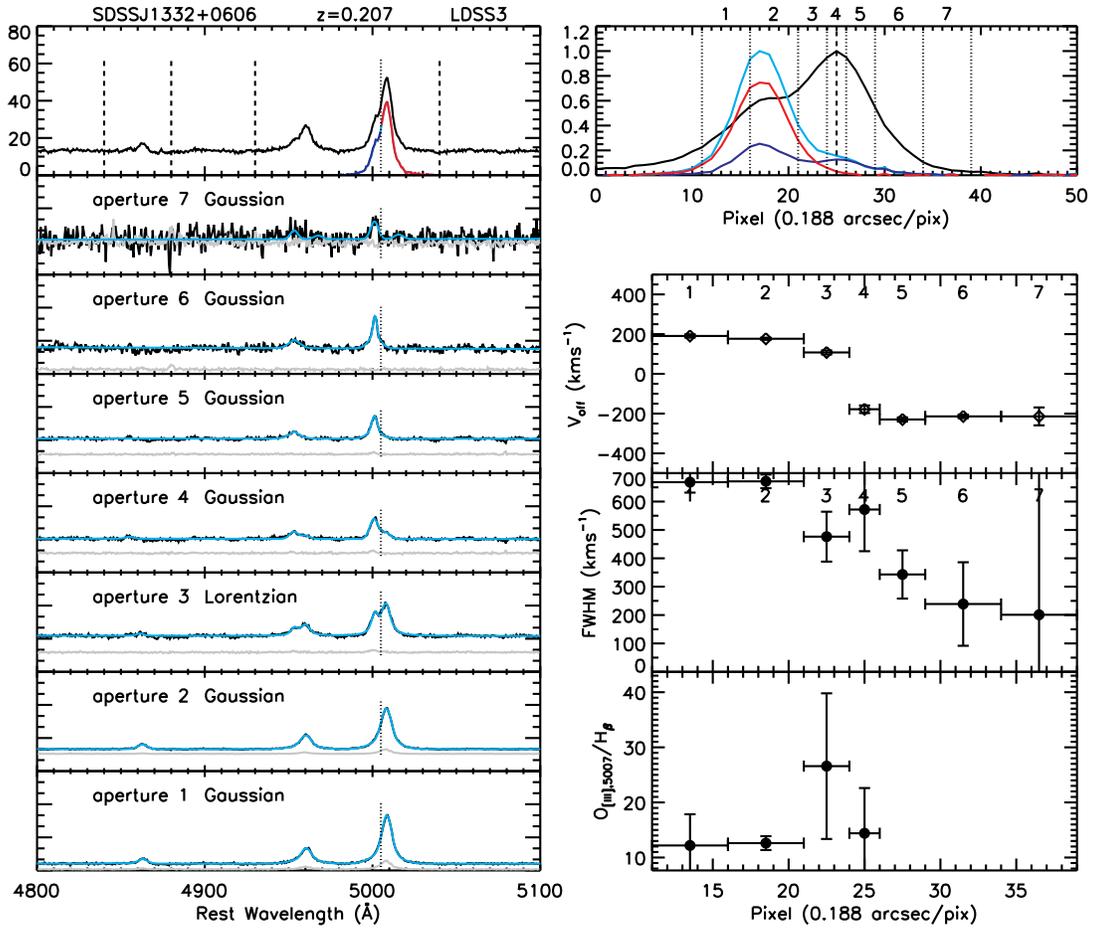}
    \caption{Diagnosis of the 2d spectrum of J1332+0606. Notation is the same as Fig.\ \ref{fig:1108+0659_diag}. The two velocity components are
    relatively well separated in space, and the stronger redshifted component has a larger line width than the weaker blueshifted component. }
    \label{fig:1332+0606_diag}
\end{figure*}

\begin{figure*}
  \centering
    \includegraphics[width=0.6\textwidth]{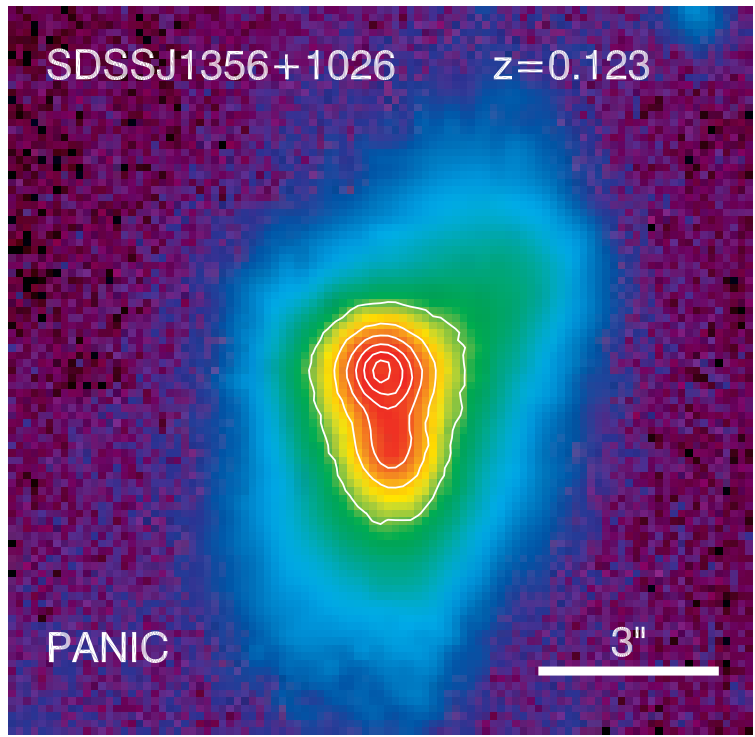}\vspace{9pt}
    \caption{SDSS J1356+1026 (binary). {\em Upper:} PANIC NIR image in $K_s$. This object has two stellar nuclei separated by $\sim 1.3$\arcsec. It
    shows a highly disturbed morphology, presumably caused by the merger. The slit spectroscopy of this object was reported in
    \citet{Greene_etal_2011}.}
    \label{fig:1356+1026}
\end{figure*}


\begin{figure*}
  \centering
    \includegraphics[width=0.6\textwidth]{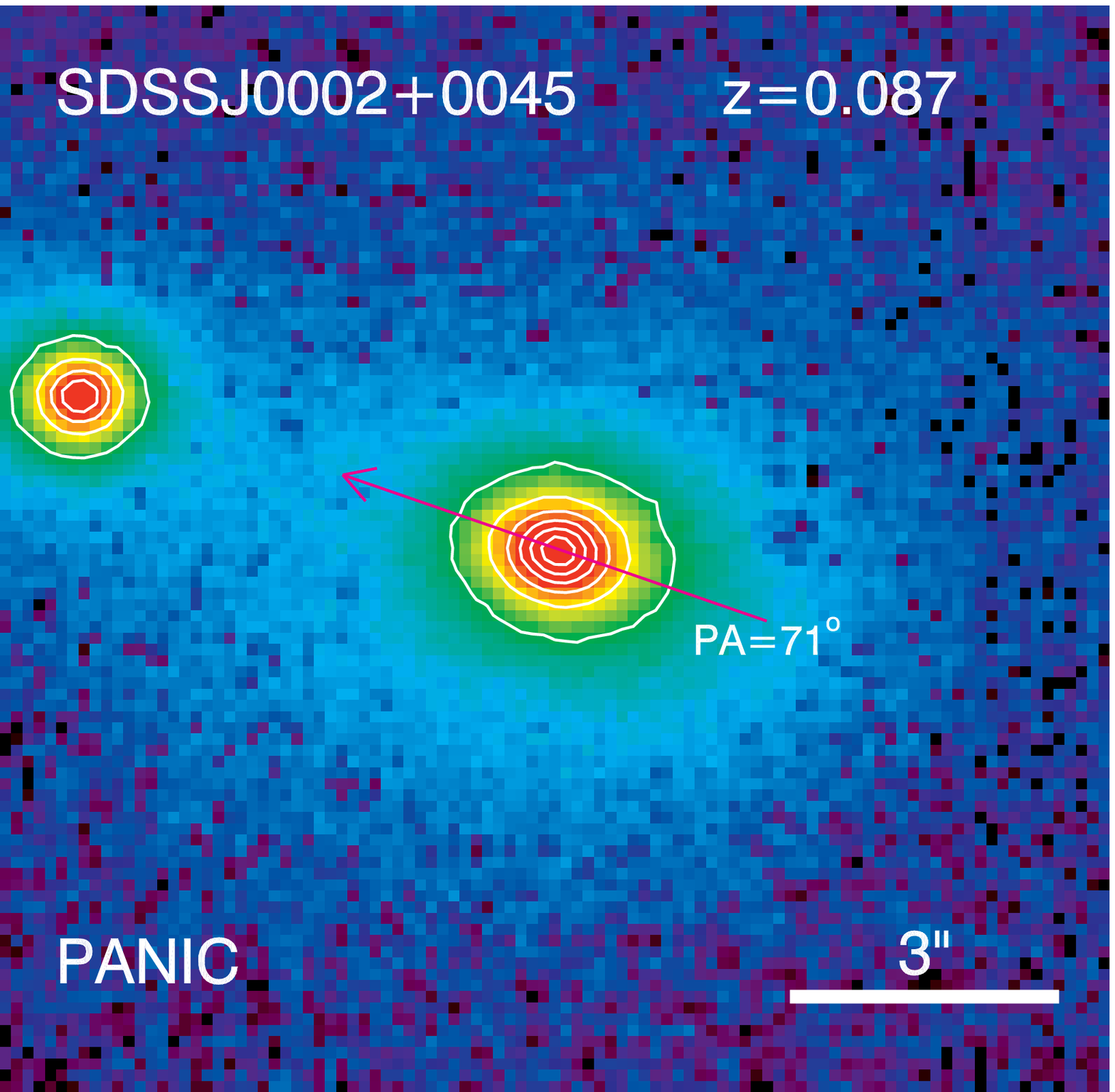}\vspace{9pt}
    \includegraphics[width=0.9\textwidth]{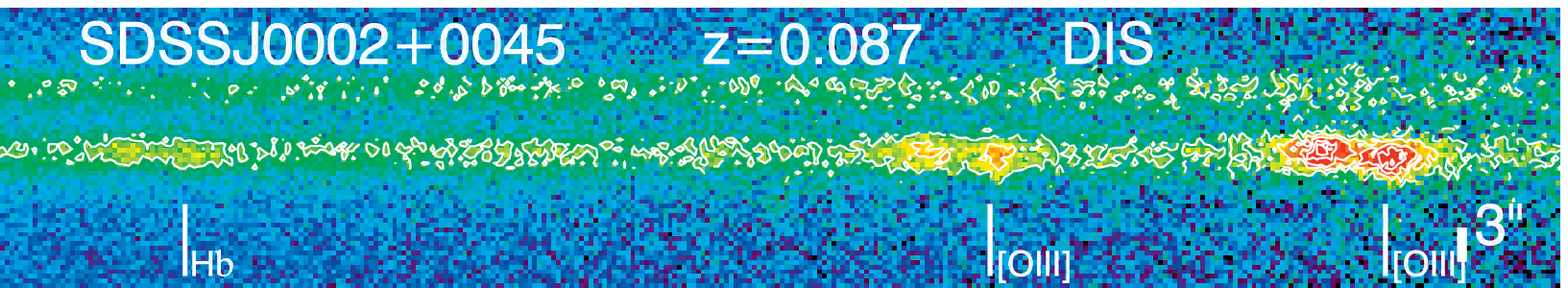}
    \caption{SDSS J0002+0045 (NLR kinematics). {\em Upper:} PANIC NIR image in $K_s$. This object shows a single nucleus, and has a companion $\sim
    5$\arcsec\ away toward the north-east (which is not covered by the SDSS fiber). {\em Bottom:} DIS 2d spectrum for the \hbeta-\OIII\ region with
    corresponding lines marked (note that the locations of these line marks are approximate). The two velocity components are spatially offset by
    $\sim 0.8$\arcsec. The north-east companion does not have detectable \OIII\ emission. Notation is the same as Fig.\ \ref{fig:1108+0659}.}
    \label{fig:0002+0045}
\end{figure*}

\begin{figure*}
  \centering
    \includegraphics[width=0.6\textwidth]{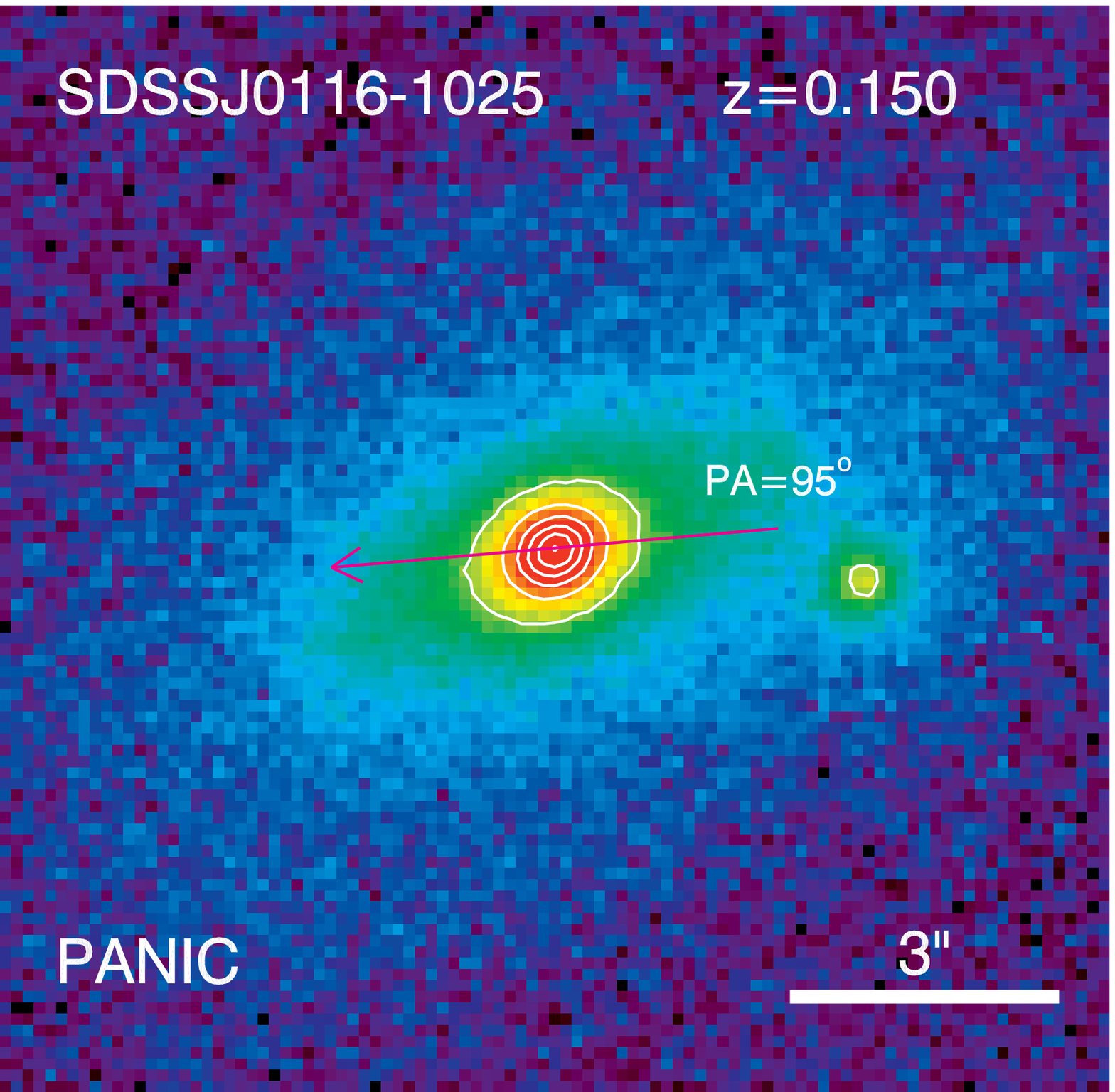}\vspace{9pt}
    \includegraphics[width=0.9\textwidth]{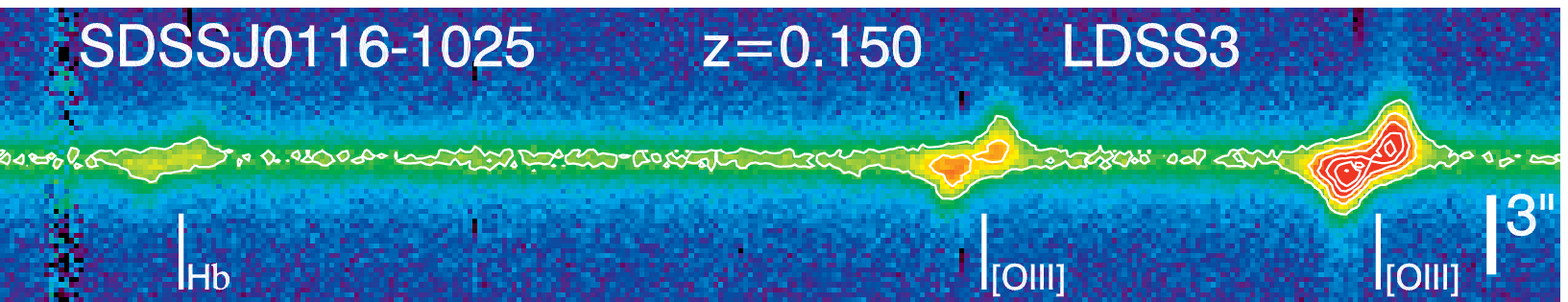}
    \caption{SDSS J0116-1025 (NLR kinematics). {\em Upper:} PANIC NIR image in $K_s$. This object has no resolved double nucleus, and it has a small
    companion $\sim 4$\arcsec\ away towards the west (which is not covered by the SDSS fiber). {\em Bottom:} LDSS3 2d spectrum for the \hbeta-\OIII\
    region with corresponding lines marked (note that the locations of these line marks are approximate). The two velocity components are spatially
    offset by $\sim 0.9$\arcsec. Notation is the same as Fig.\ \ref{fig:1108+0659}.}
    \label{fig:0116-1025}
\end{figure*}

\newpage
\begin{figure*}
  \centering
    \includegraphics[width=0.9\textwidth]{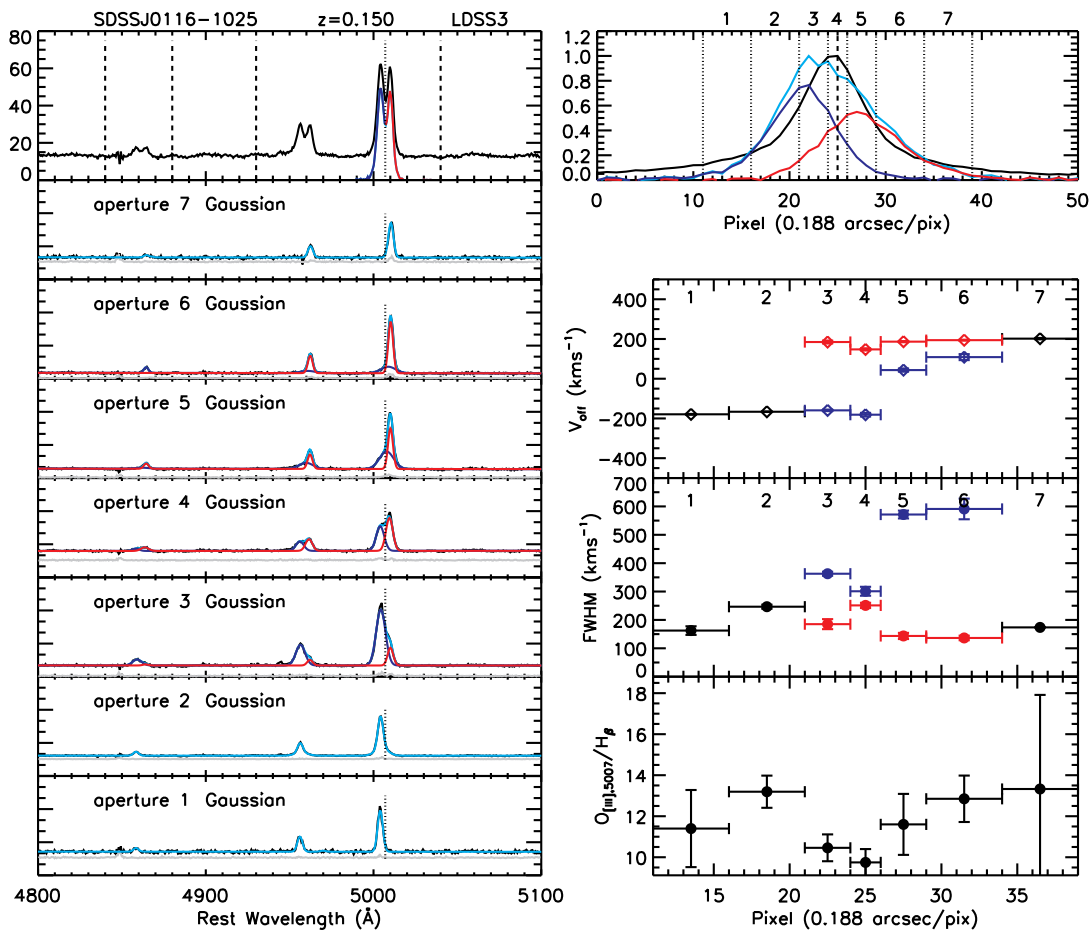}
    \caption{Diagnosis of the 2d spectrum of J0116$-$1025. Notation is the same as Fig.\ \ref{fig:1108+0659_diag}. The \OIII/\hbeta\ flux ratio is
    almost independent of position (within $\sim 0.15$ dex), but the \OIII\ line width seems to increase towards the center of the continuum emission
    except for the blueshifted component in bins 5 and 6. No obvious velocity gradient is seen for either of the two \OIII\ components, and the data
    are consistent with a flat rotation curve with $V_c\sin i\sim 200\ {\rm km\,s^{-1}}$ on each side of the continuum. }
    \label{fig:0116-1025_diag}
\end{figure*}

\begin{figure*}
  \centering
    \includegraphics[width=0.6\textwidth]{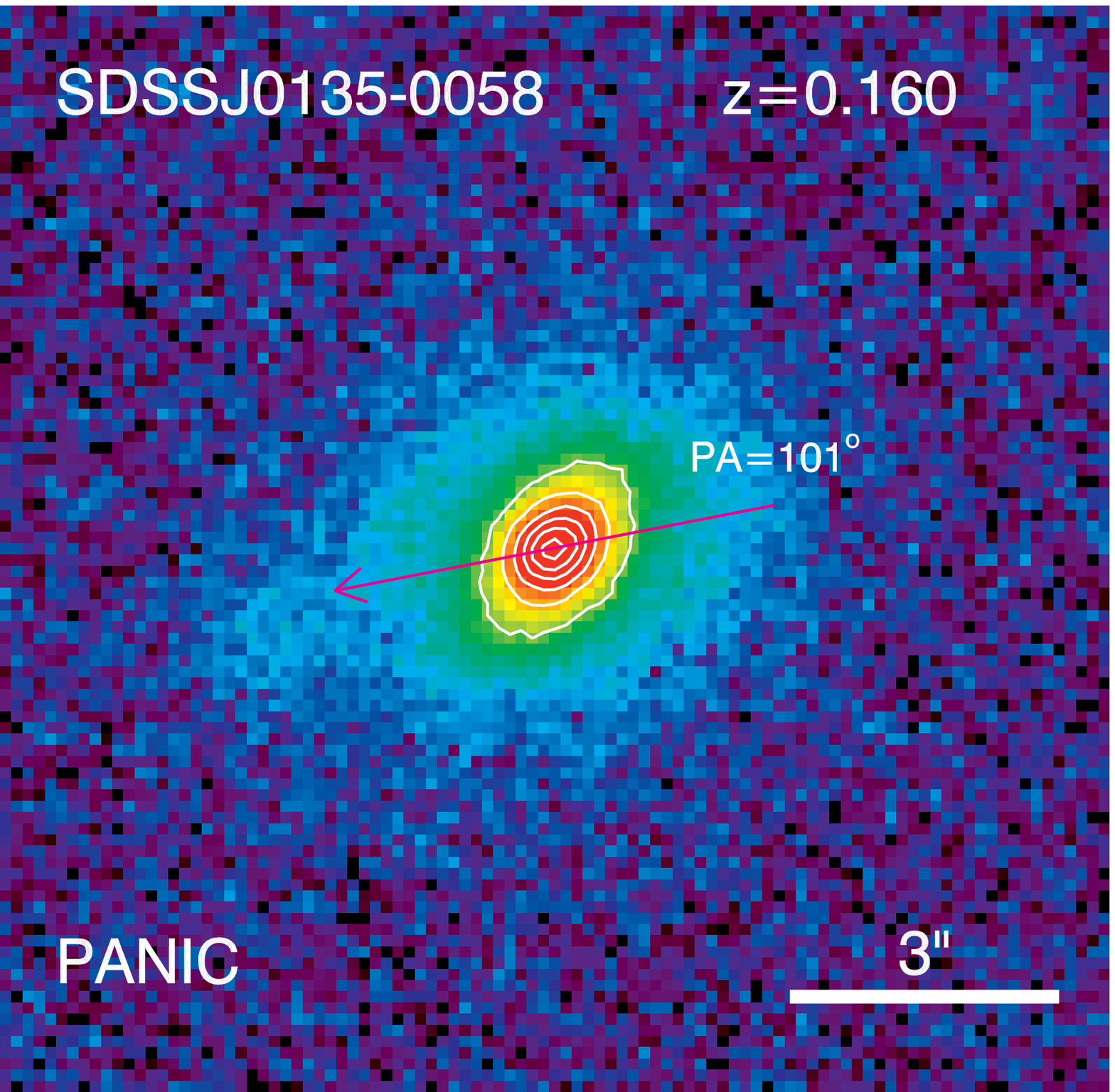}\vspace{9pt}
    \includegraphics[width=0.9\textwidth]{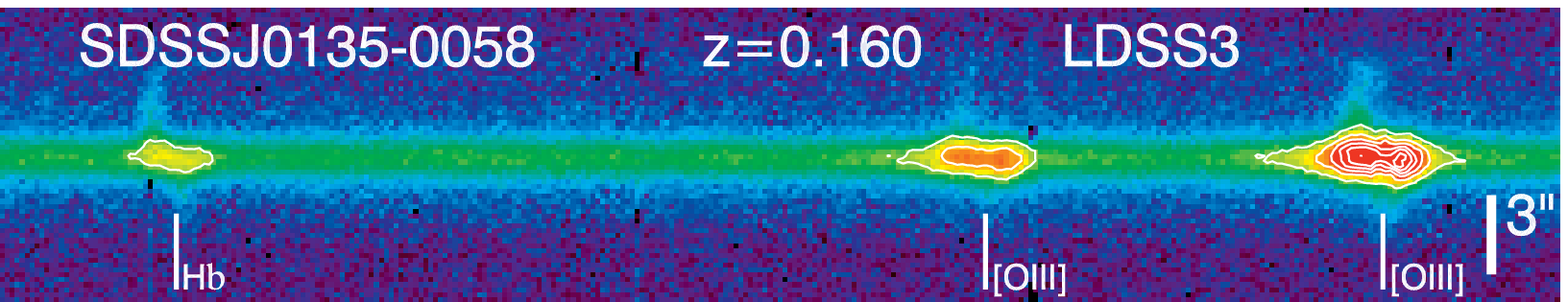}
    \caption{SDSS J0135$-$0058 (NLR kinematics). {\em Upper:} PANIC NIR image in $K_s$. This object has no resolved double nucleus. There are tidal
    features towards the south-east, which is covered by our slit observation. {\em Bottom:} LDSS3 2d spectrum for the \hbeta-\OIII\ region with
    corresponding lines marked (note that the locations of these line marks are approximate). The two velocity components are spatially offset by
    $\sim 0.2$\arcsec. Notation is the same as Fig.\ \ref{fig:1108+0659}.}
    \label{fig:0135-0058}
\end{figure*}

\newpage
\begin{figure*}
  \centering
    \includegraphics[width=0.9\textwidth]{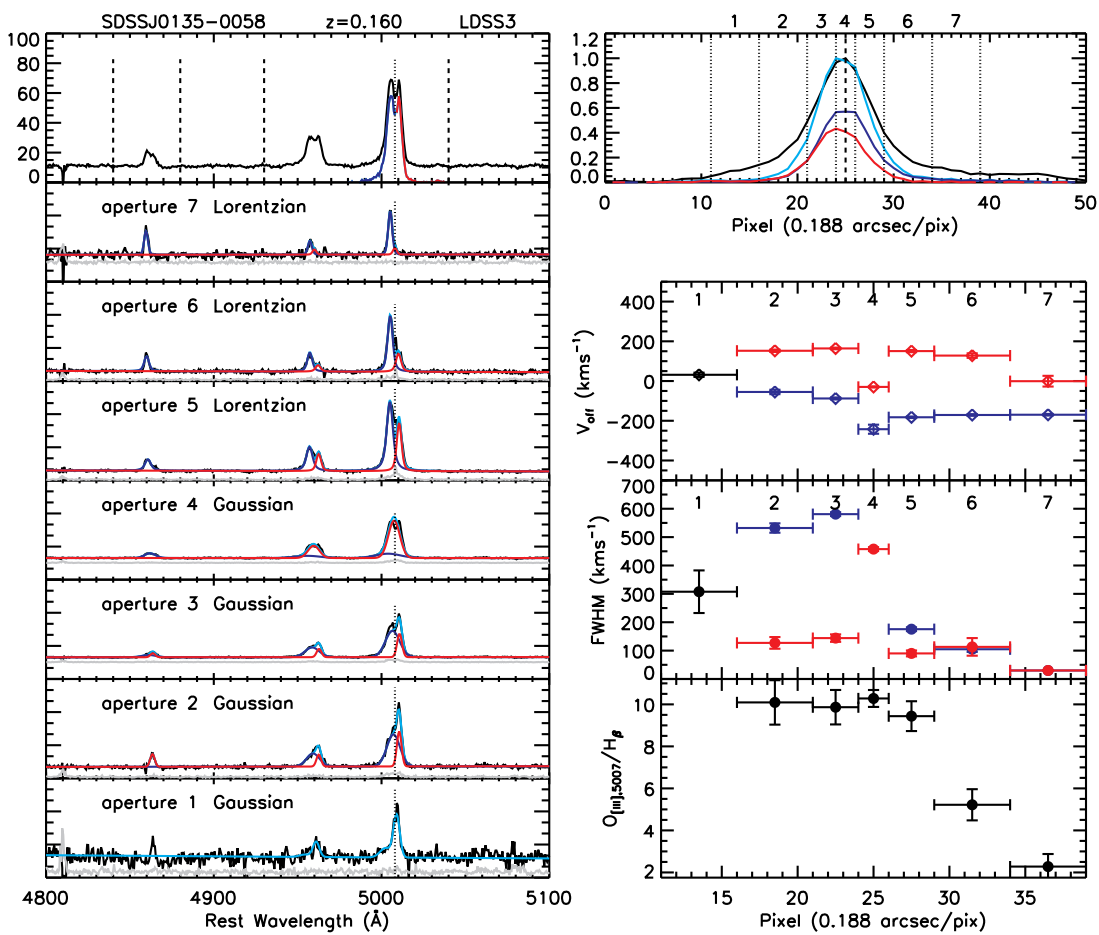}
    \caption{Diagnosis of the 2d spectrum of J0135$-$0058. Notation is the same as Fig.\ \ref{fig:1108+0659_diag}. The \OIII/\hbeta\ flux ratio is
    almost constant until the outermost apertures (6 and 7), where it drops rapidly as it is now tracing the faint tidal feature seen in the 2d
    spectrum. There is a slight velocity gradient for the blue- and red-shifted components. Deblending of the emission lines is ambiguous in bins 3
    and 4.}
    \label{fig:0135-0058_diag}
\end{figure*}

\newpage
\begin{figure*}
  \centering
    \includegraphics[width=0.6\textwidth]{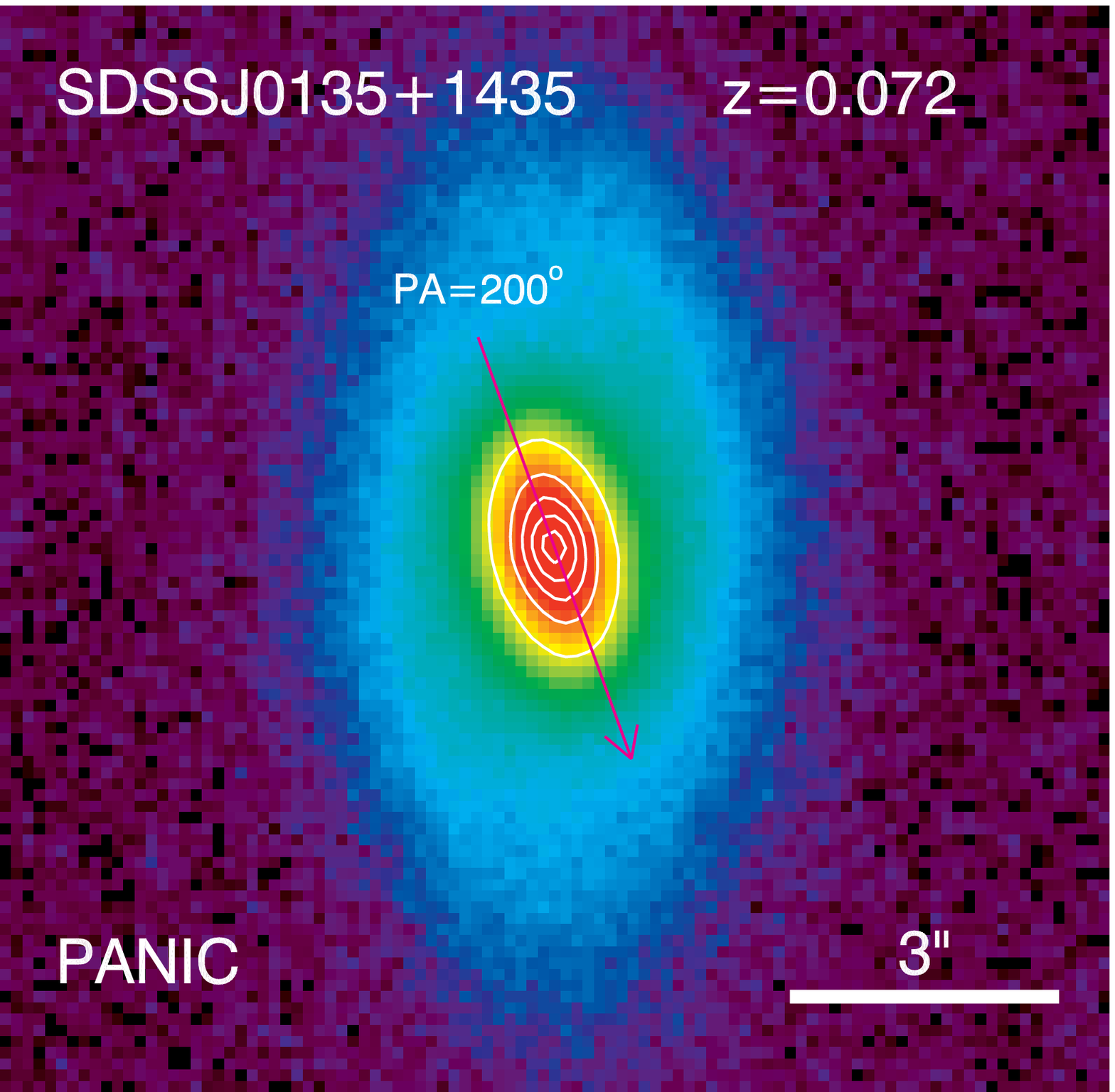}\vspace{9pt}
    \includegraphics[width=0.9\textwidth]{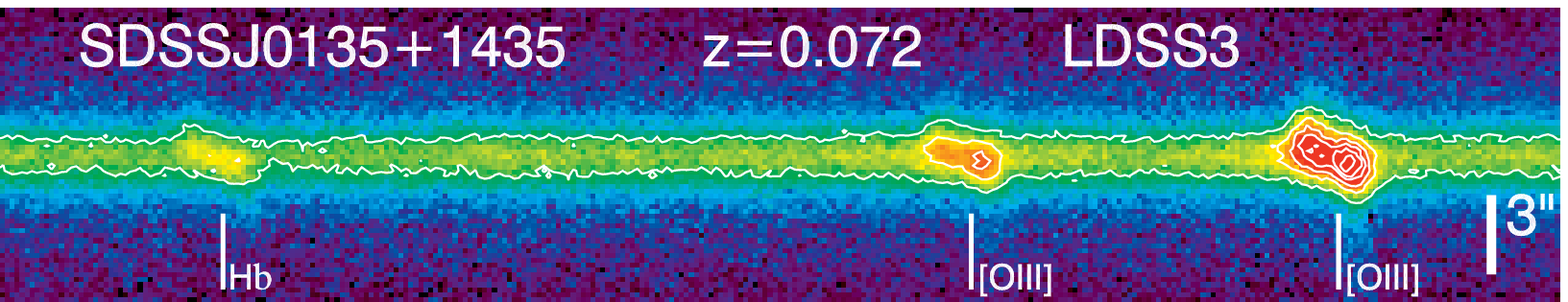}
    \caption{SDSS J0135+1435 (NLR kinematics). {\em Upper:} PANIC NIR image in $K_s$. This object has no resolved double nucleus. {\em Bottom:} LDSS3
    2d spectrum for the \hbeta-\OIII\ region with corresponding lines marked (note that the locations of these line marks are approximate). The two
    velocity components are spatially offset by $\sim 0.8$\arcsec. Velocity gradients can be seen in the 2d spectrum. Notation is the same as Fig.\
    \ref{fig:1108+0659}.}
    \label{fig:0135+1435}
\end{figure*}

\clearpage
\begin{figure*}
  \centering
    \includegraphics[width=0.9\textwidth]{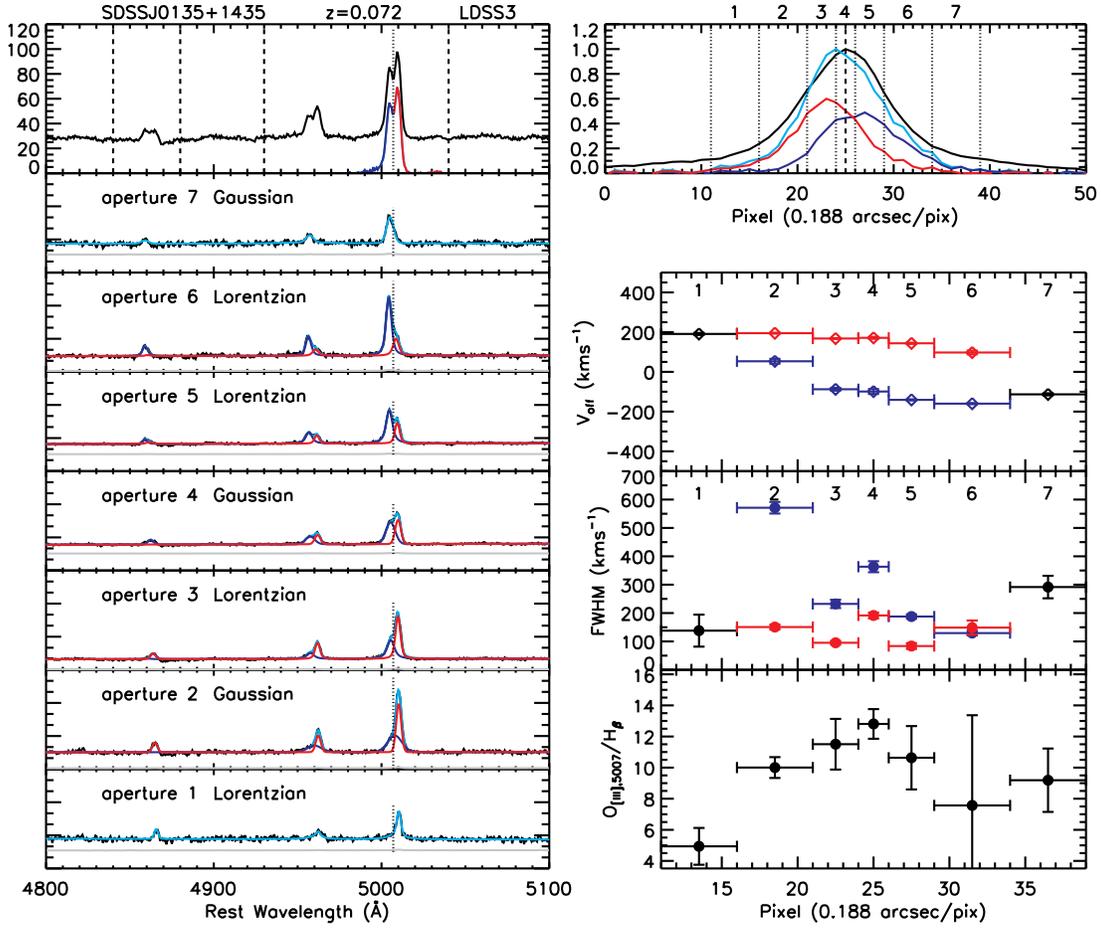}
    \caption{Diagnosis of the 2d spectrum of J0135+1435. Notation is the same as Fig.\ \ref{fig:1108+0659_diag}. We detect clear velocity gradients
    for the blueshifted and redshifted velocity components, and possible increasing velocity dispersion towards the center of the continuum. The
    \OIII/\hbeta\ flux ratio also increases towards the center of the continuum in general. }
    \label{fig:0135+1435_diag}
\end{figure*}

\begin{figure*}
  \centering
    \includegraphics[width=0.6\textwidth]{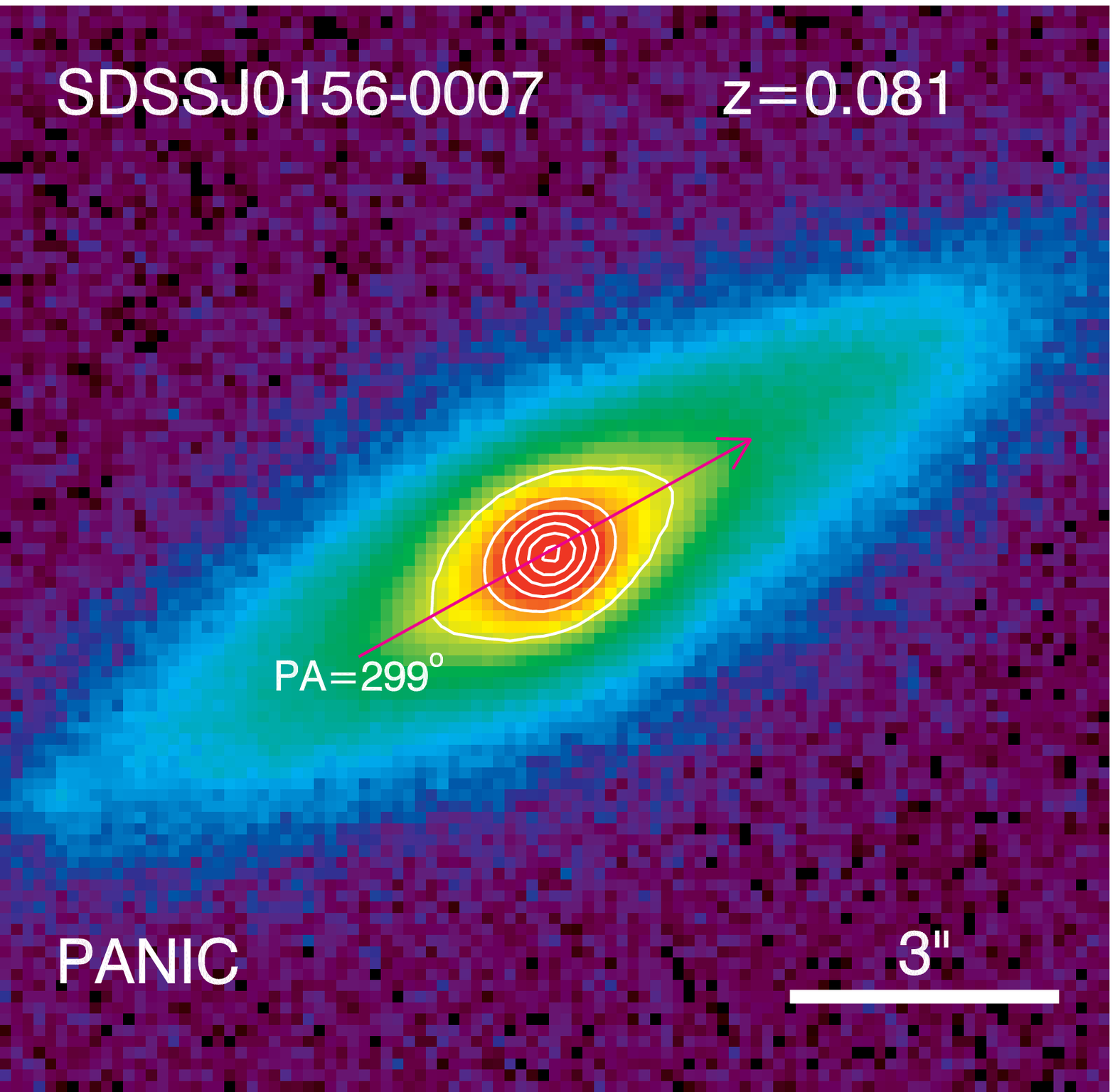}\vspace{9pt}
    \includegraphics[width=0.9\textwidth]{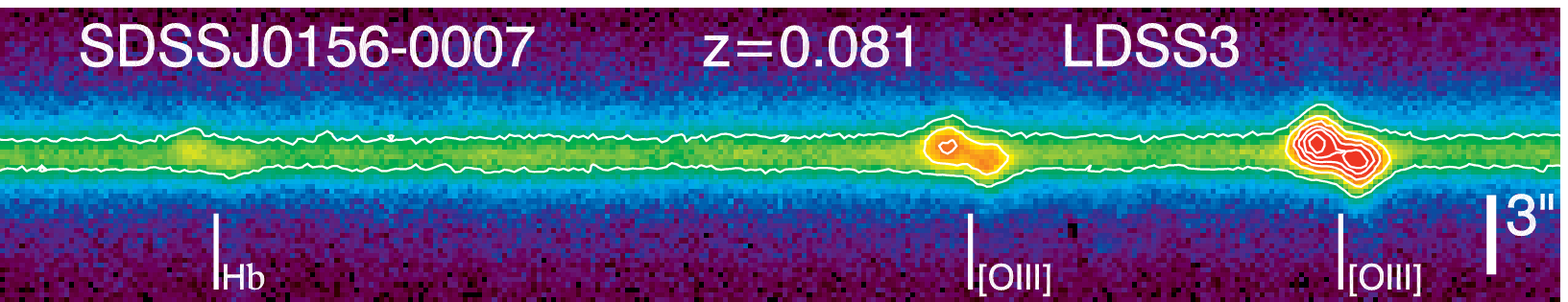}
    \caption{SDSS J0156$-$0007 (NLR kinematics). {\em Upper:} PANIC NIR image in $K_s$. This object has a single nucleus, and it has a large disk
    component. {\em Bottom:} LDSS3 2d spectrum for the \hbeta-\OIII\ region with corresponding lines marked (note that the locations of these line
    marks are approximate). The two velocity components are spatially offset by $\sim 0.6$\arcsec. Notation is the same as Fig.\
    \ref{fig:1108+0659}.}
    \label{fig:0156-0007}
\end{figure*}

\newpage
\begin{figure*}
  \centering
    \includegraphics[width=0.9\textwidth]{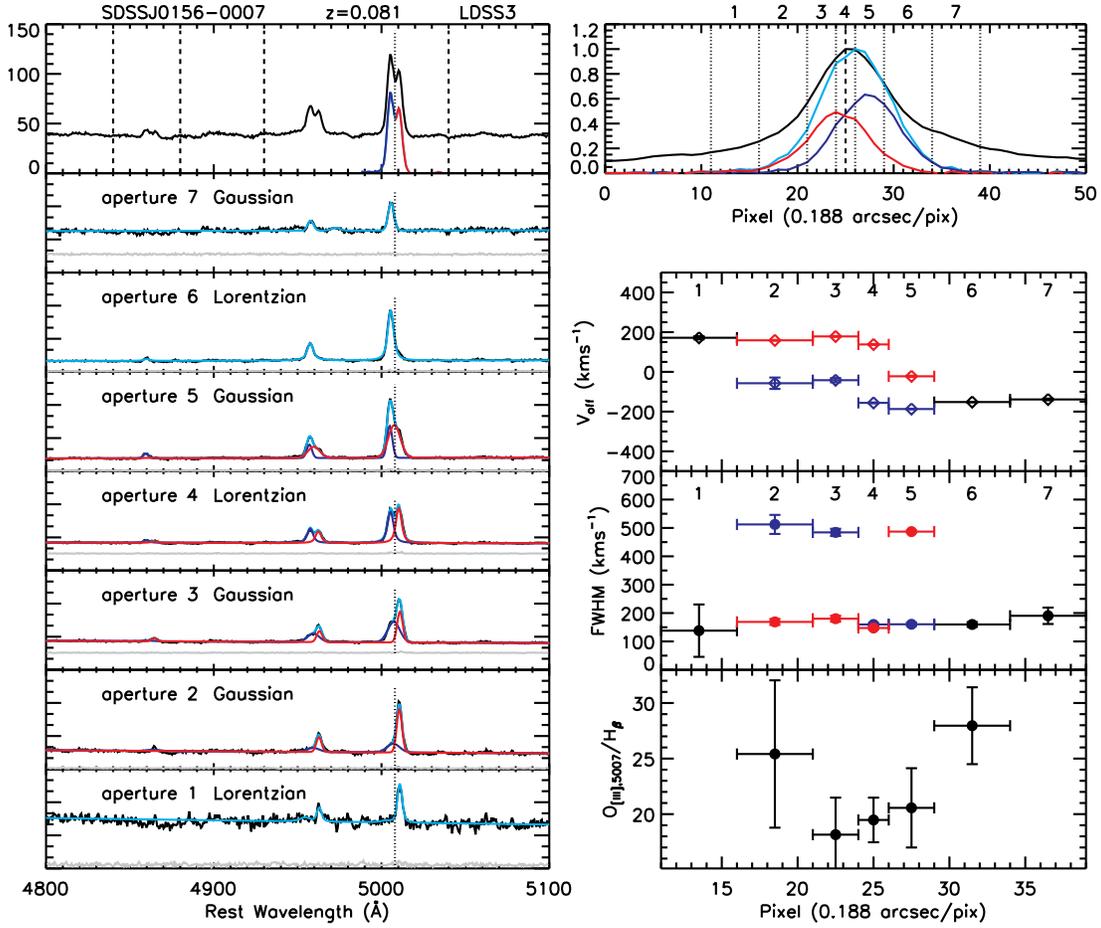}
    \caption{Diagnosis of the 2d spectrum of J0156$-$0007. Notation is the same as Fig.\ \ref{fig:1108+0659_diag}. Deblending of the two velocity
    components is ambiguous in several spatial bins. The \OIII/\hbeta\ flux ratio varies within $\sim 0.2$ dex in bins for which we have
    measurements.
    Weak velocity gradients can been seen for both \OIII\ components. }
    \label{fig:0156-0007_diag}
\end{figure*}

\begin{figure*}
  \centering
    \includegraphics[width=0.442\textwidth]{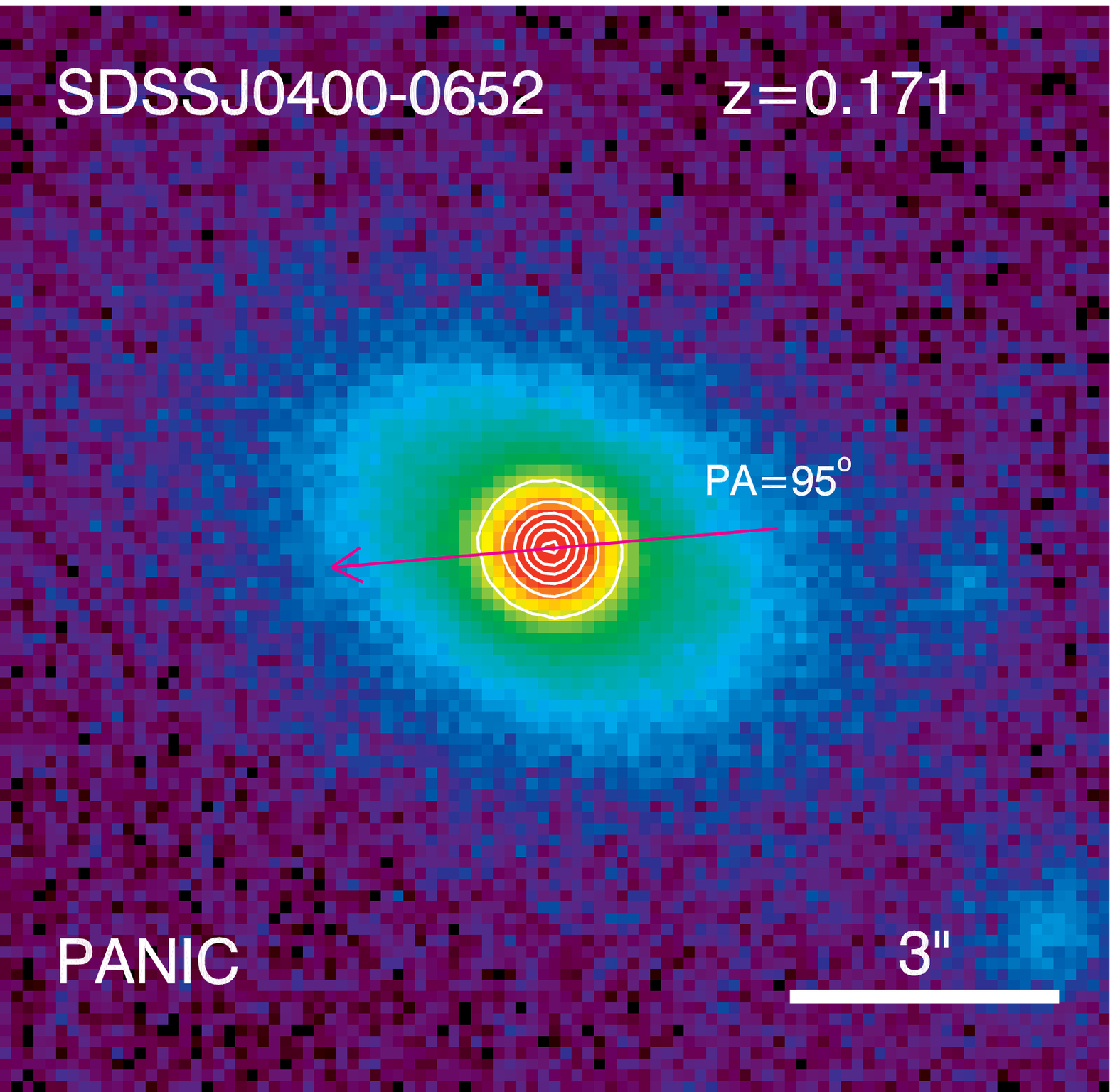}\hspace{5pt}
    \includegraphics[width=0.442\textwidth]{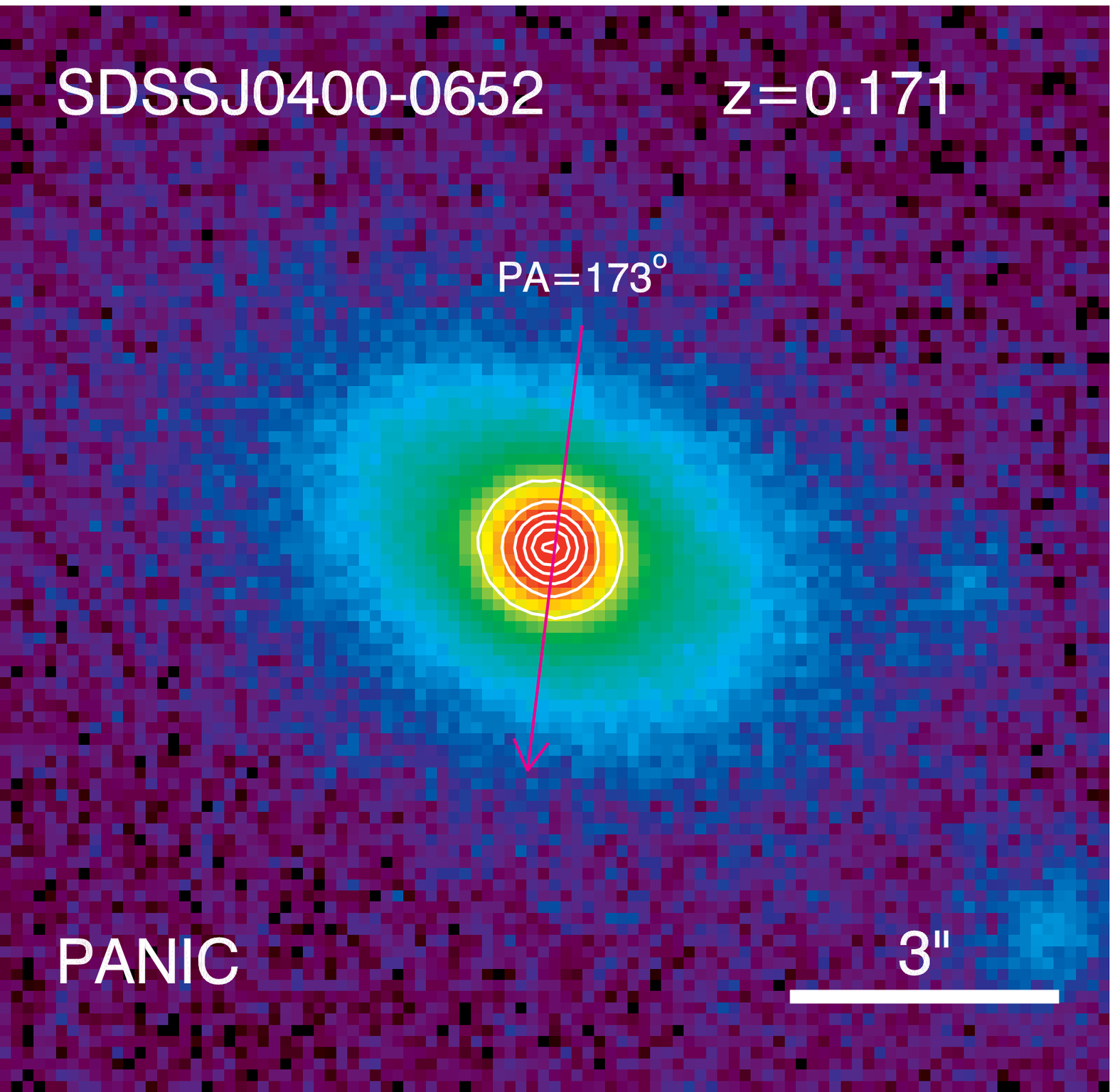}\vspace{9pt}
    \includegraphics[width=0.9\textwidth]{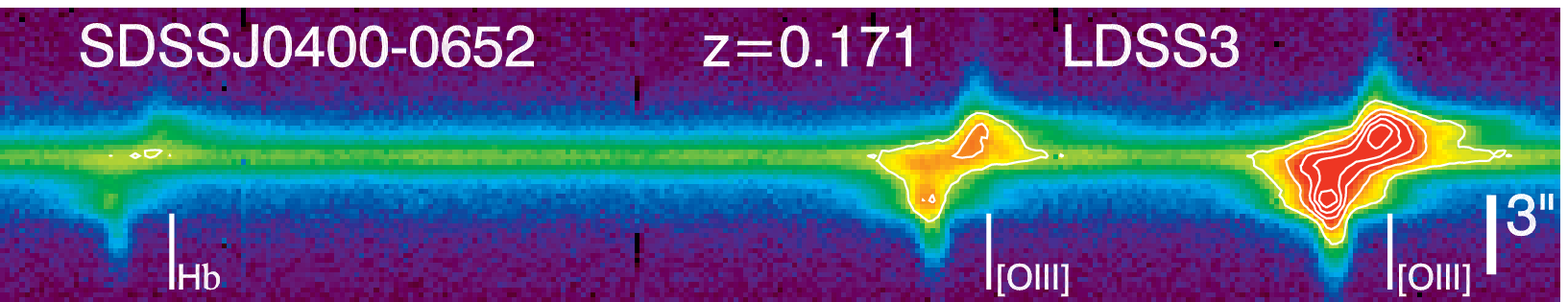}\vspace{5pt}
    \includegraphics[width=0.9\textwidth]{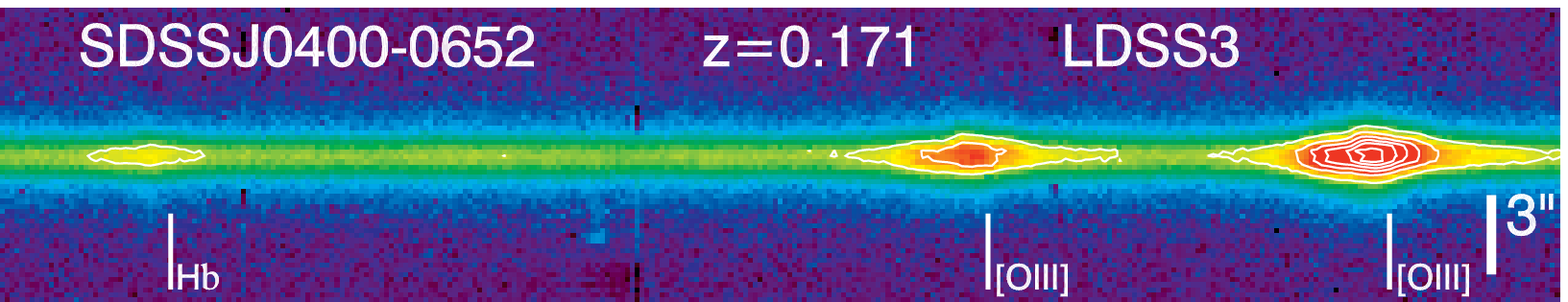}
    \caption{SDSS J0400$-$0652 (NLR kinematics). {\em Upper:} PANIC NIR image in $K_s$. This object has no resolved double nucleus. Two slit
    positions
    are shown. {\em Bottom:} LDSS3 2d spectrum for the \hbeta-\OIII\ region with corresponding lines marked (note that the locations of these line
    marks are approximate). The 2d spectrum shows a rich kinematic structure at one slit position, but not at the other slit position. Notation is
    the
    same as Fig.\ \ref{fig:1108+0659}.}
    \label{fig:0400-0652}
\end{figure*}

\newpage
\begin{figure*}
  \centering
  \subfigure{
    \includegraphics[width=0.9\textwidth]{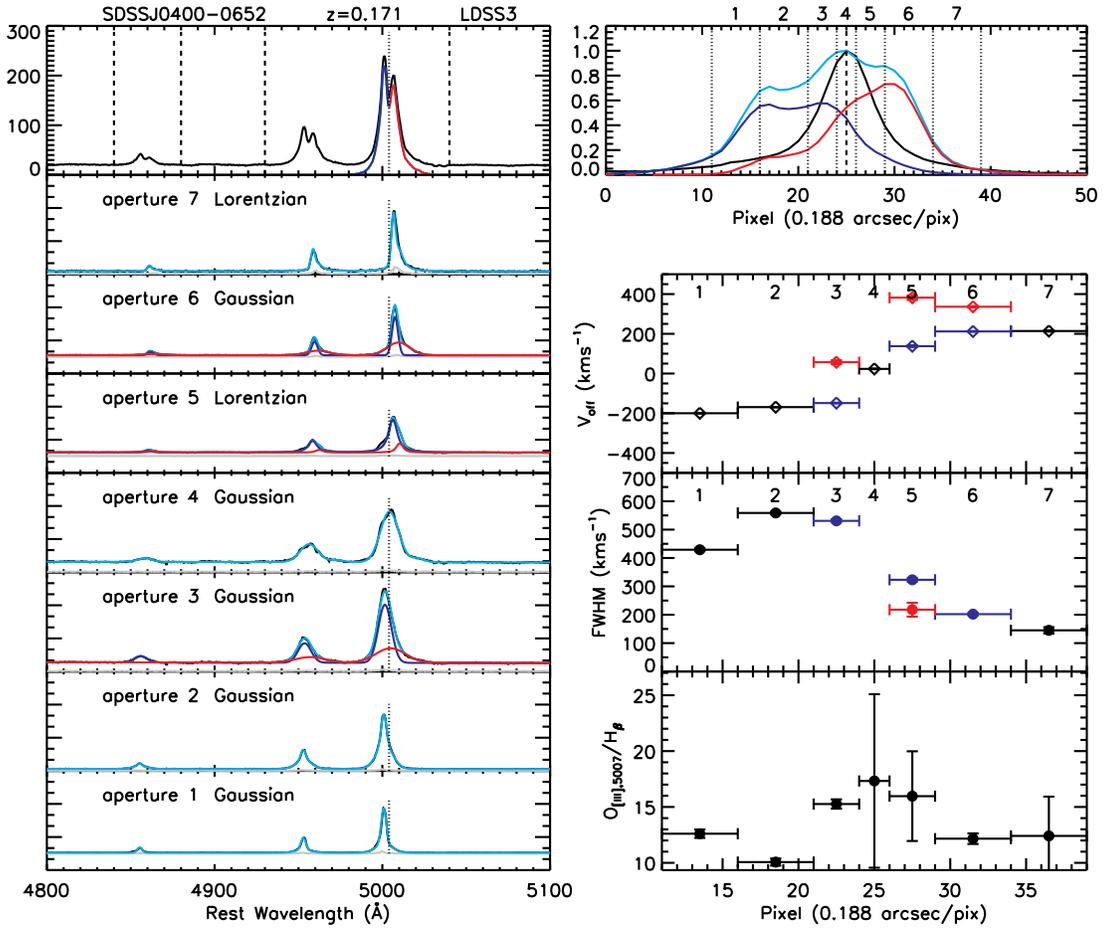}}
    \caption{Diagnosis of the 2d spectra of J0400$-$0652. Notation is the same as Fig.\ \ref{fig:1108+0659_diag}. The upper panel is for the slit
    position PA$=95^\circ$ and the bottom panel is for the slit position PA$=173^\circ$. The two slit positions show very different distributions of
    the line emission. Deblending of the two velocity components is ambiguous in most spatial bins because of the complicated kinematics as seen in
    the 2d spectrum.}
    \label{fig:0400-0652_diag}
\end{figure*}

\addtocounter{figure}{-1}
\begin{figure*}
  \centering
  \subfigure{
    \includegraphics[width=0.9\textwidth]{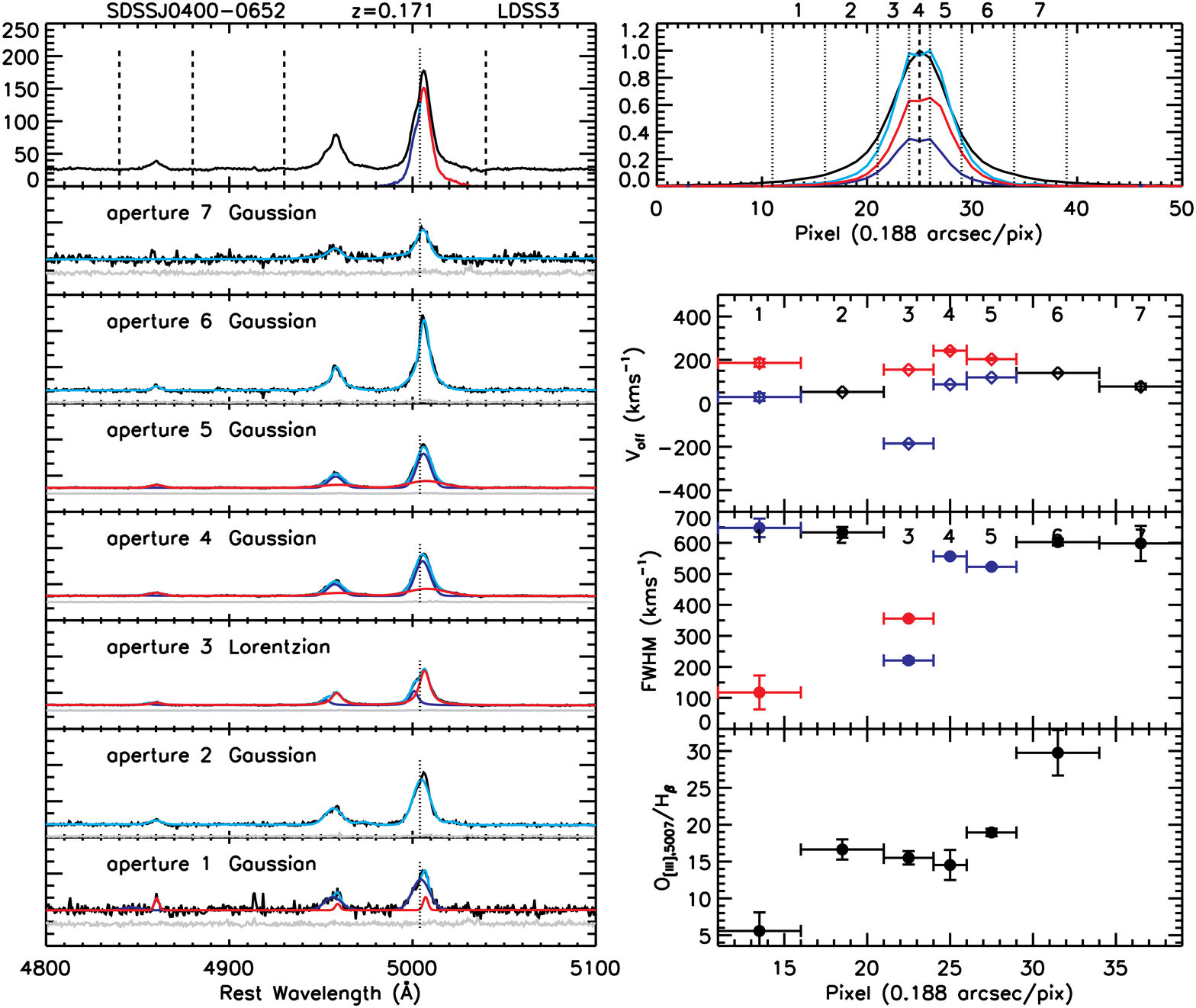}}
    \caption{Continued.}
    \label{fig:0400-0652_diag}
\end{figure*}

\begin{figure*}
  \centering
    \includegraphics[width=0.6\textwidth]{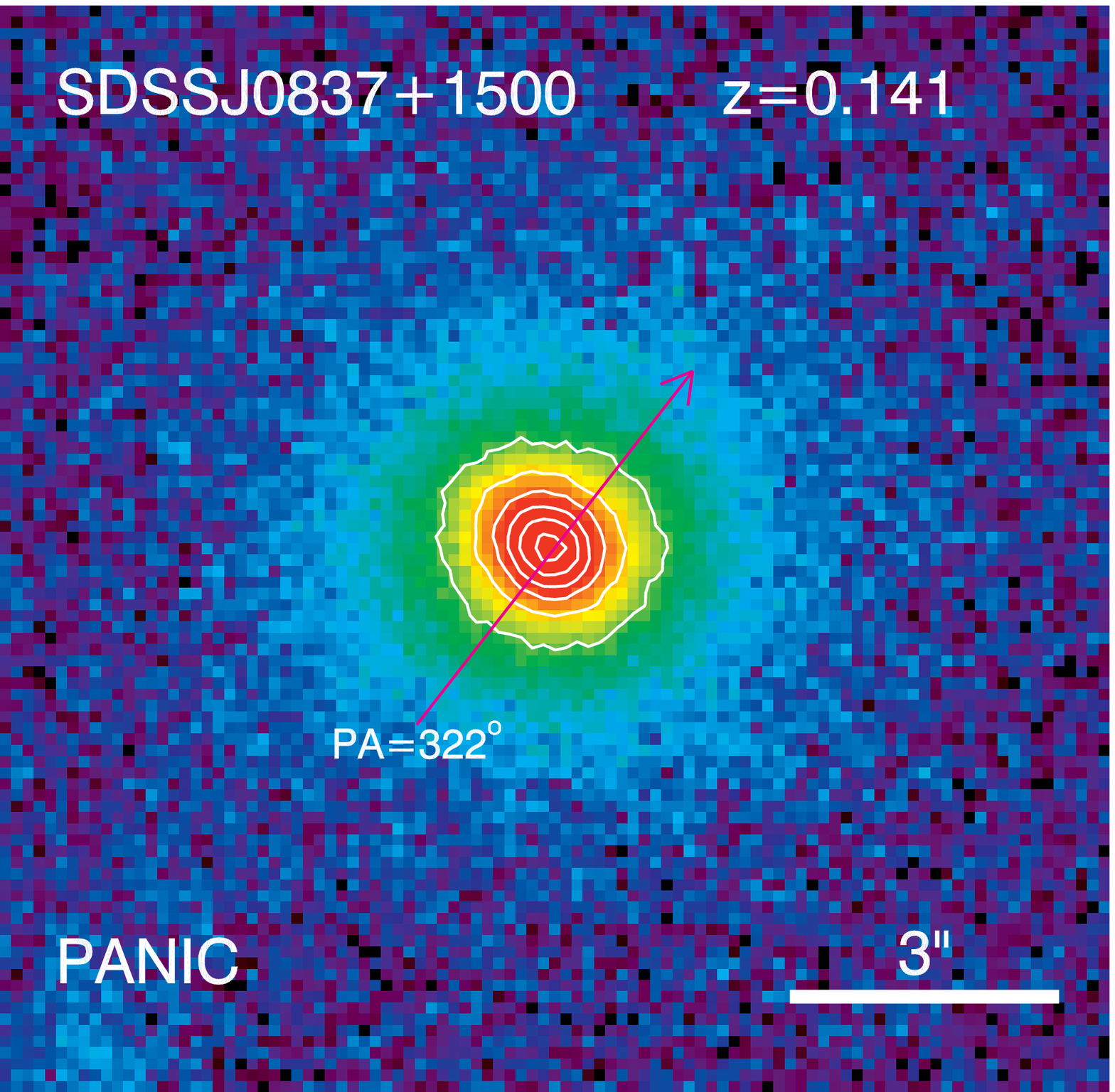}\vspace{9pt}
    \includegraphics[width=0.9\textwidth]{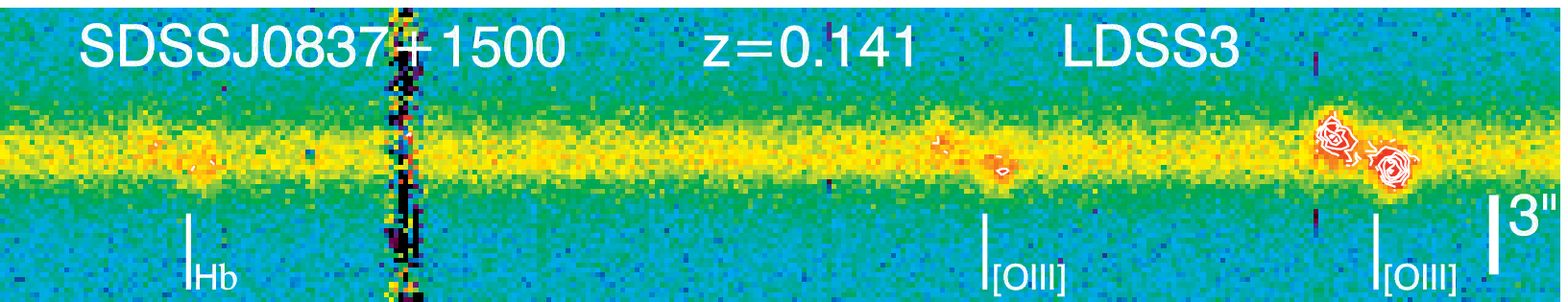}
    \caption{SDSS J0837+1500 (NLR kinematics). {\em Upper:} PANIC NIR image in $K_s$. This object has no resolved double nucleus. {\em Bottom:} LDSS3
    2d spectrum for the \hbeta-\OIII\ region with corresponding lines marked (note that the locations of these line marks are approximate). The two
    velocity components are spatially offset by $\sim 1.3$\arcsec. Notation is the same as Fig.\ \ref{fig:1108+0659}.}
    \label{fig:0837+1500}
\end{figure*}

\newpage
\begin{figure*}
  \centering
    \includegraphics[width=0.9\textwidth]{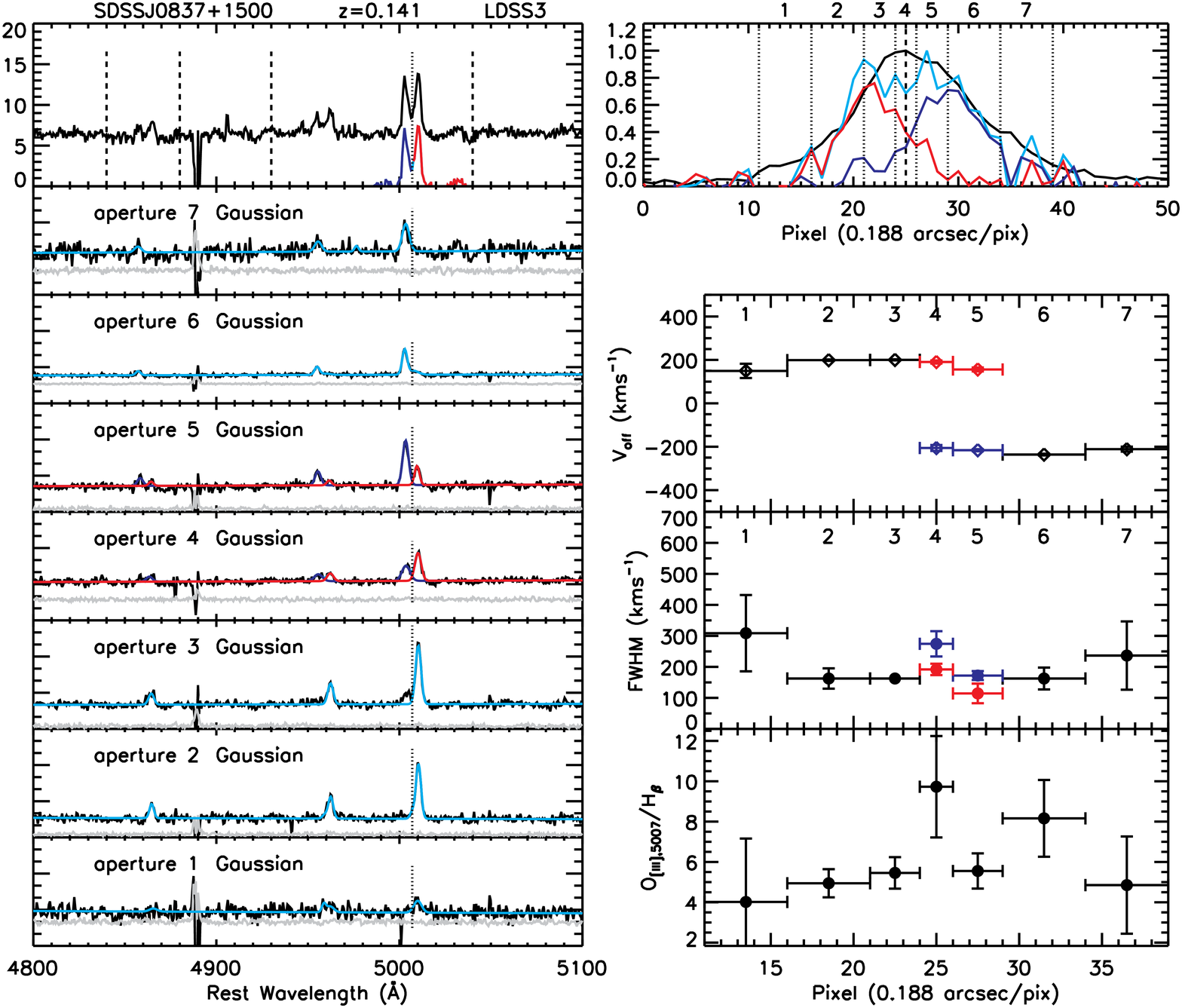}
    \caption{Diagnosis of the 2d spectrum of J0837+1500. Notation is the same as Fig.\ \ref{fig:1108+0659_diag}. The two \OIII\ components are
    comparable in line strength and symmetric about the center of the continuum. No obvious trends are seen in the velocity offset and line width of
    the two \OIII\ components. There is some indication that the \OIII/\hbeta\ flux ratio increases towards the center of the continuum emission,
    which may suggest that the ionizing source is coincident with the center of the continuum.}
    \label{fig:0837+1500_diag}
\end{figure*}

\begin{figure*}
  \centering
    \includegraphics[width=0.6\textwidth]{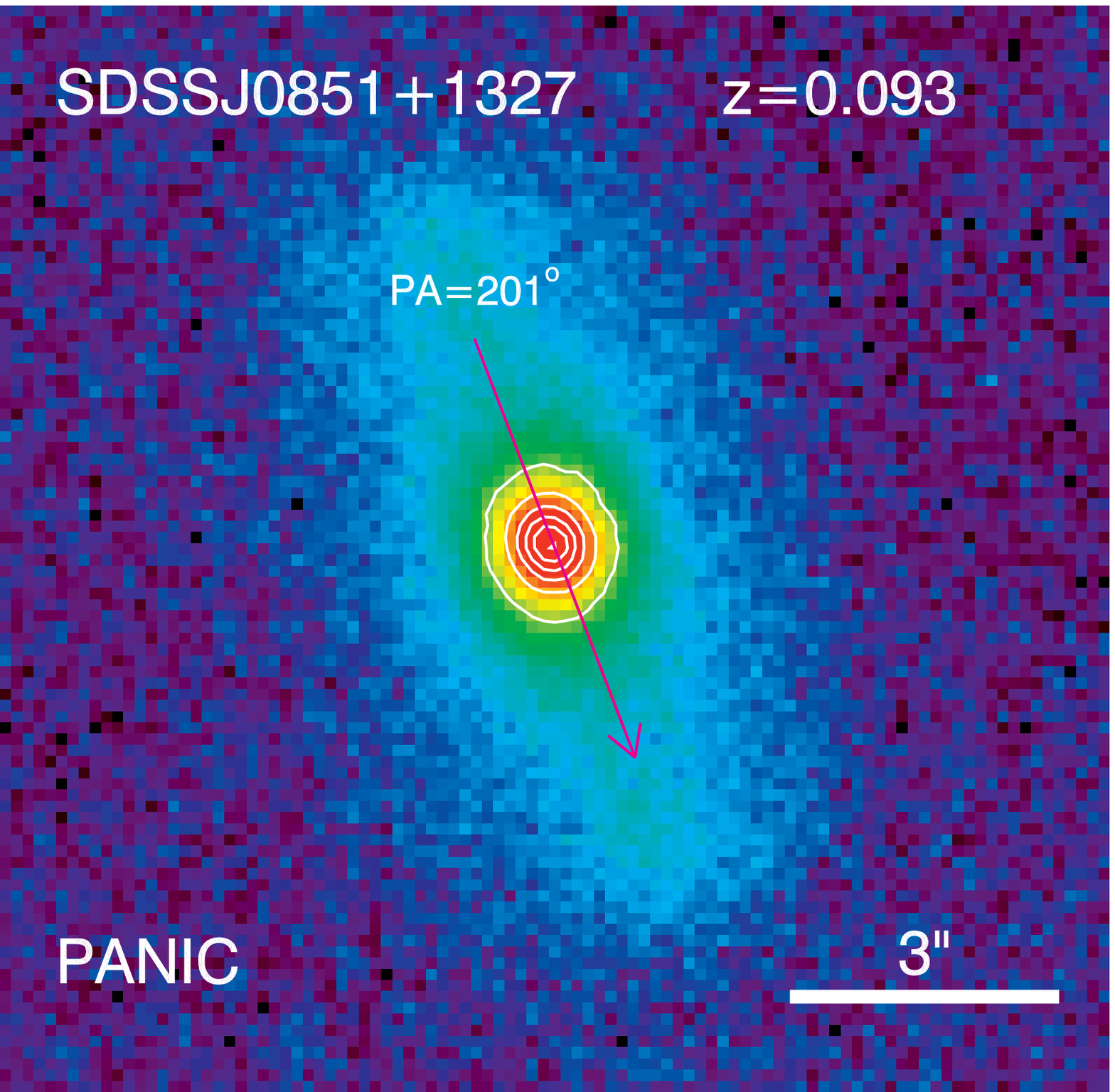}\vspace{9pt}
    \includegraphics[width=0.9\textwidth]{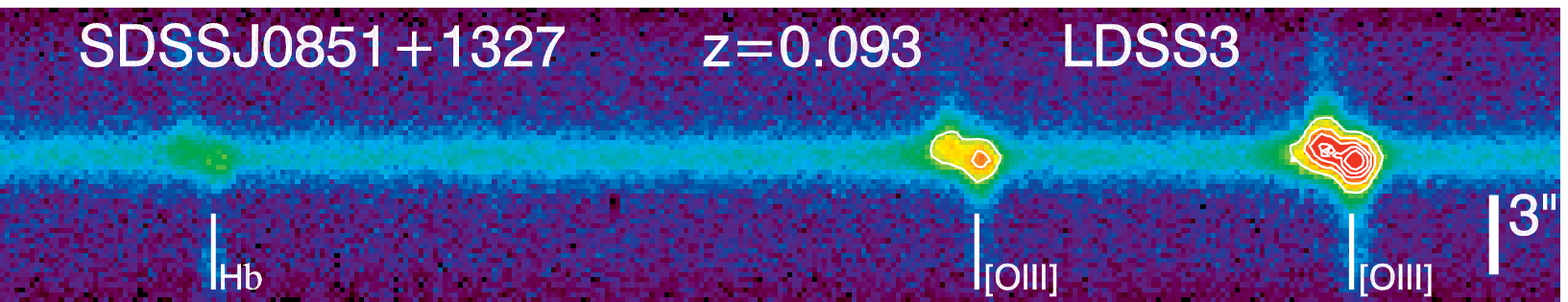}
    \caption{SDSS J0851+1327 (NLR kinematics). {\em Upper:} PANIC NIR image in $K_s$. This object has no resolved double nucleus. {\em Bottom:} LDSS3
    2d spectrum for the \hbeta-\OIII\ region with corresponding lines marked (note that the locations of these line marks are approximate). The two
    velocity components are spatially offset by $\sim 0.6$\arcsec. Notation is the same as Fig.\ \ref{fig:1108+0659}.}
    \label{fig:0851+1327}
\end{figure*}

\newpage
\begin{figure*}
  \centering
    \includegraphics[width=0.9\textwidth]{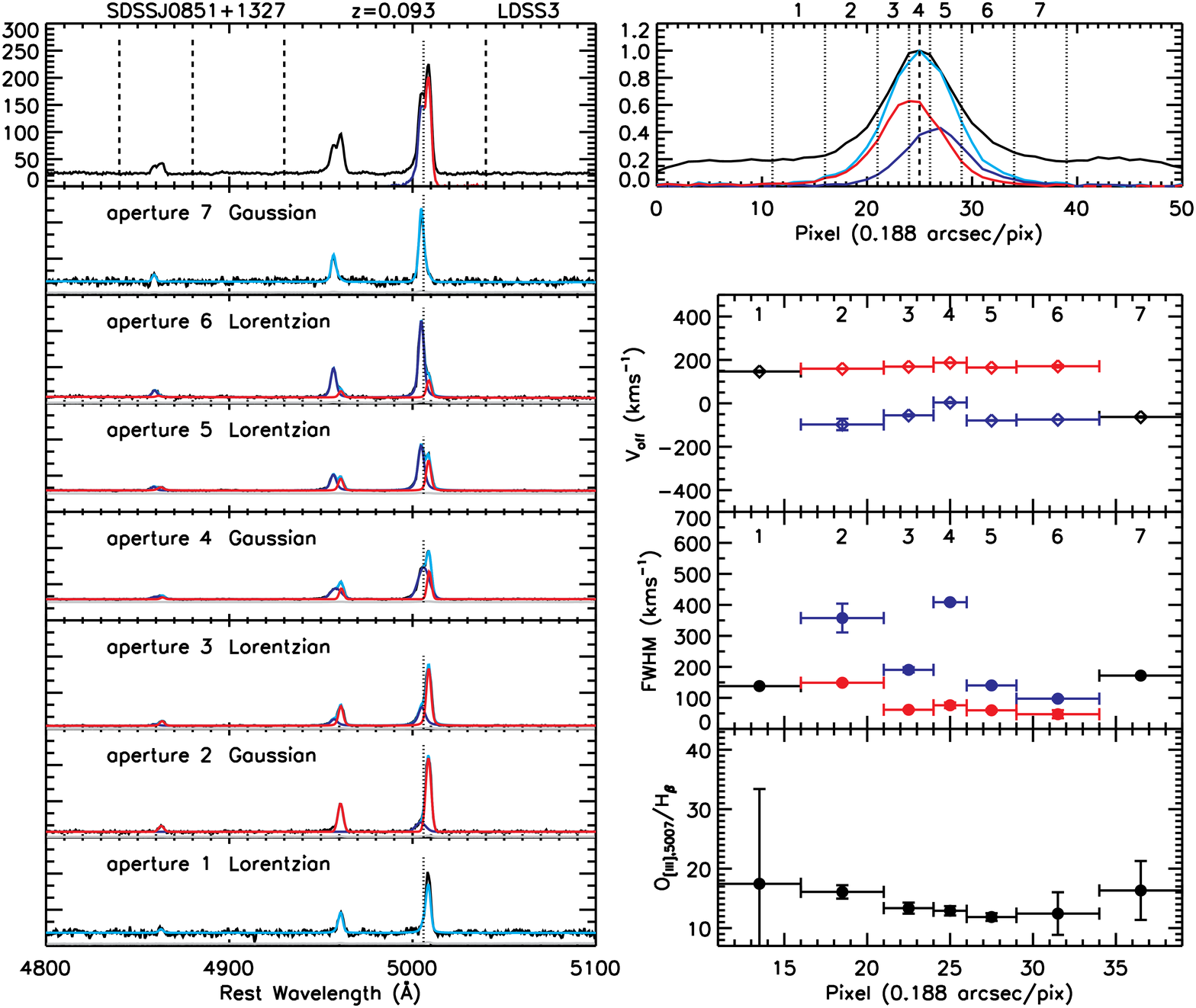}
    \caption{Diagnosis of the 2d spectrum of J0851+1327. Notation is the same as Fig.\ \ref{fig:1108+0659_diag}. This object has a giant disk
    component seen in the optical image and 2d spectrum, which is also seen in the extended distribution of the continuum emission in the upper-right
    panel. No obvious trends are seen for the velocity offset and line with of the two \OIII\ components. The \OIII/\hbeta\ flux is constant within
    $\sim 0.1$ dex.}
    \label{fig:0851+1327_diag}
\end{figure*}

\begin{figure*}
  \centering
    \includegraphics[width=0.6\textwidth]{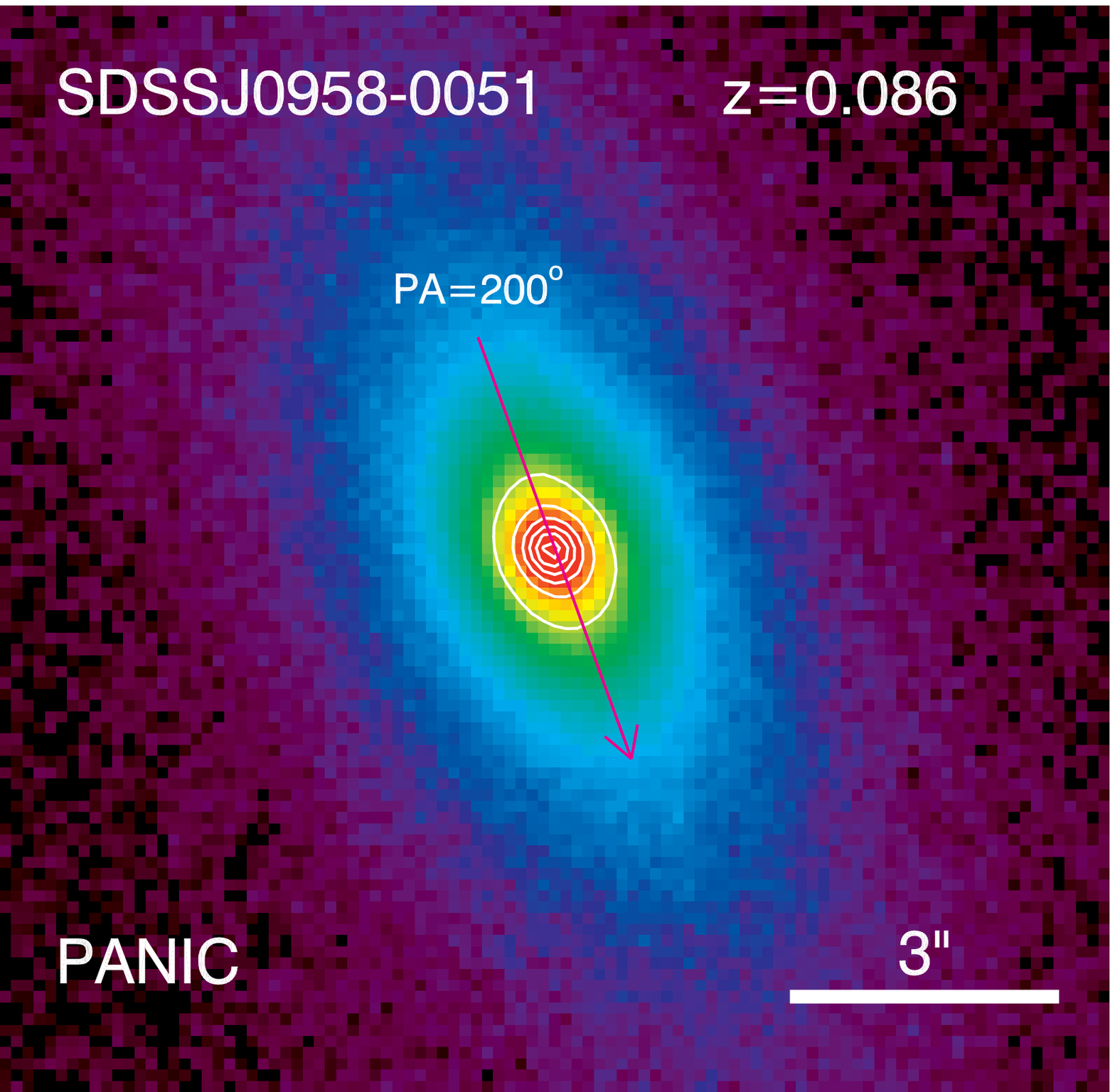}\vspace{9pt}
    \includegraphics[width=0.9\textwidth]{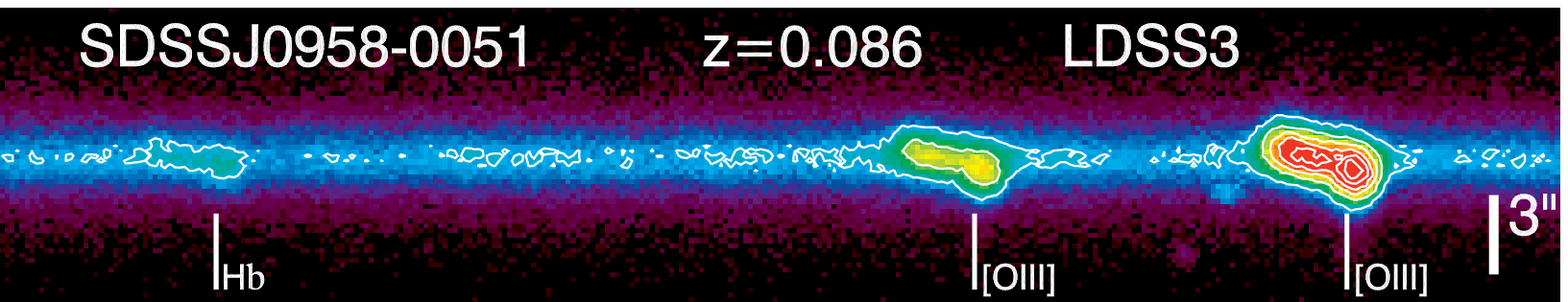}
    \caption{SDSS J0958$-$0051 (NLR kinematics). {\em Upper:} PANIC NIR image in $K_s$. This object has no resolved double nucleus, and it has a
    large
    disk component. {\em Bottom:} LDSS3 2d spectrum for the \hbeta-\OIII\ region with corresponding lines marked (note that the locations of these
    line marks are approximate). The two velocity components are spatially offset by $\sim 0.8$\arcsec. There are clear velocity gradients in the
    line
    emission. Notation is the same as Fig.\ \ref{fig:1108+0659}.}
    \label{fig:0958-0051}
\end{figure*}

\newpage
\begin{figure*}
  \centering
    \includegraphics[width=0.9\textwidth]{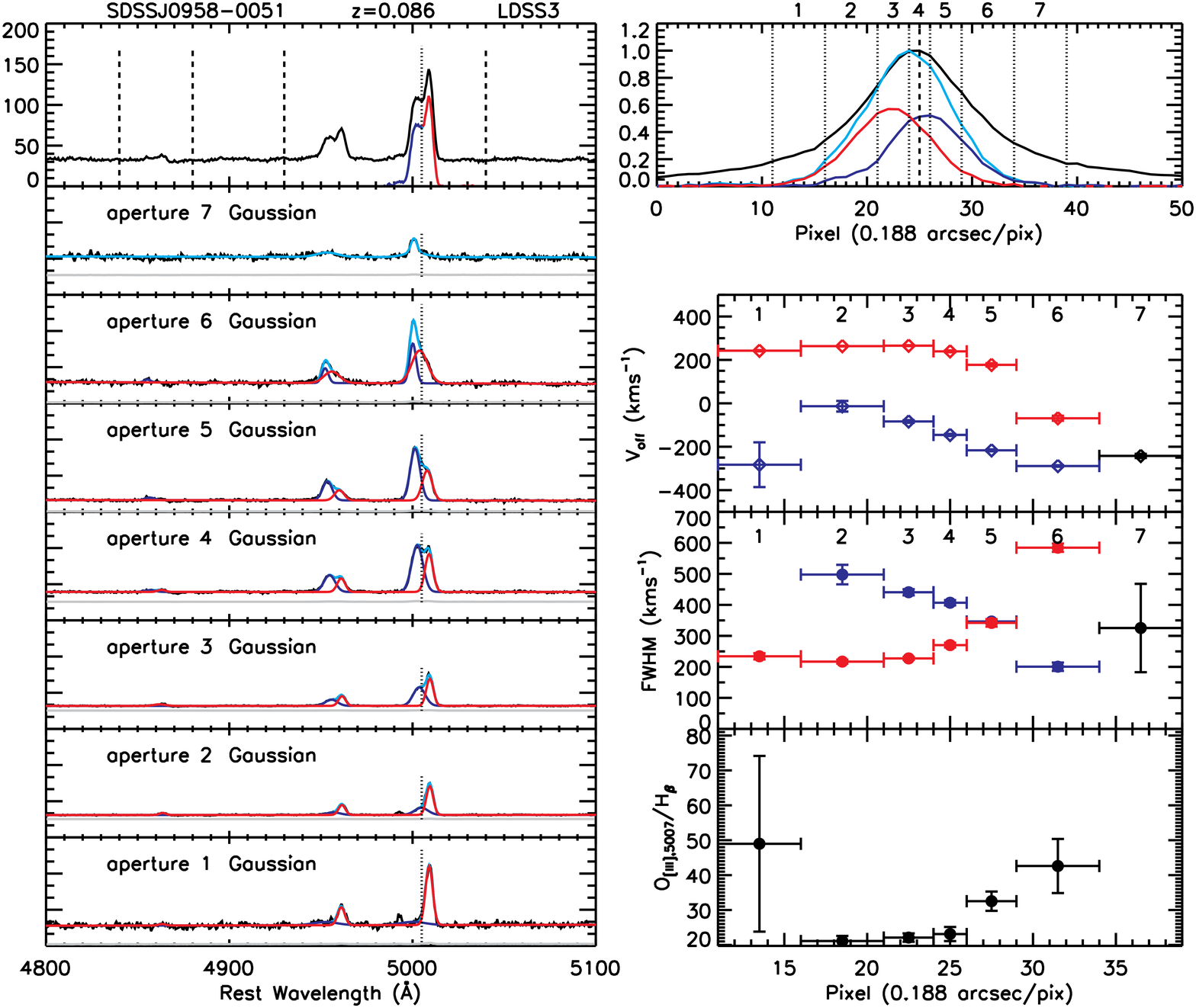}
    \caption{Diagnosis of the 2d spectrum. Notation is the same as Fig.\ \ref{fig:1108+0659_diag}. Velocity gradients can been seen for both the
    blueshifted and redshifted components, which is indicative of rotation. The FWHM of the two velocity components show peculiar patterns, which do
    not seem to be caused by failed deblending, and is suggestive of more complicated kinematics in this system. The \OIII/\hbeta\ flux ratio seems
    to
    rise towards the outermost bins, which may by caused by density stratification. }
    \label{fig:0958-0051_diag}
\end{figure*}

\begin{figure*}
  \centering
    \includegraphics[width=0.6\textwidth]{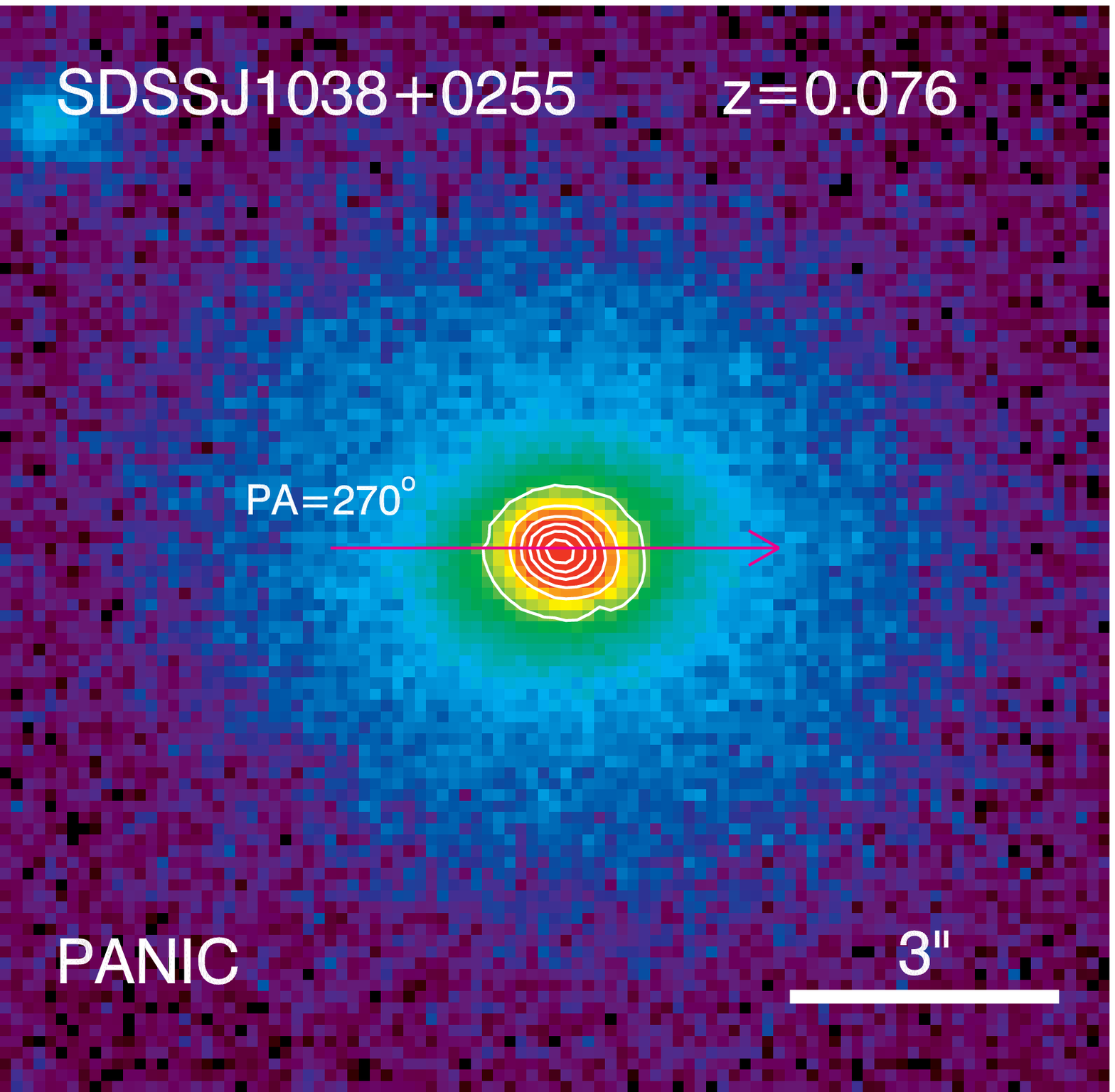}\vspace{9pt}
    \includegraphics[width=0.9\textwidth]{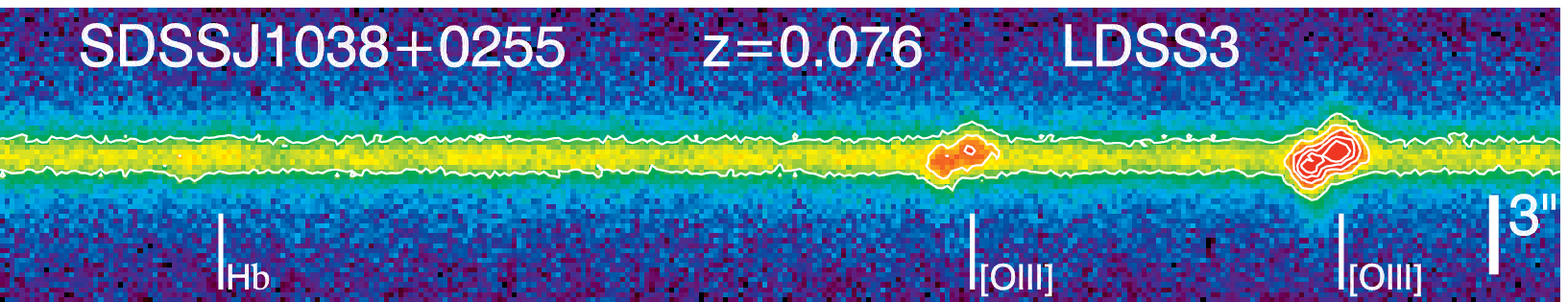}
    \caption{SDSS J1038+0255 (NLR kinematics). {\em Upper:} PANIC NIR image in $K_s$. This object has no resolved double nucleus. {\em Bottom:} LDSS3
    2d spectrum for the \hbeta-\OIII\ region with corresponding lines marked (note that the locations of these line marks are approximate). The two
    velocity components are spatially offset by $\sim 0.9$\arcsec. Notation is the same as Fig.\ \ref{fig:1108+0659}.}
    \label{fig:1038+0255}
\end{figure*}

\newpage
\begin{figure*}
  \centering
    \includegraphics[width=0.9\textwidth]{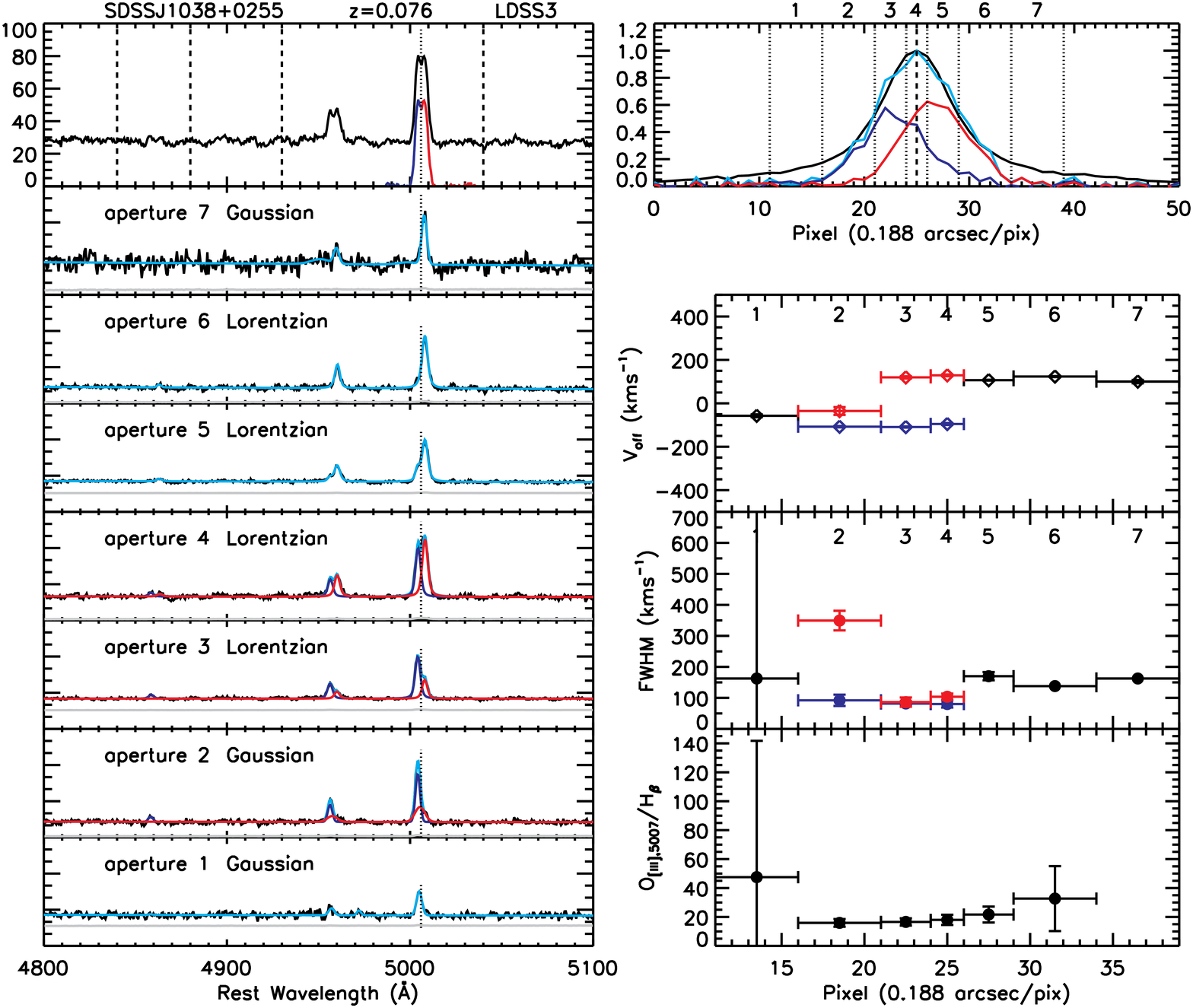}
    \caption{Diagnosis of the 2d spectrum of J1038+0255. Notation is the same as Fig.\ \ref{fig:1108+0659_diag}. Deblending of the two velocity
    components is not successful in over half the spatial bins.}
    \label{fig:1038+0255_diag}
\end{figure*}

\begin{figure*}
  \centering
    \includegraphics[width=0.6\textwidth]{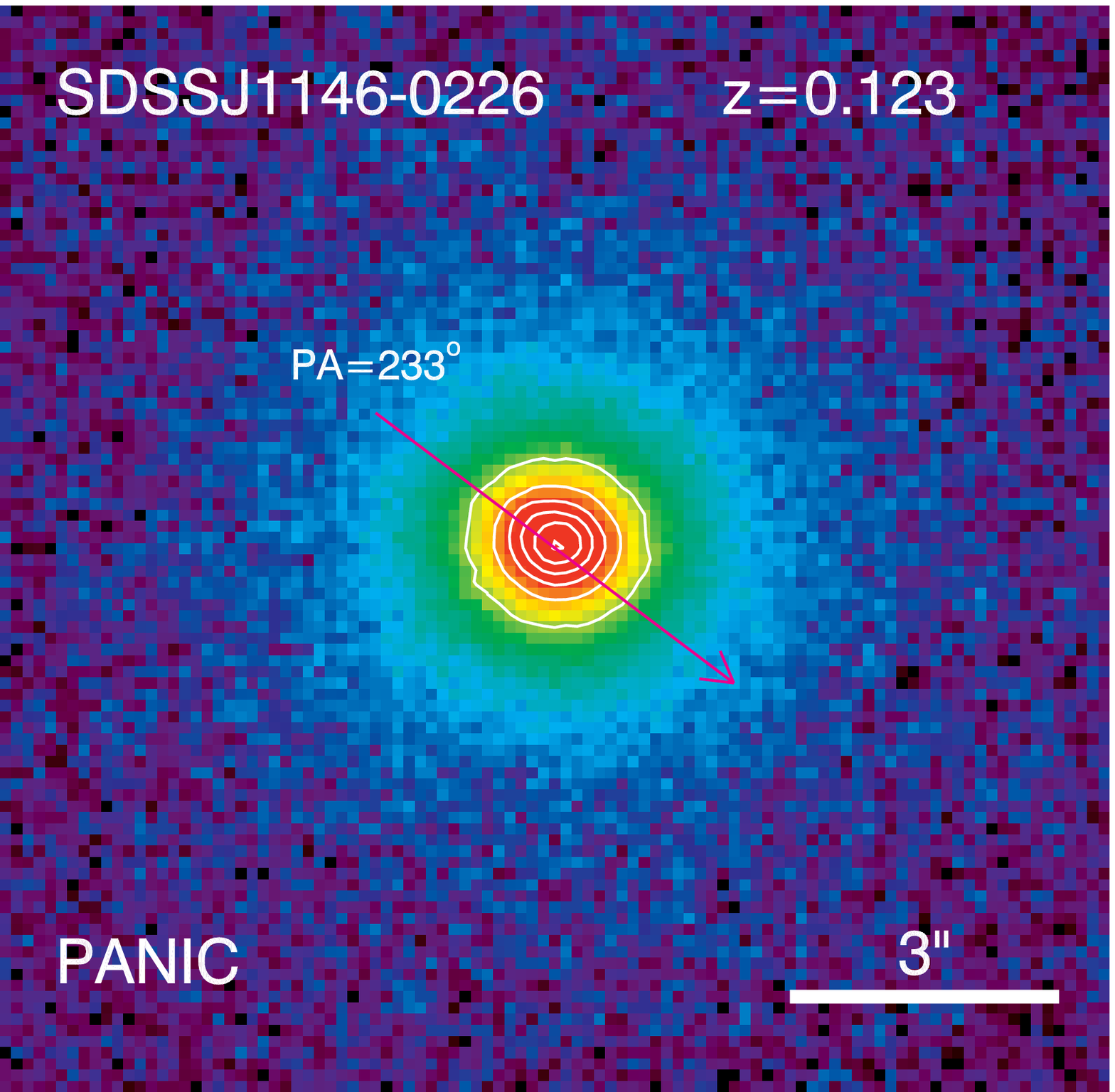}\vspace{9pt}
    \includegraphics[width=0.9\textwidth]{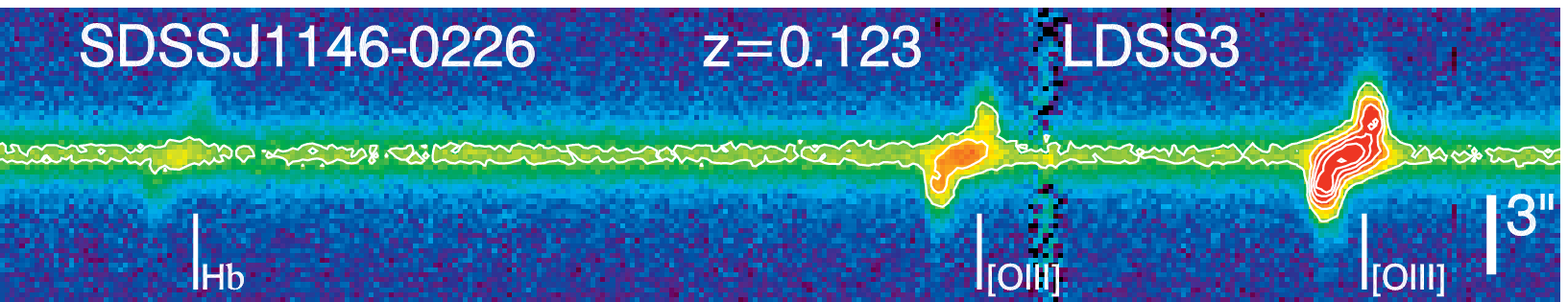}
    \caption{SDSS J1146$-$0226 (NLR kinematics). {\em Upper:} PANIC NIR image in $K_s$. This object has no resolved double nucleus. {\em Bottom:}
    LDSS3 2d spectrum for the \hbeta-\OIII\ region with corresponding lines marked (note that the locations of these line marks are approximate).
    This
    object shows a clear rotation curve. The two velocity components are spatially offset by $\sim 1.3$\arcsec\ measured from the centroids of the
    blue- and red-shifted emission. Notation is the same as Fig.\ \ref{fig:1108+0659}.}
    \label{fig:1146-0226}
\end{figure*}

\newpage
\begin{figure*}
  \centering
    \includegraphics[width=0.9\textwidth]{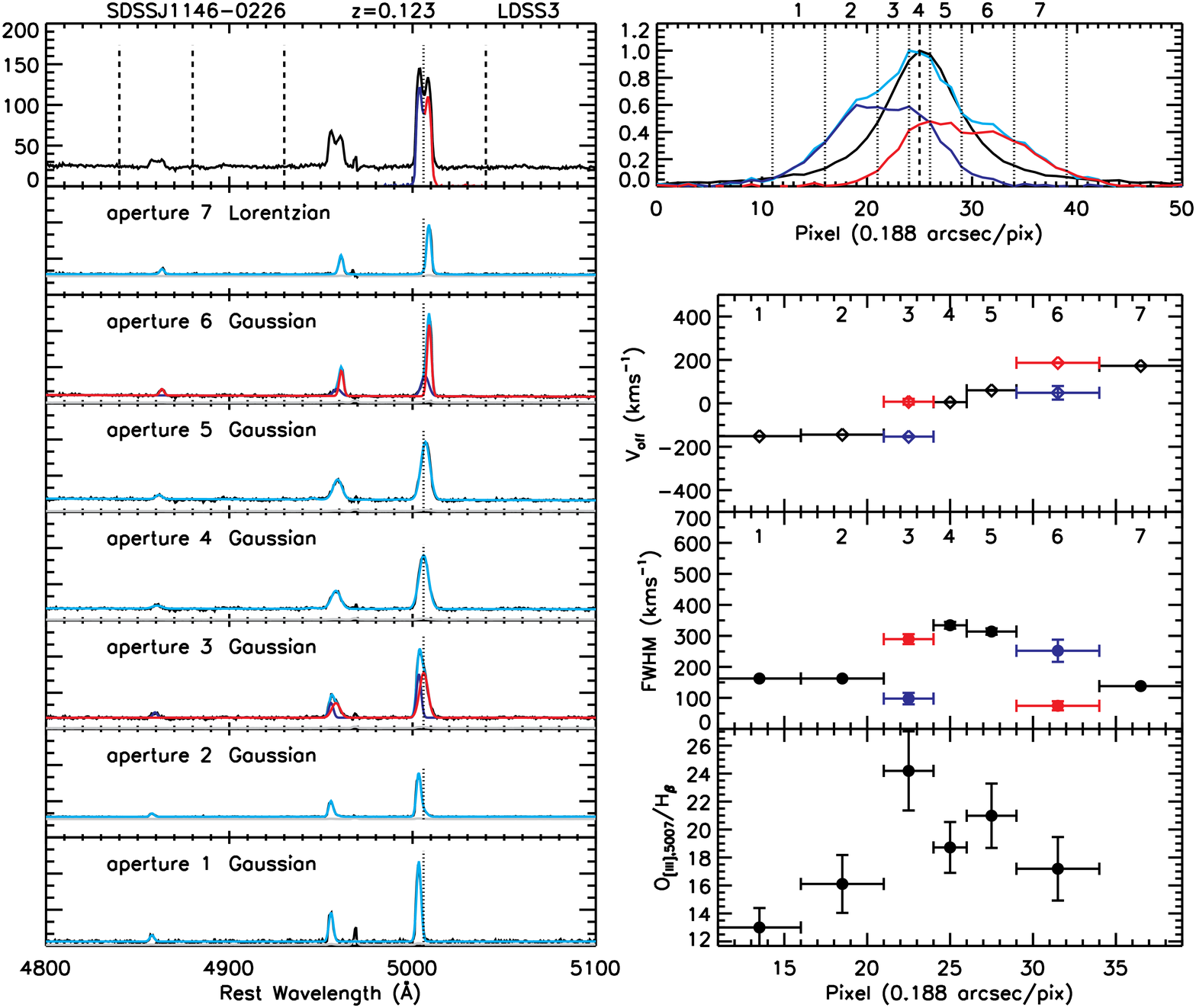}
    \caption{Diagnosis of the 2d spectrum of J1146$-$0226. Notation is the same as Fig.\ \ref{fig:1108+0659_diag}. Deblending is unsuccessful in the
    central bins as the velocity difference between the two components is too small there. This can be seen clearly in the 2d spectrum. The line
    width
    and the \OIII/\hbeta\ flux ratio seem to increase towards the center of the continuum, which is consistent with the ionizing source being at the
    center of the galaxy. }
    \label{fig:1146-0226_diag}
\end{figure*}

\begin{figure*}
  \centering
    \includegraphics[width=0.6\textwidth]{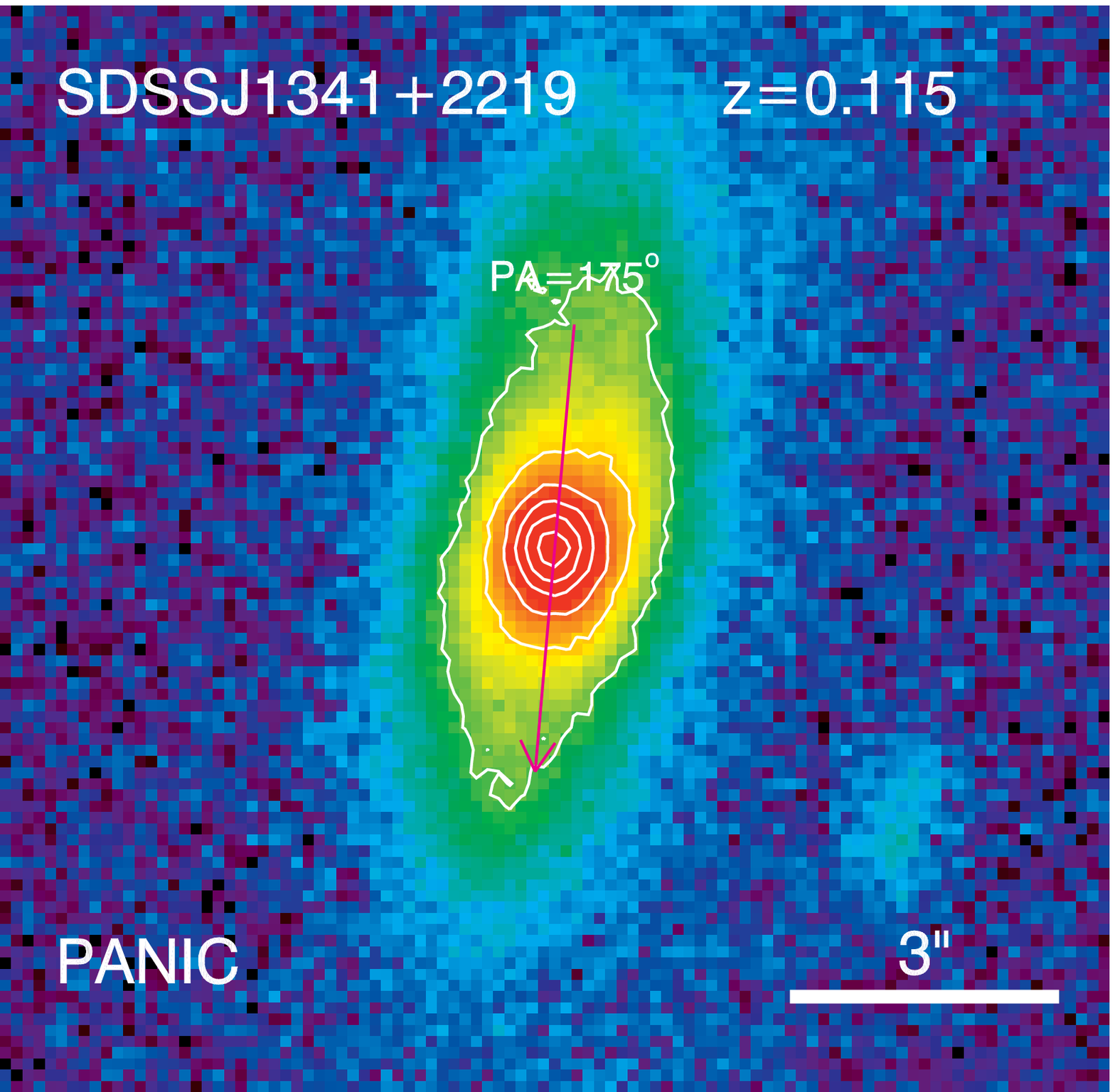}\vspace{9pt}
    \includegraphics[width=0.9\textwidth]{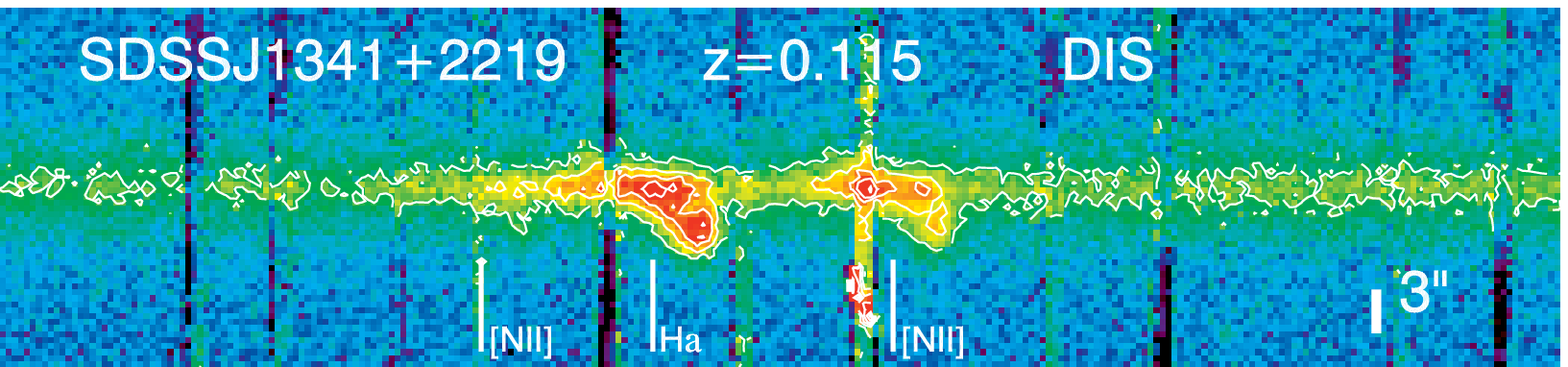}
    \caption{SDSS J1341+2219 (NLR kinematics). {\em Upper:} PANIC NIR image in $K_s$. This object has no resolved double nucleus, although it has a
    disturbed disk component. {\em Bottom:} DIS 2d spectrum for the \halpha\ region with corresponding lines marked (note that the locations of these
    line marks are approximate). The DIS spectrum for the \hbeta-\OIII\ region has poor quality. The two velocity components are spatially offset by
    $\sim 1.7$\arcsec\ measured from the peaks of the blueshifted and redshifted \halpha\ emission. Notation is the same as Fig.\
    \ref{fig:1108+0659}.}
    \label{fig:1341+2219}
\end{figure*}

\clearpage
\begin{figure*}
  \centering
    \includegraphics[width=0.6\textwidth]{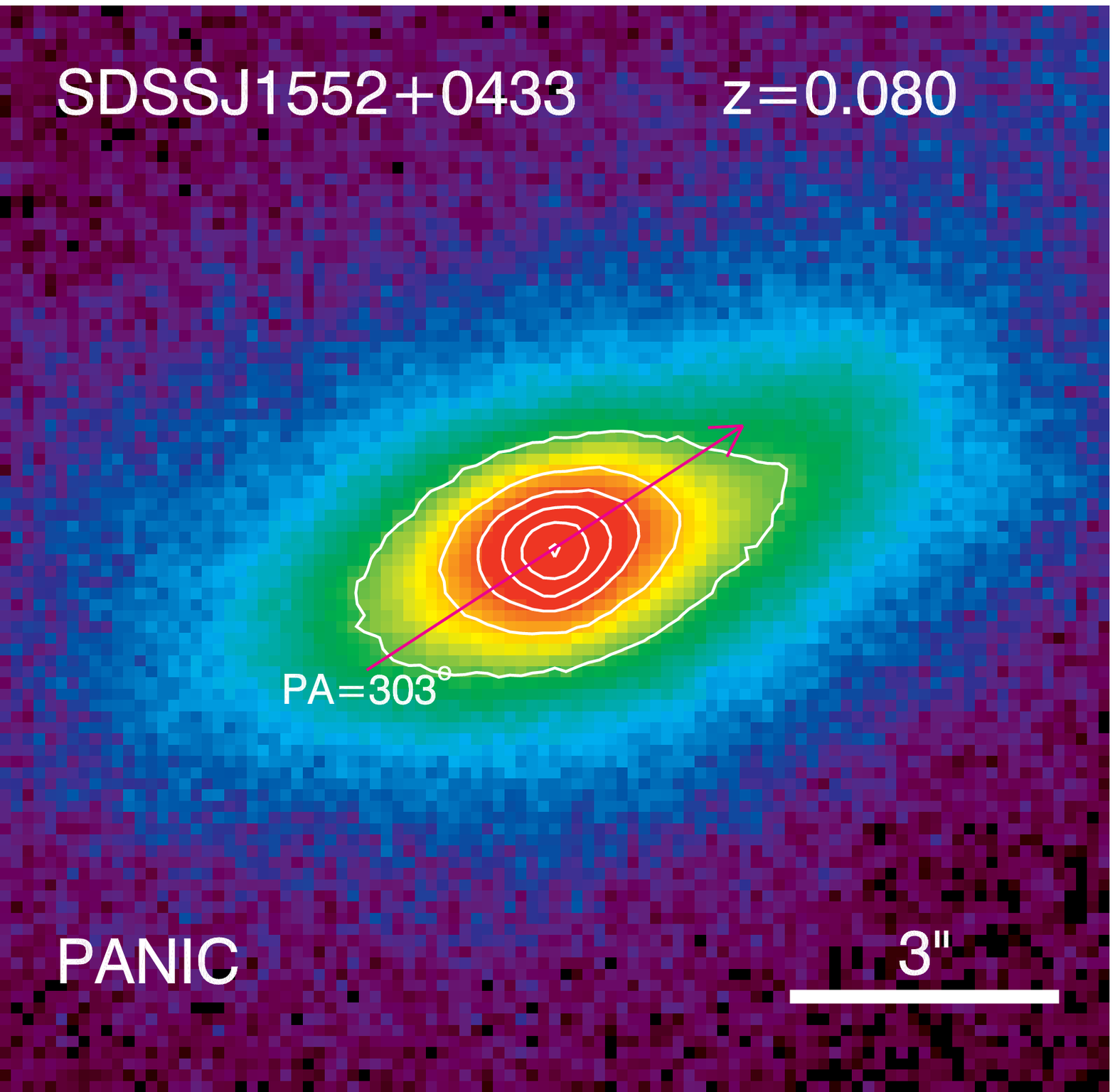}\vspace{9pt}
    \includegraphics[width=0.9\textwidth]{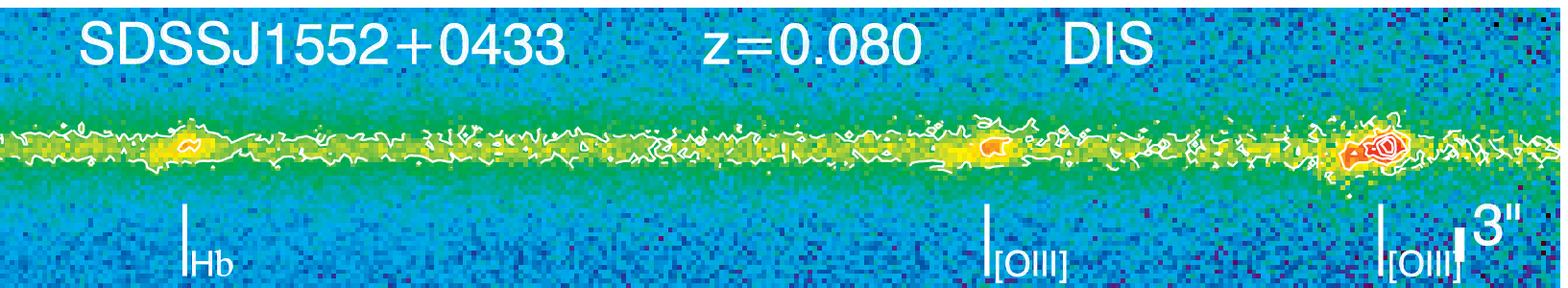}
  \caption{SDSS J1552+0433 (NLR kinematics). {\em Upper:} PANIC NIR image in $K_s$. This object has no resolved double nucleus, although it has a
  disturbed disk component in the optical and in the NIR. {\em Bottom:} DIS 2d spectrum for the \hbeta-\OIII\ region with corresponding lines marked
  (note that the locations of these line marks are approximate). The two velocity components are spatially offset by $\sim 1.2$\arcsec. Notation is
  the same as Fig.\ \ref{fig:1108+0659}.}
    \label{fig:1552+0433}
\end{figure*}

\begin{figure*}
  \centering
    \includegraphics[width=0.6\textwidth]{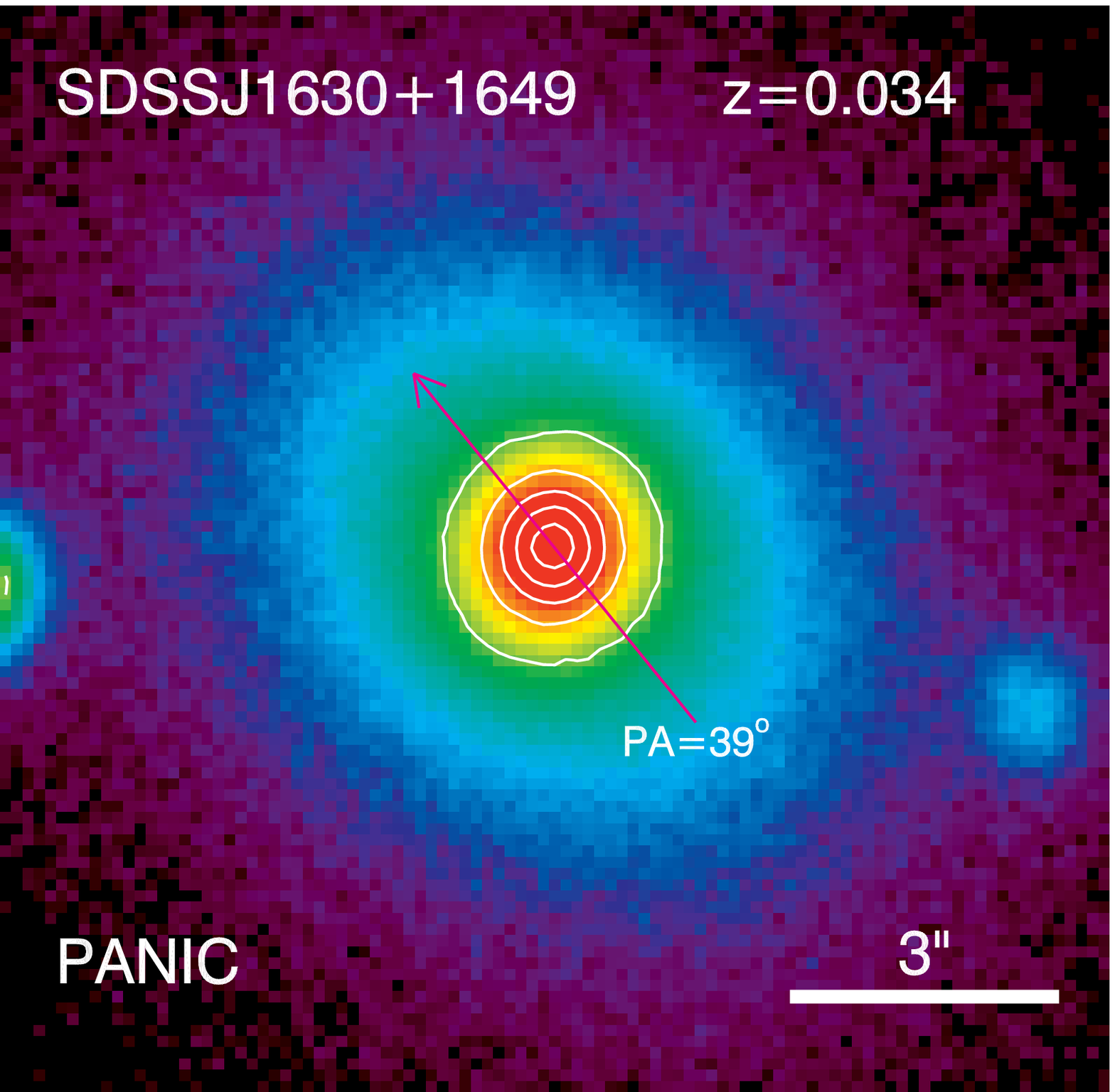}\vspace{9pt}
    \includegraphics[width=0.9\textwidth]{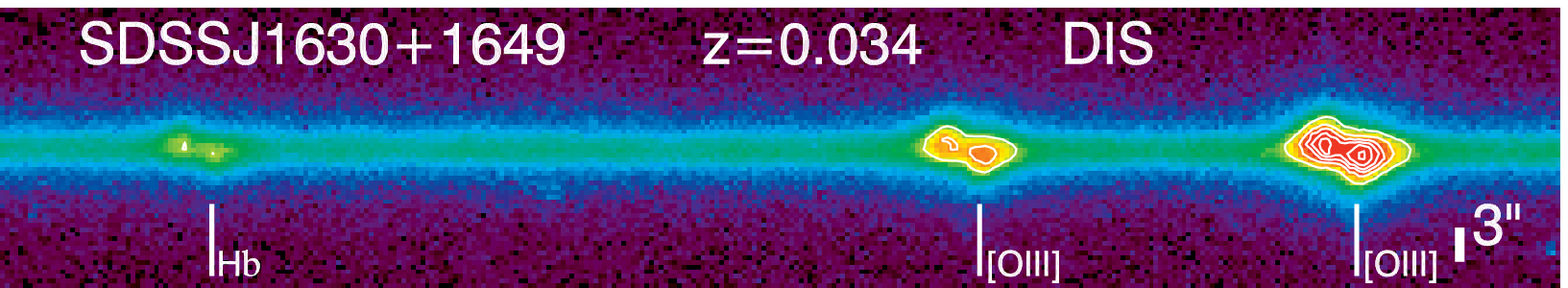}\vspace{9pt}
    \includegraphics[width=0.9\textwidth]{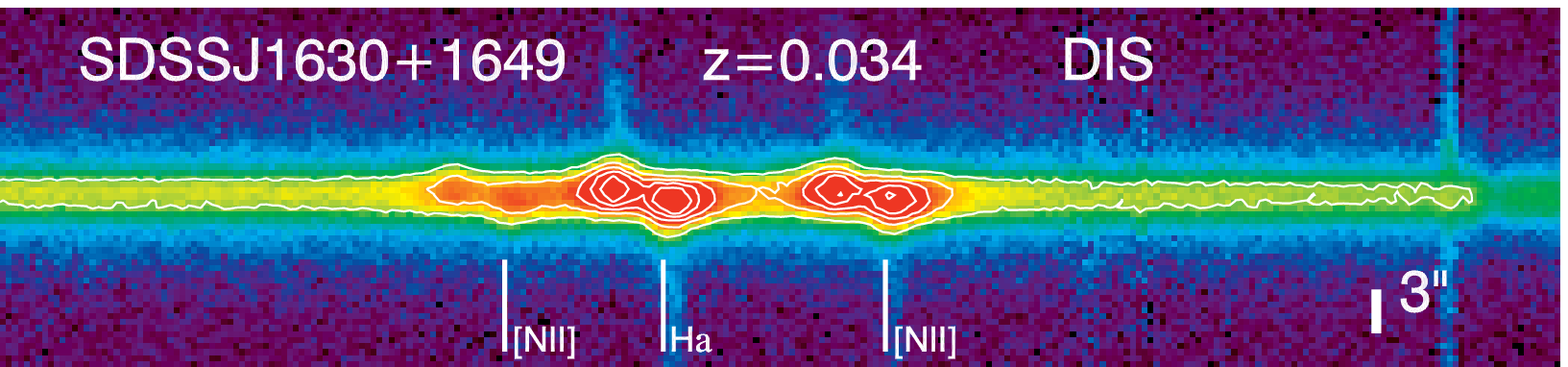}
    \caption{SDSS J1630+1649 (NLR kinematics). {\em Upper:} PANIC NIR image in $K_s$. This object has no resolved double nucleus, and it has a faint
    companion $\sim 5$\arcsec\ away towards the east. {\em Middle and Bottom:} DIS 2d spectrum for the \hbeta-\OIII\ and \halpha\ regions with
    corresponding lines marked (note that the locations of these line marks are approximate). The two \OIII\ components are spatially offset by $\sim
    0.8$\arcsec. Notation is the same as Fig.\ \ref{fig:1108+0659}.}
    \label{fig:1630+1649}
\end{figure*}

\newpage
\begin{figure*}
  \centering
    \includegraphics[width=0.9\textwidth]{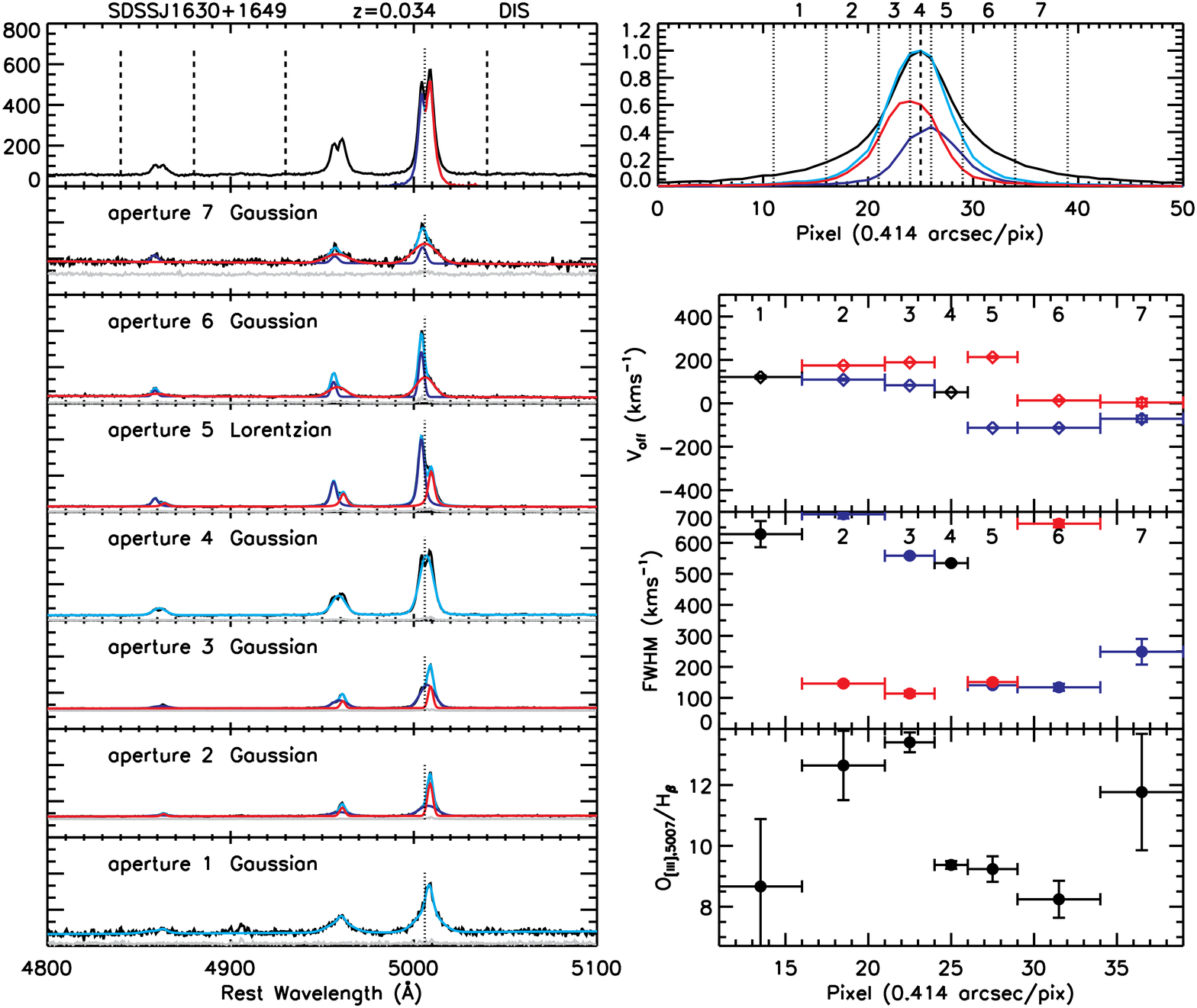}
    \caption{Diagnosis of the 2d spectrum of J1630+1649. Notation is the same as Fig.\ \ref{fig:1108+0659_diag}. Deblending of the two velocity
    components is ambiguous in most spatial bins. The \OIII/\hbeta\ flux ratio varies within $\sim 0.2$ dex.}
    \label{fig:1630+1649_diag}
\end{figure*}

\begin{figure*}
  \centering
    \includegraphics[width=0.6\textwidth]{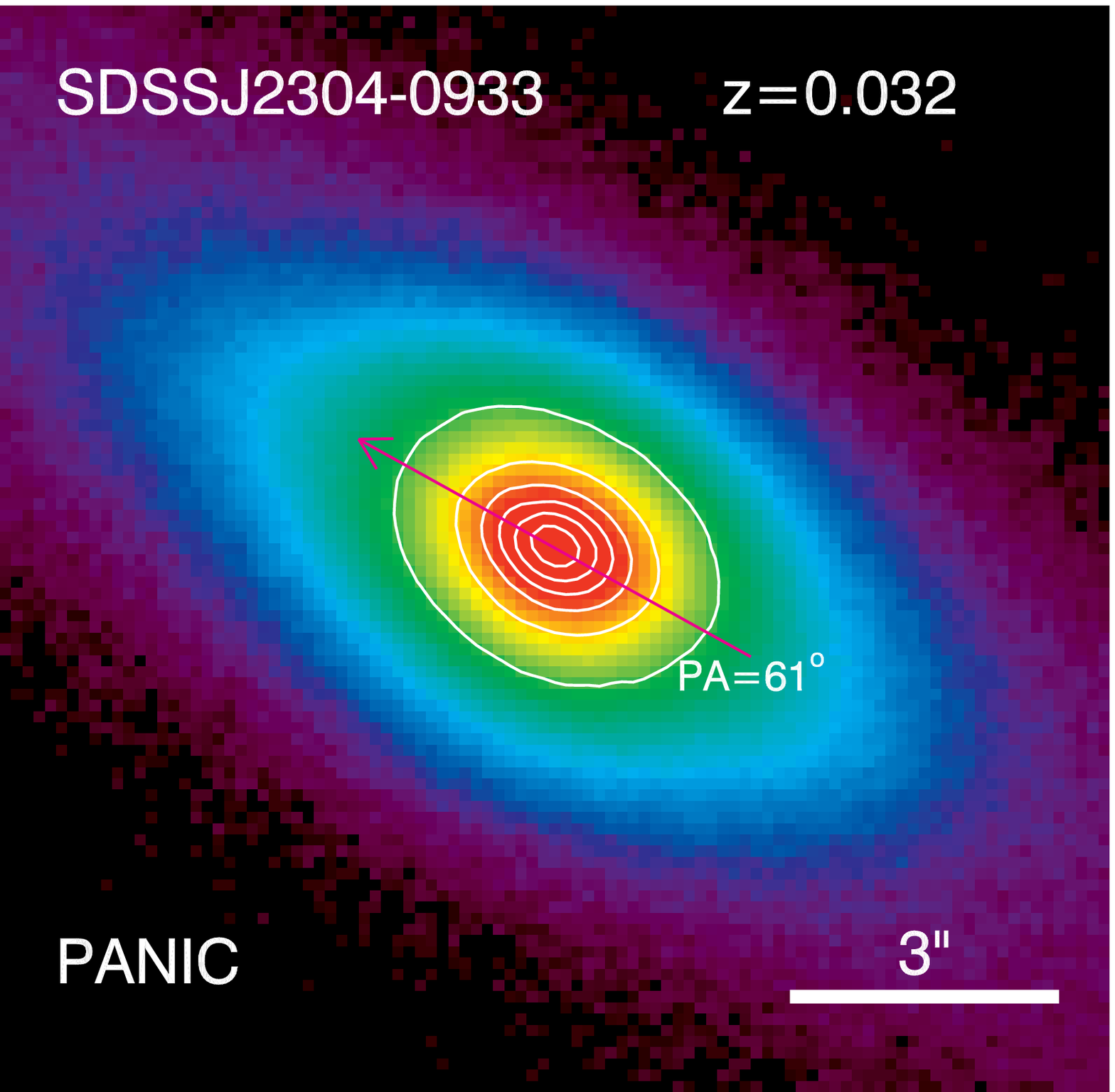}\vspace{9pt}
    \includegraphics[width=0.9\textwidth]{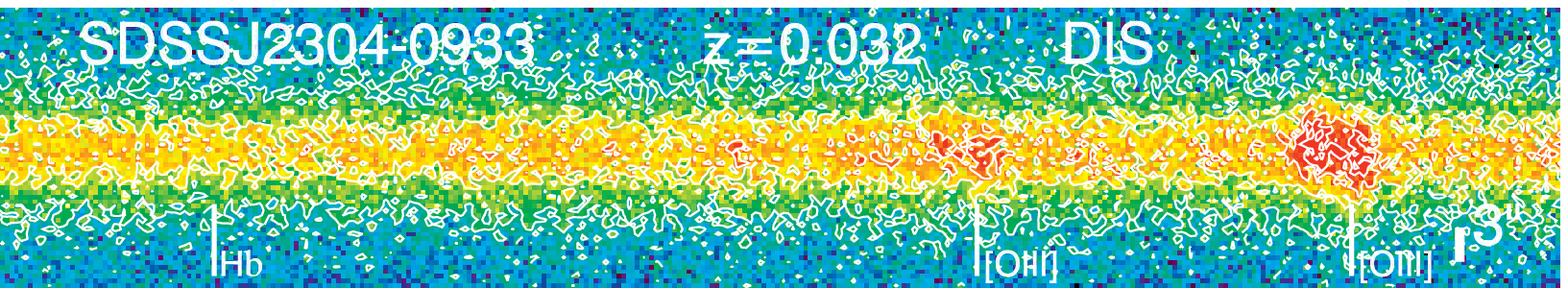}
    \caption{SDSS J2304$-$0933 (NLR kinematics). {\em Upper:} PANIC NIR image in $K_s$. This object has no resolved double nucleus, and a large disk
    component is apparent. {\em Bottom:} DIS 2d spectrum for the \hbeta-\OIII\ region with corresponding lines marked (note that the locations of
    these line marks are approximate). The two velocity components are spatially offset by $\sim 0.8$\arcsec. Notation is the same as Fig.\
    \ref{fig:1108+0659}.}
    \label{fig:2304-0933}
\end{figure*}


\clearpage
\begin{figure*}
  \centering
    \subfigure{
    \includegraphics[width=0.2\textwidth]{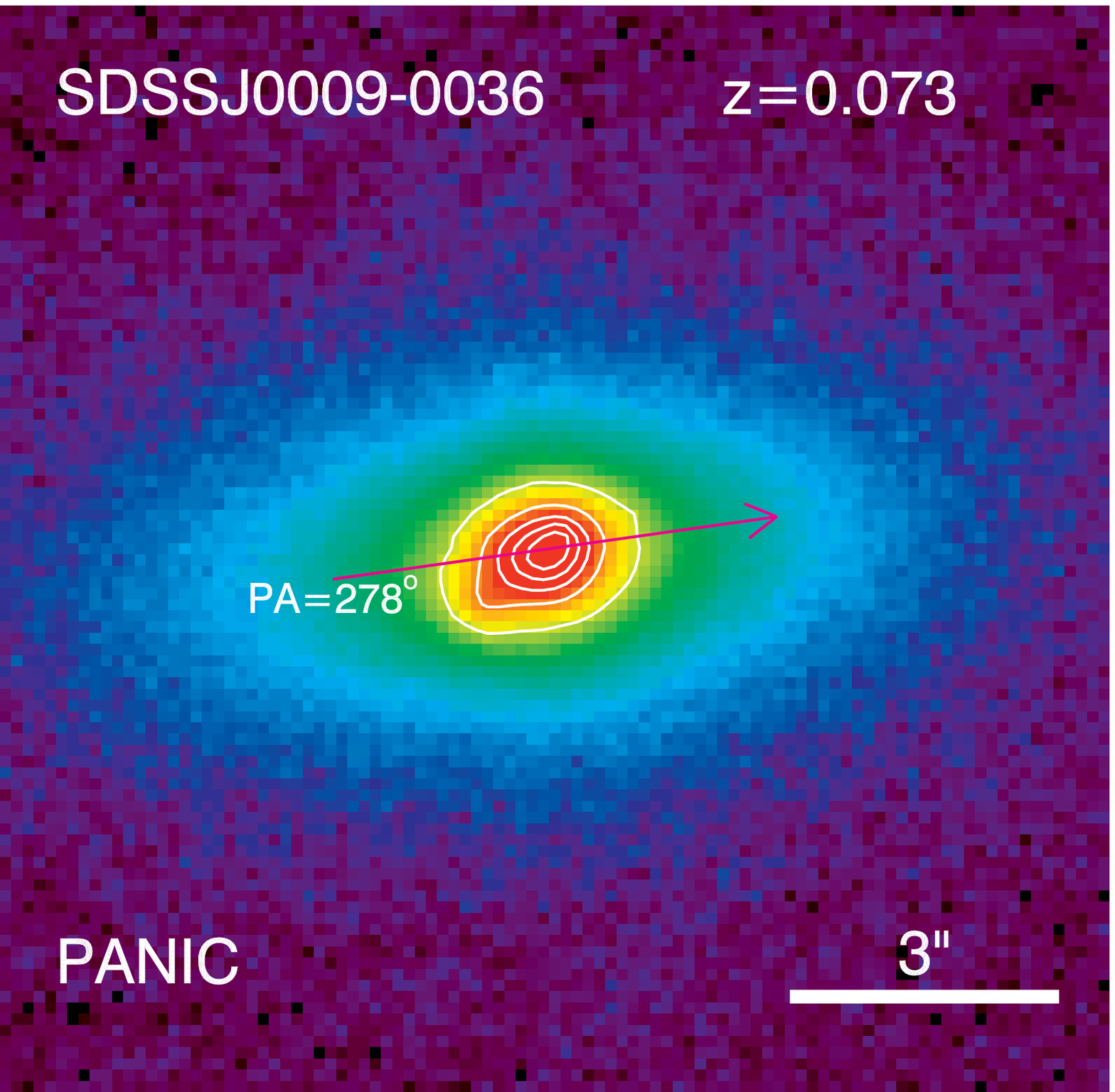}\hspace{9pt}
    \includegraphics[width=0.7\textwidth]{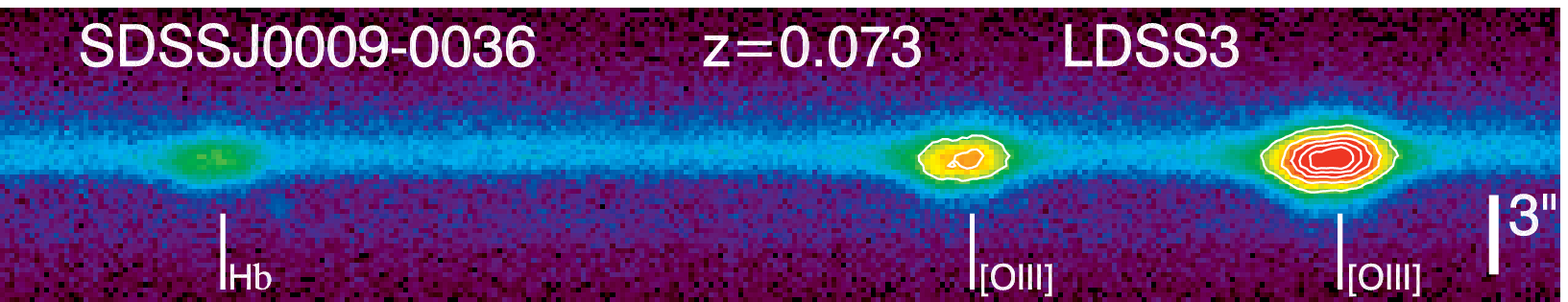}}
    \subfigure{
    \includegraphics[width=0.2\textwidth]{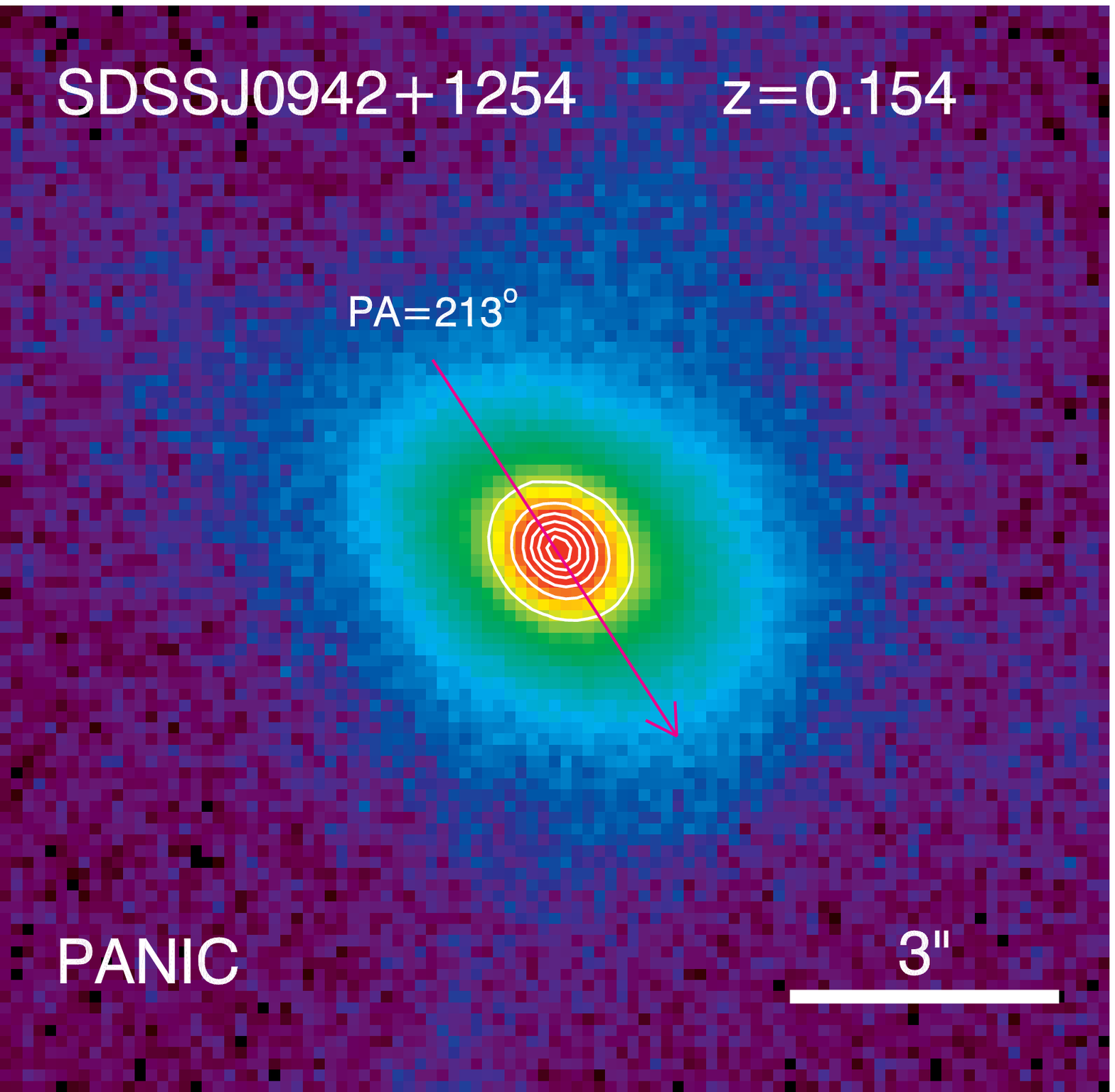}\hspace{9pt}
    \includegraphics[width=0.7\textwidth]{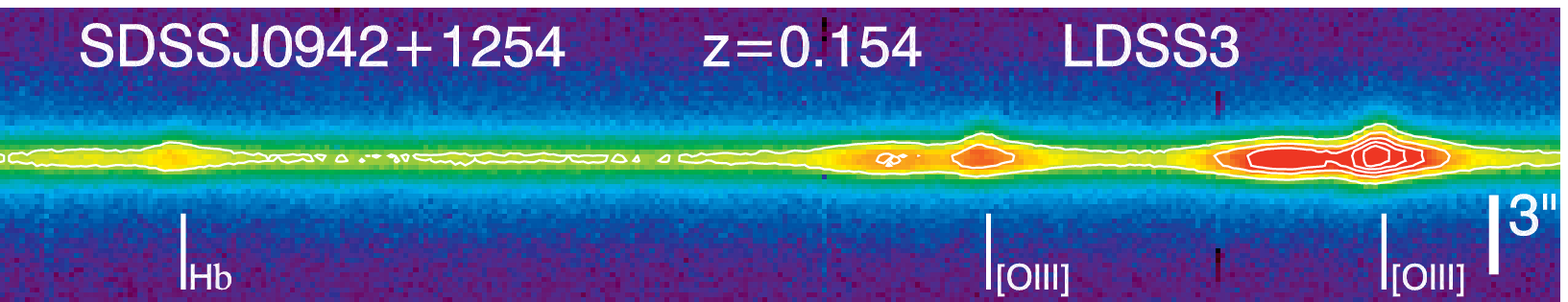}}
    \subfigure{
    \includegraphics[width=0.2\textwidth]{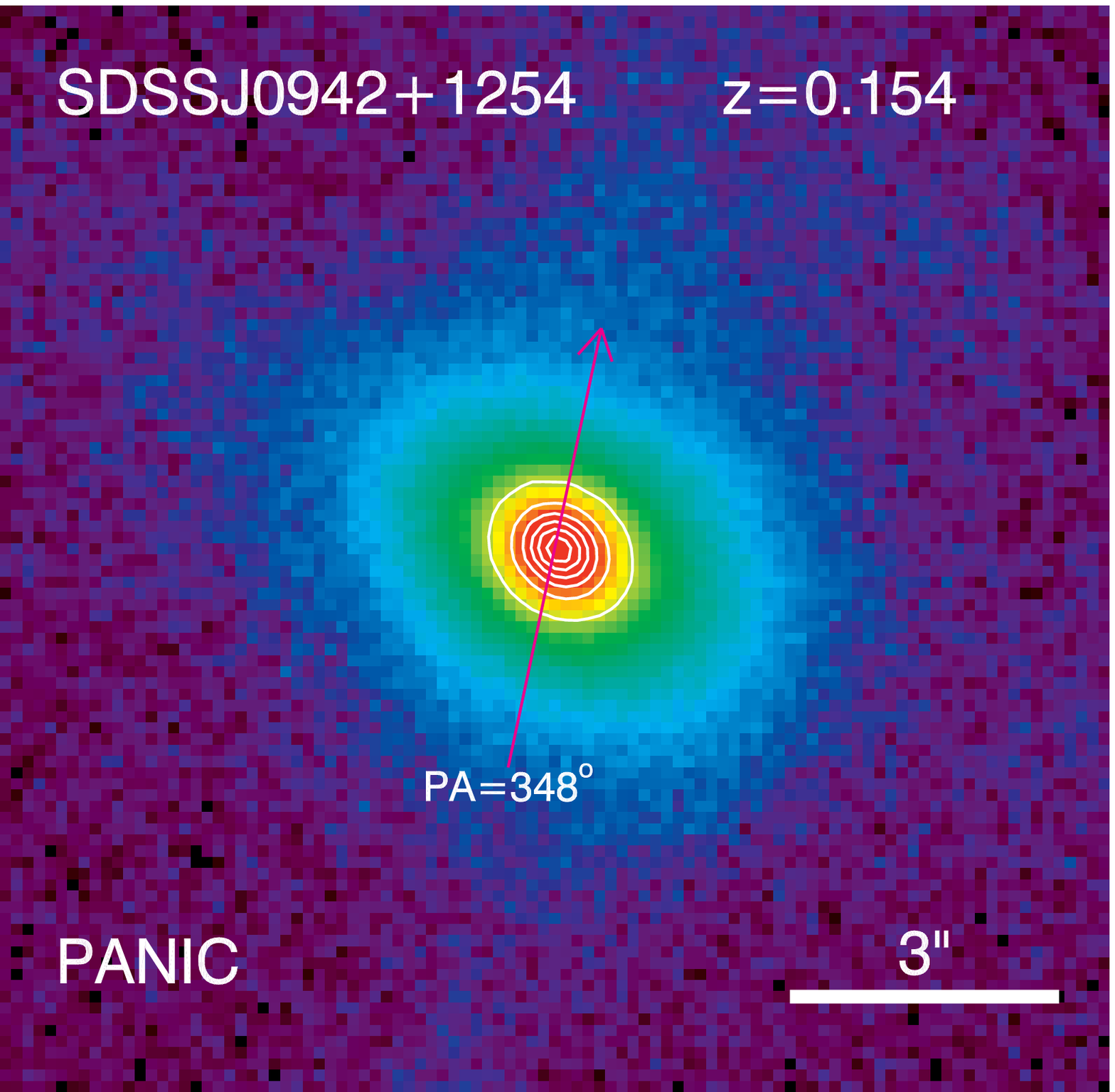}\hspace{9pt}
    \includegraphics[width=0.7\textwidth]{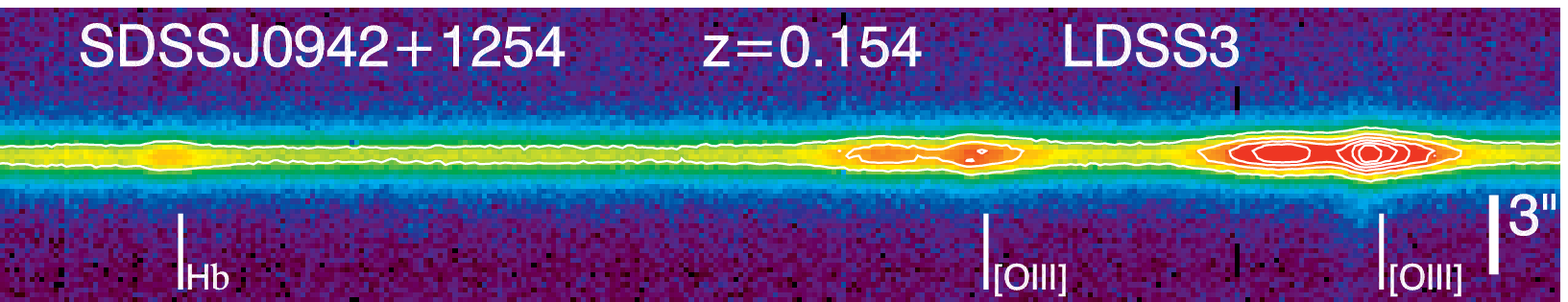}}
    \subfigure{
    \includegraphics[width=0.2\textwidth]{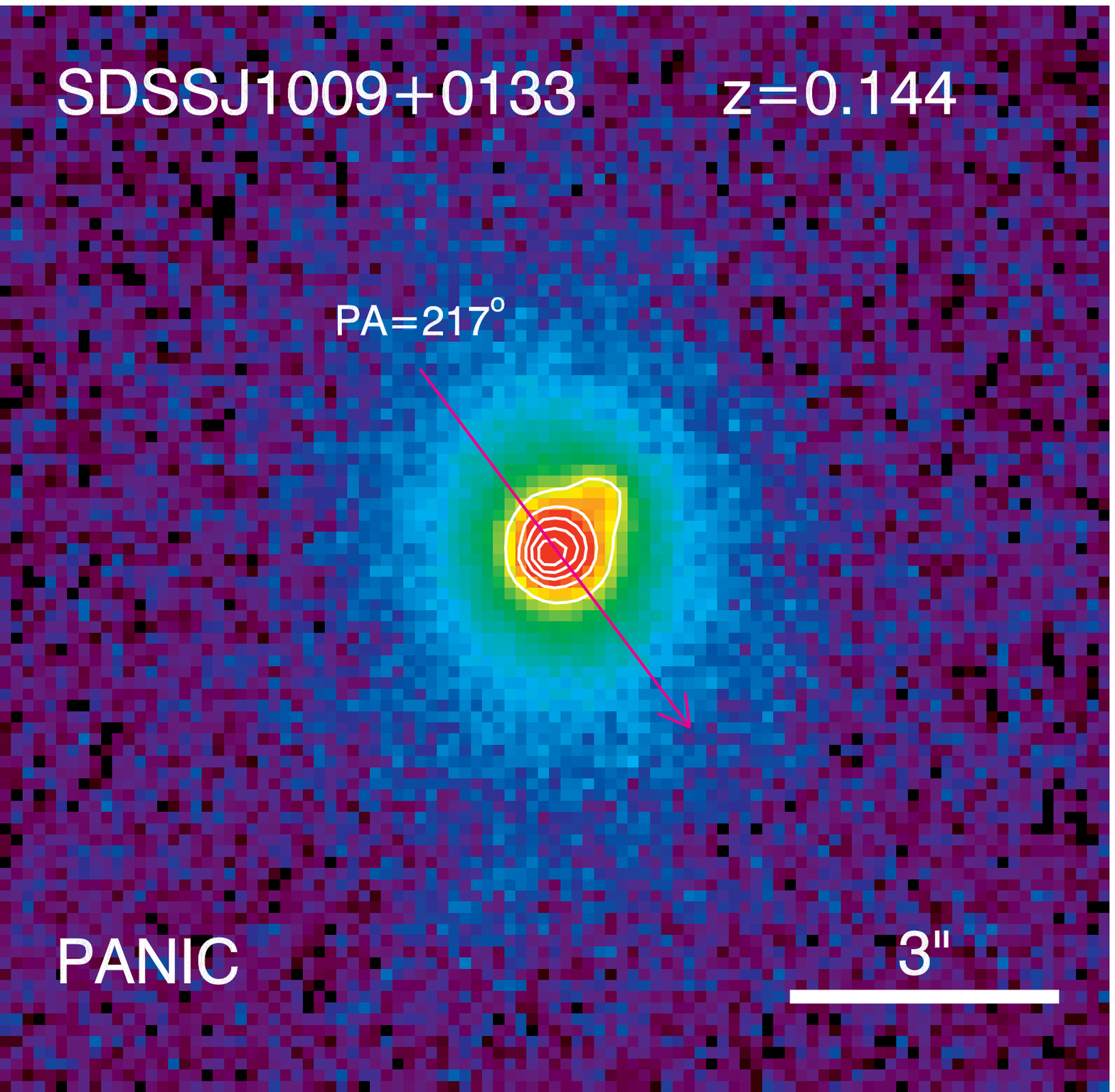}\hspace{9pt}
    \includegraphics[width=0.7\textwidth]{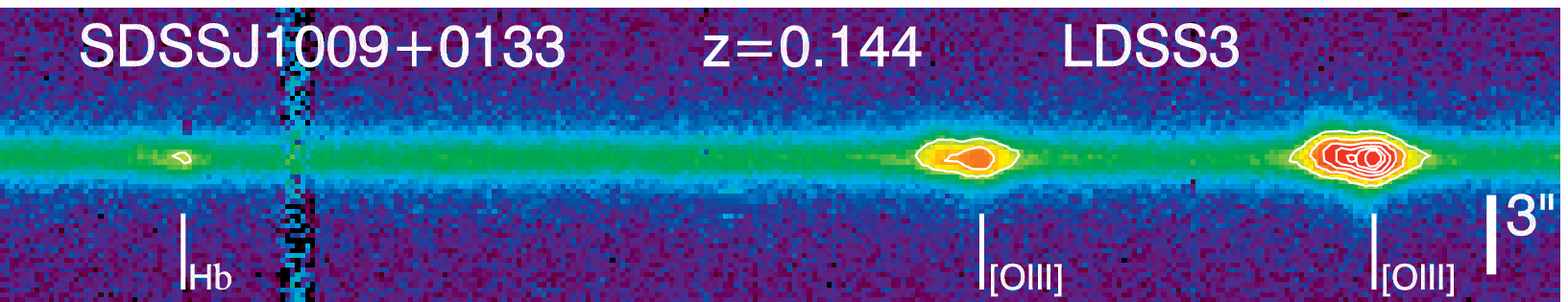}}
    \subfigure{
    \includegraphics[width=0.2\textwidth]{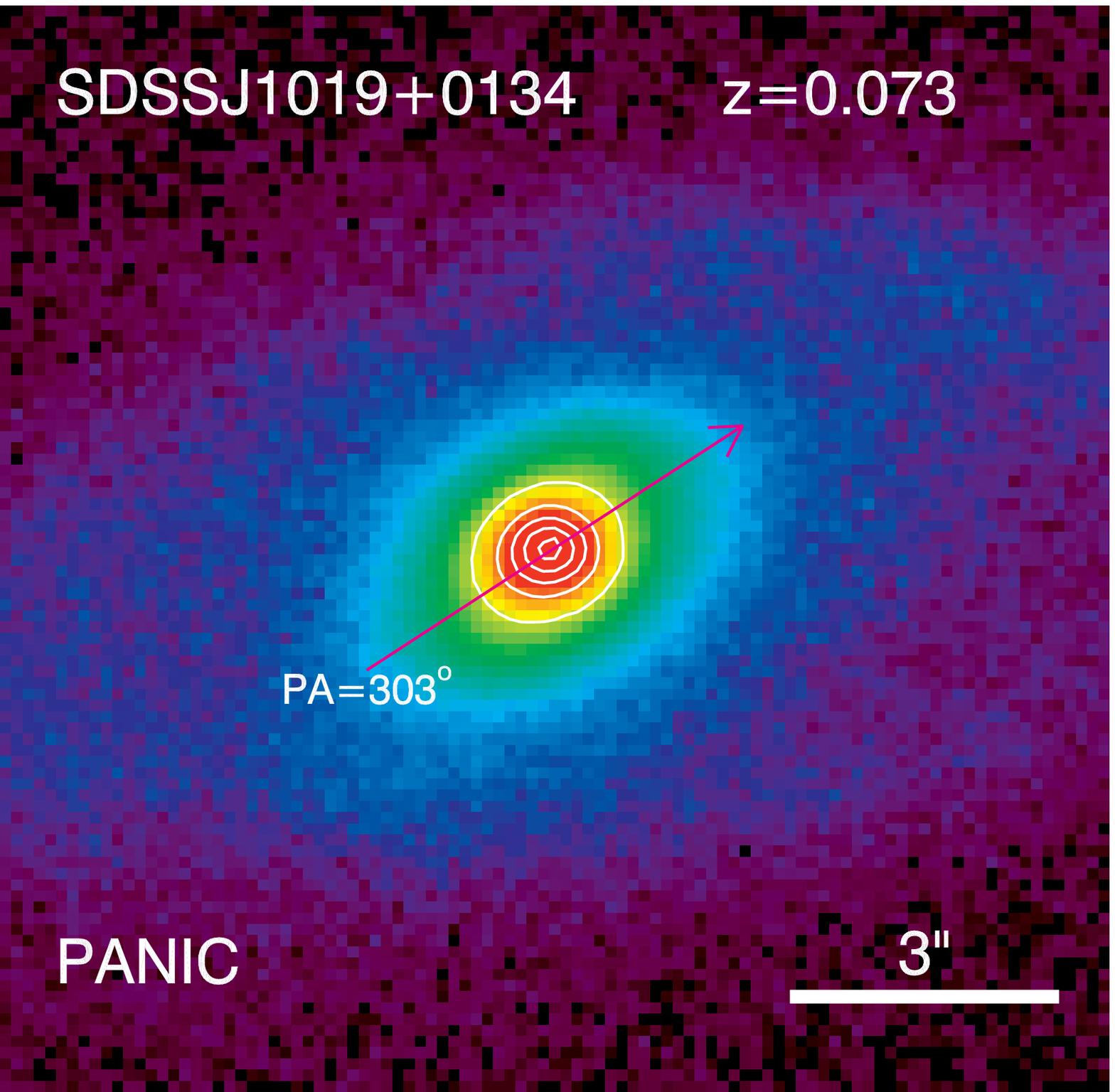}\hspace{9pt}
    \includegraphics[width=0.7\textwidth]{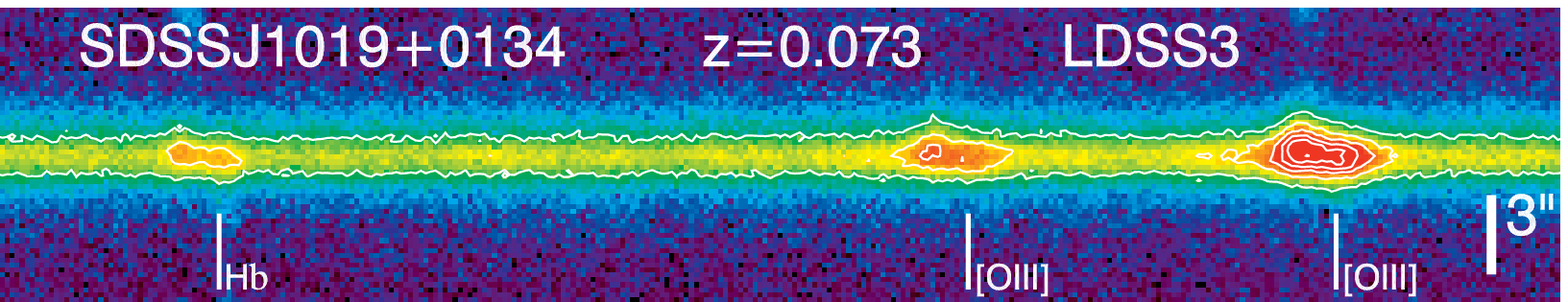}}
    \subfigure{
    \includegraphics[width=0.2\textwidth]{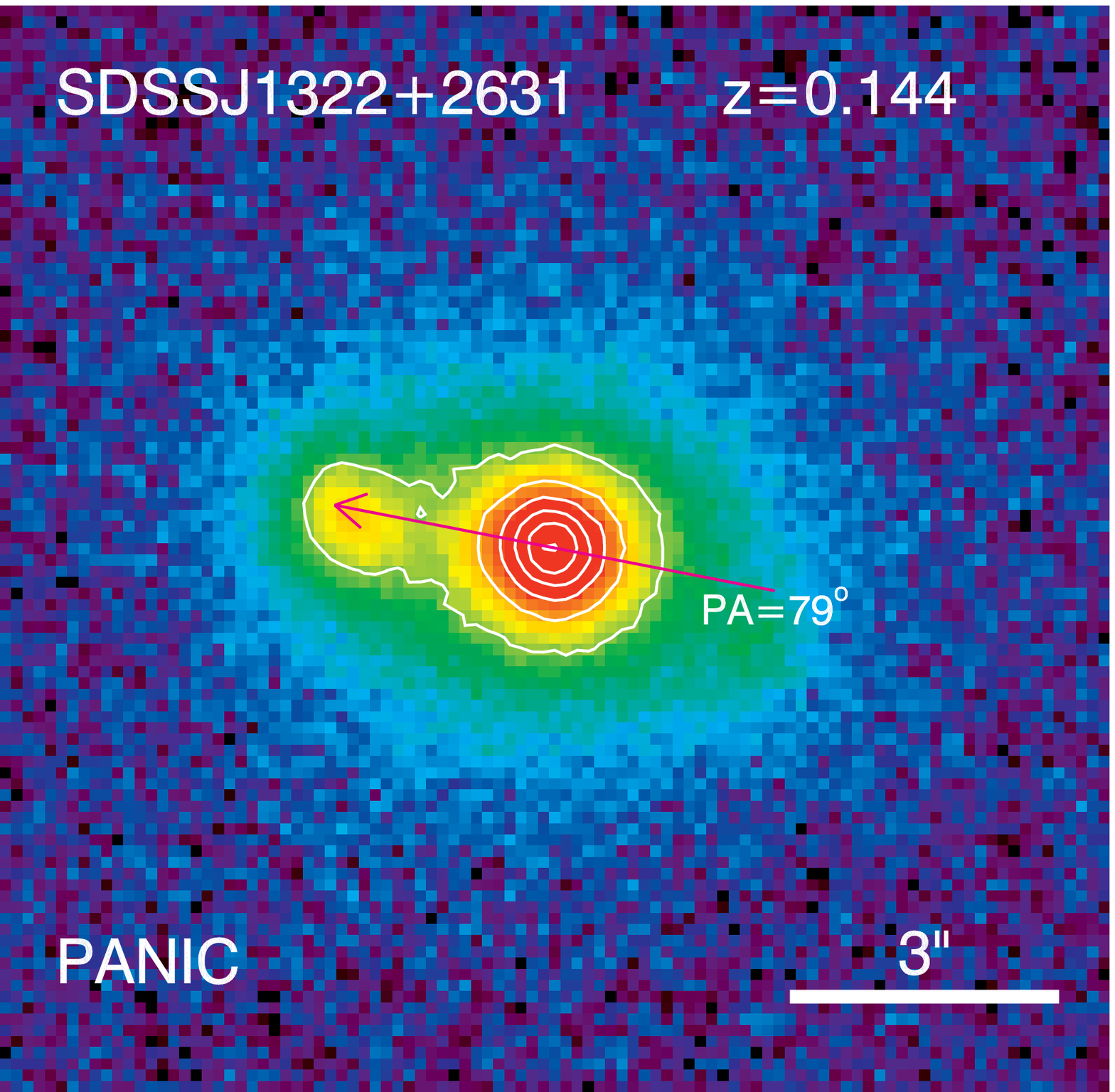}\hspace{9pt}
    \includegraphics[width=0.7\textwidth]{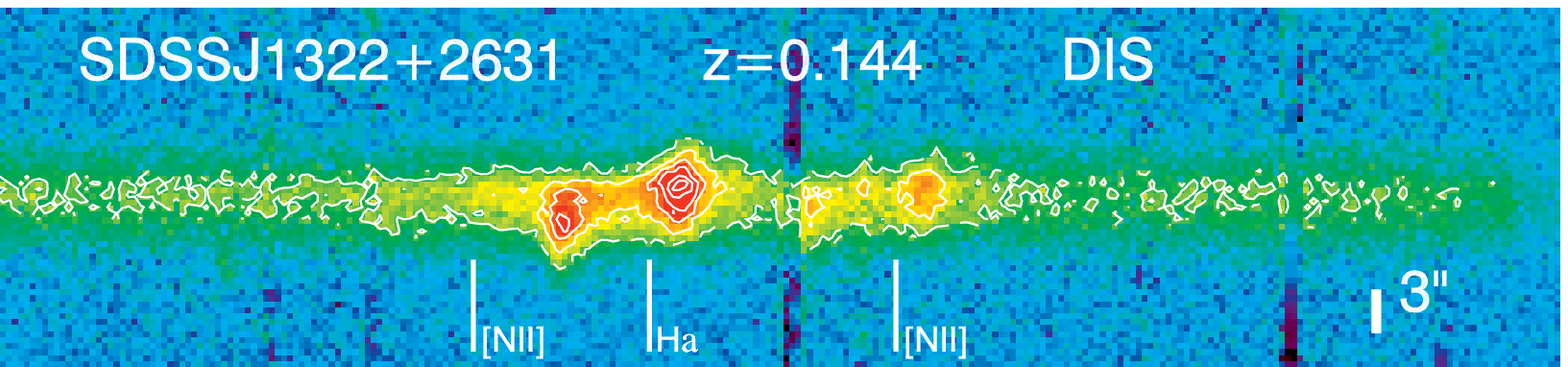}}
  \caption{PANIC NIR images and 2d spectra for objects classified as ambiguous cases. In all objects except for SDSS J1322+2631, the spatial offset
  between the two emission line velocity components is smaller than the typical resolution of our NIR imaging.
  }
  \label{fig:amb1}
\end{figure*}

\addtocounter{figure}{-1}
\begin{figure*}
\addtocounter{subfigure}{1}
\centering
    \subfigure{
    \includegraphics[width=0.2\textwidth]{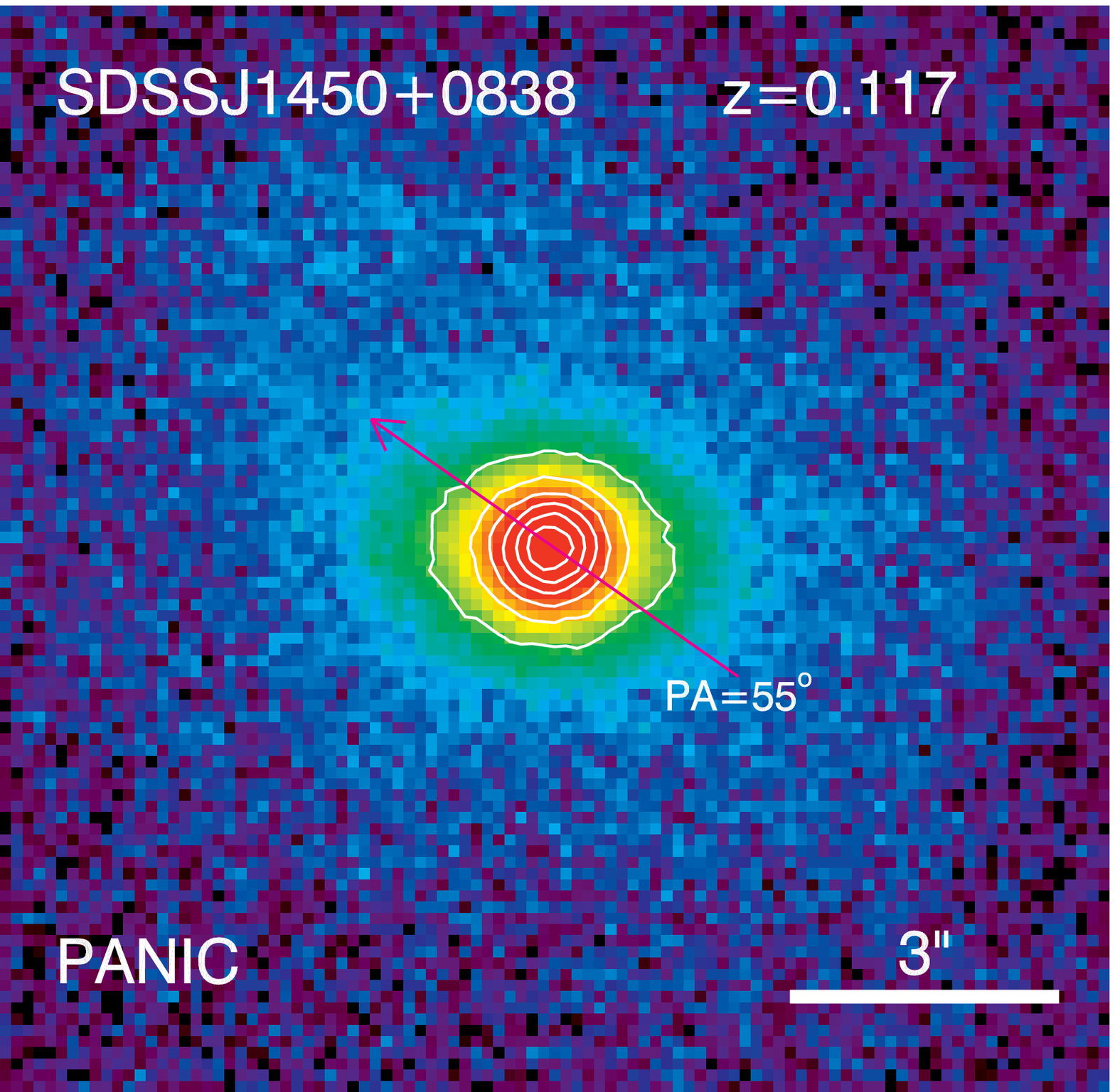}\hspace{9pt}
    \includegraphics[width=0.7\textwidth]{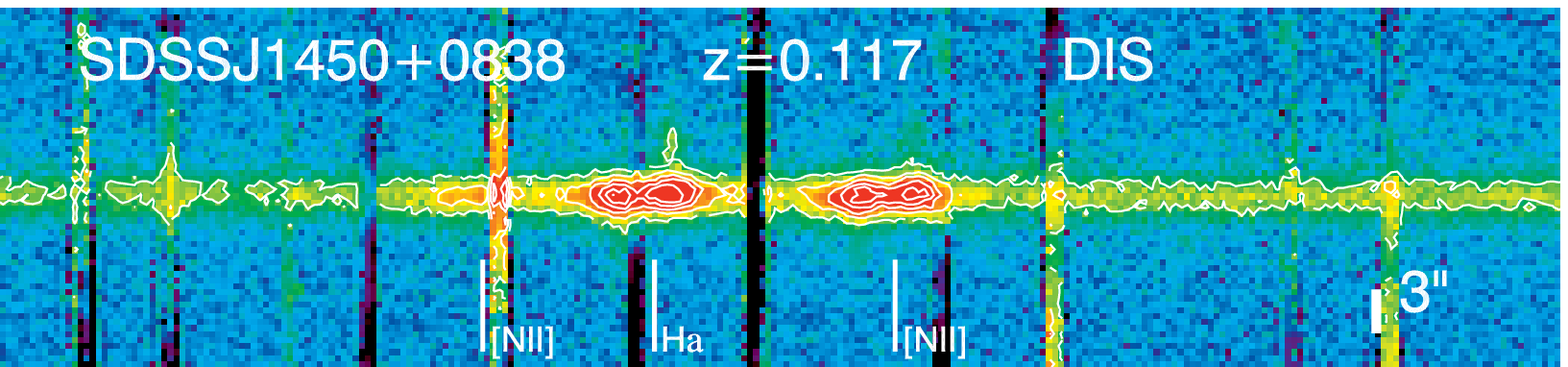}}
    \subfigure{
    \includegraphics[width=0.2\textwidth]{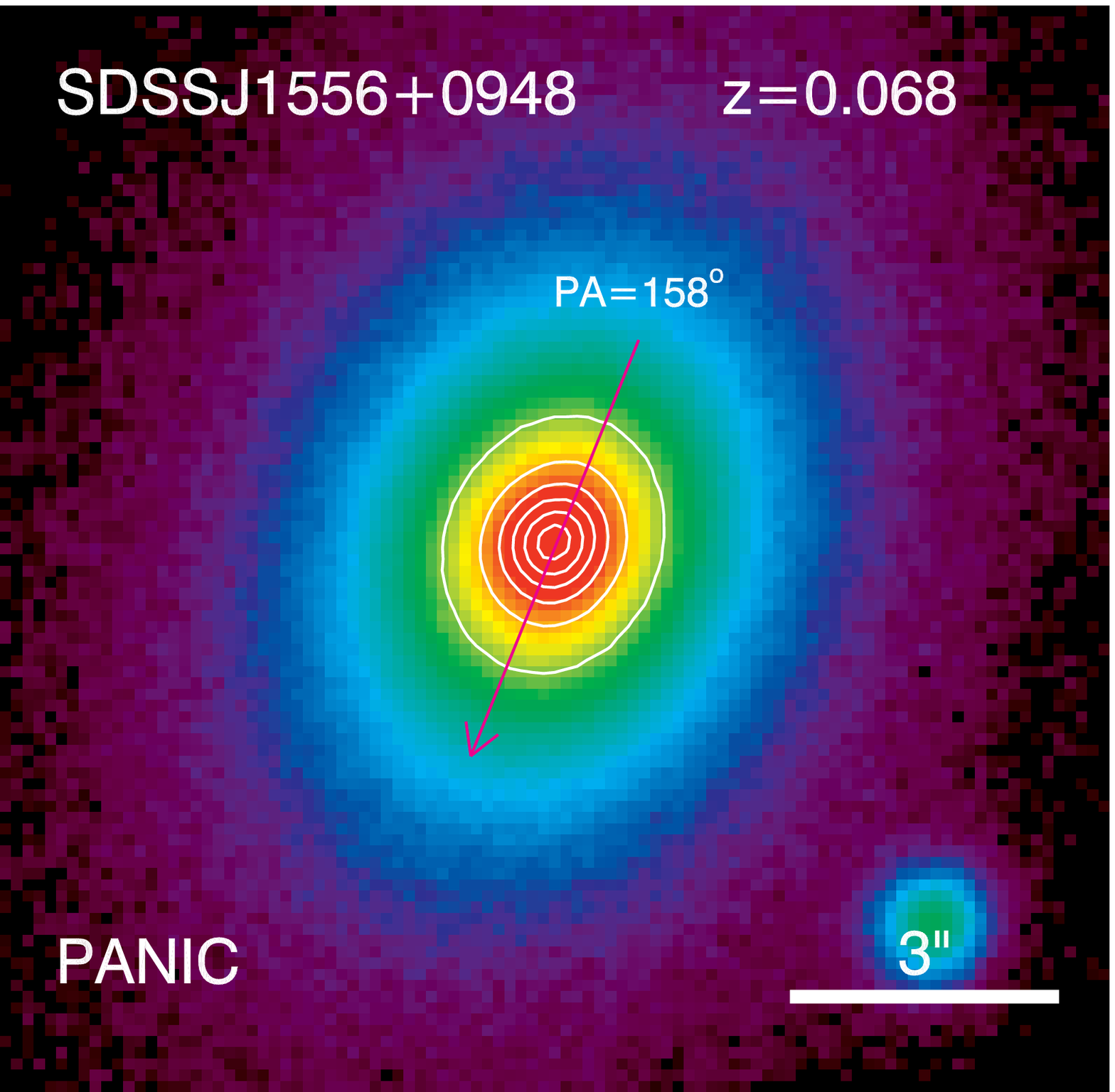}\hspace{9pt}
    \includegraphics[width=0.7\textwidth]{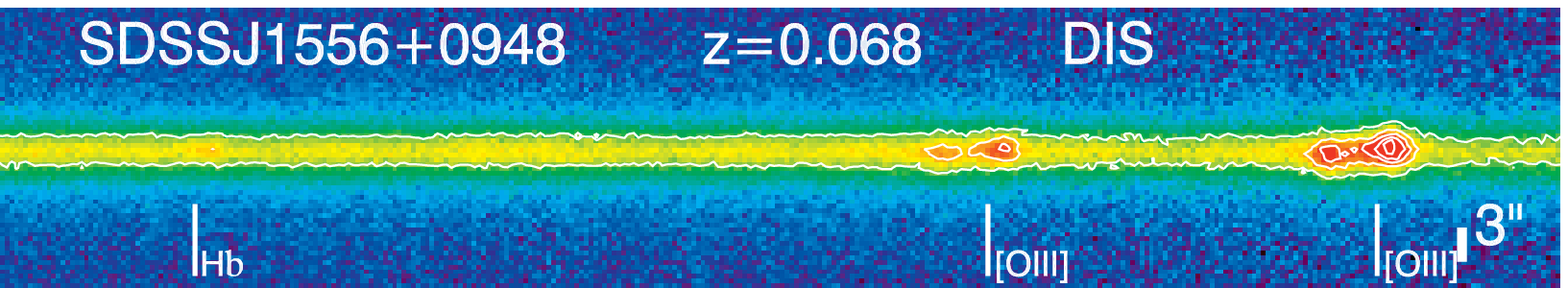}}
    \subfigure{
    \includegraphics[width=0.2\textwidth]{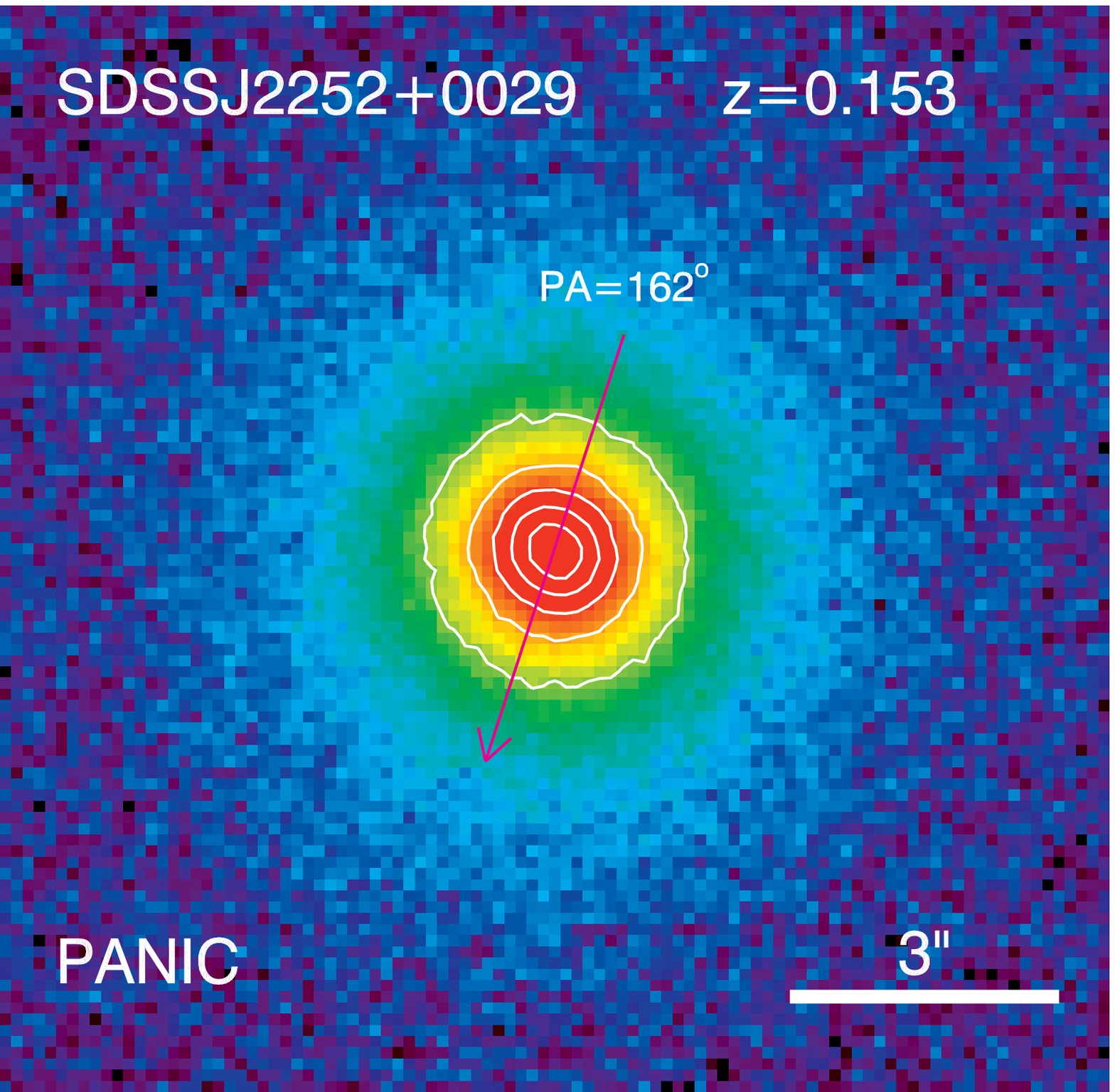}\hspace{9pt}
    \includegraphics[width=0.7\textwidth]{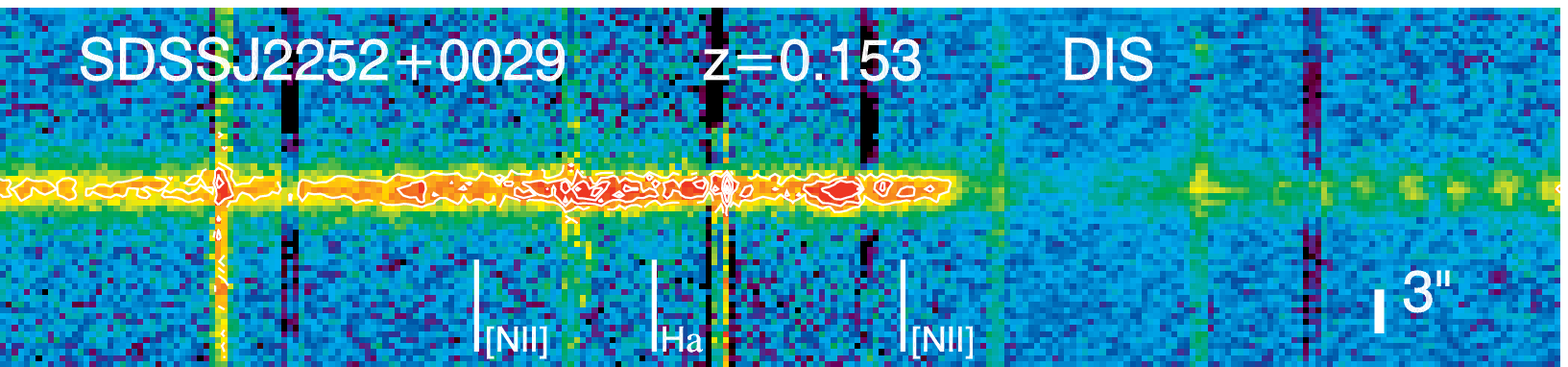}}
    \subfigure{
    \includegraphics[width=0.2\textwidth]{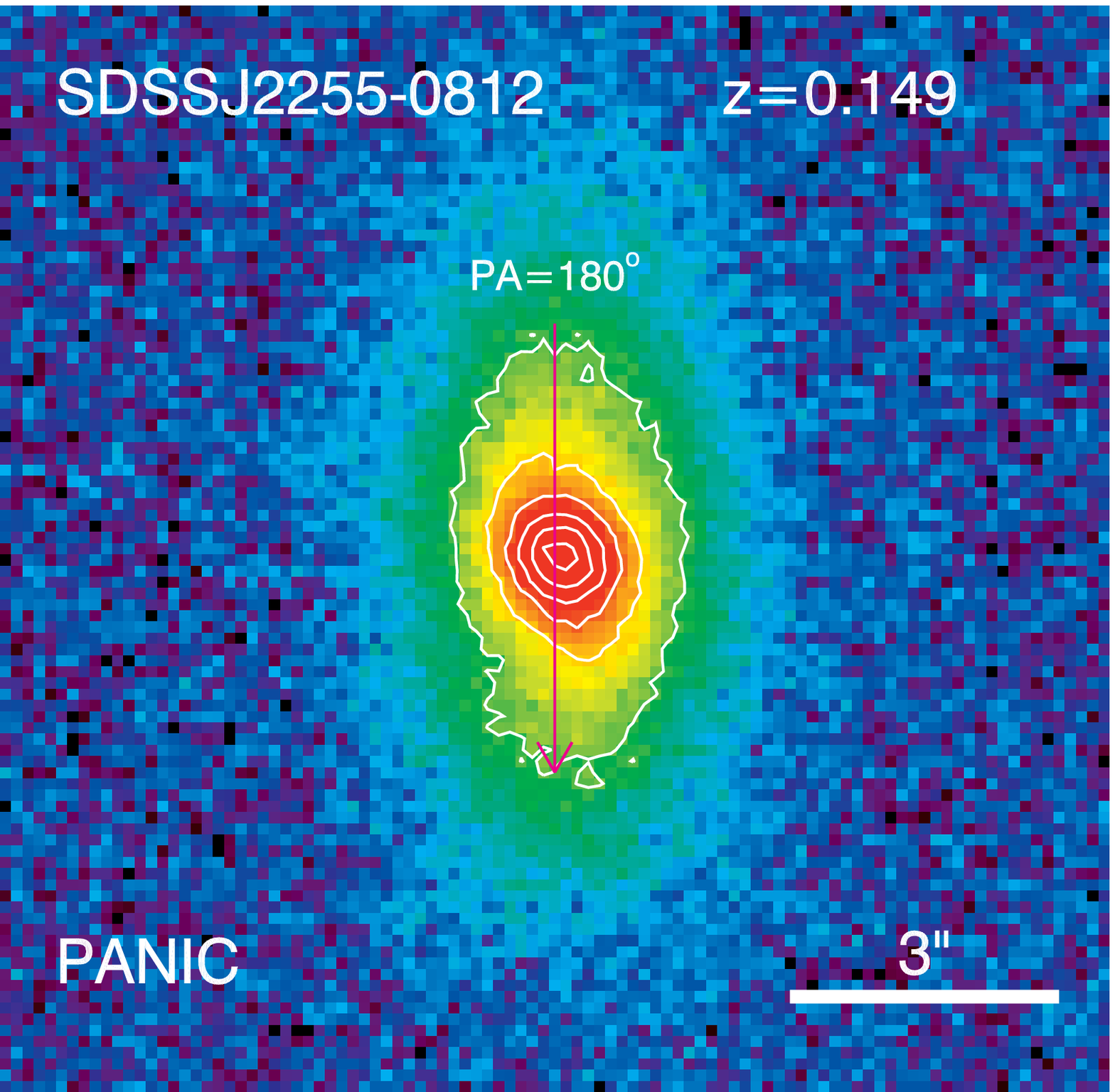}\hspace{9pt}
    \includegraphics[width=0.7\textwidth]{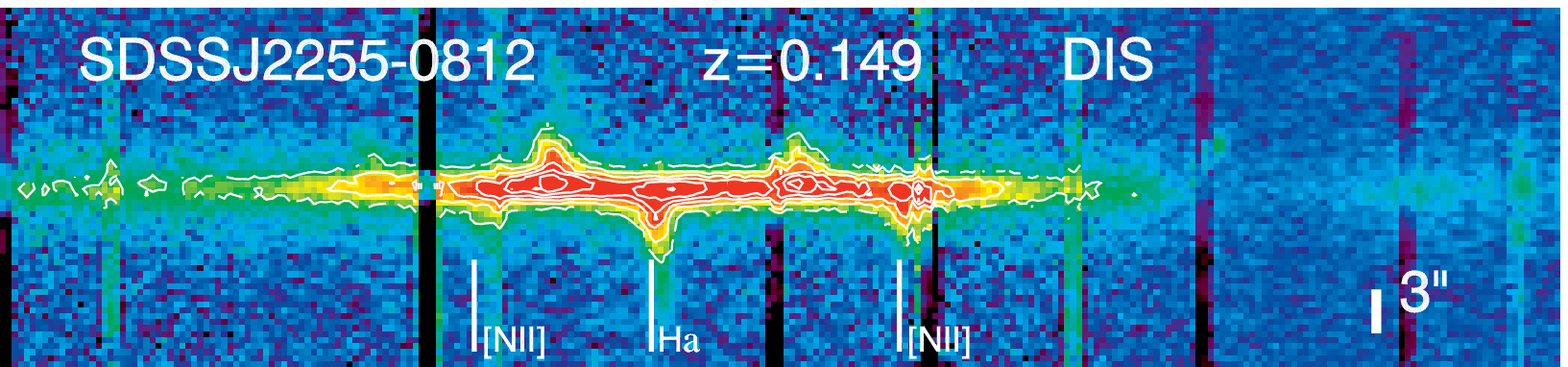}}
    \subfigure{
    \includegraphics[width=0.2\textwidth]{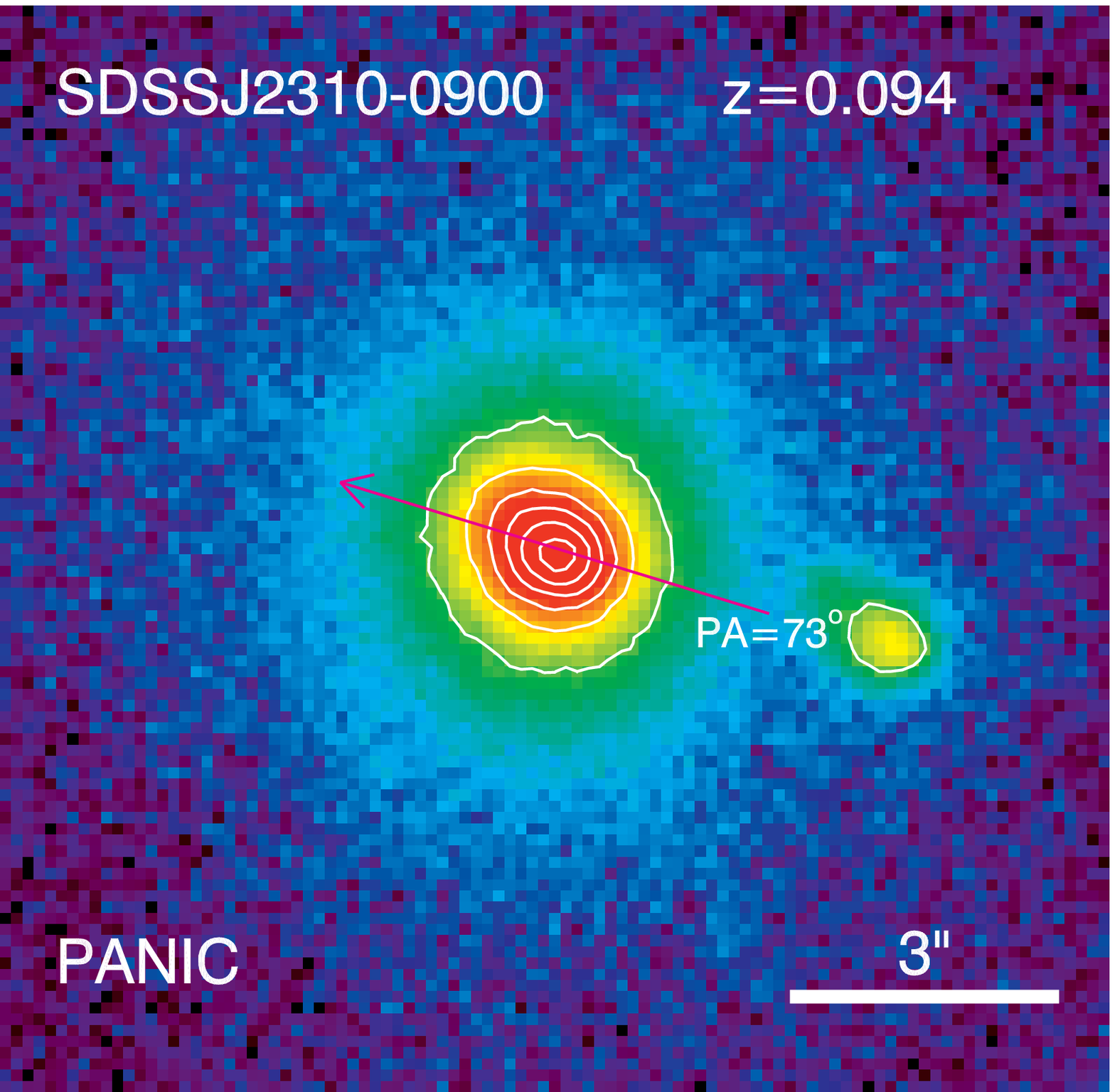}\hspace{9pt}
    \includegraphics[width=0.7\textwidth]{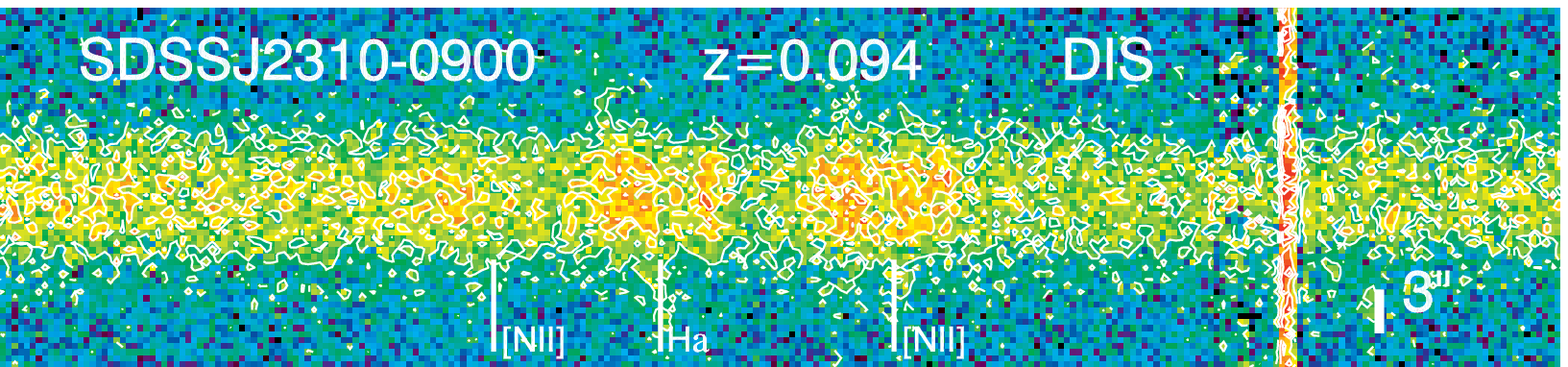}}
    \subfigure{
    \includegraphics[width=0.2\textwidth]{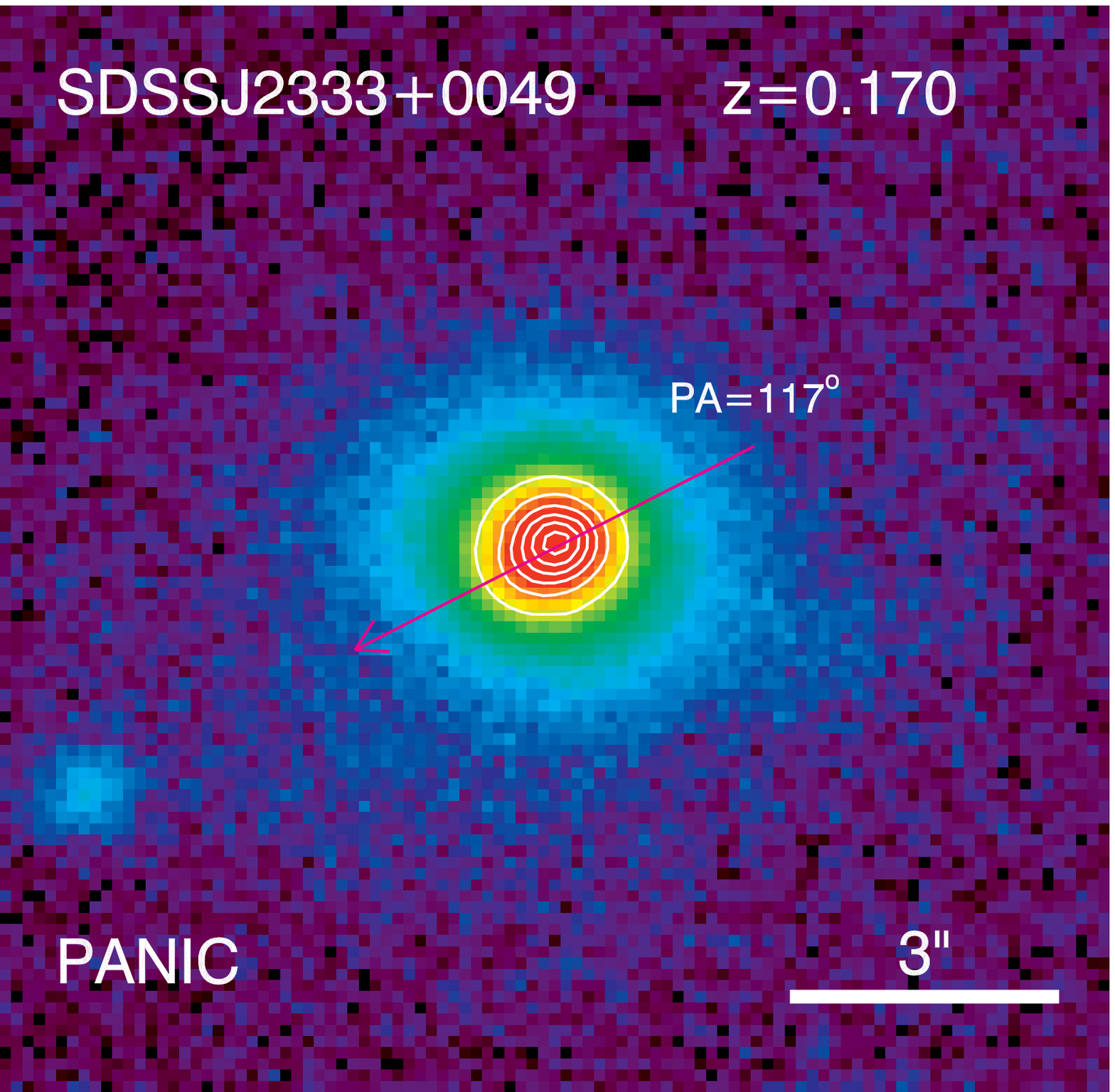}\hspace{9pt}
    \includegraphics[width=0.7\textwidth]{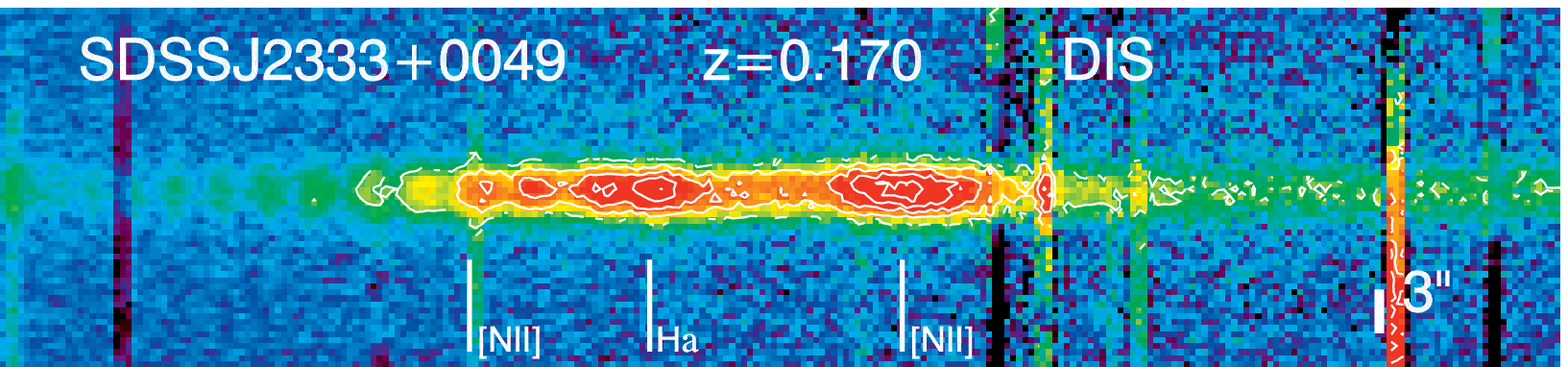}}
    \caption{Continued. }
    \label{fig:amb2}
\end{figure*}

\begin{figure*}
  \centering
    \includegraphics[width=0.32\textwidth]{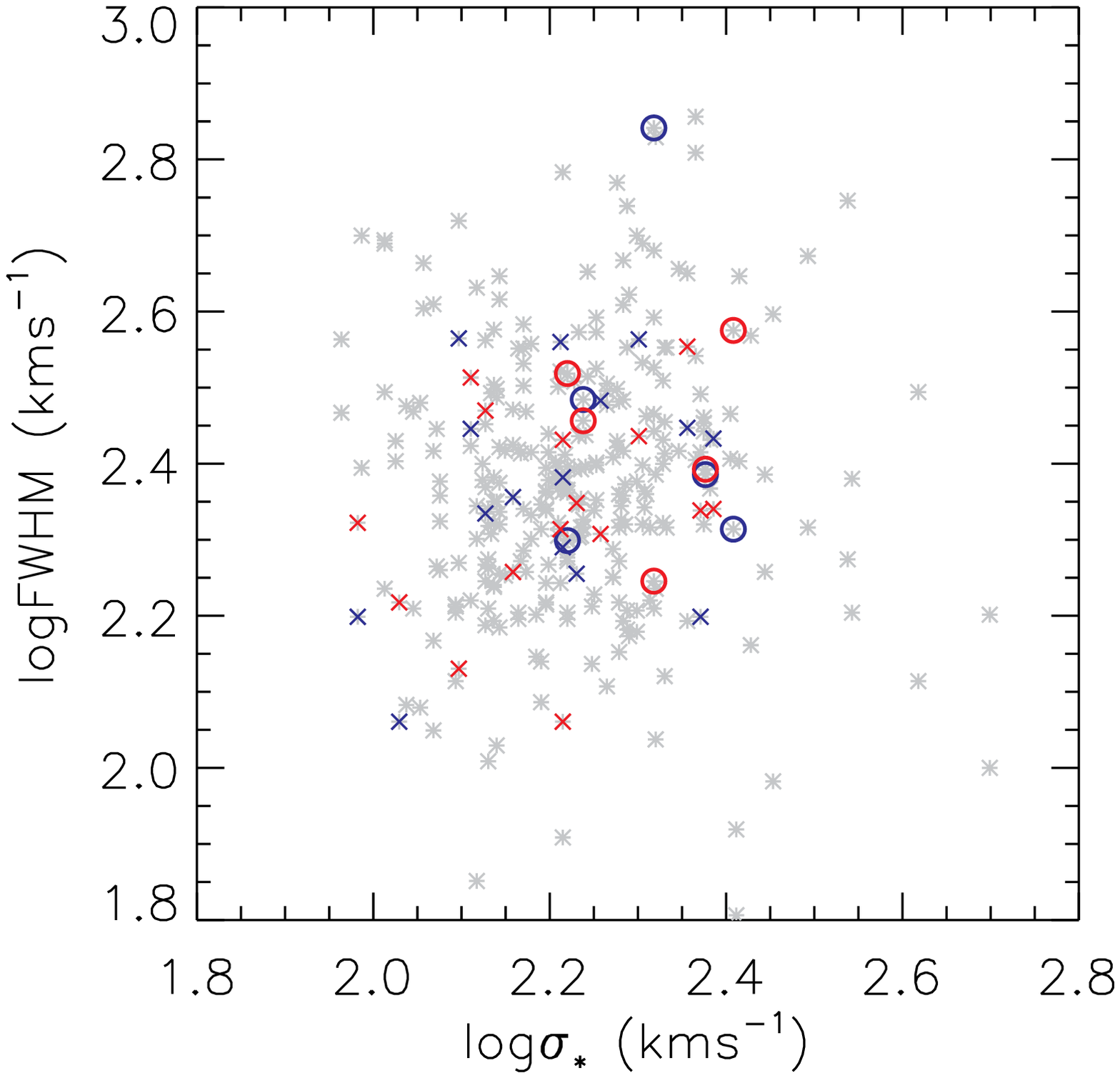}
    \includegraphics[width=0.32\textwidth]{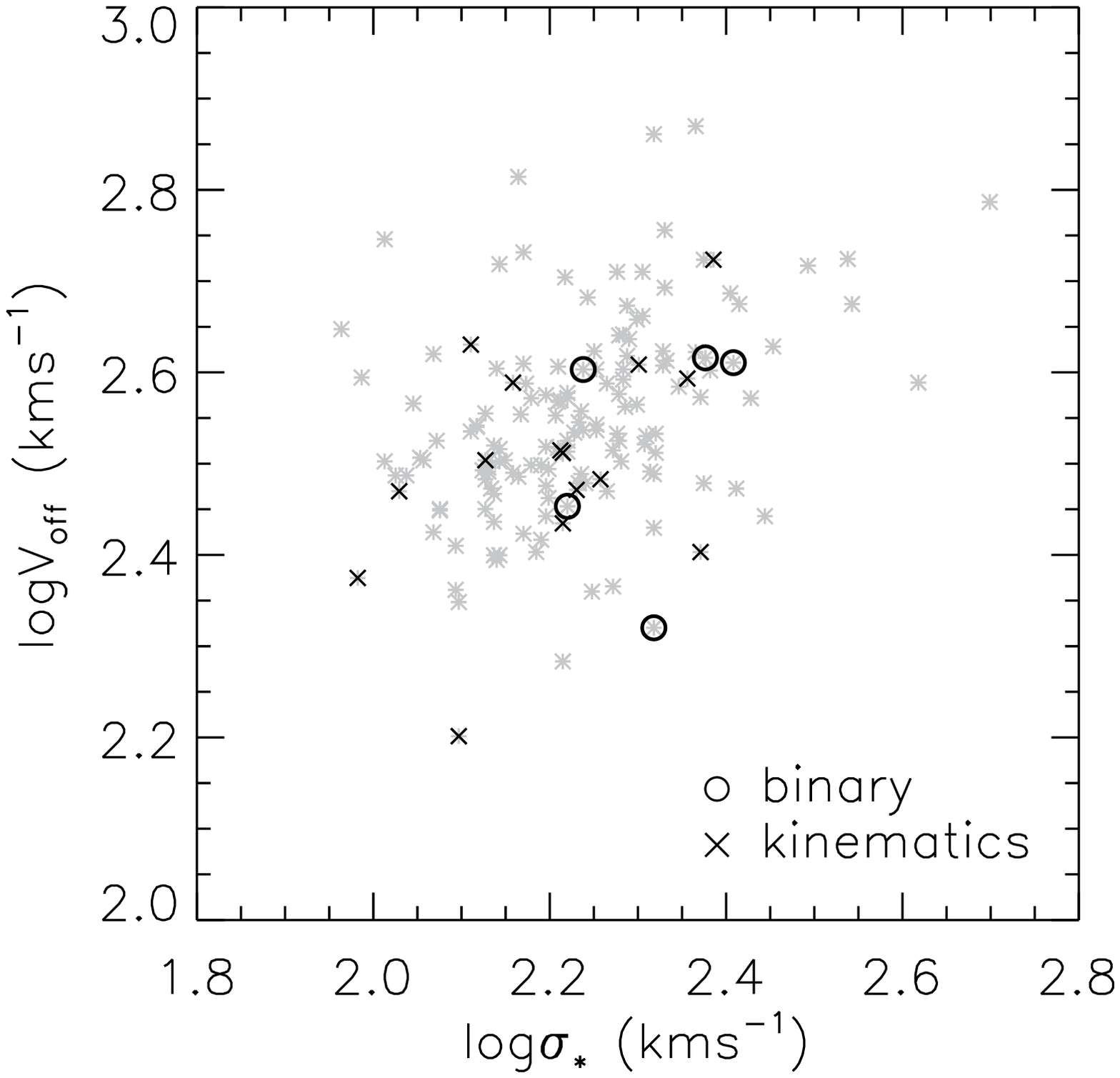}
    \includegraphics[width=0.32\textwidth]{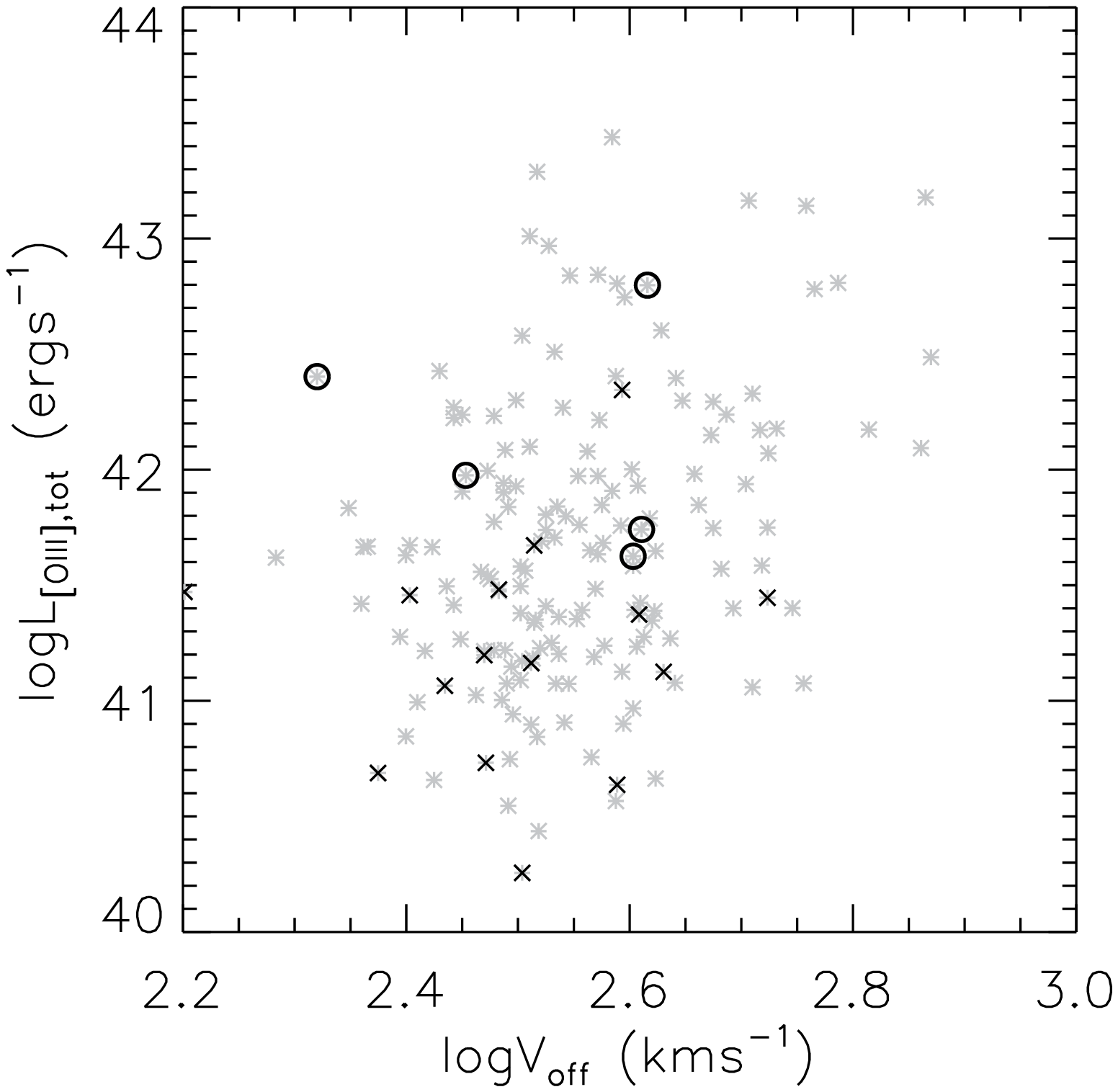}
    \caption{Bulk properties of the five objects classified as binary AGNs (circles) compared with those classified as NLR kinematics in single AGNs
    (crosses). In all panels, the gray points are for all objects in the double-peaked sample of \citet{Liu_etal_2010a}. {\em Left:} Distributions in the plane of FWHM and stellar velocity dispersion $\sigma_*$. The stellar velocity dispersions are for
    the
    whole galaxy within the 3\arcsec\ diameter fiber and the FWHMs are for the two velocity components (the blueshifted component is indicated in
    blue
    and the redshifted component is indicated in red). Measurements are taken from \citet{Liu_etal_2010a}. {\em Middle:} Distributions in the plane
    of
    $\sigma_*$ versus velocity offset between the two components, $V_{\rm off}$. {\em Right:} Distributions in the plane of the total \OIIIb\
    luminosity vs $V_{\rm off}$. }
    \label{fig:bulk_prop}
\end{figure*}

\begin{figure*}
  \centering
    \includegraphics[width=0.9\textwidth]{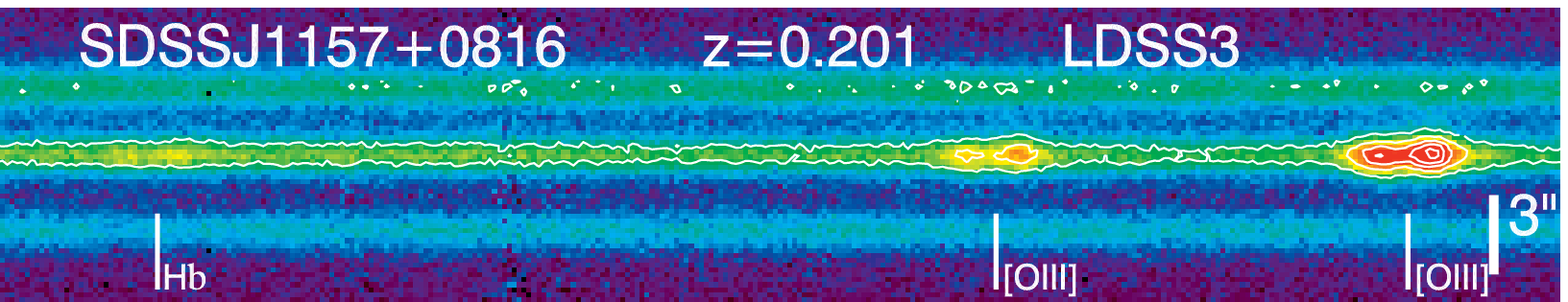}
    \caption{A slit spectrum for J1157+0816, which shows three nuclei in the optical SDSS image \citep{Smith_etal_2010} and in the NIR image
    \citep{Fu_etal_2010}. The slit was placed across the three nuclei. Only the central nucleus has \OIII\ emission, which is responsible for the
    double-peaked \OIII\ profile seen in the SDSS spectrum. }
    \label{fig:1157+0816}
\end{figure*}

\begin{figure*}
  \centering
    \subfigure{
    \includegraphics[width=0.3\textwidth]{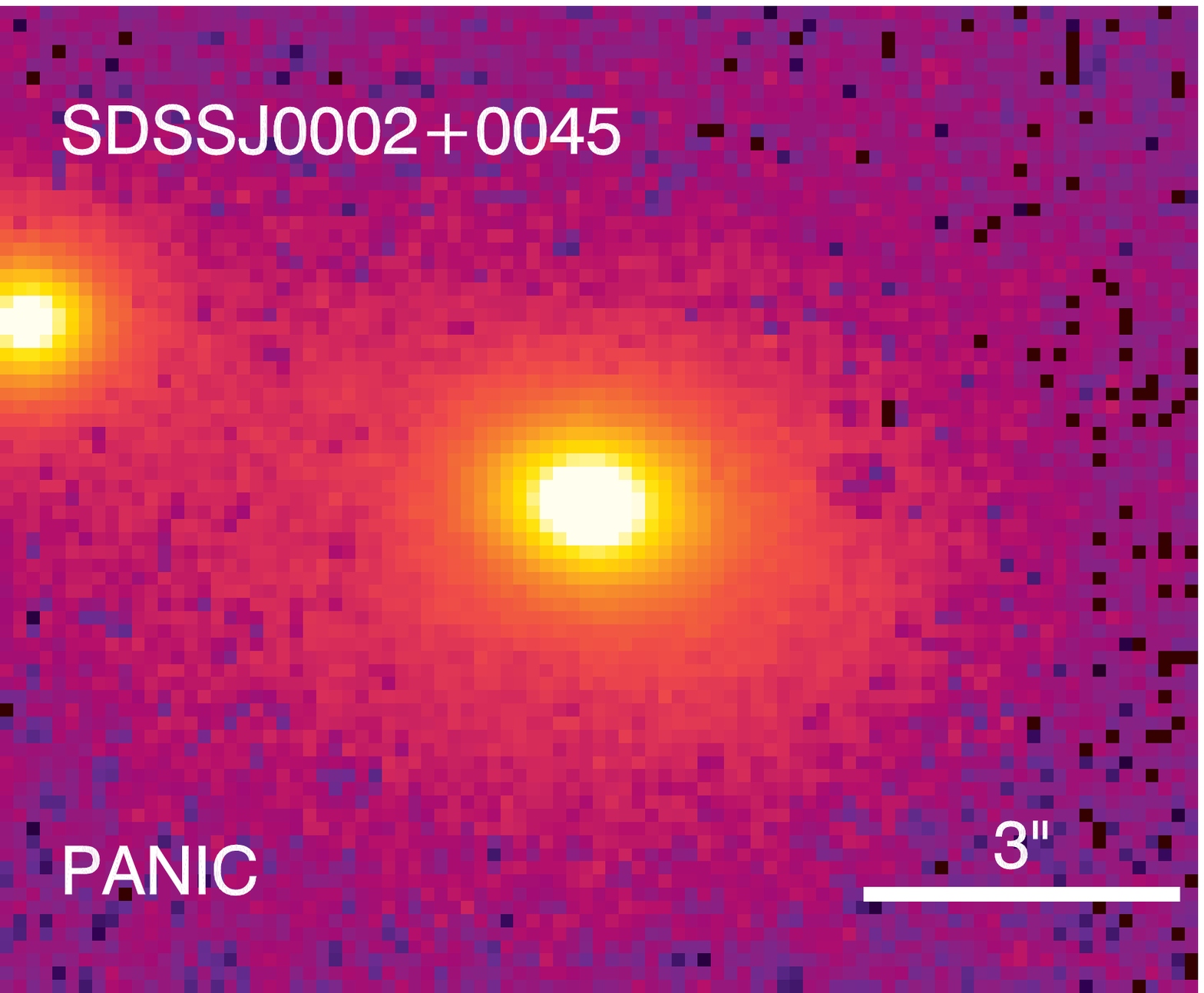}
    \includegraphics[width=0.3\textwidth]{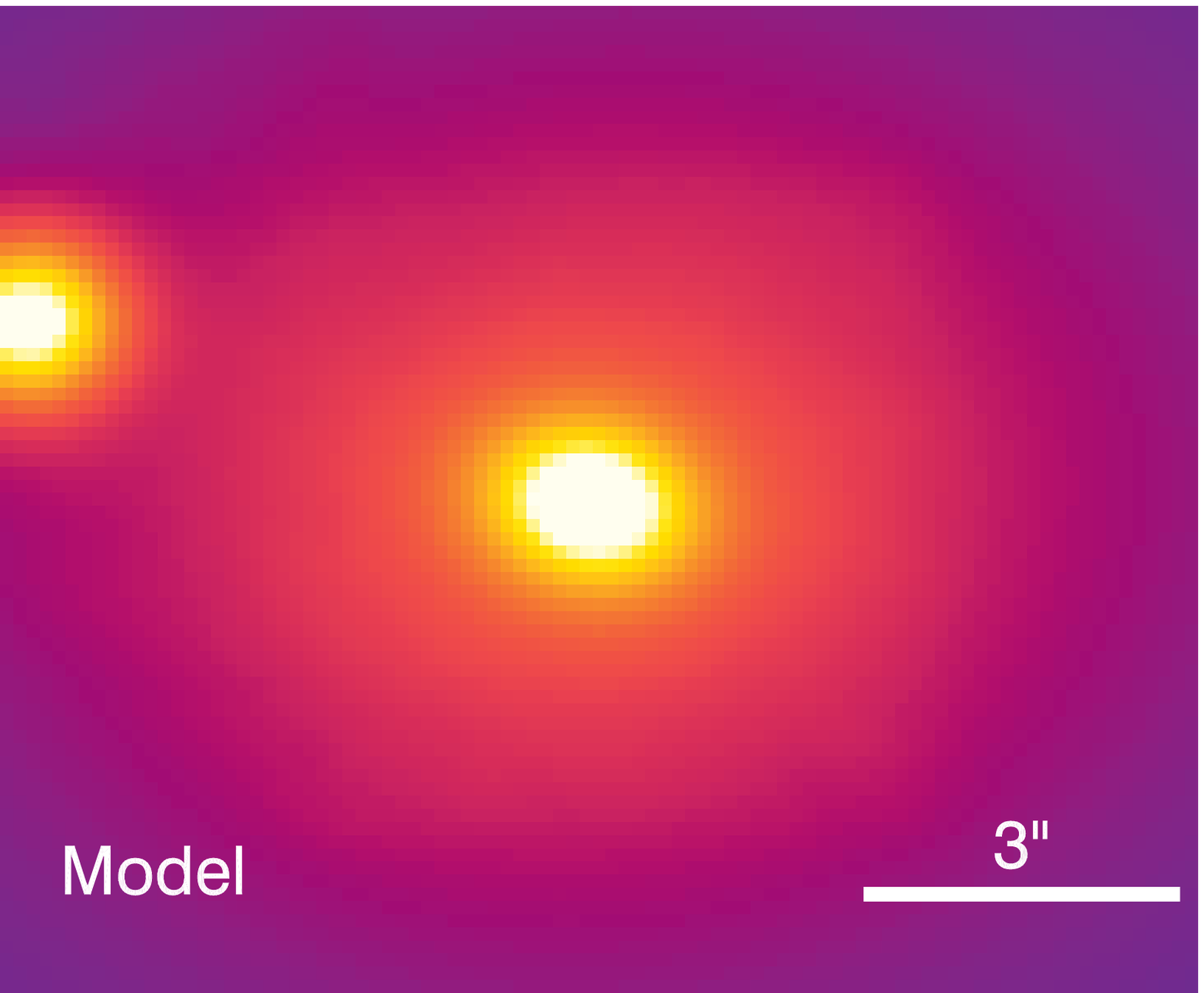}
    \includegraphics[width=0.3\textwidth]{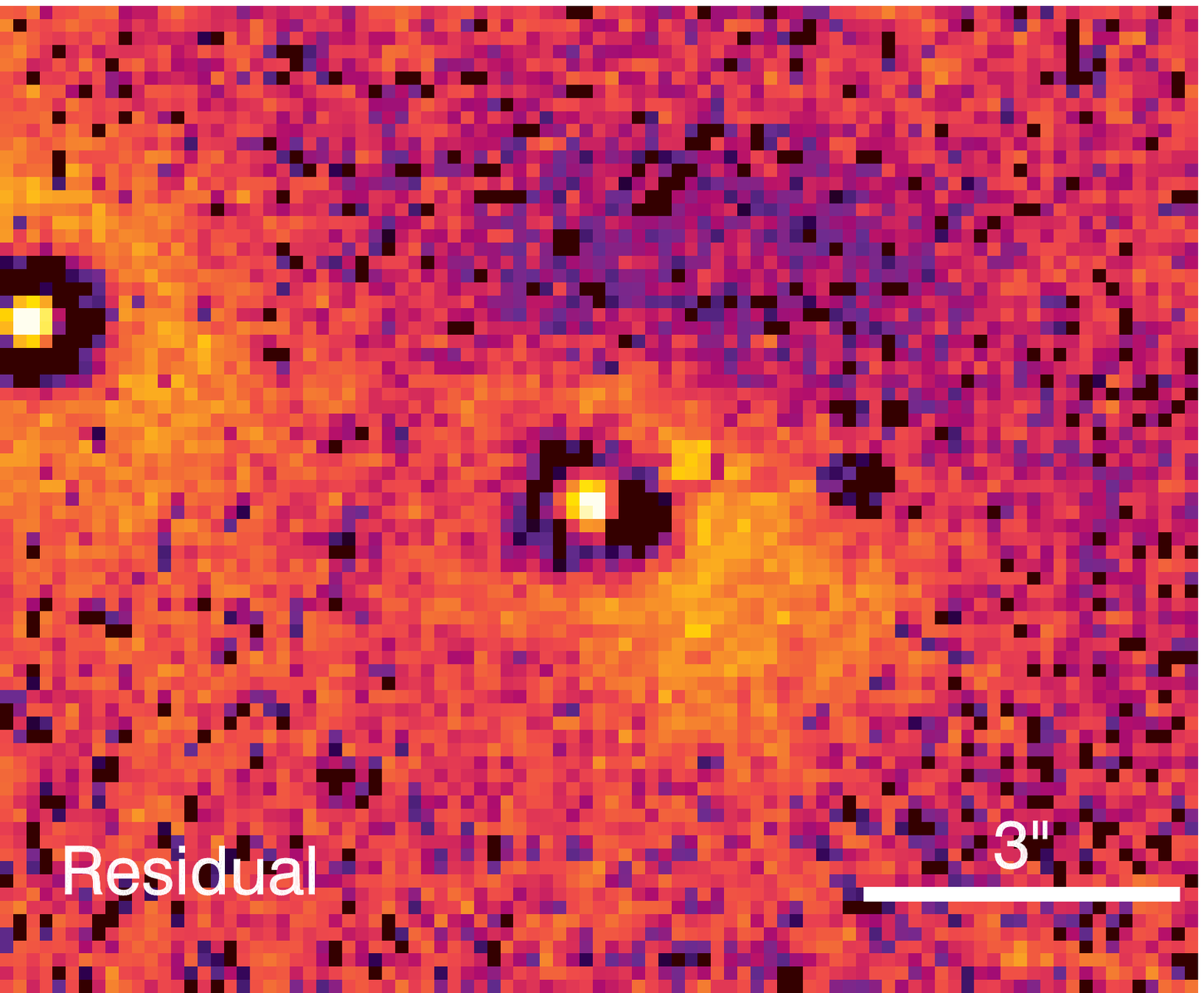}}
    \subfigure{
    \includegraphics[width=0.3\textwidth]{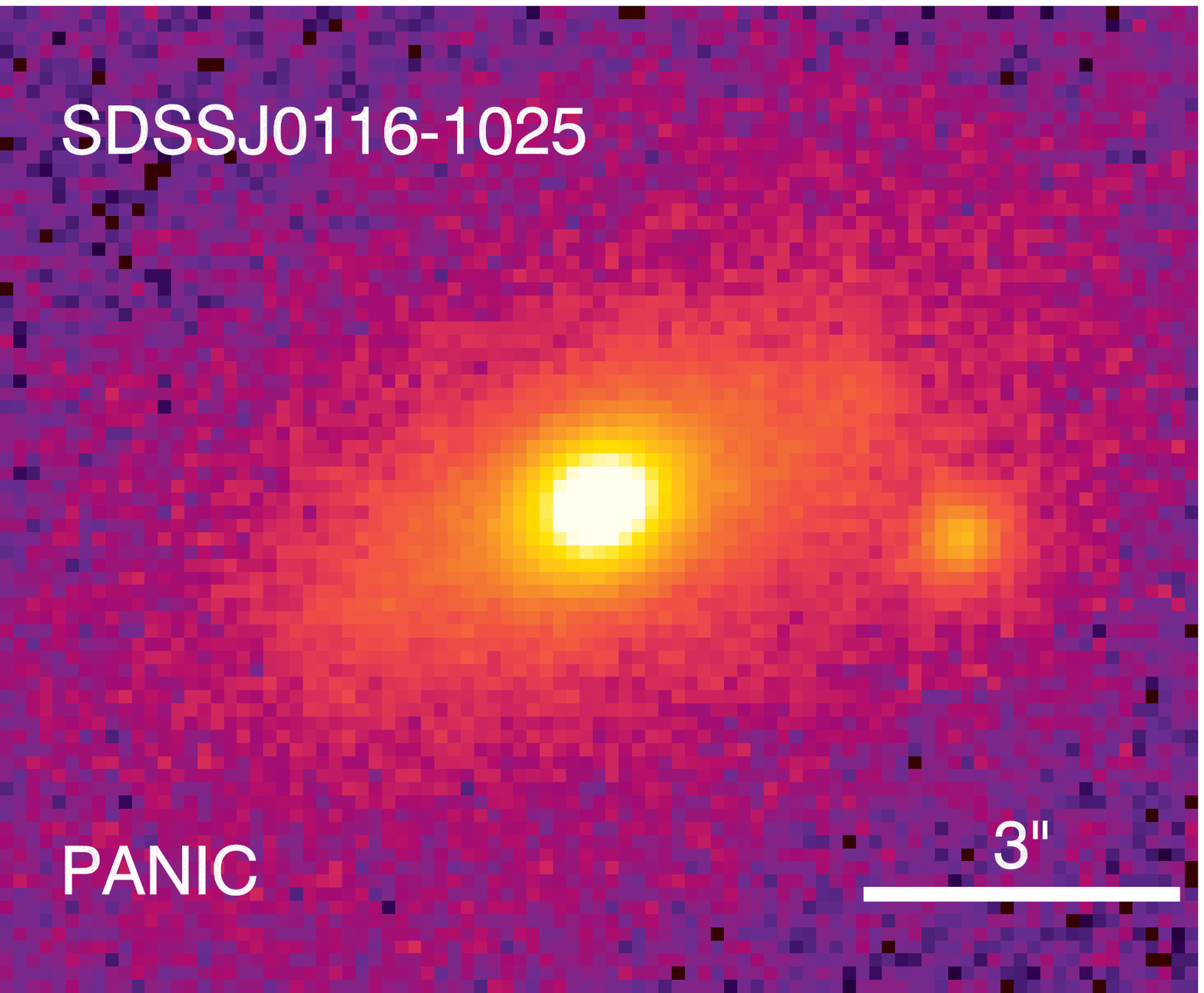}
    \includegraphics[width=0.3\textwidth]{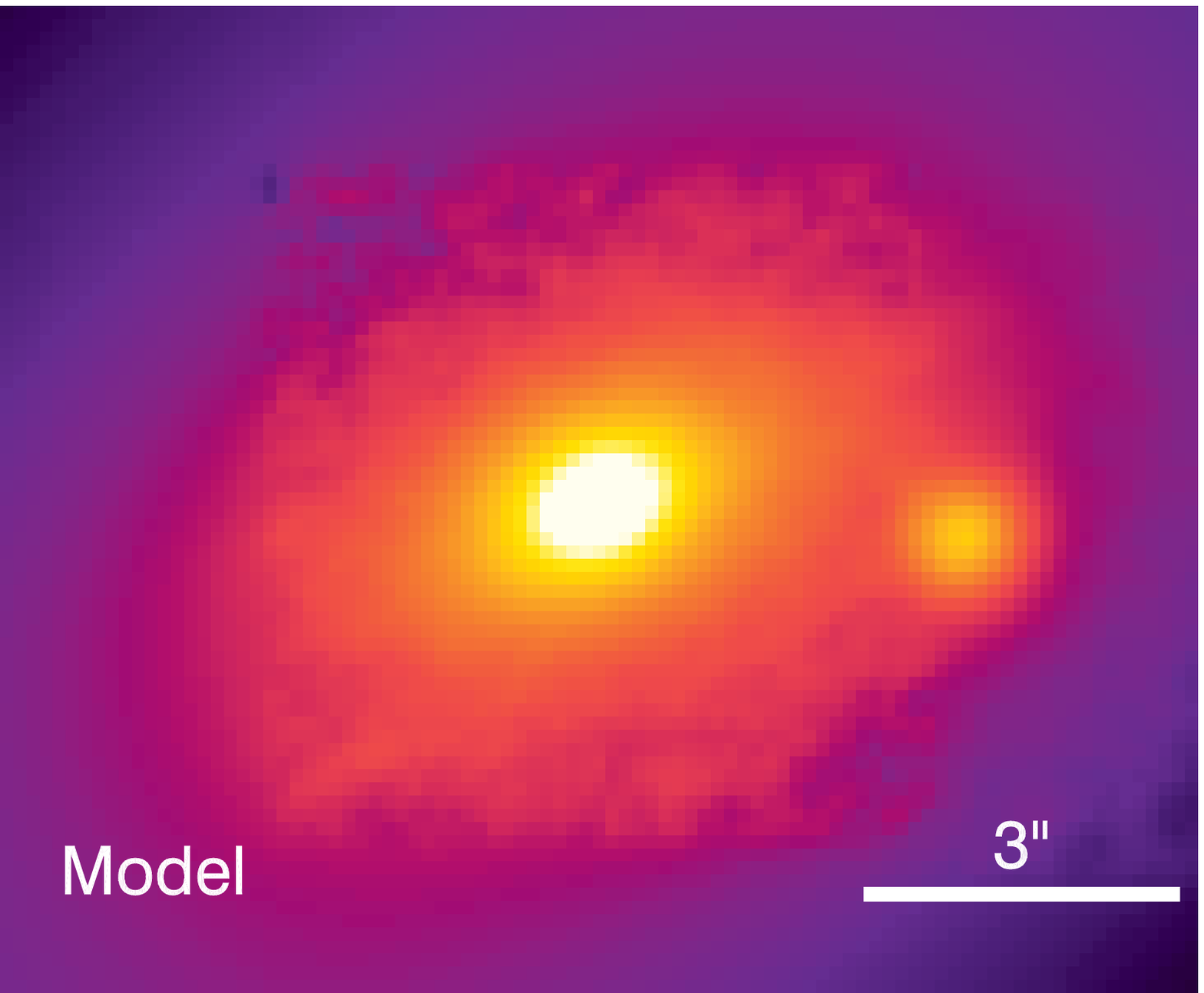}
    \includegraphics[width=0.3\textwidth]{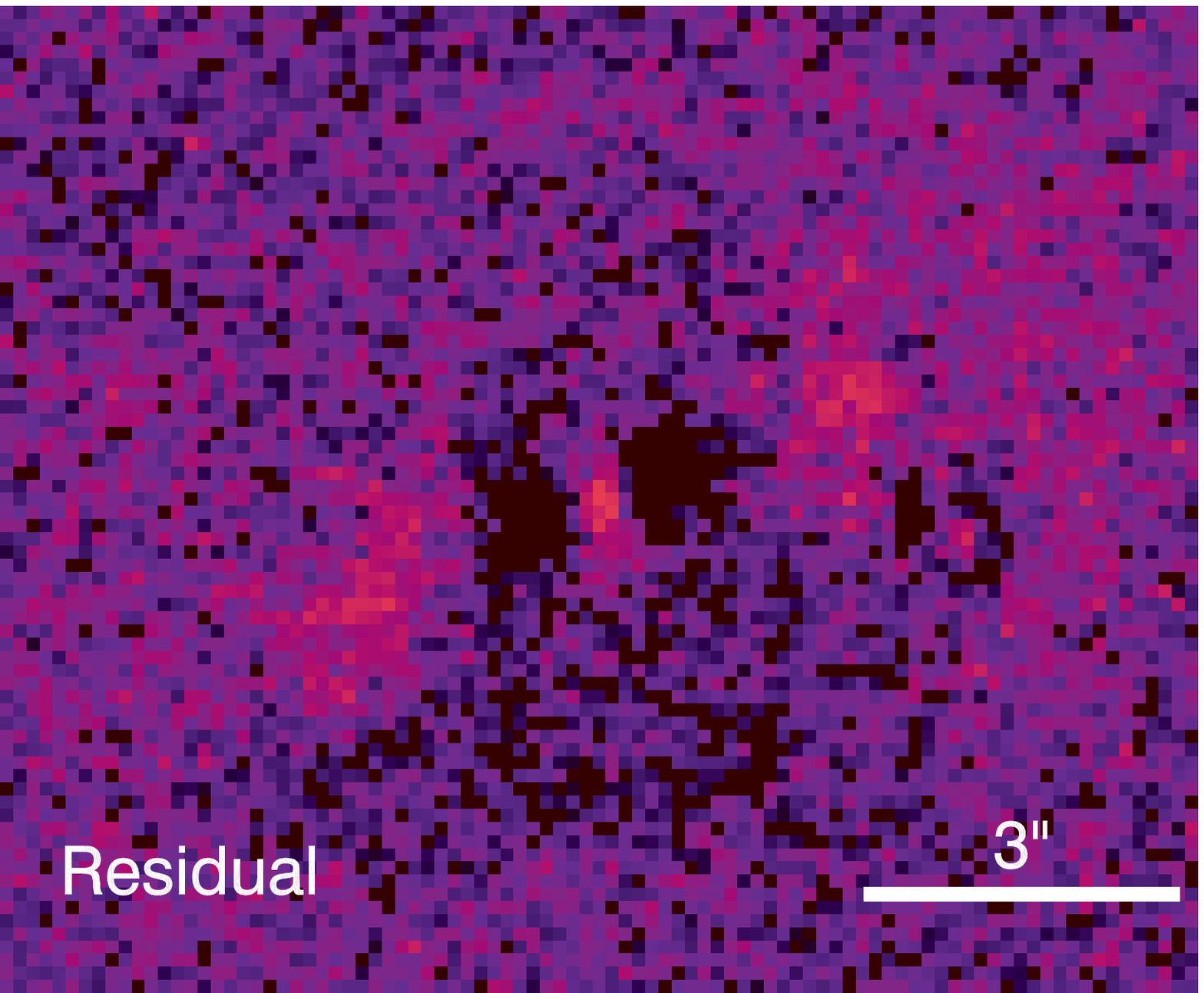}}
    \subfigure{
    \includegraphics[width=0.3\textwidth]{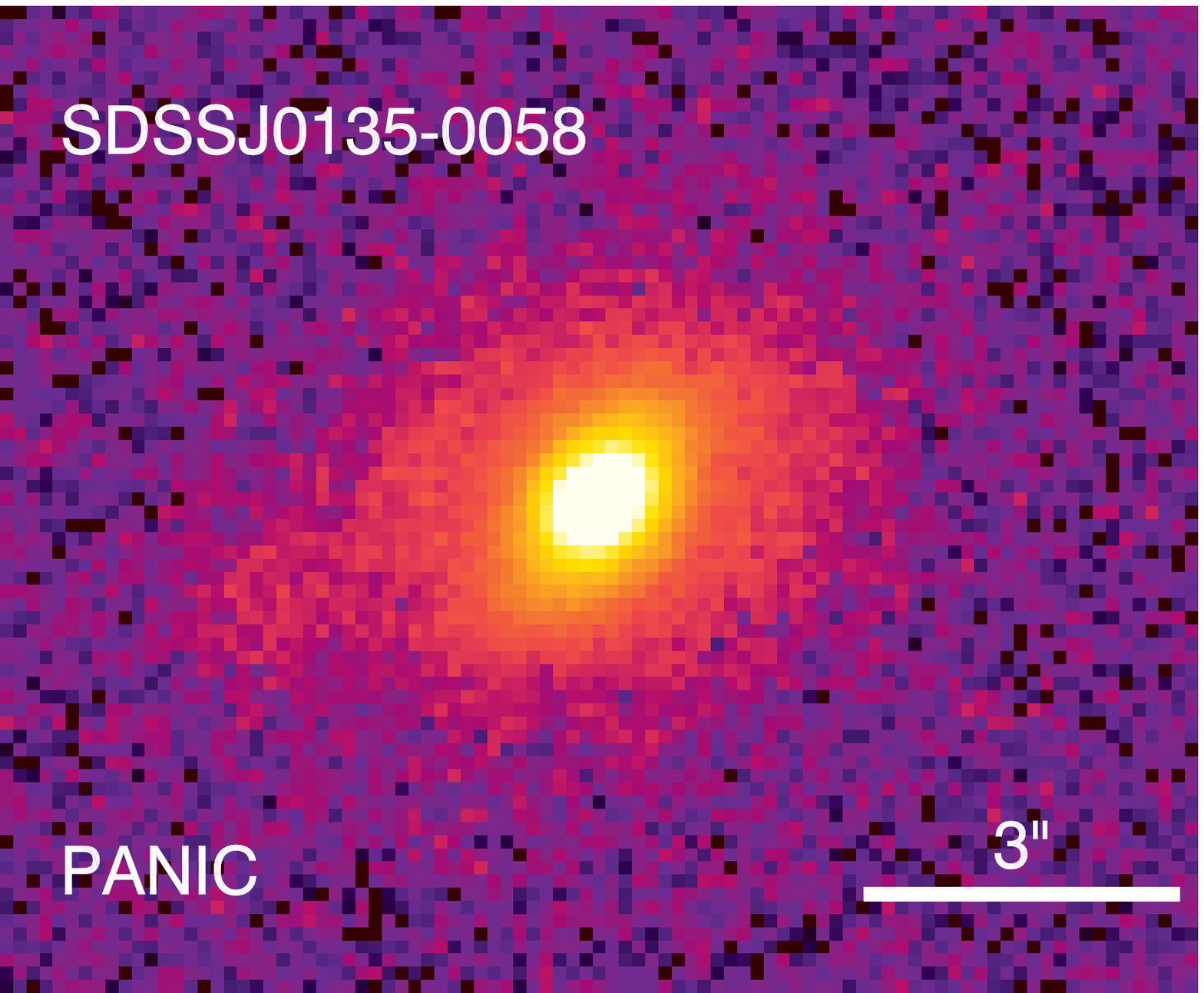}
    \includegraphics[width=0.3\textwidth]{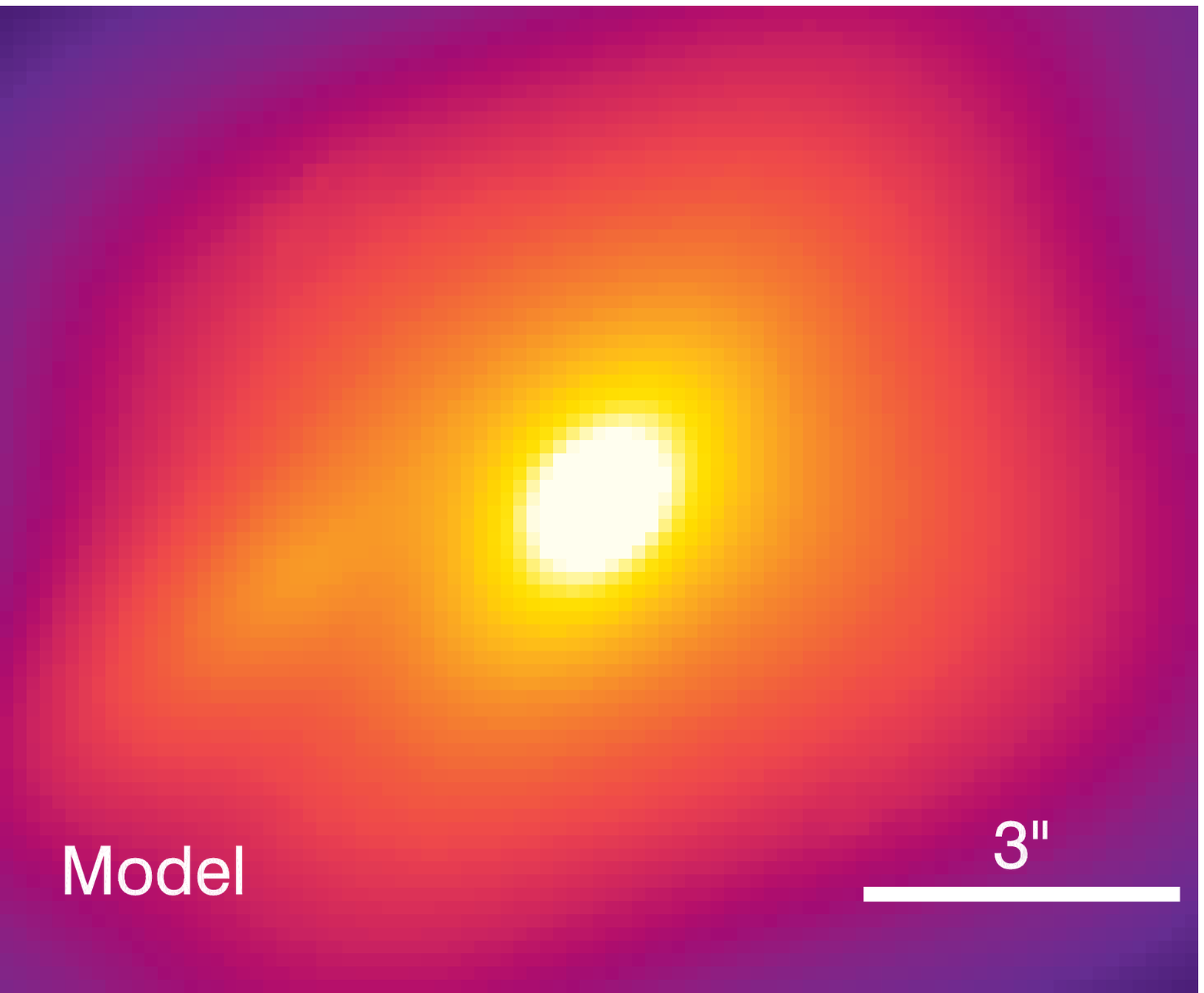}
    \includegraphics[width=0.3\textwidth]{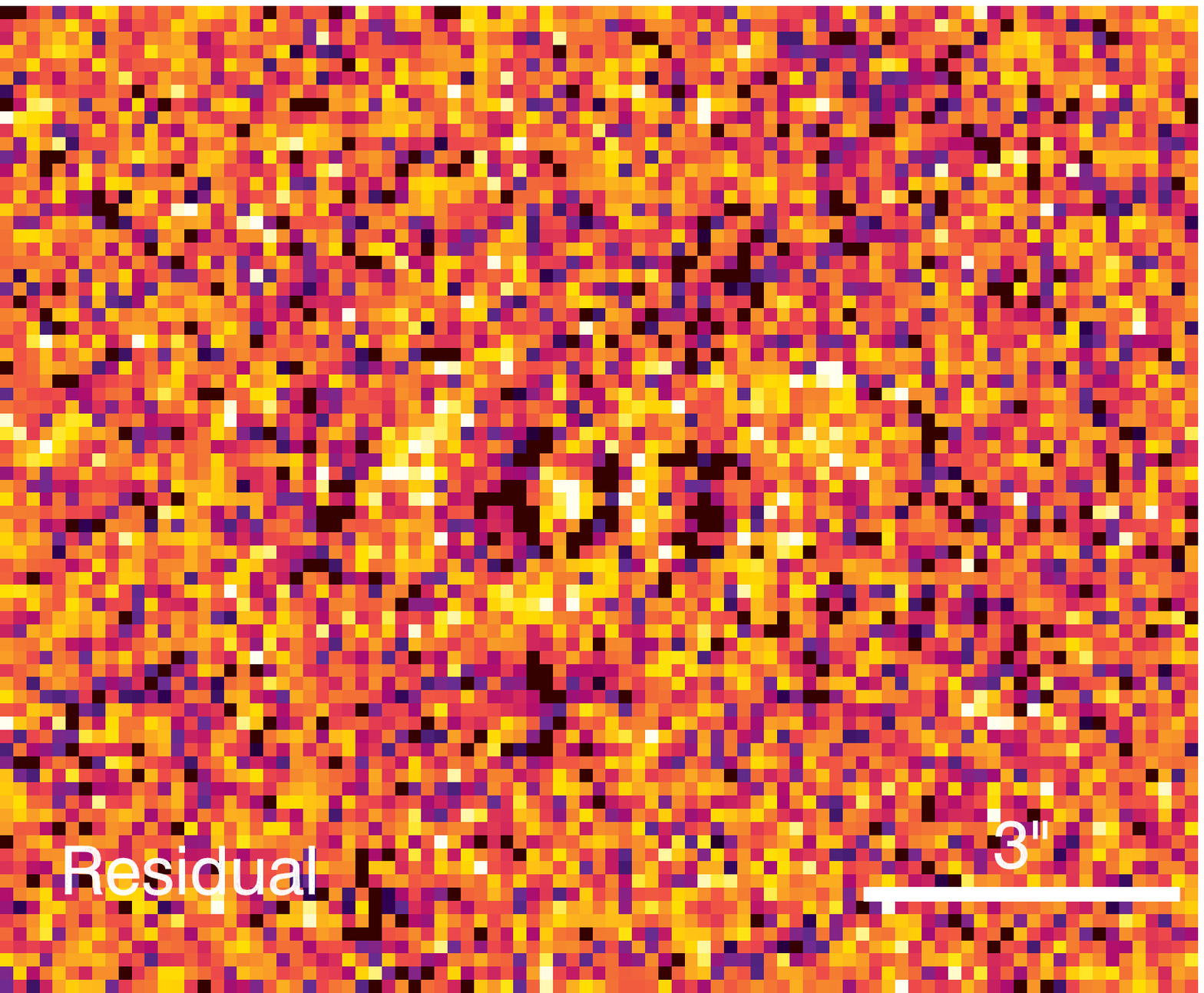}}
    \subfigure{
    \includegraphics[width=0.3\textwidth]{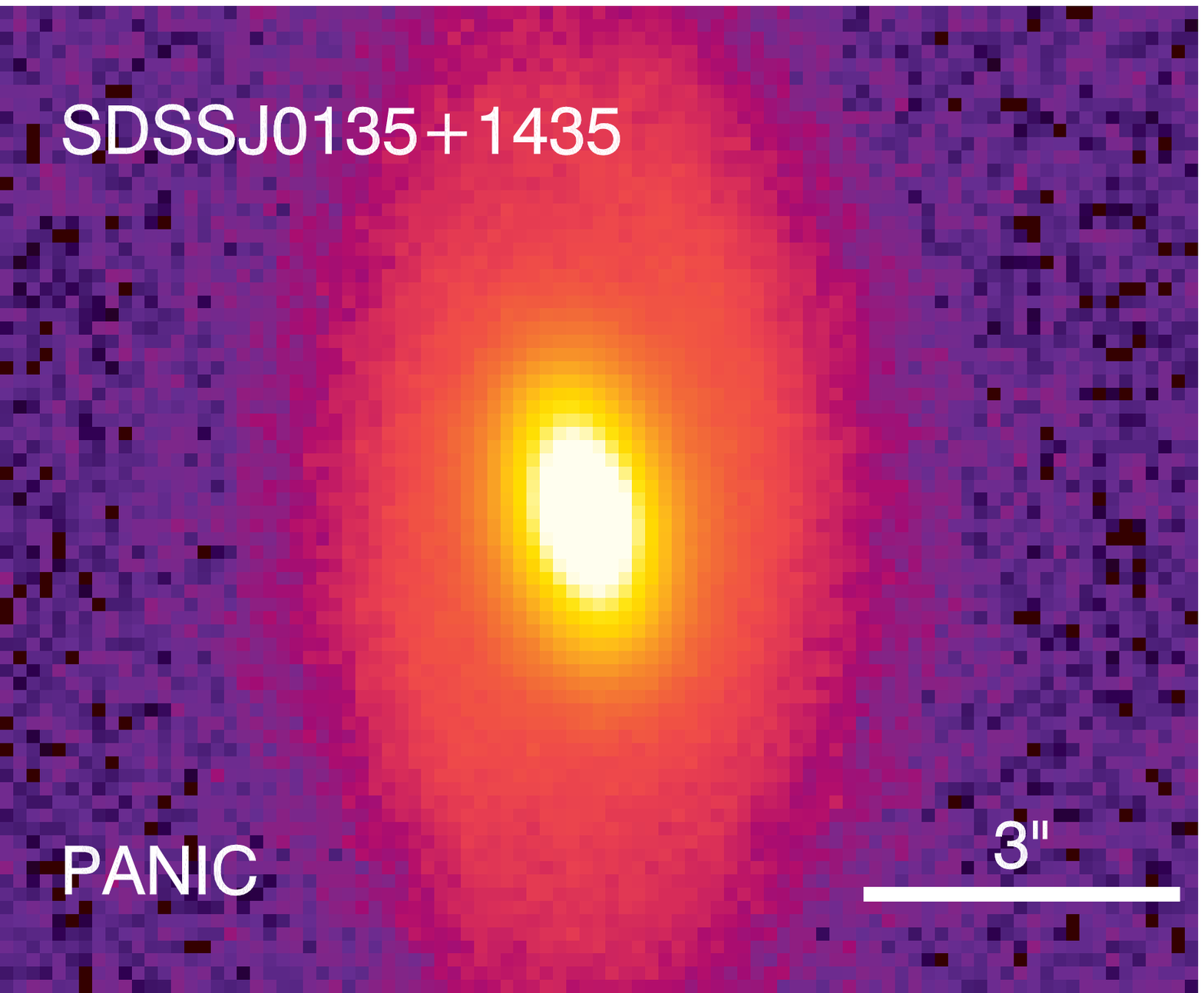}
    \includegraphics[width=0.3\textwidth]{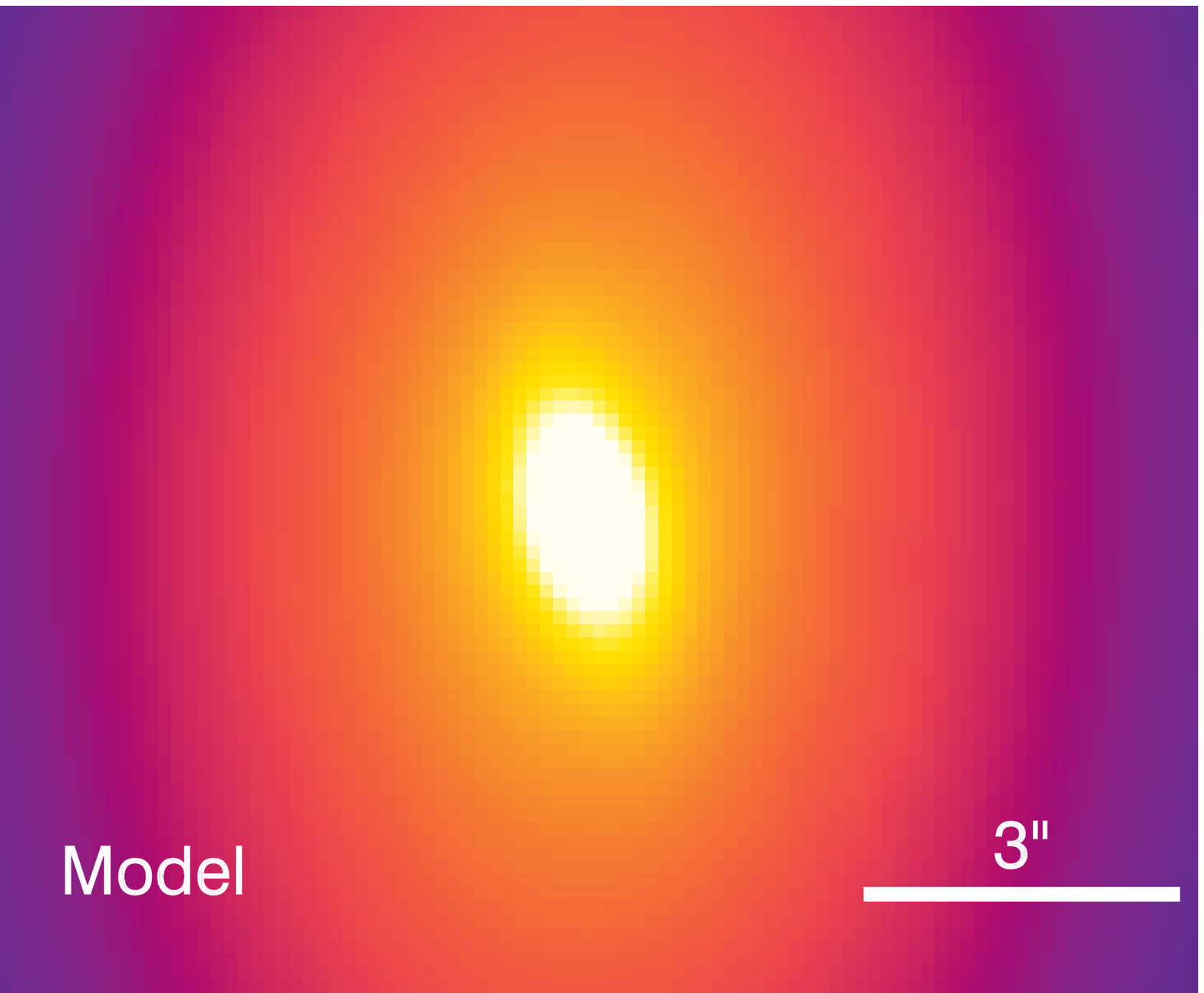}
    \includegraphics[width=0.3\textwidth]{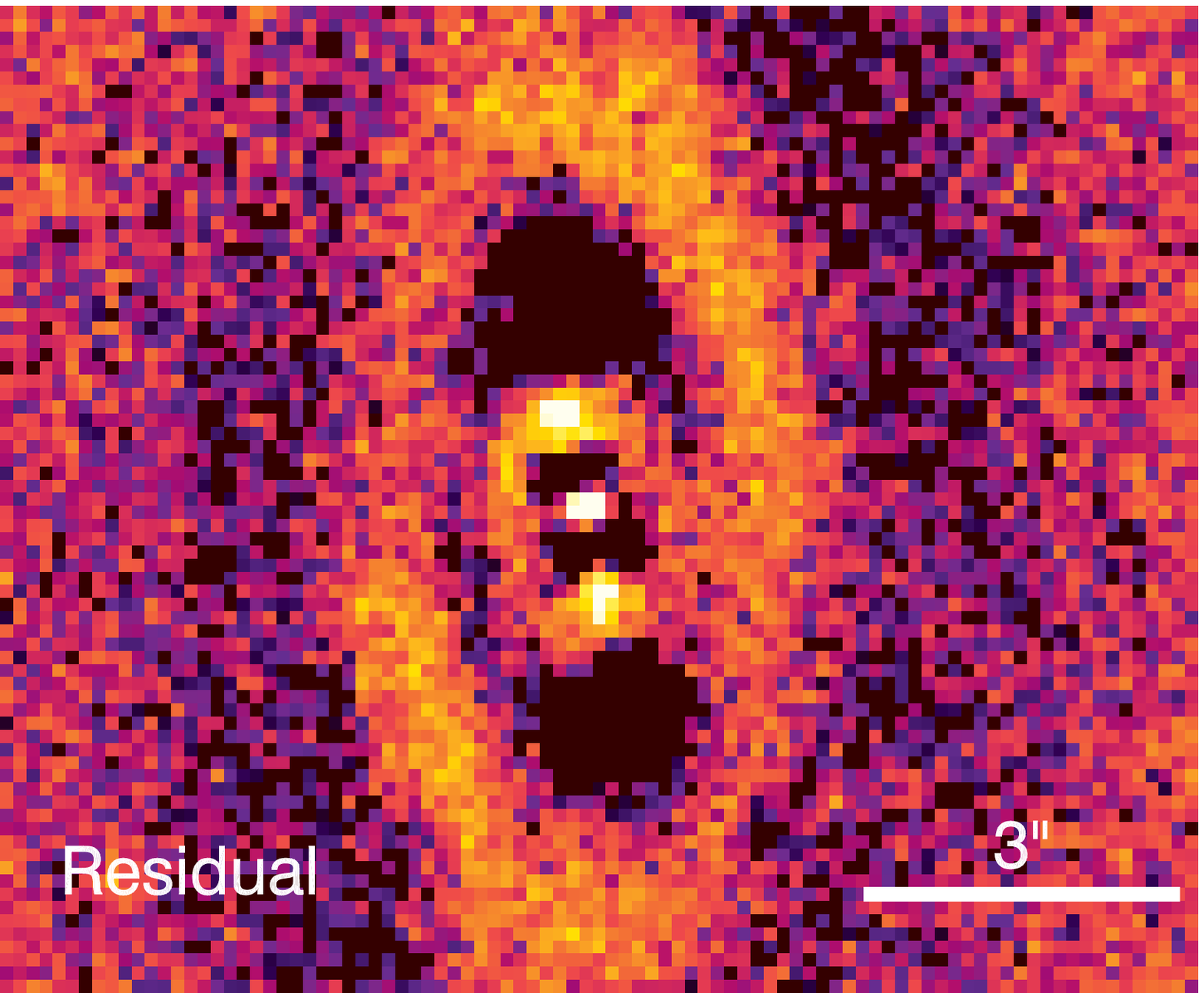}}
    \subfigure{
    \includegraphics[width=0.3\textwidth]{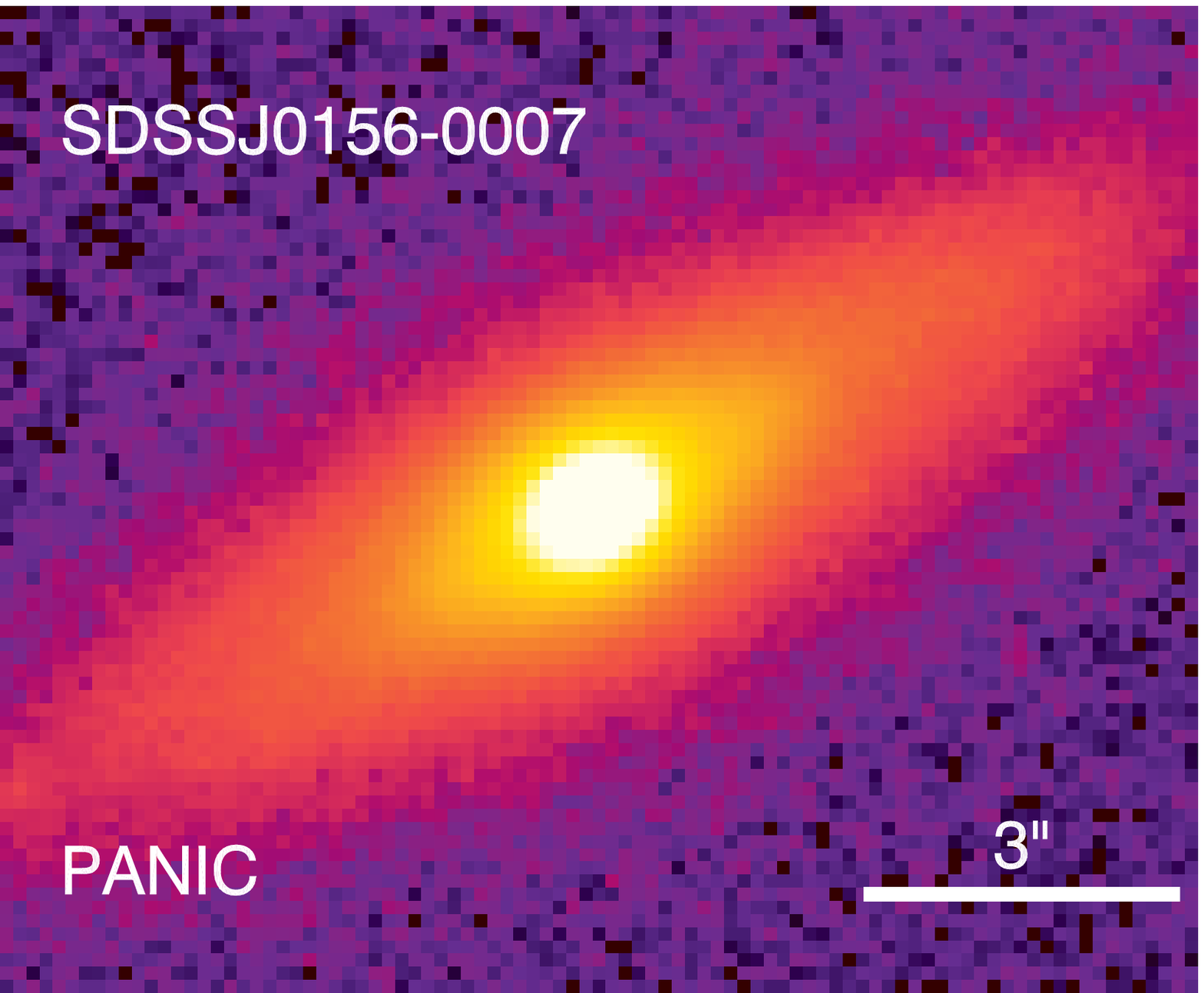}
    \includegraphics[width=0.3\textwidth]{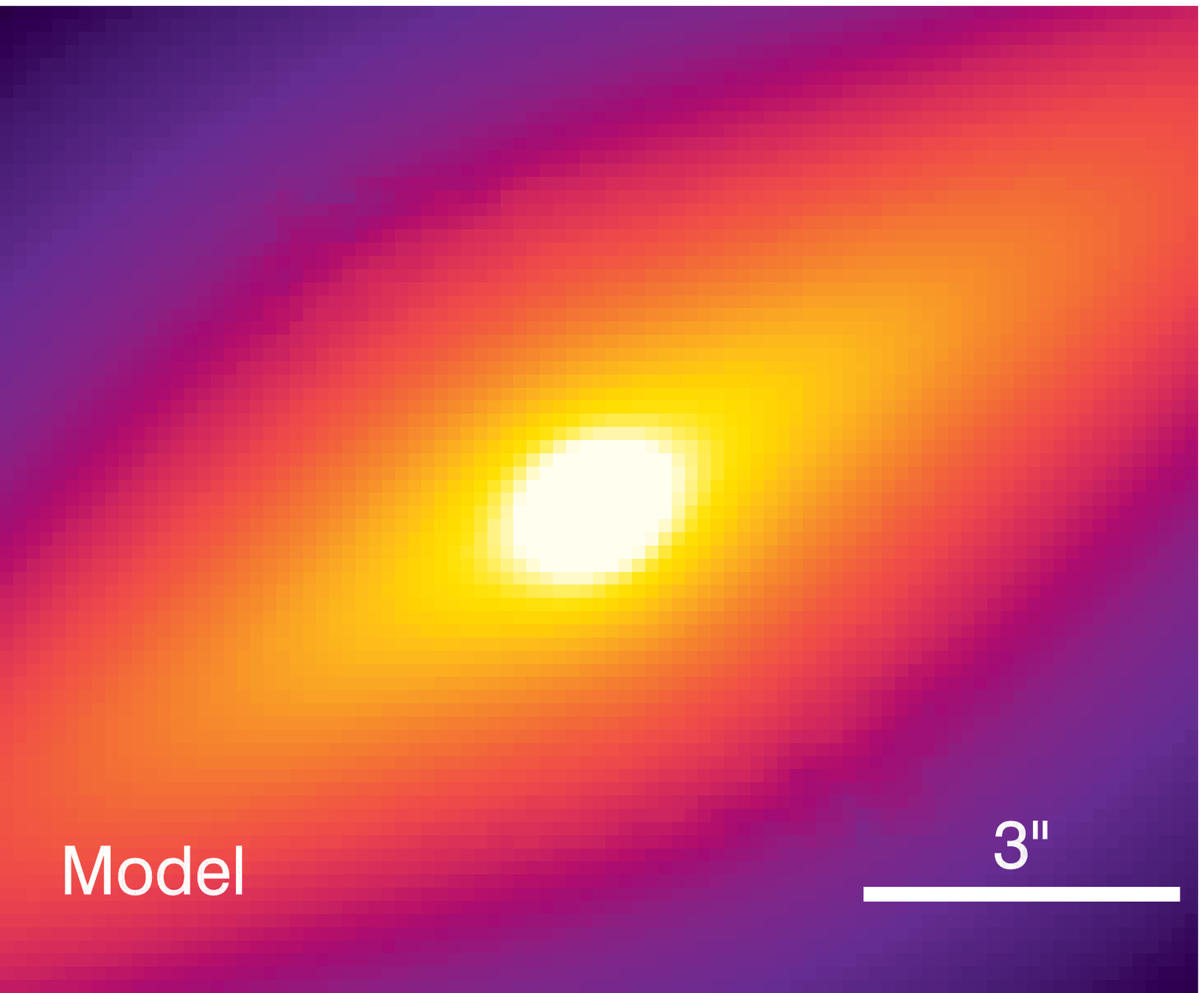}
    \includegraphics[width=0.3\textwidth]{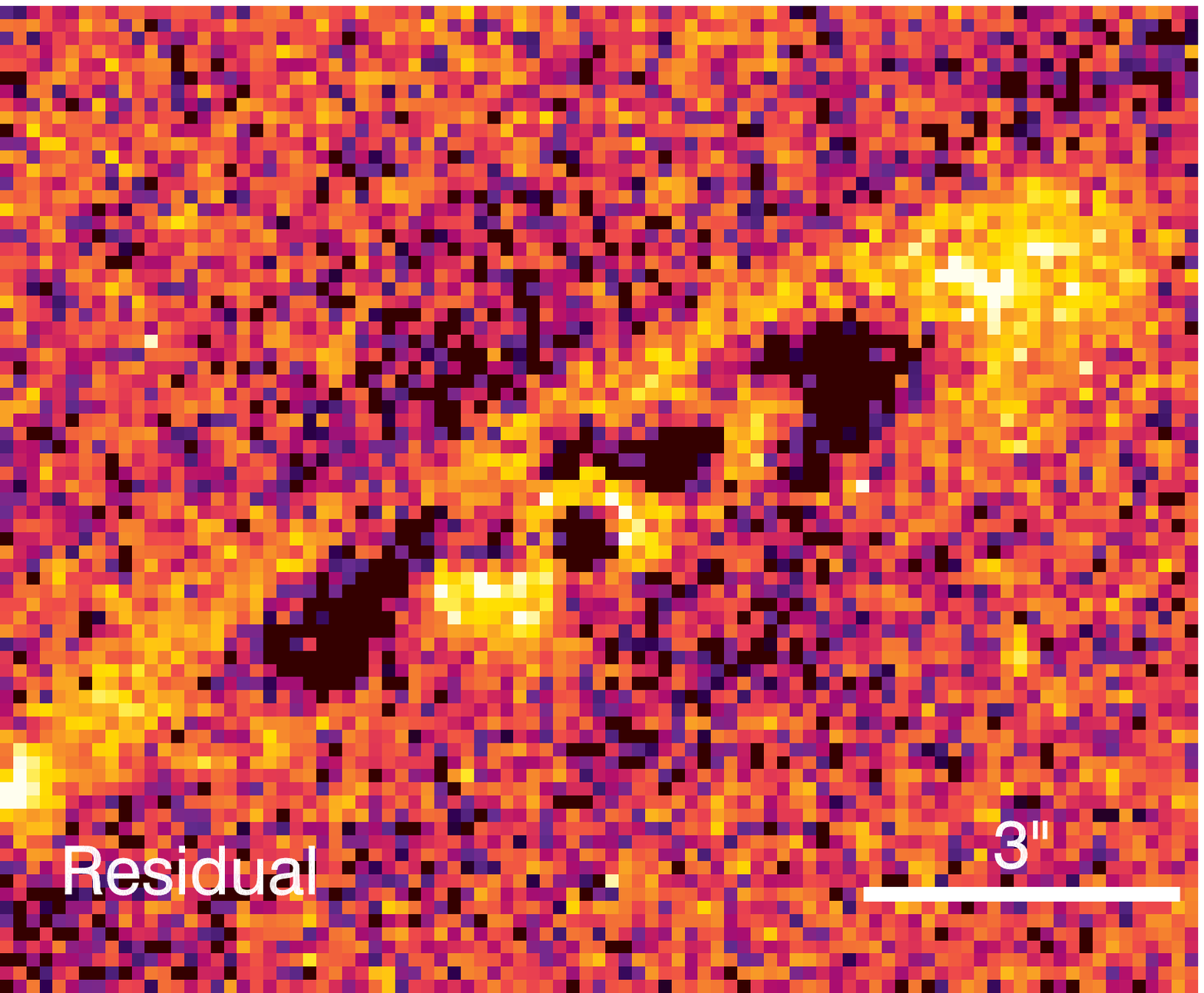}}
    \caption{\footnotesize Model fits of the NIR surface brightness for objects classified as NLR kinematics of single AGNs (\S\ref{sec:kinematics})
    using {\sevenrm GALFIT}. The first column shows the data, the second column shows the models and the last column shows the residuals. Note that
    the residual maps have been re-stretched to enhance the contrast and the residuals are not important compared to the observed fluxes (the
    apparent
    excess seen in some of the residual maps has values less than $5\%$ of the
    observed fluxes at the corresponding locations). }
    \label{fig:NLR_galfit}
\end{figure*}

\addtocounter{figure}{-1}
\begin{figure*}
\addtocounter{subfigure}{1}
  \centering
    \subfigure{
    \includegraphics[width=0.3\textwidth]{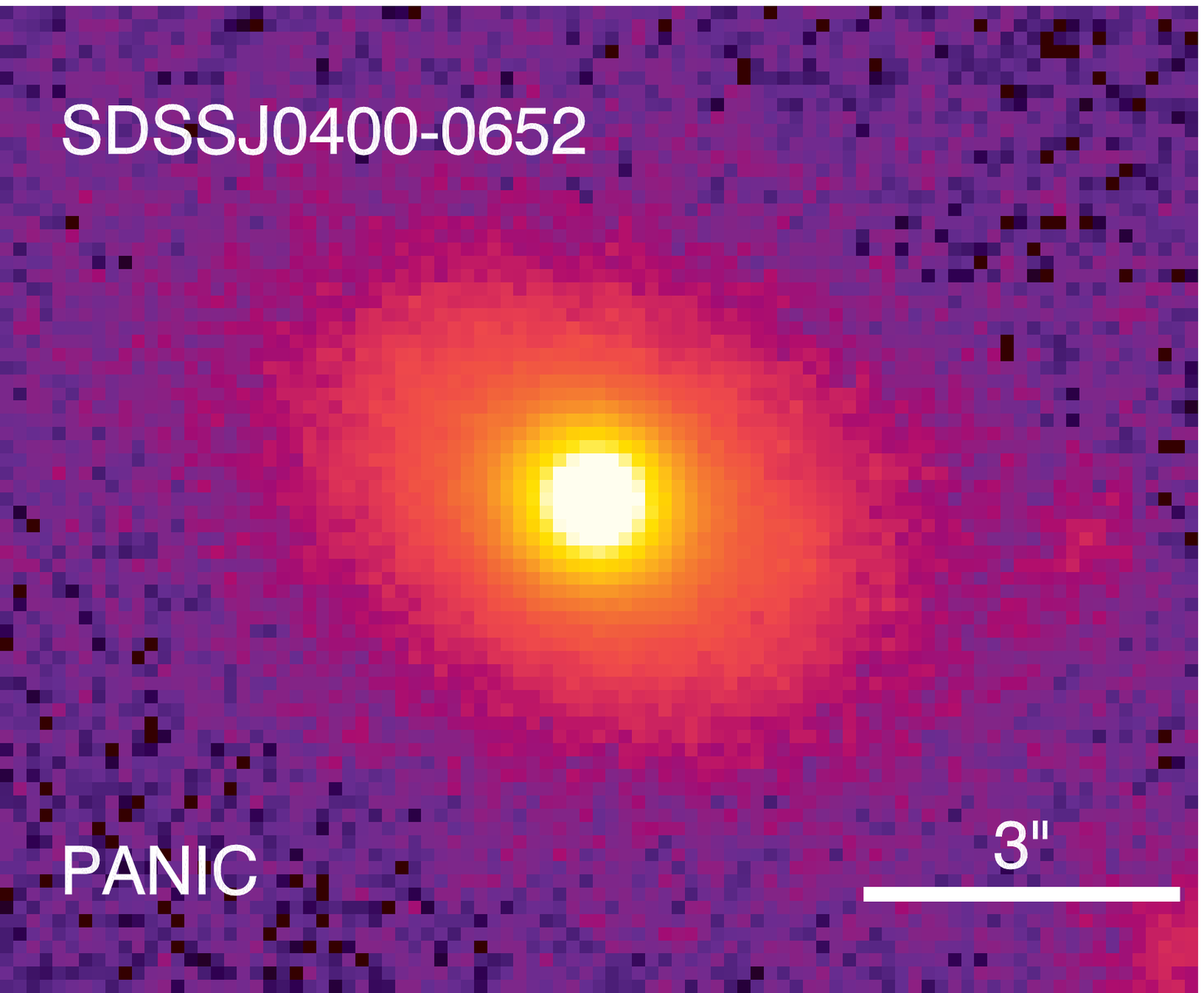}
    \includegraphics[width=0.3\textwidth]{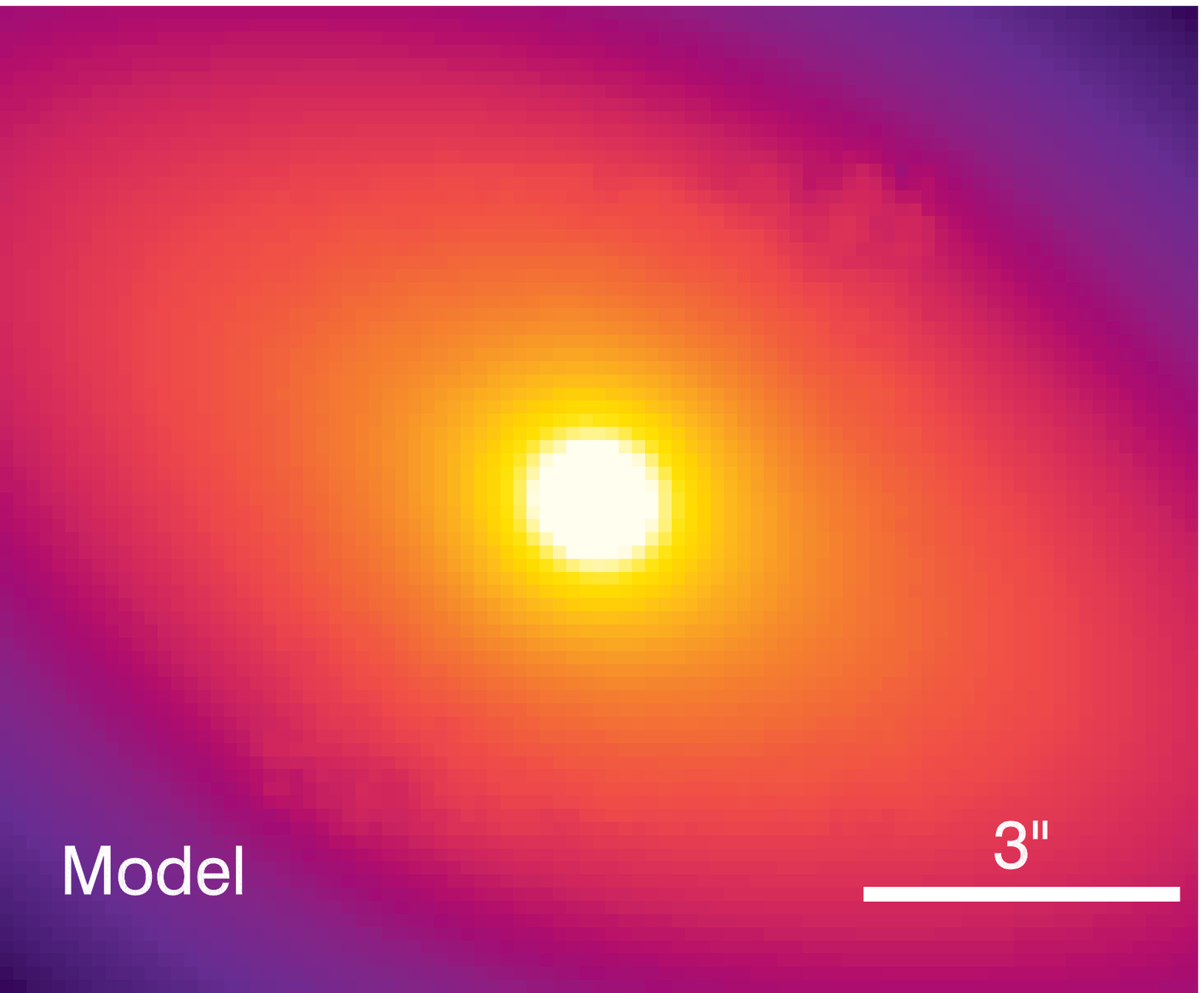}
    \includegraphics[width=0.3\textwidth]{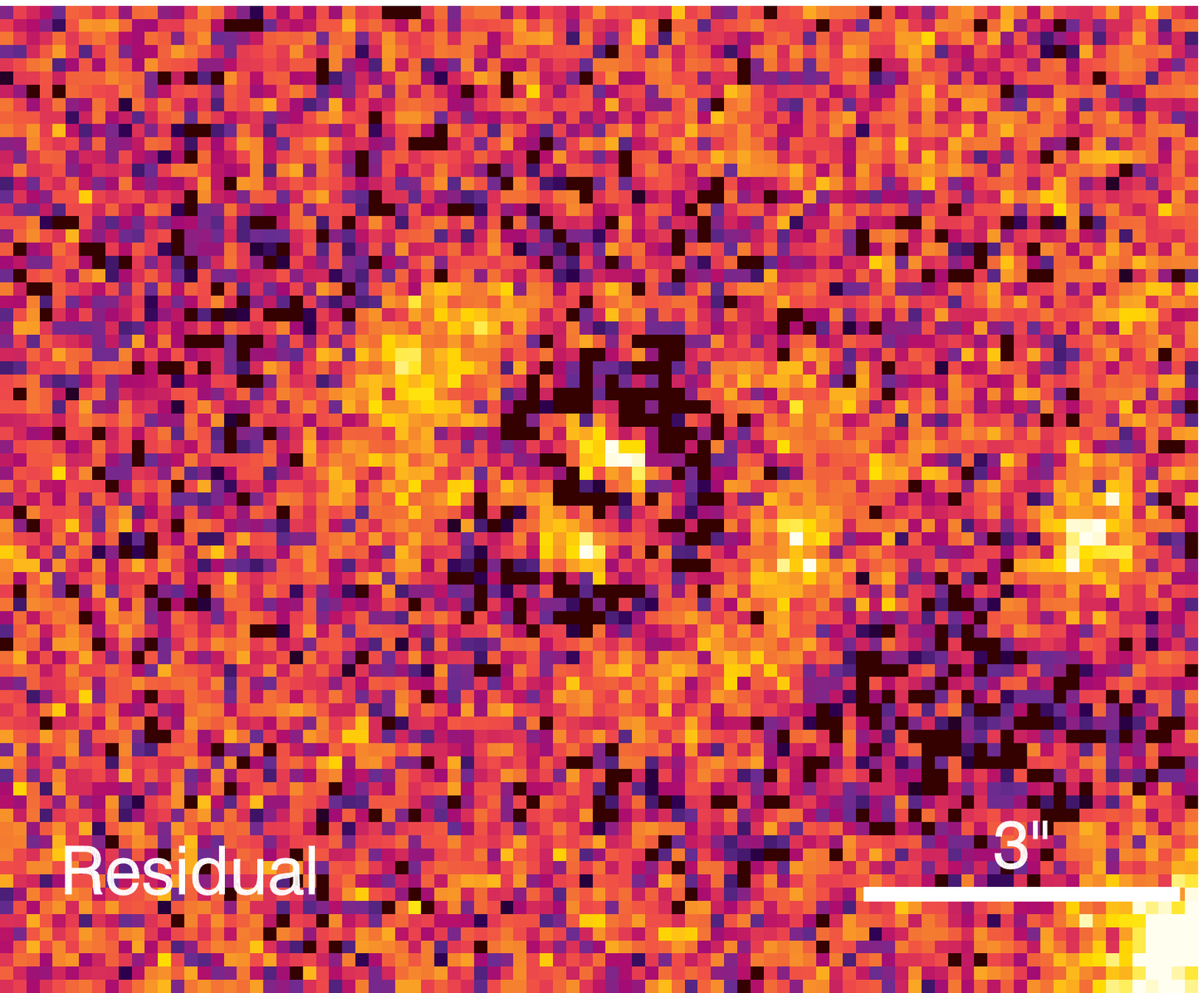}}
    \subfigure{
    \includegraphics[width=0.3\textwidth]{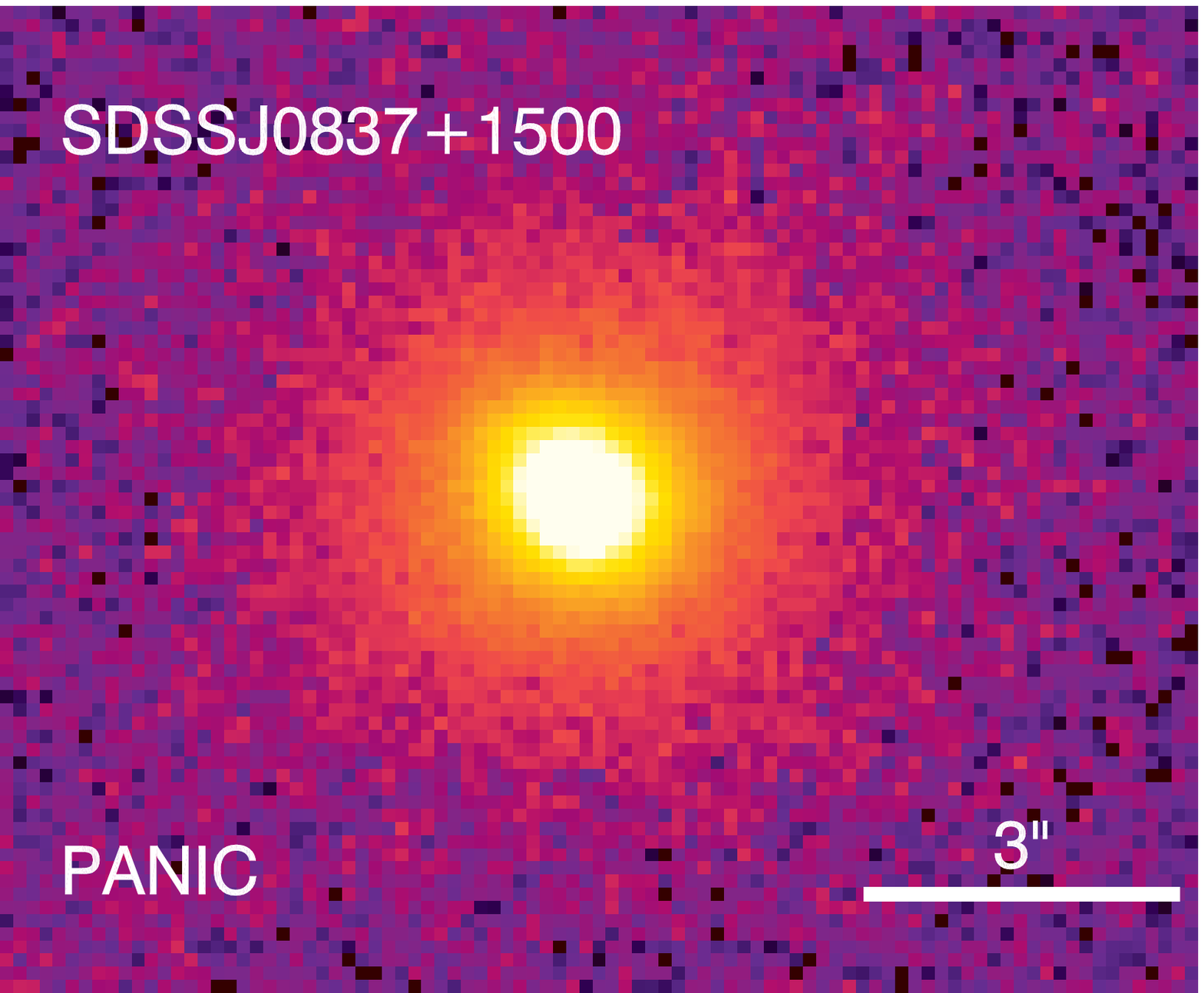}
    \includegraphics[width=0.3\textwidth]{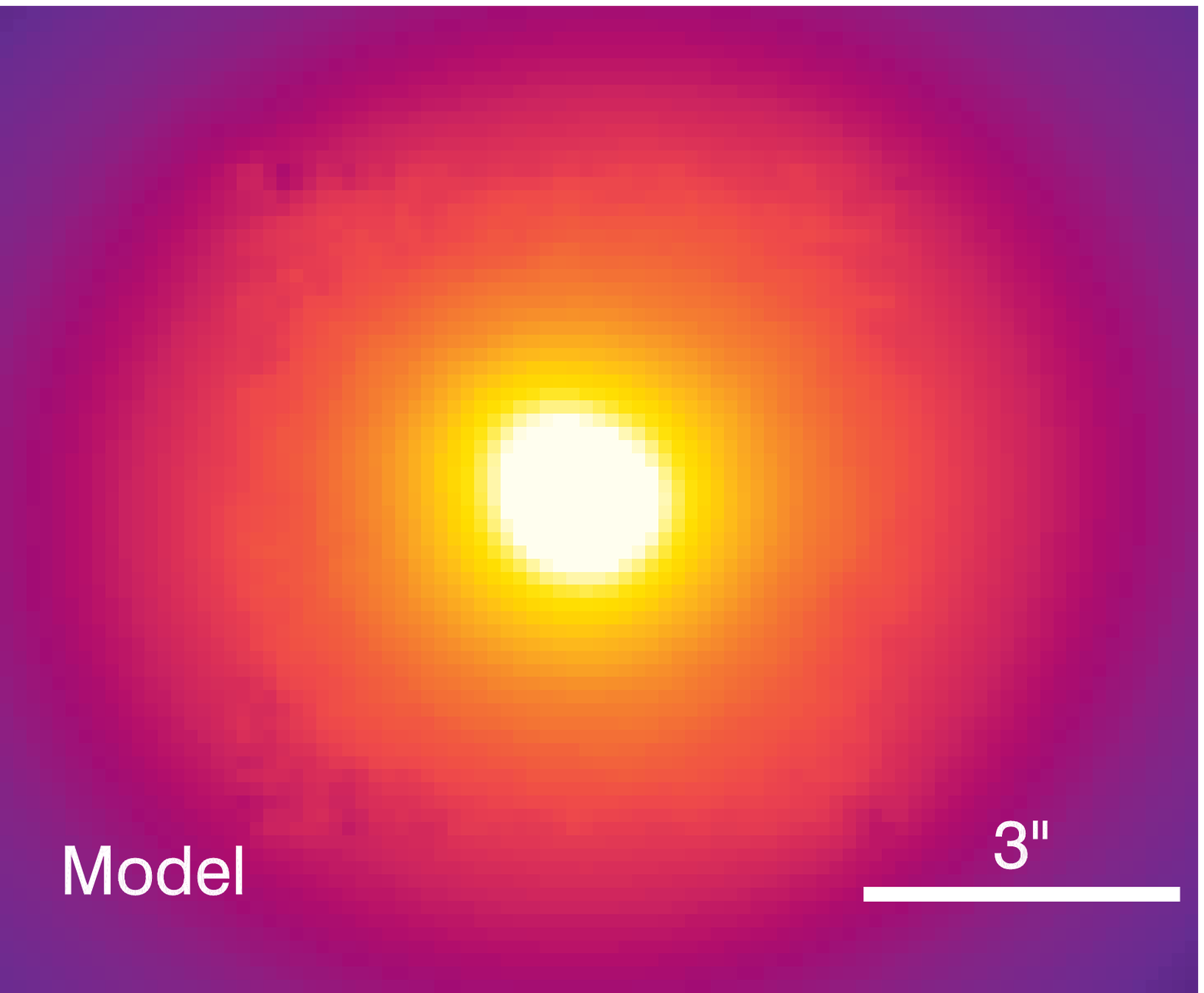}
    \includegraphics[width=0.3\textwidth]{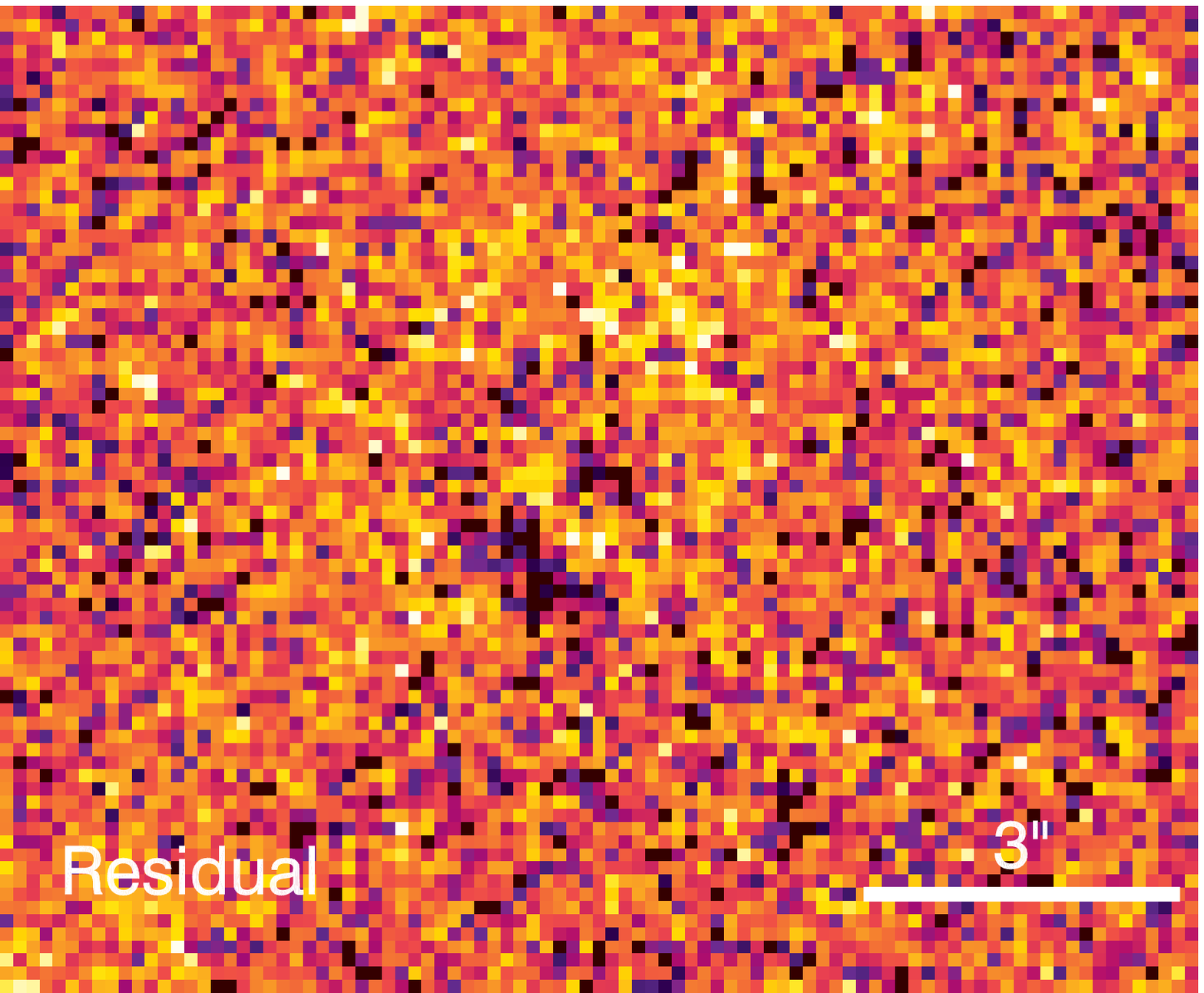}}
    \subfigure{
    \includegraphics[width=0.3\textwidth]{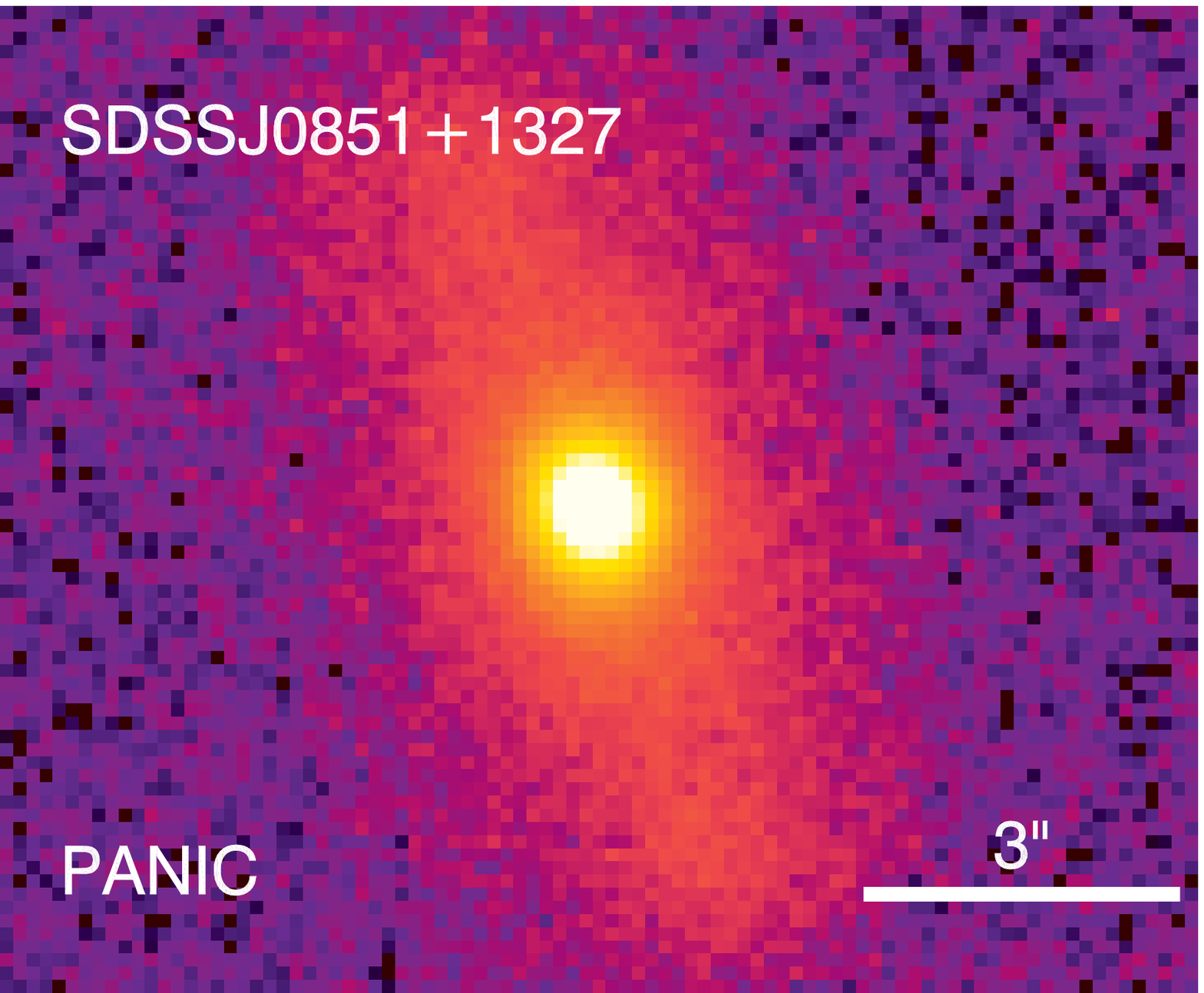}
    \includegraphics[width=0.3\textwidth]{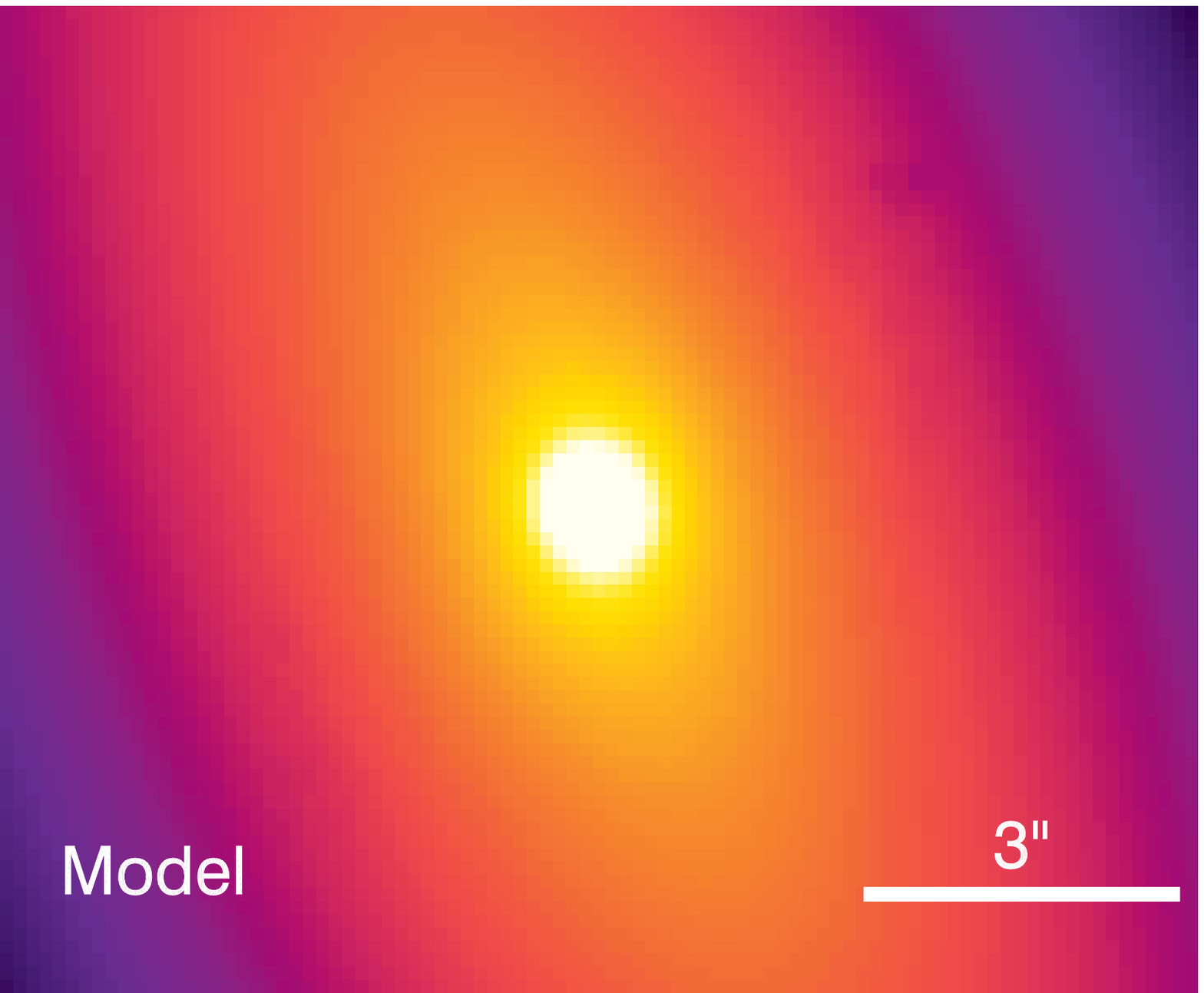}
    \includegraphics[width=0.3\textwidth]{0116-1025_galfit_residual.eps}}
    \subfigure{
    \includegraphics[width=0.3\textwidth]{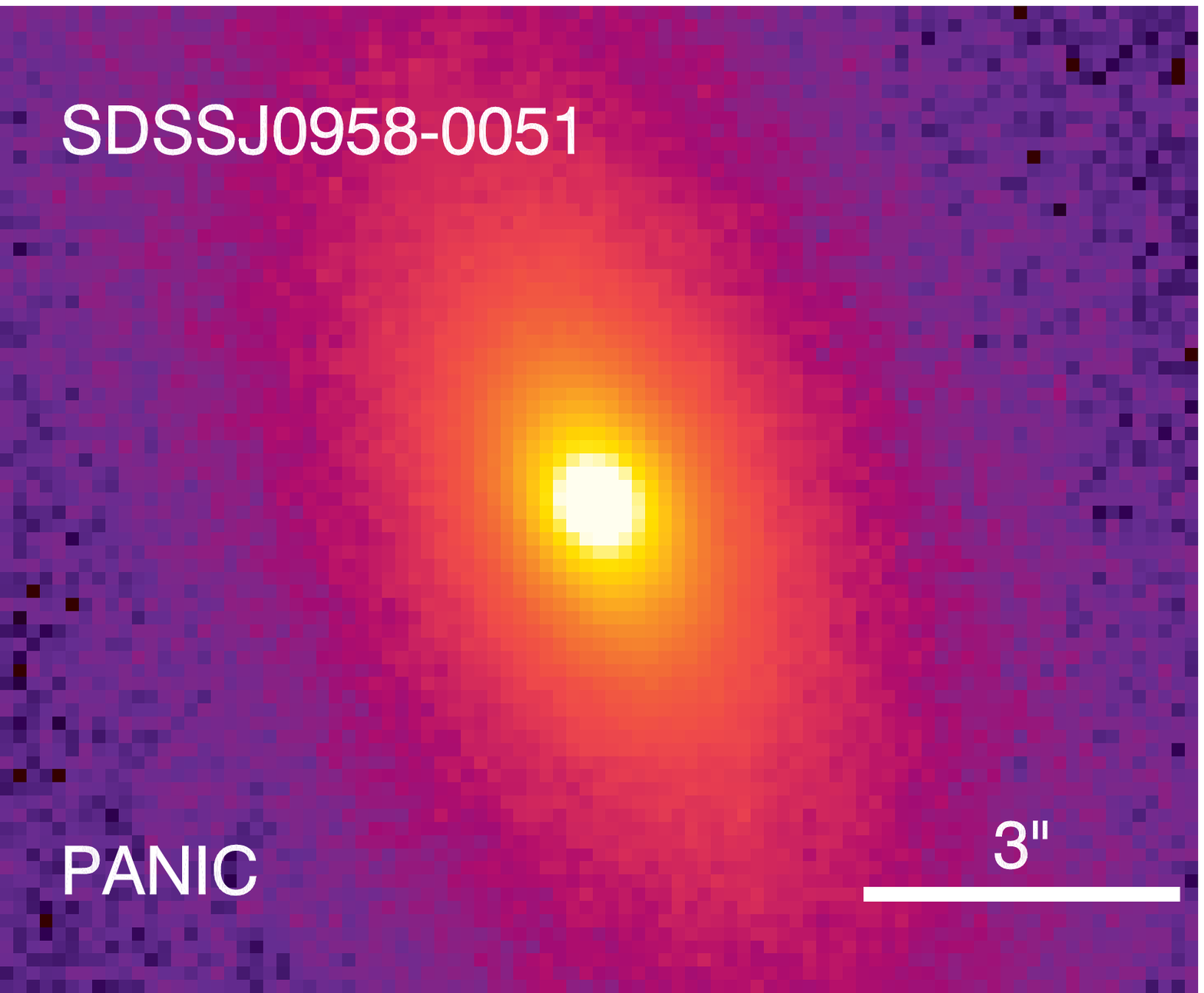}
    \includegraphics[width=0.3\textwidth]{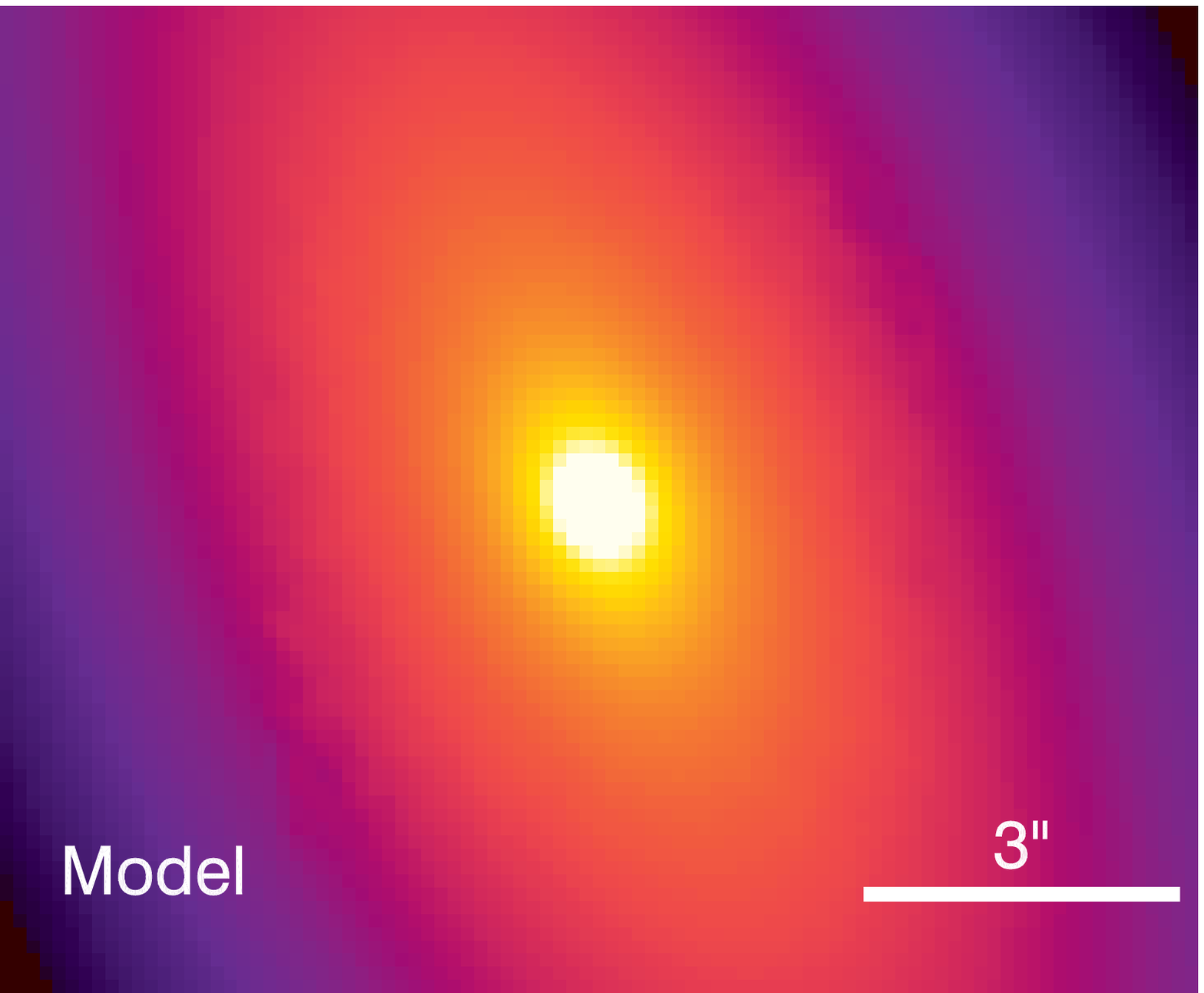}
    \includegraphics[width=0.3\textwidth]{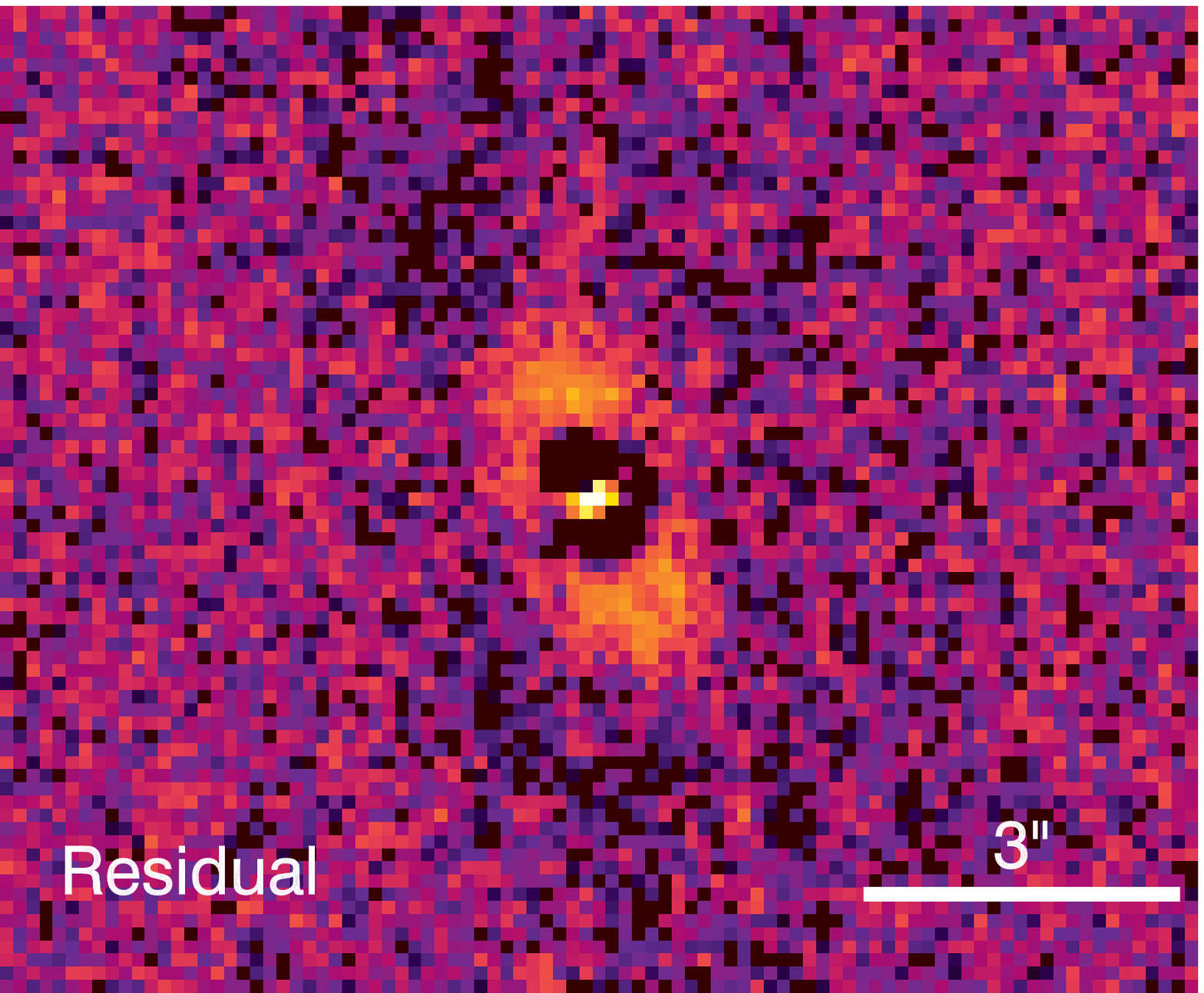}}
    \subfigure{
    \includegraphics[width=0.3\textwidth]{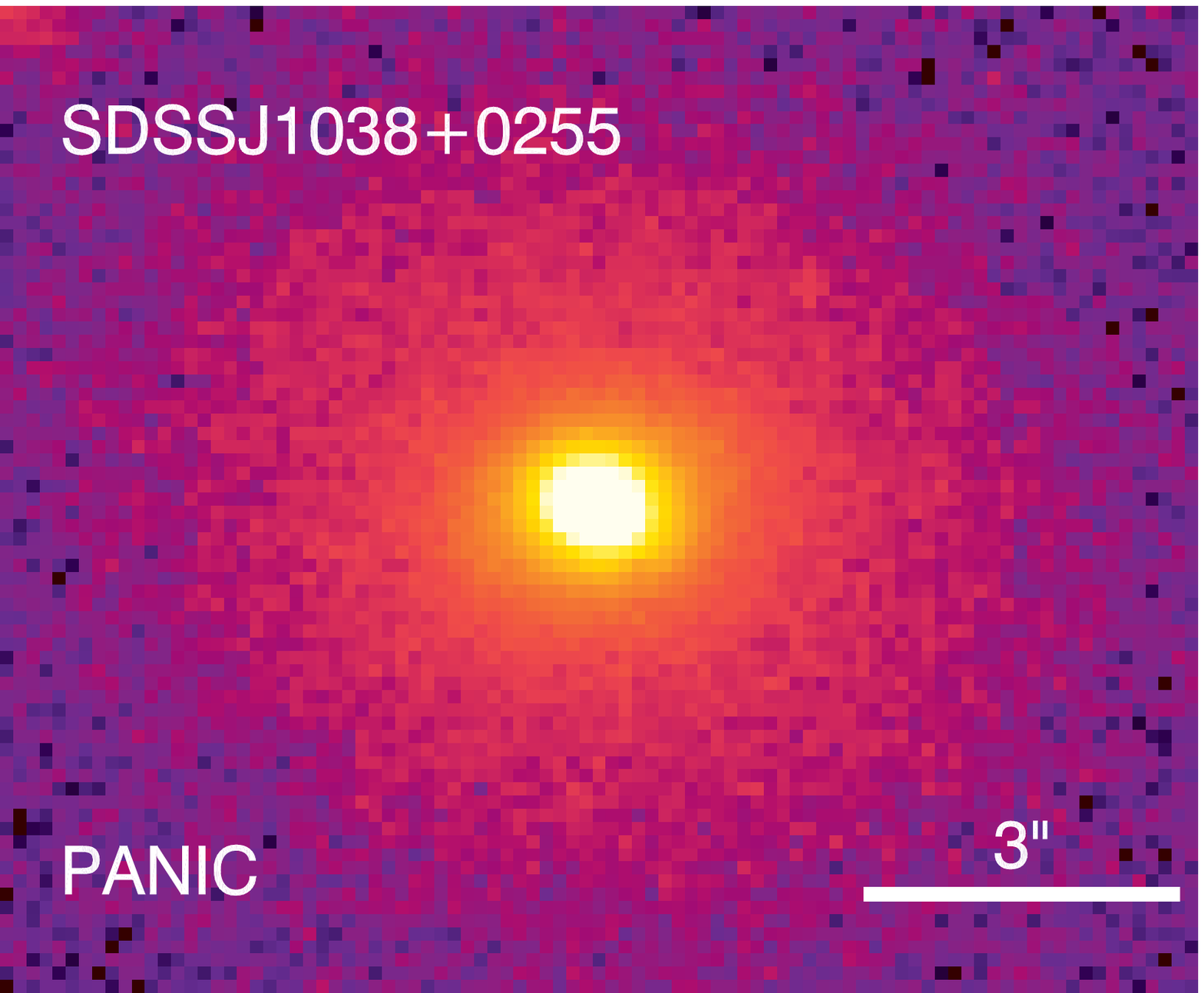}
    \includegraphics[width=0.3\textwidth]{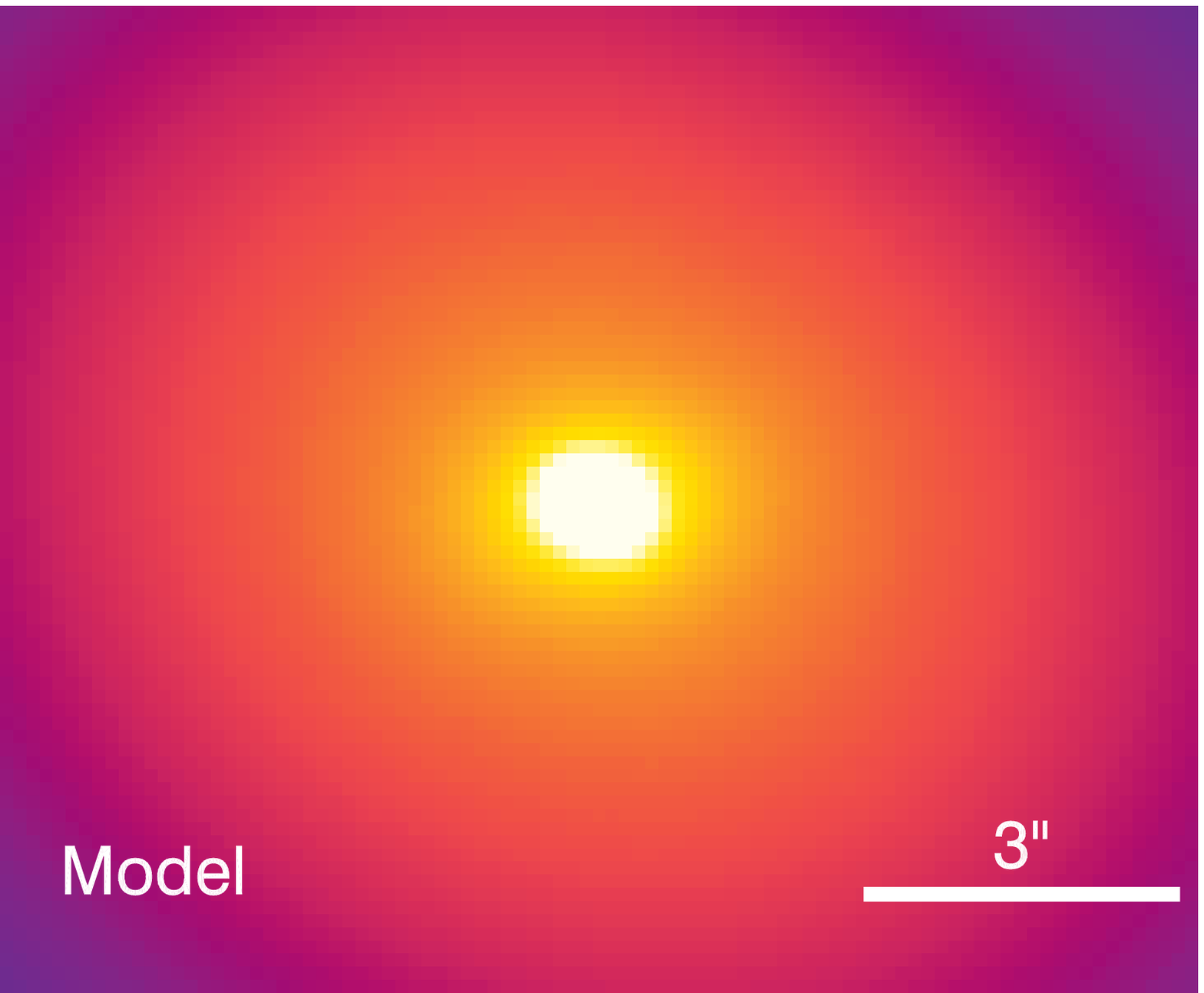}
    \includegraphics[width=0.3\textwidth]{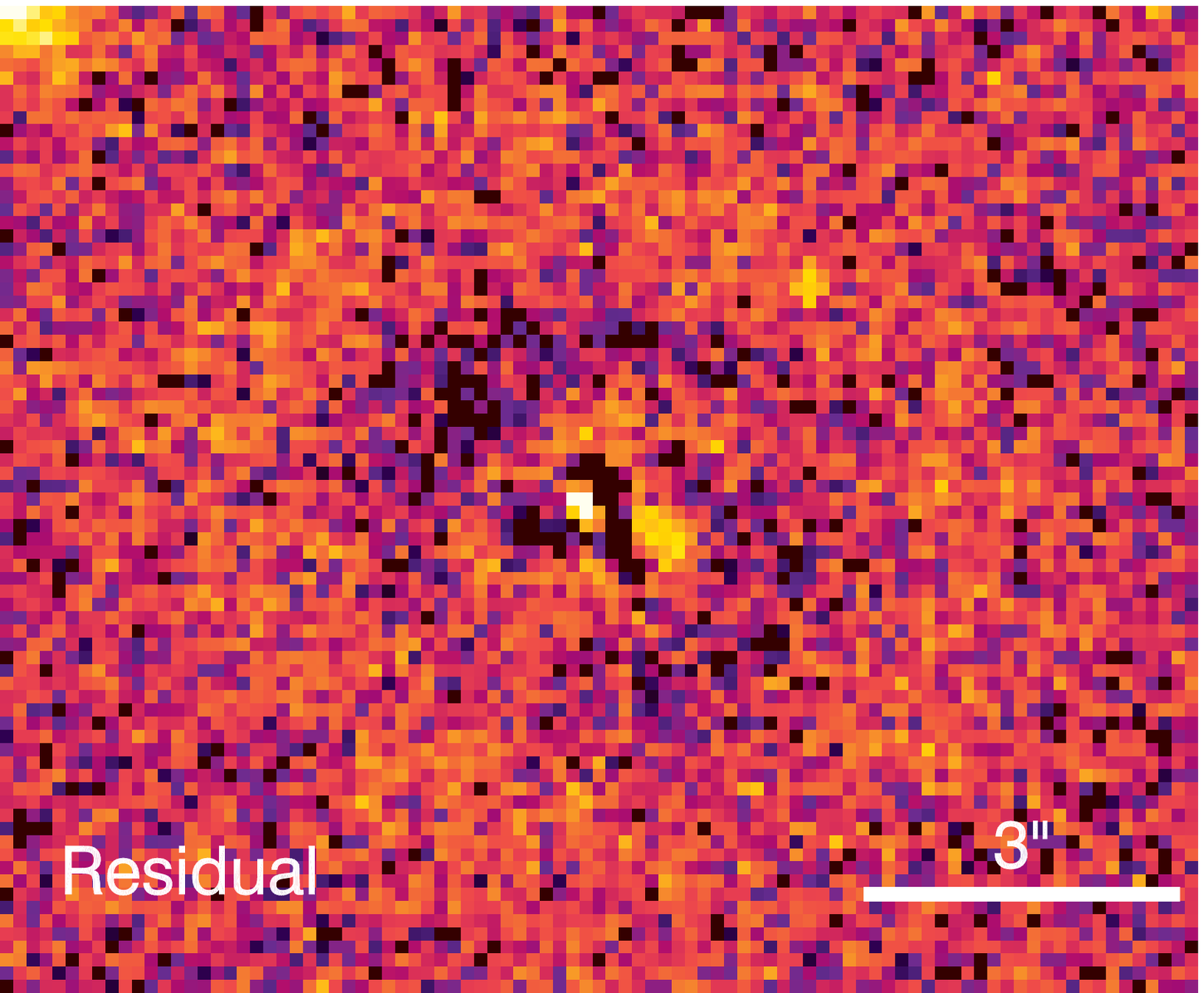}}
\caption{Continued. }
    \label{fig:NLR_galfit}
\end{figure*}

\addtocounter{figure}{-1}
\begin{figure*}
\addtocounter{subfigure}{1}
  \centering
    \subfigure{
    \includegraphics[width=0.3\textwidth]{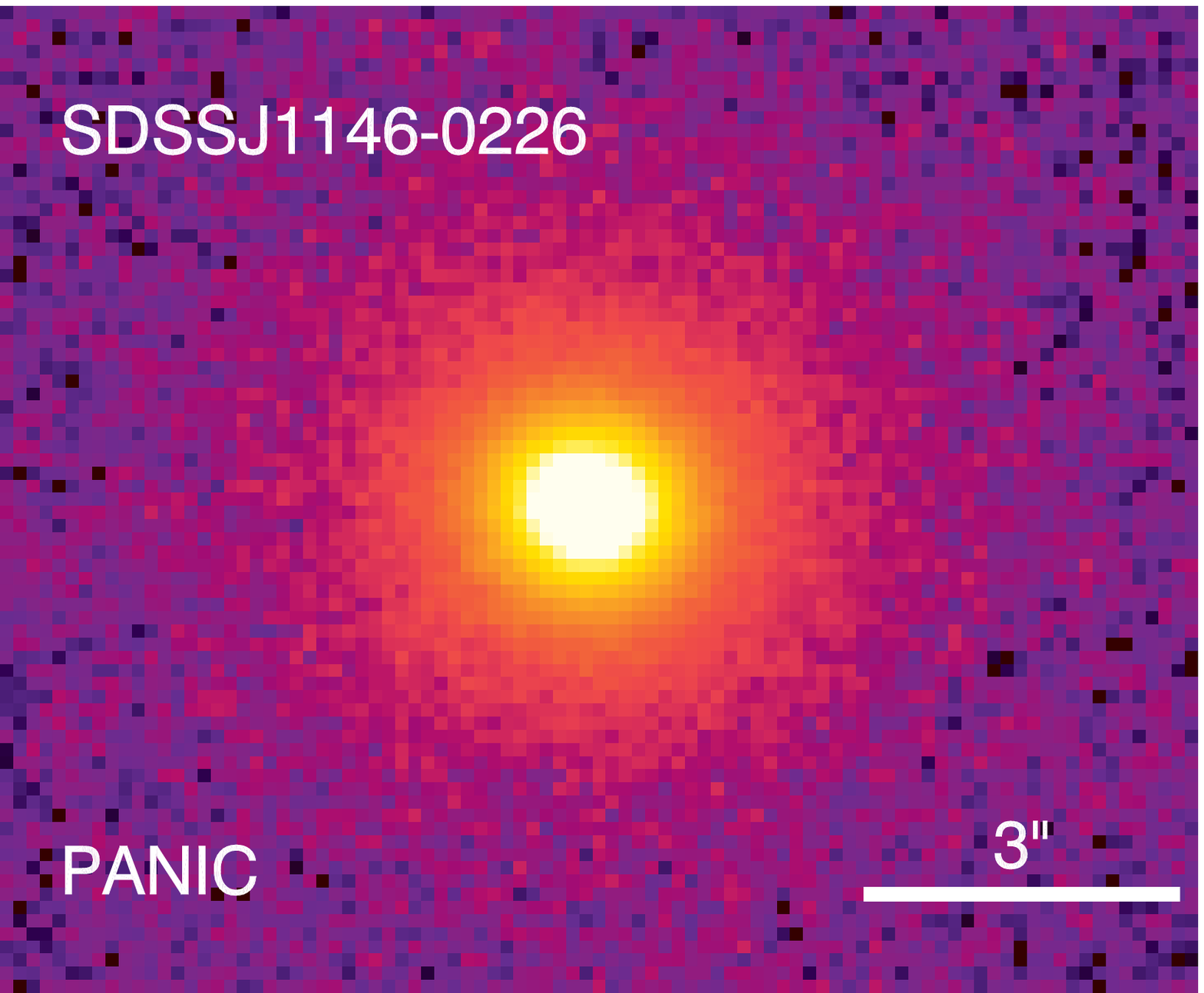}
    \includegraphics[width=0.3\textwidth]{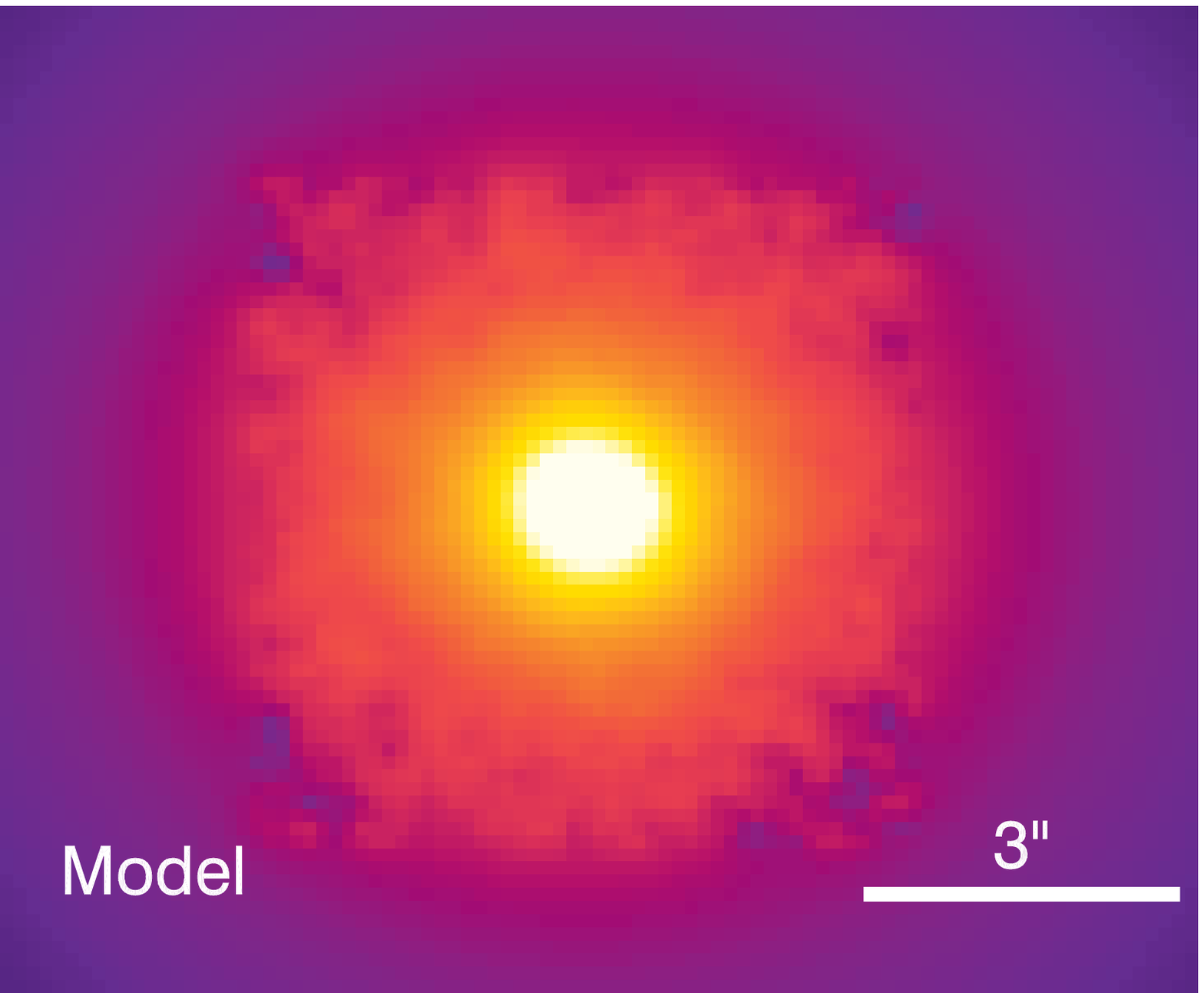}
    \includegraphics[width=0.3\textwidth]{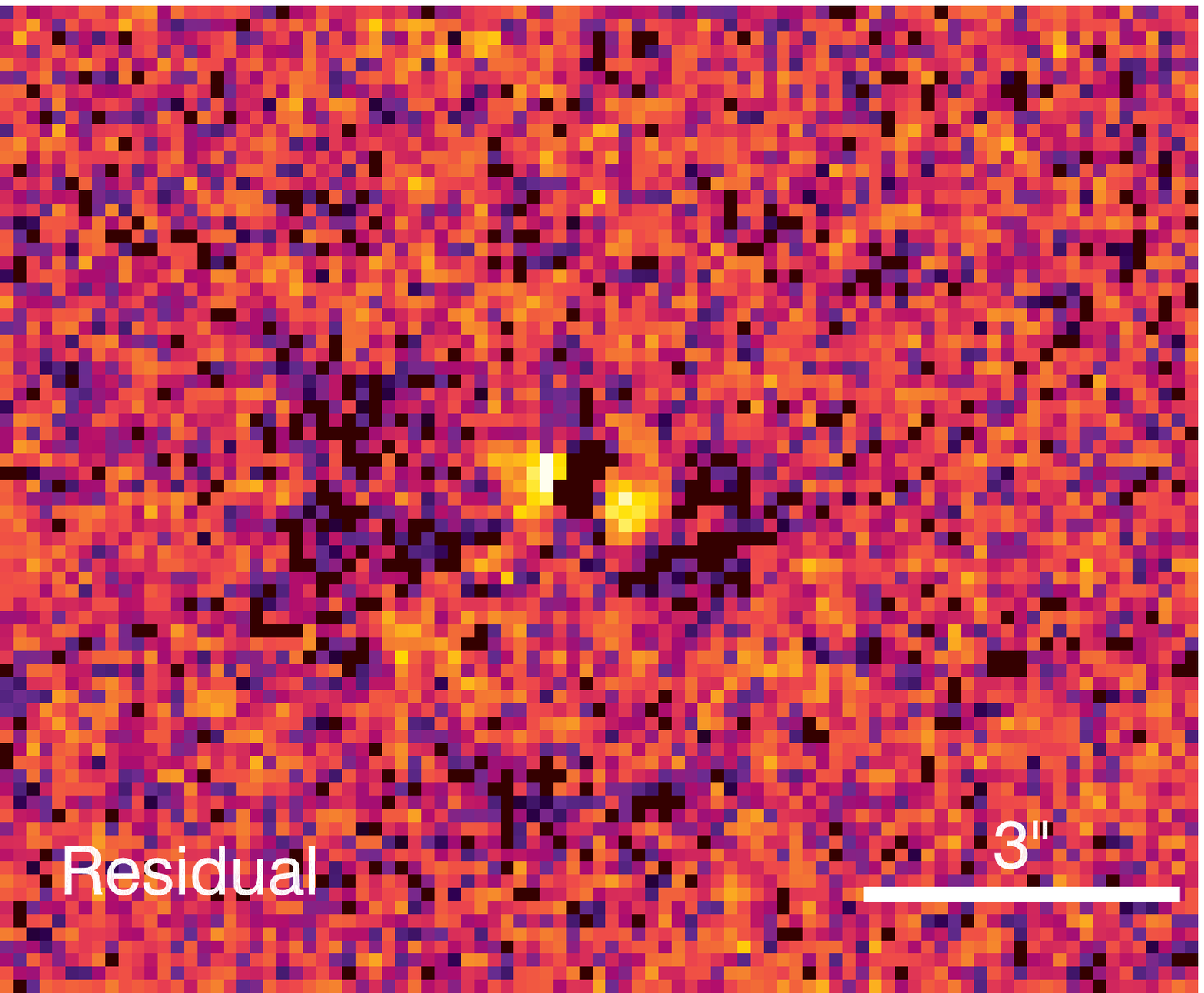}}
    \subfigure{
    \includegraphics[width=0.3\textwidth]{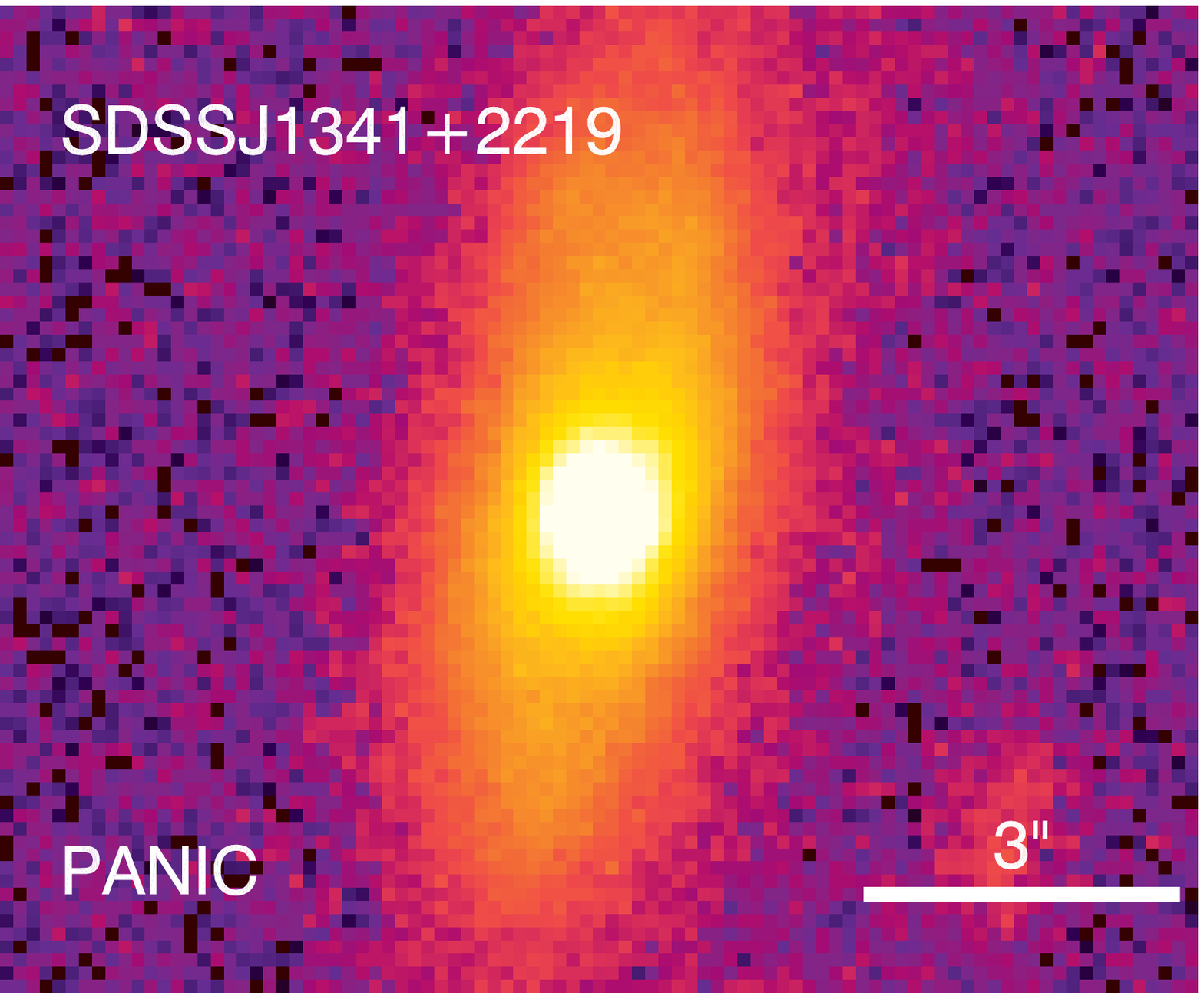}
    \includegraphics[width=0.3\textwidth]{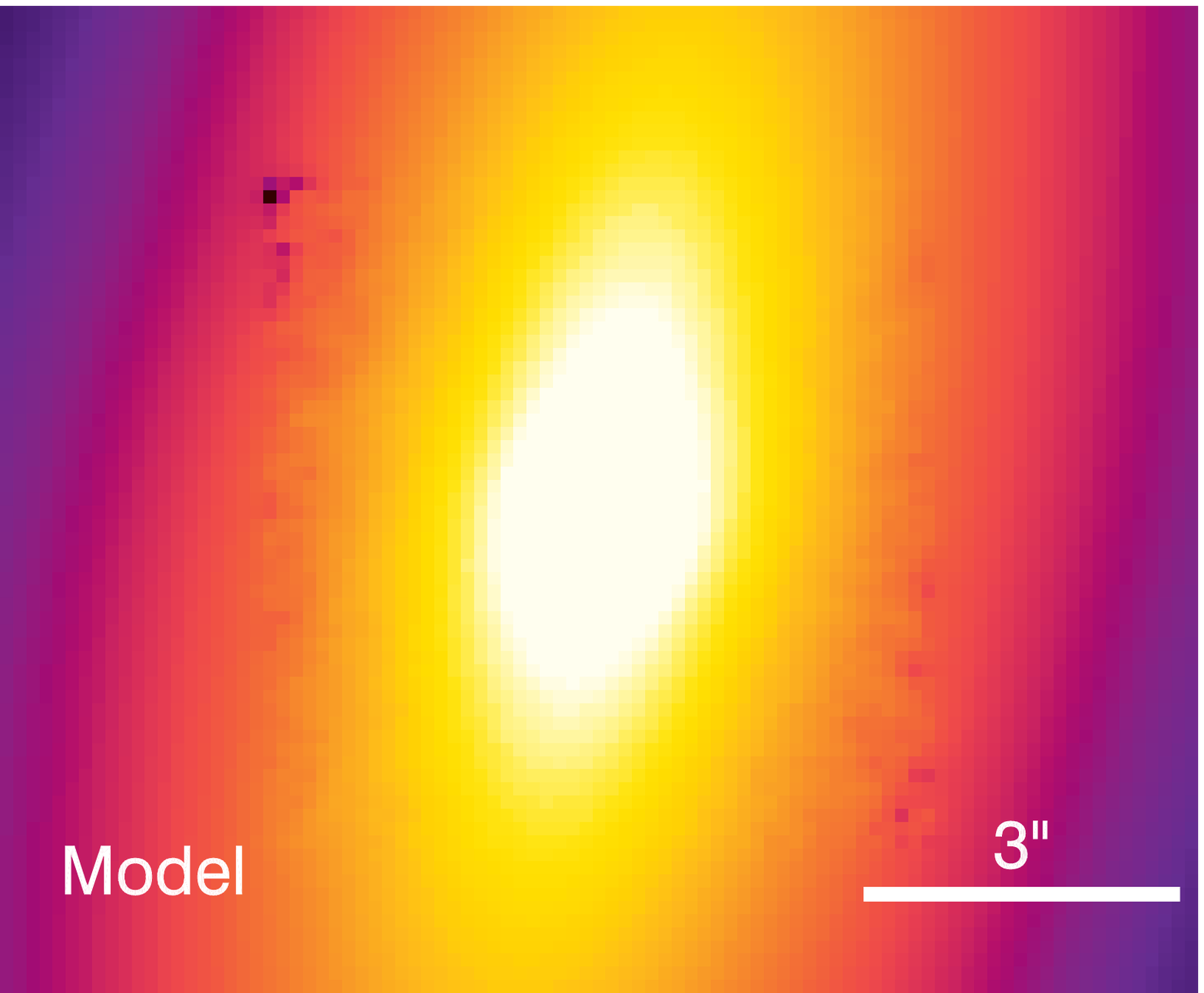}
    \includegraphics[width=0.3\textwidth]{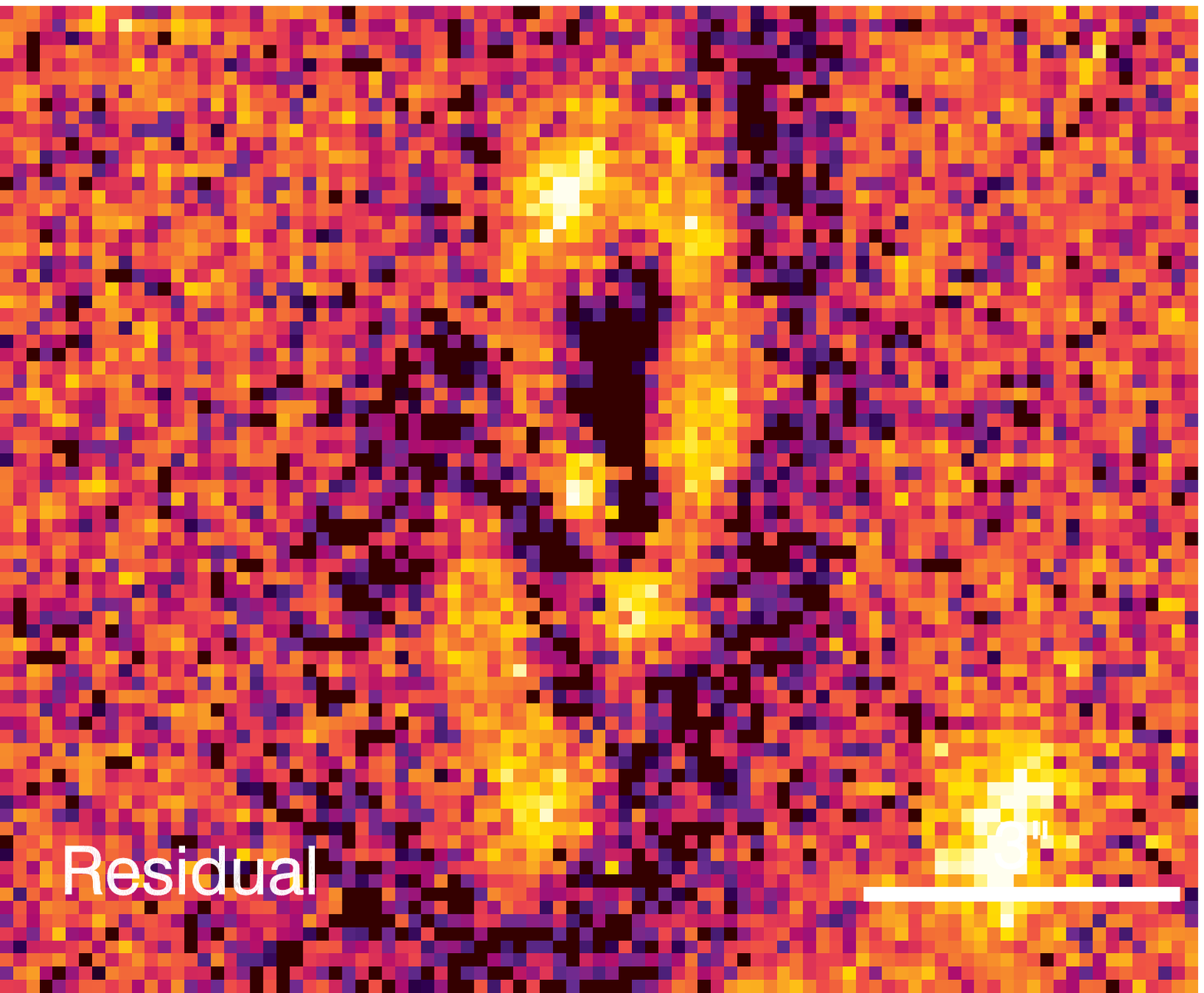}}
    \subfigure{
    \includegraphics[width=0.3\textwidth]{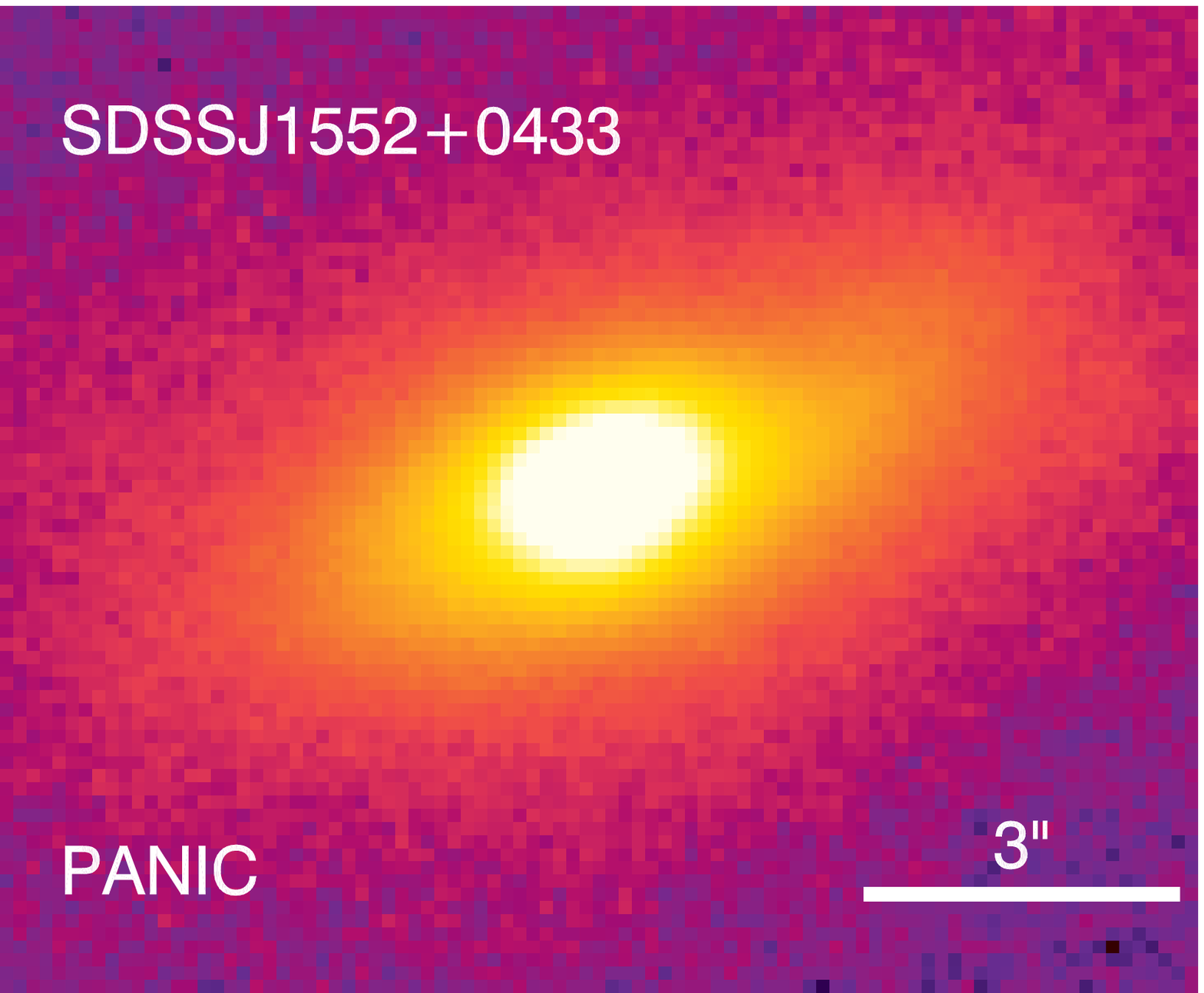}
    \includegraphics[width=0.3\textwidth]{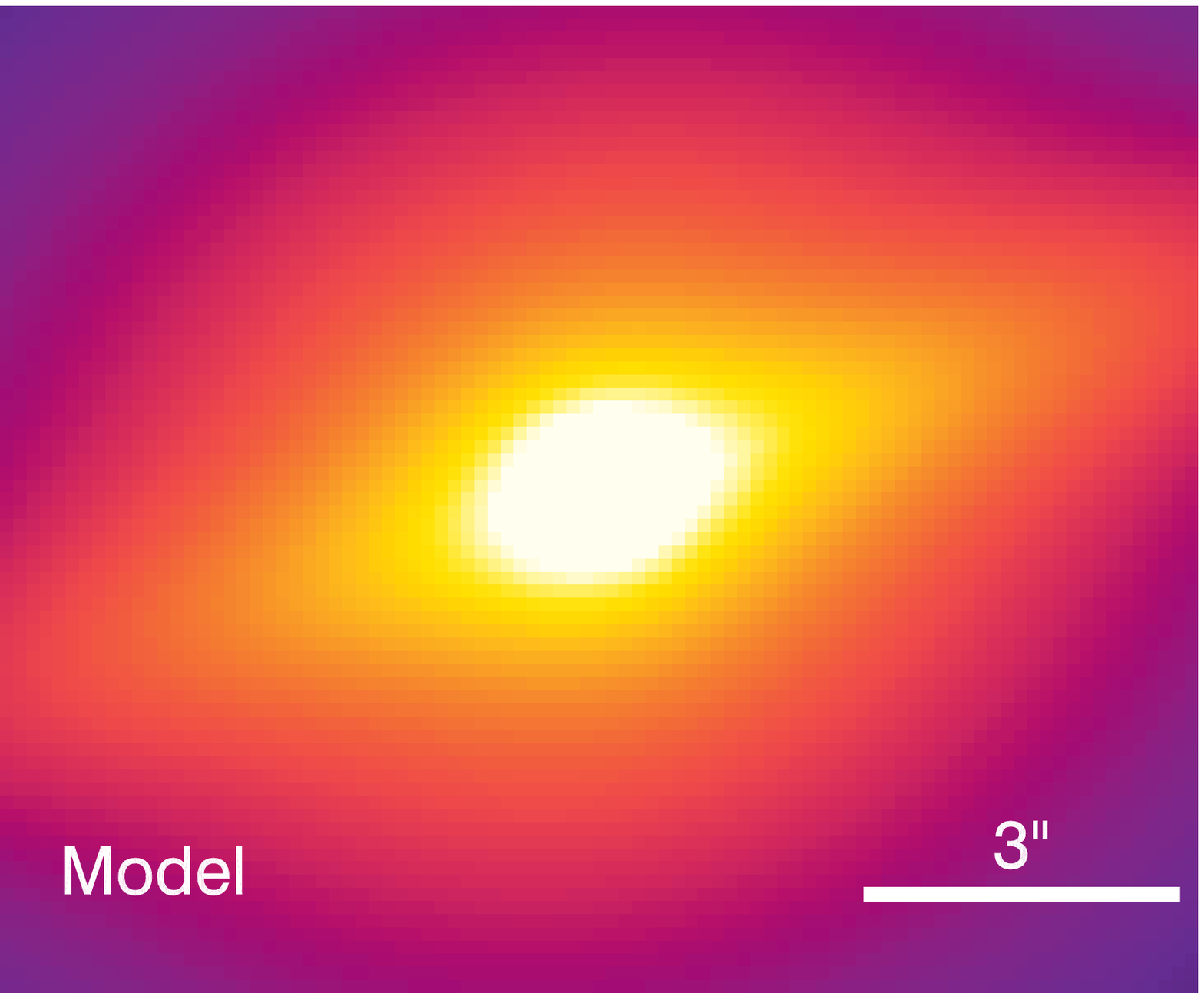}
    \includegraphics[width=0.3\textwidth]{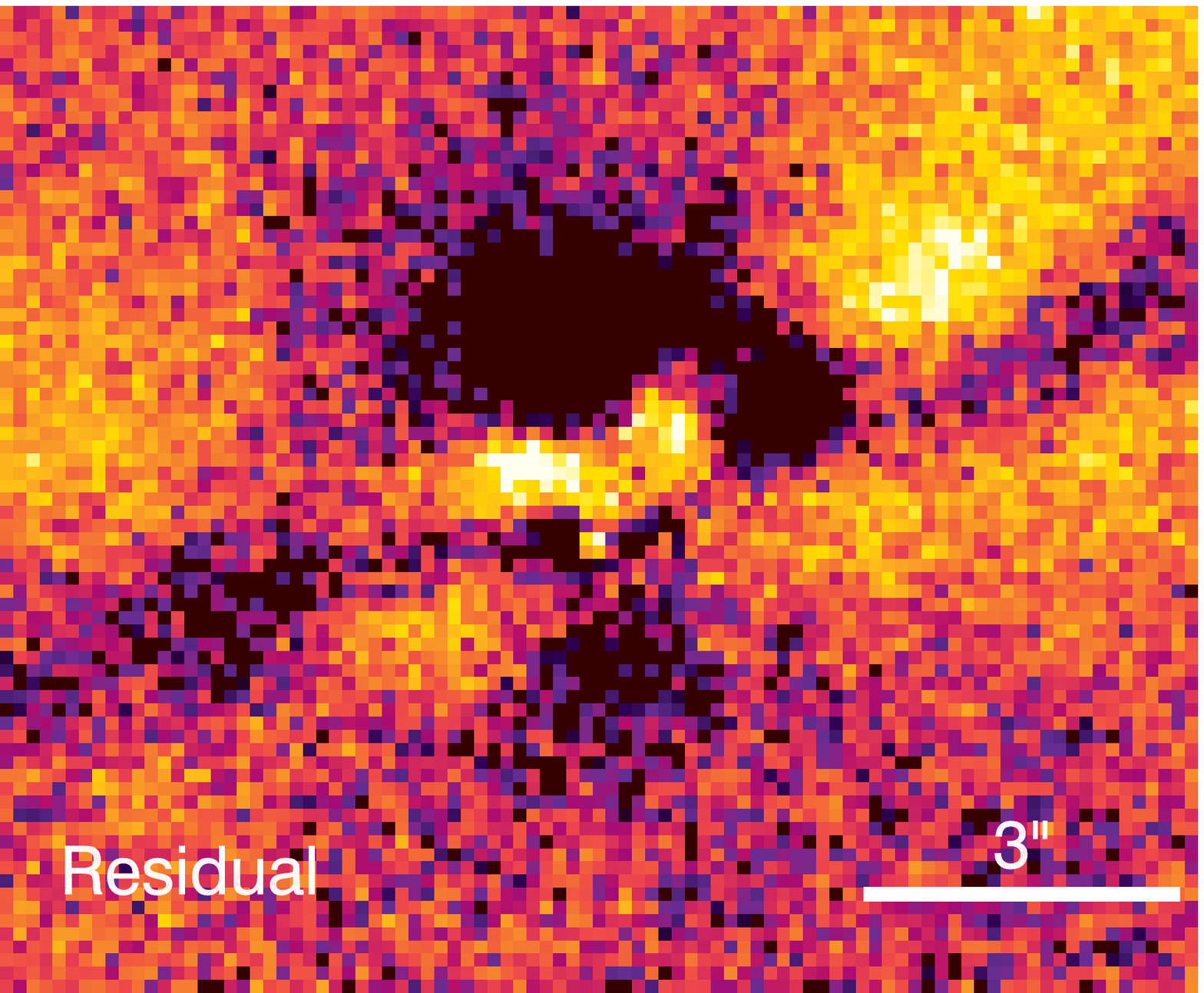}}
    \subfigure{
    \includegraphics[width=0.3\textwidth]{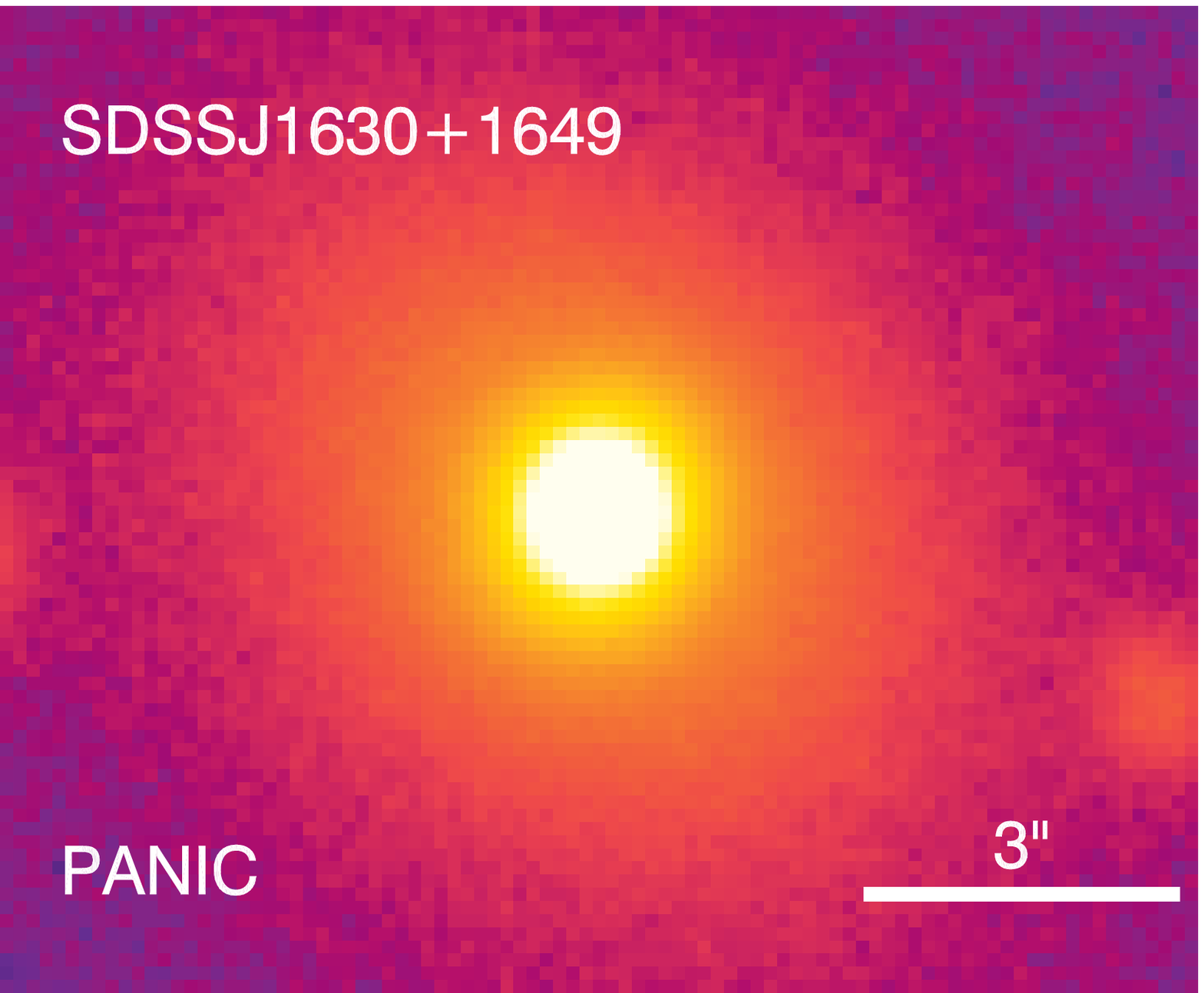}
    \includegraphics[width=0.3\textwidth]{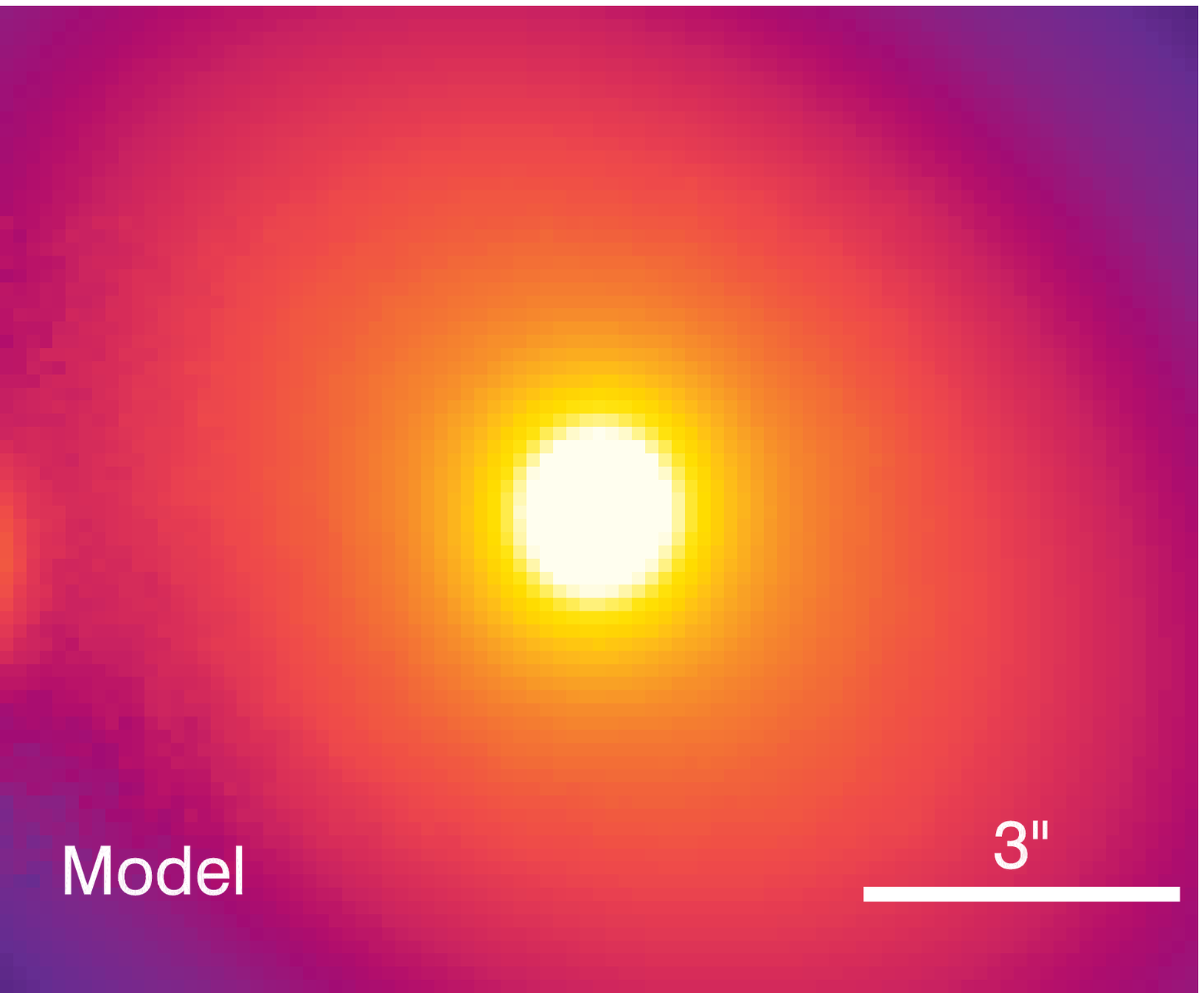}
    \includegraphics[width=0.3\textwidth]{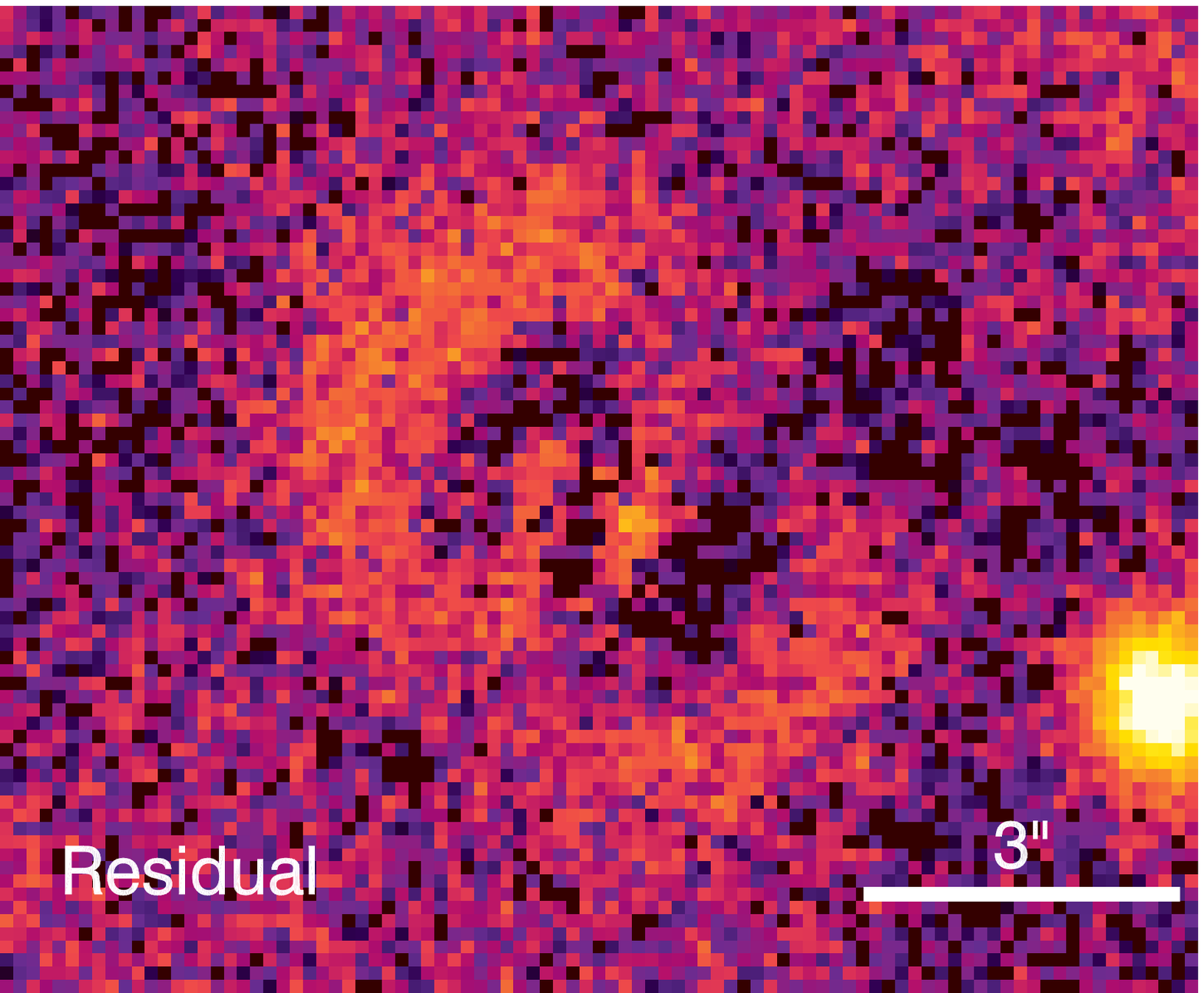}}
    \subfigure{
    \includegraphics[width=0.3\textwidth]{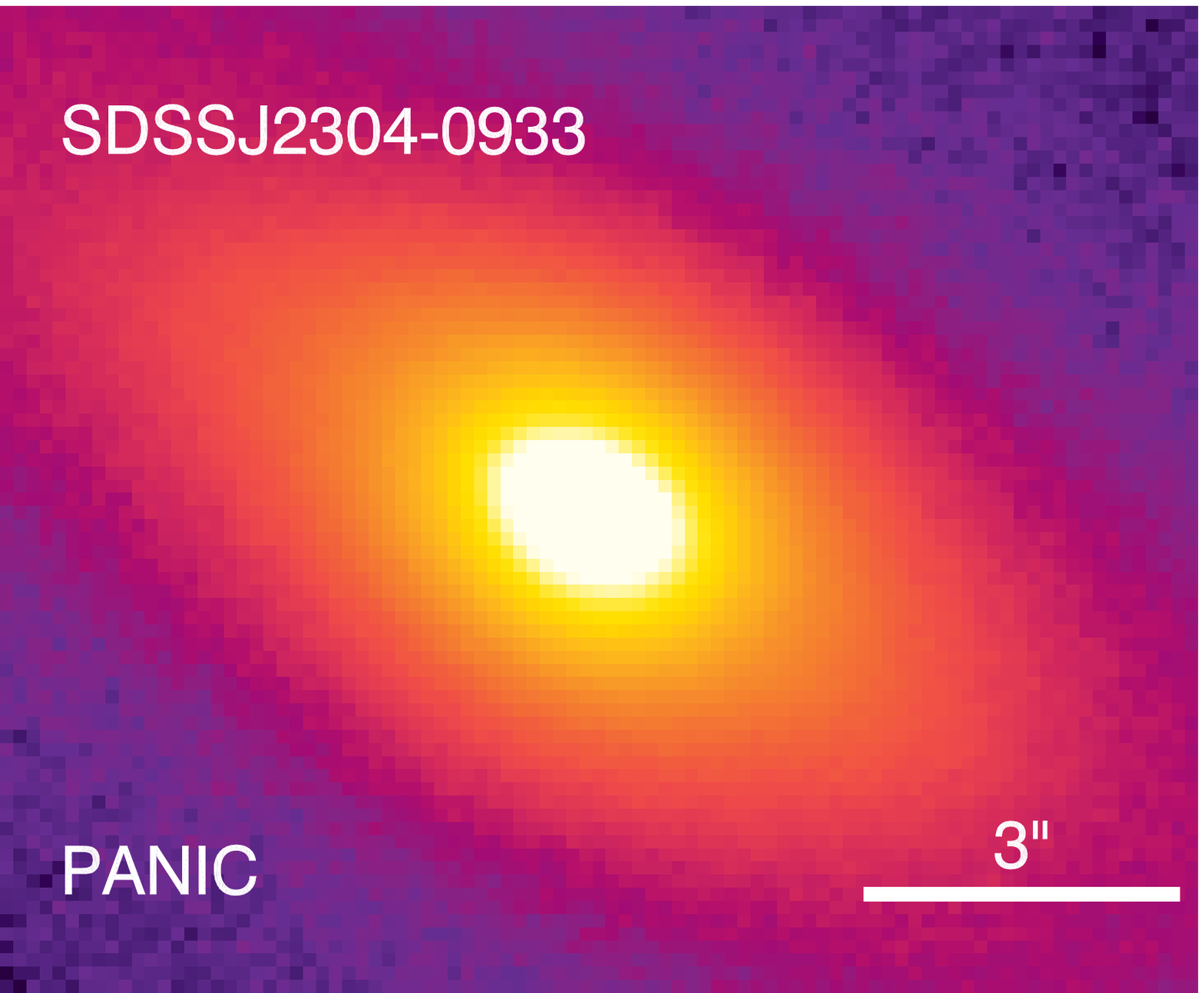}
    \includegraphics[width=0.3\textwidth]{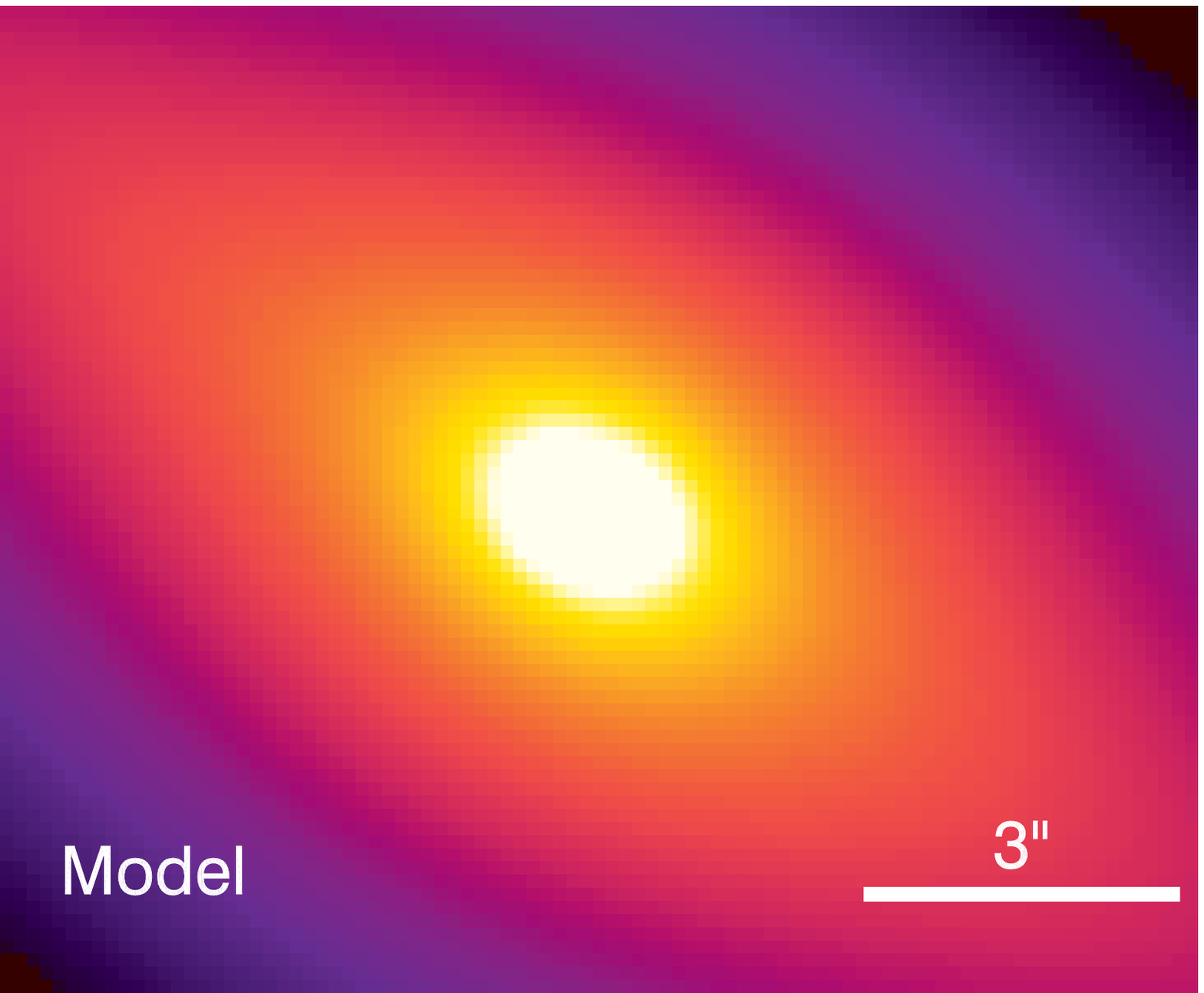}
    \includegraphics[width=0.3\textwidth]{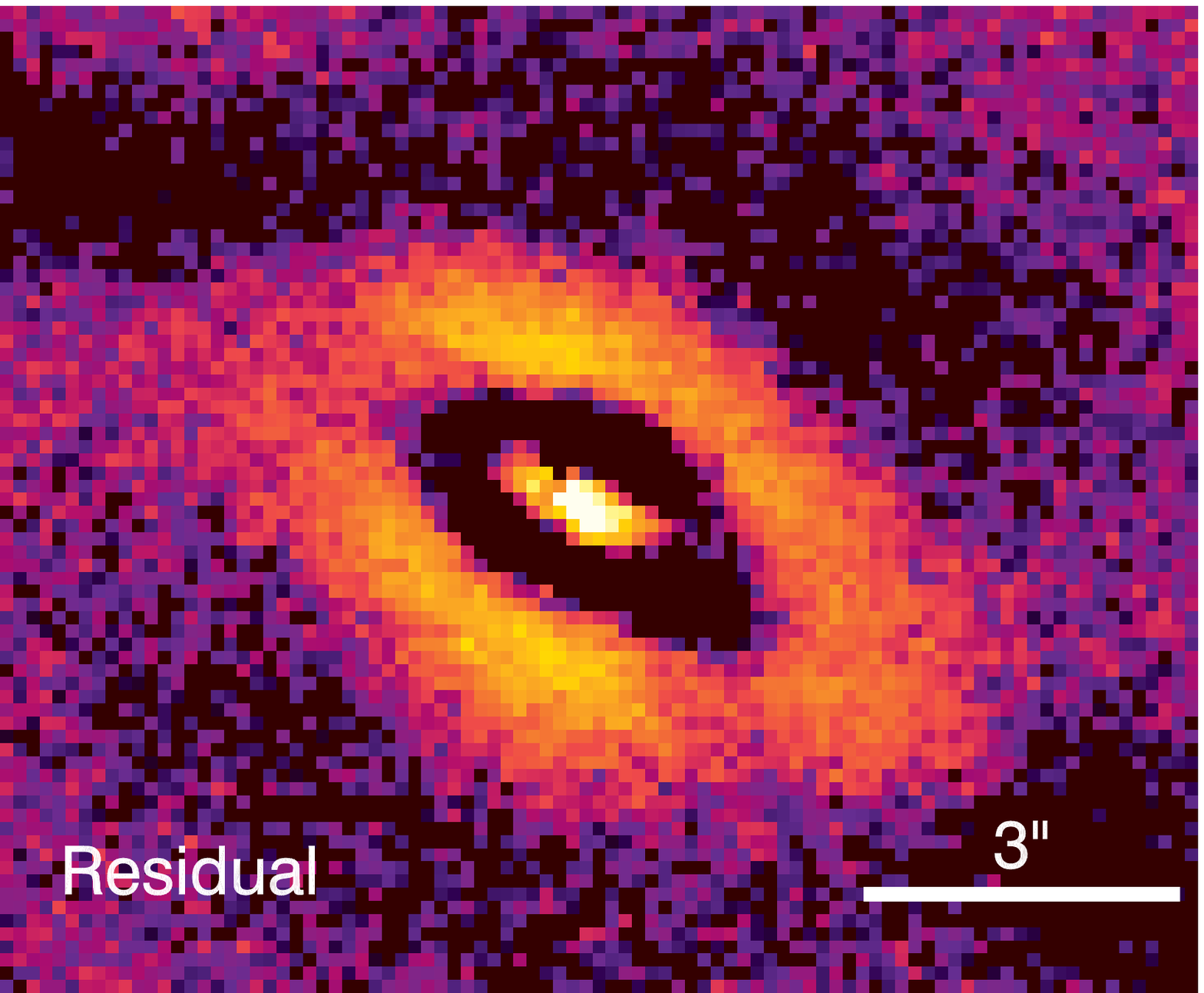}}
\caption{Continued. }
    \label{fig:NLR_galfit}
\end{figure*}

\begin{figure*}
  \centering
    \subfigure{
    \includegraphics[width=0.3\textwidth]{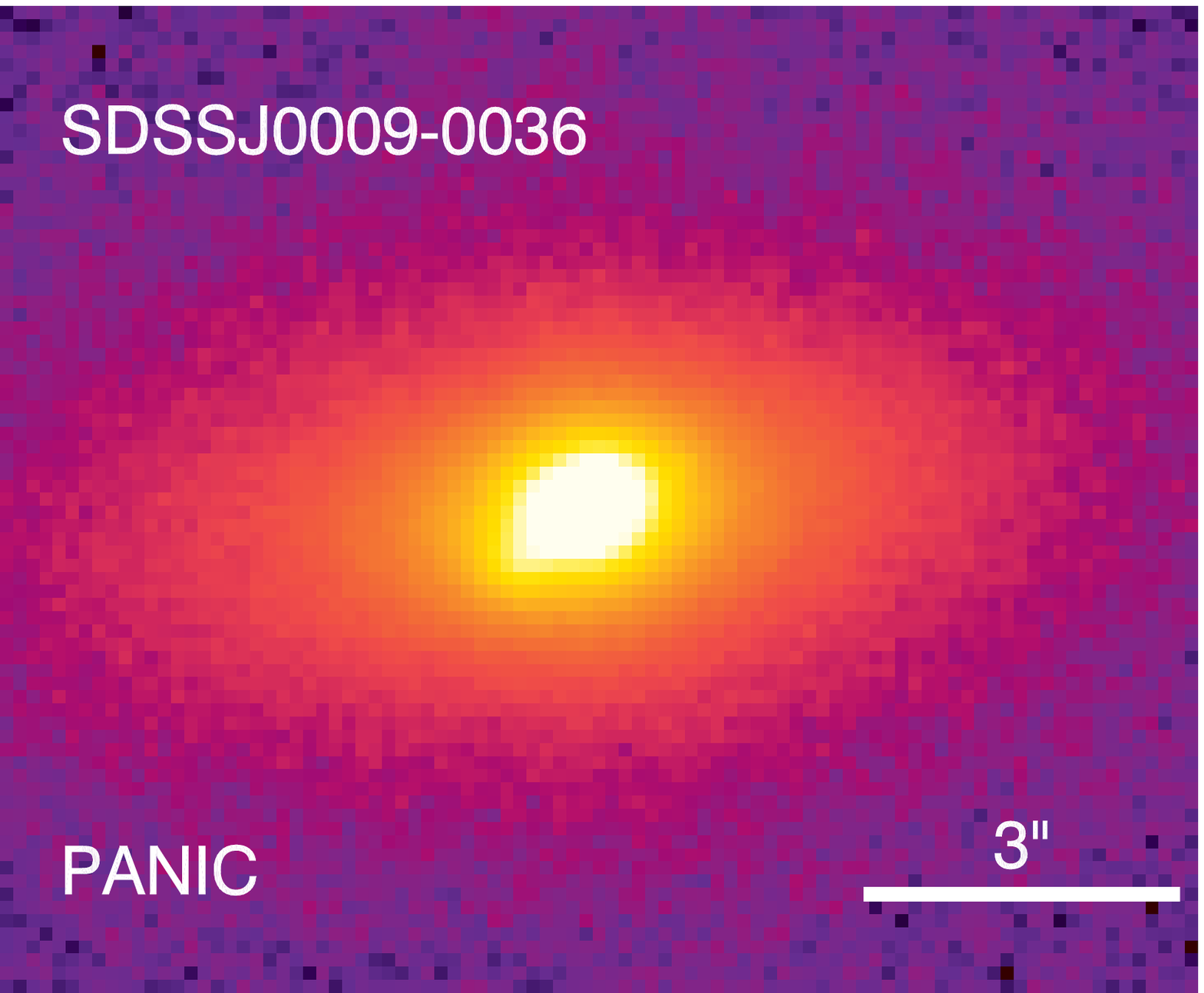}
    \includegraphics[width=0.3\textwidth]{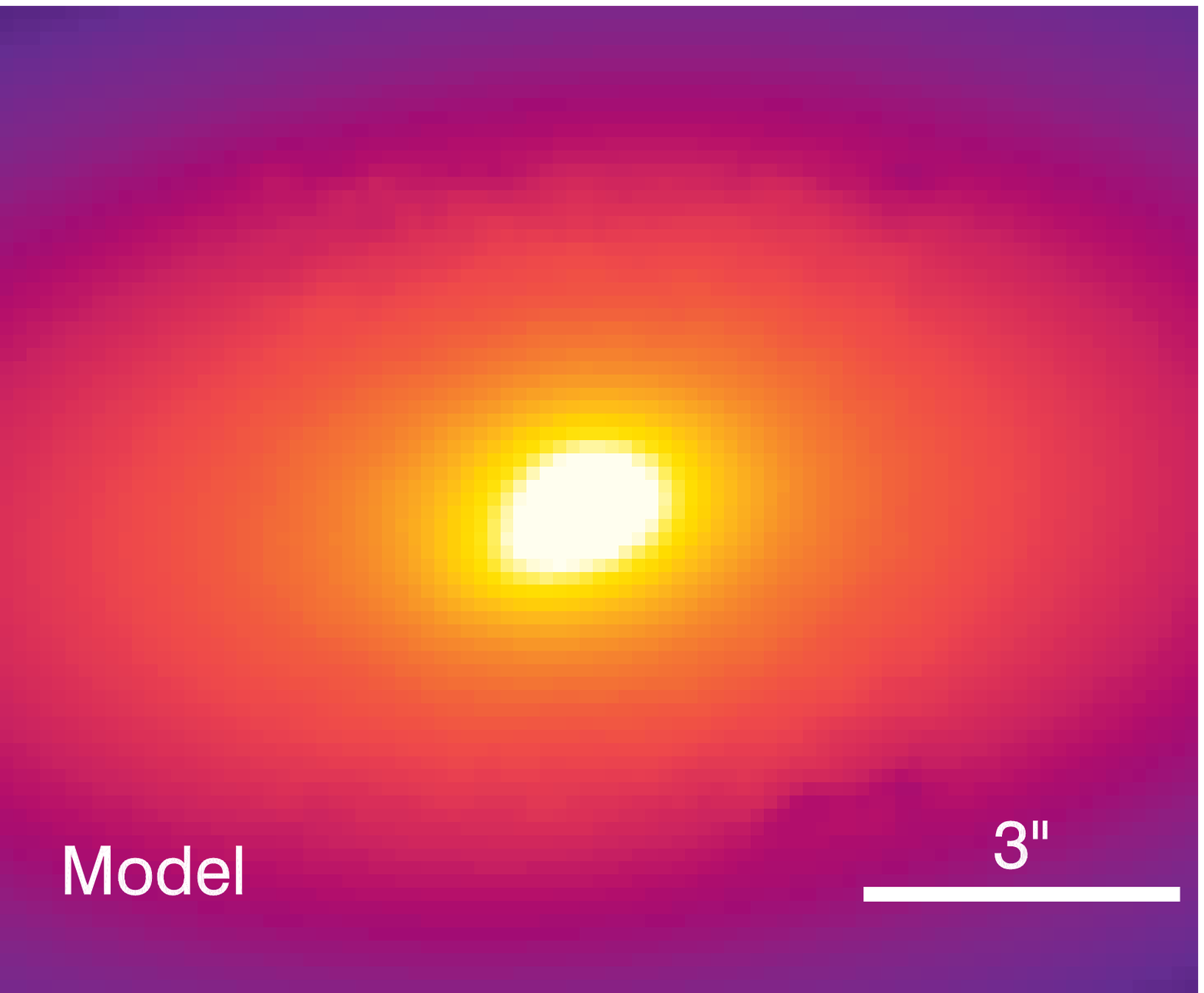}
    \includegraphics[width=0.3\textwidth]{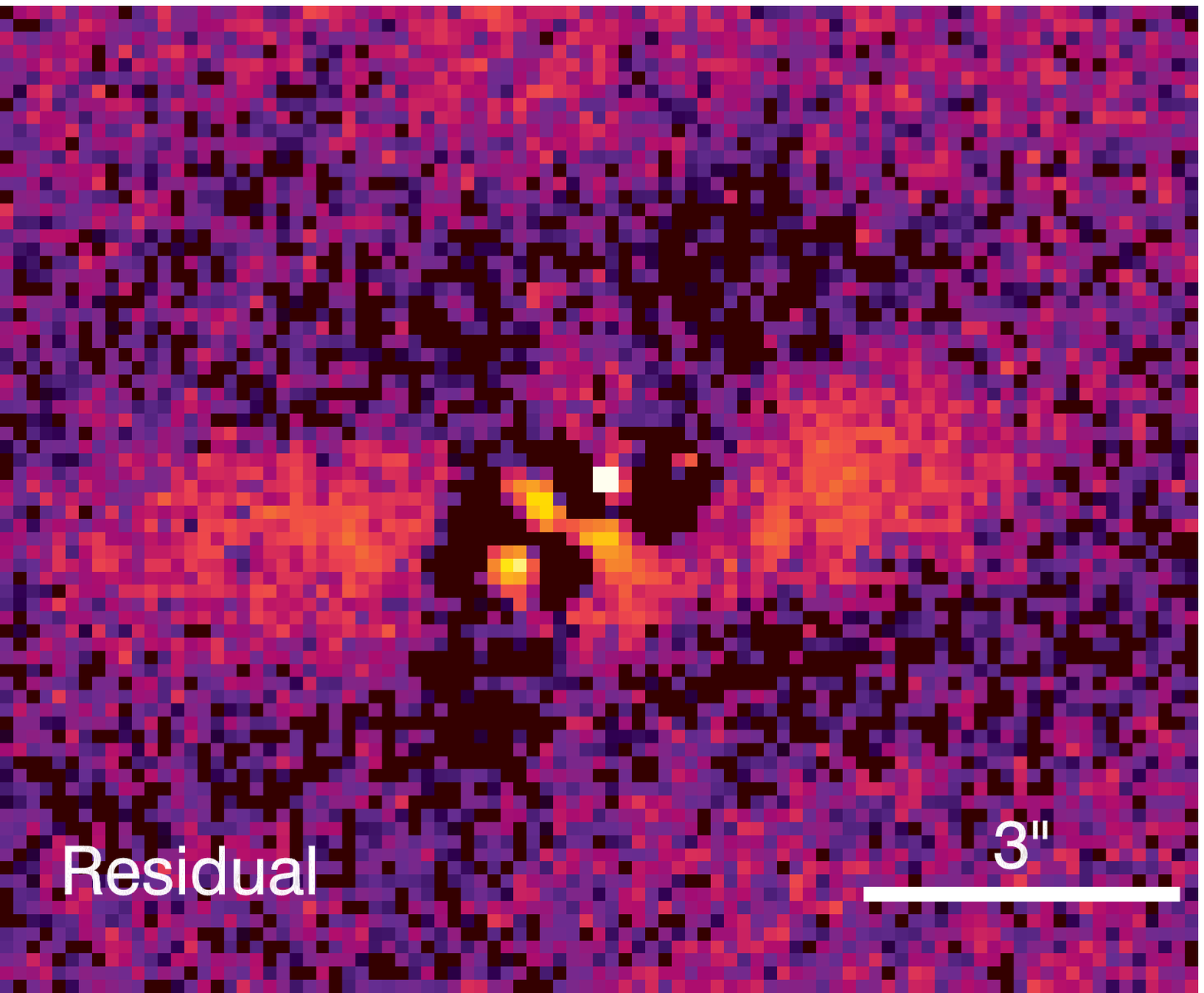}}
    \subfigure{
    \includegraphics[width=0.3\textwidth]{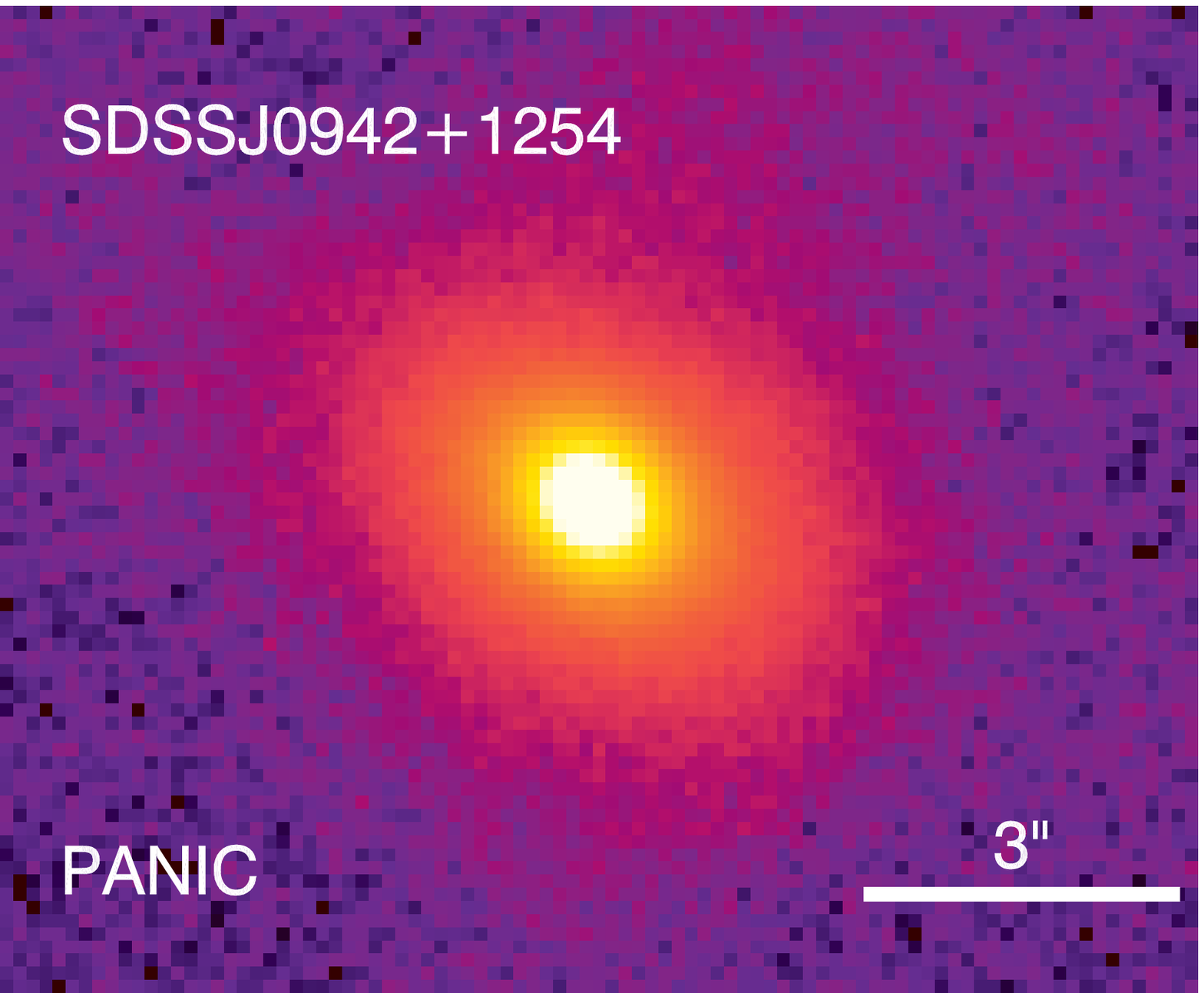}
    \includegraphics[width=0.3\textwidth]{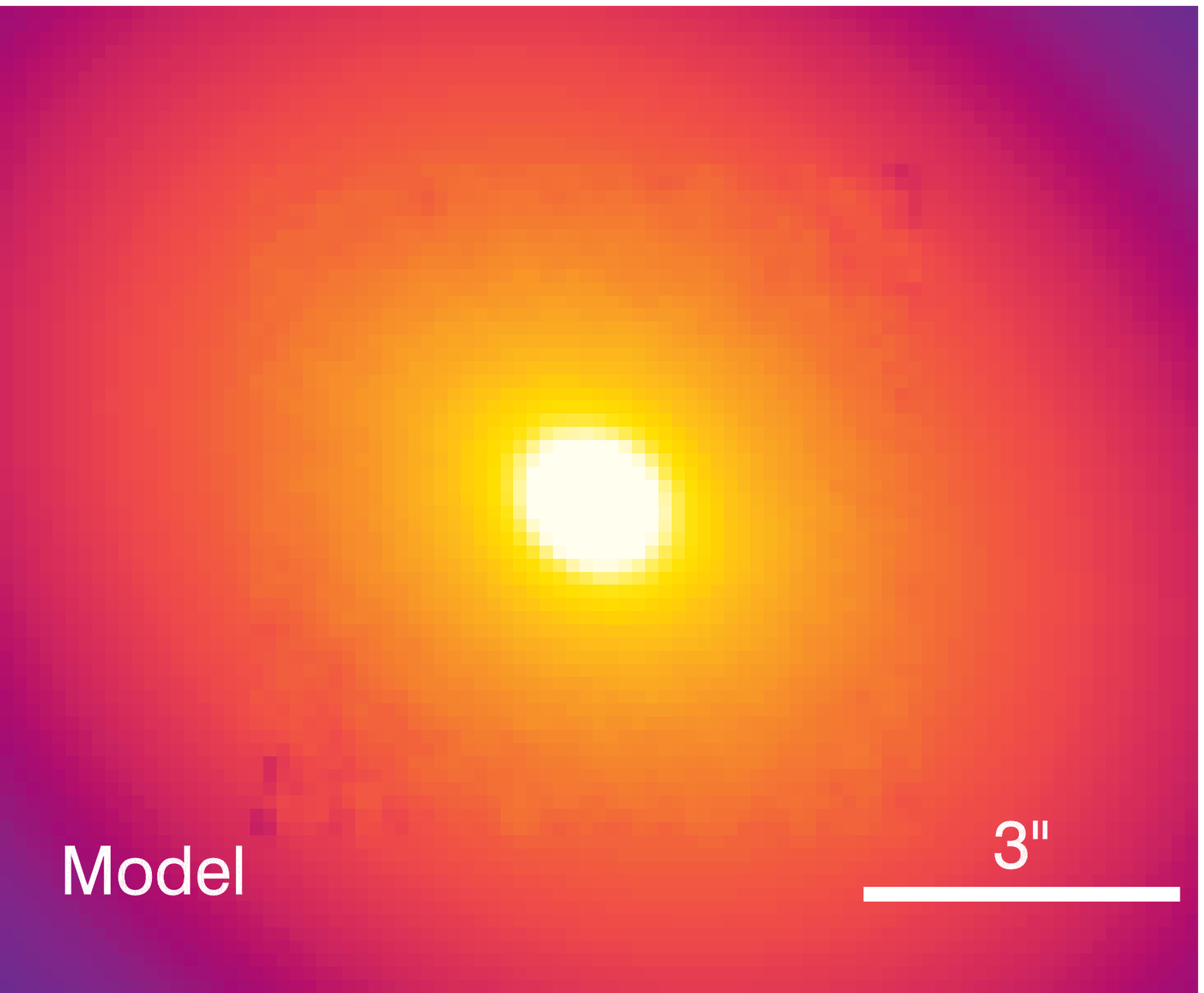}
    \includegraphics[width=0.3\textwidth]{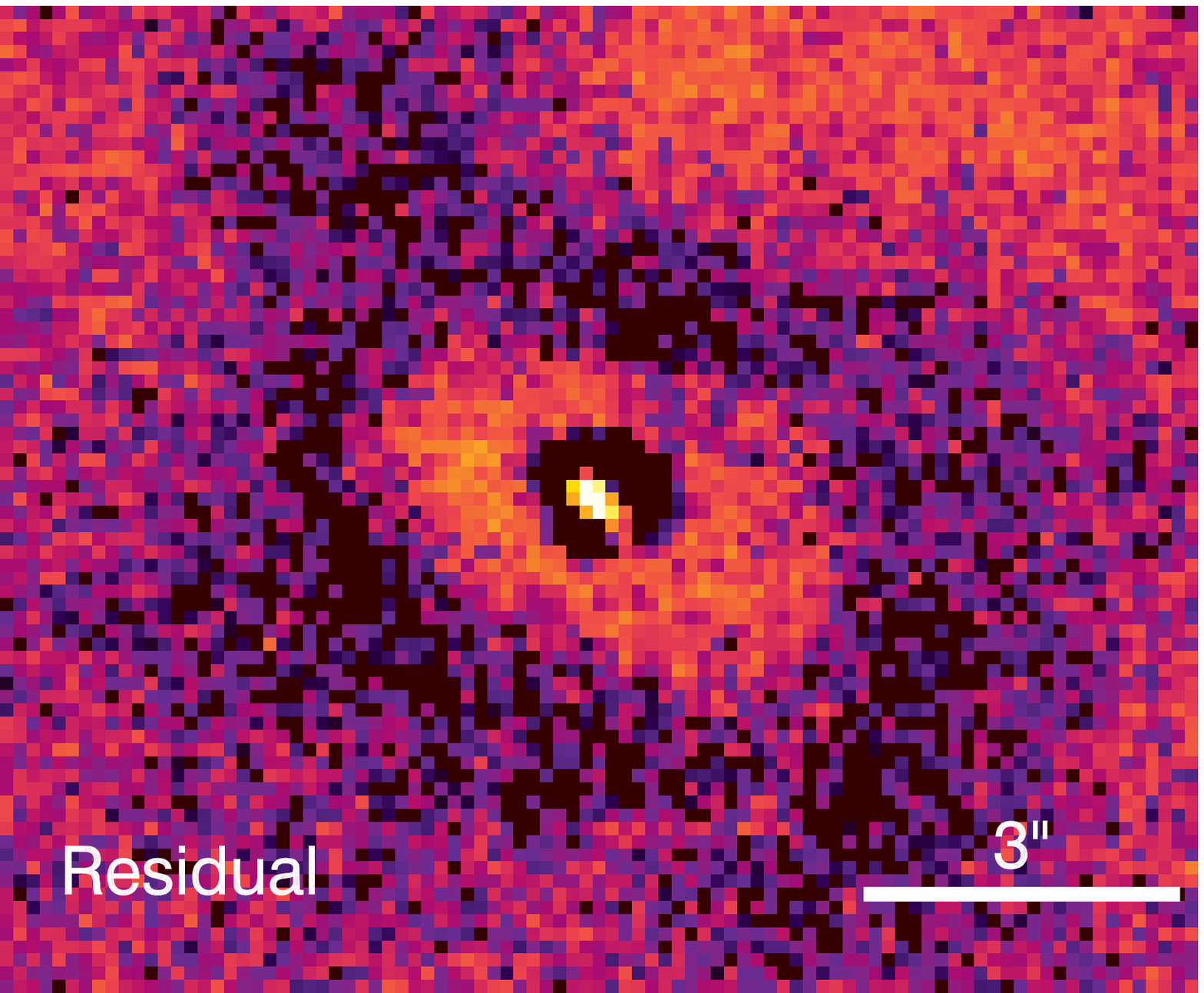}}
    \subfigure{
    \includegraphics[width=0.3\textwidth]{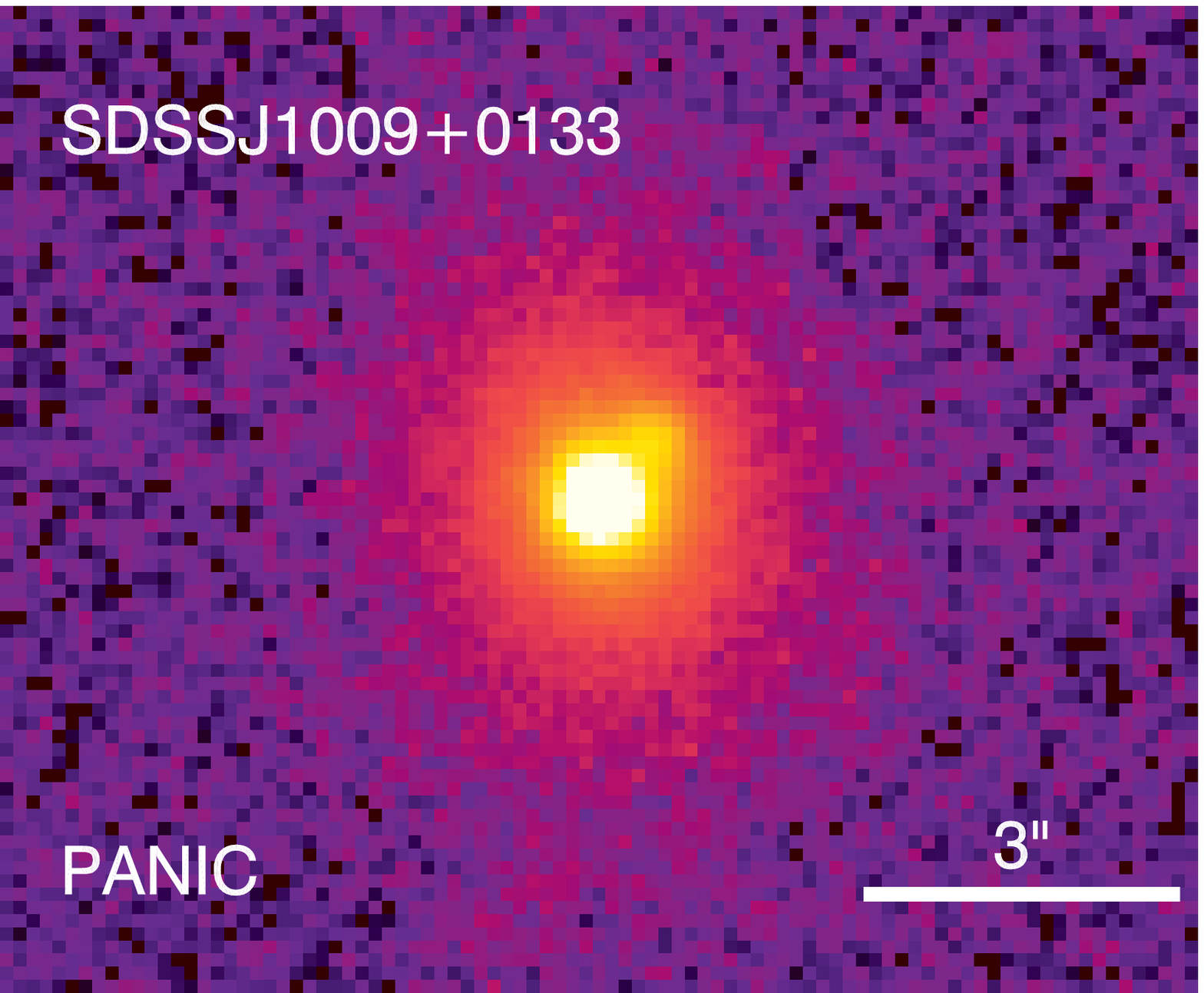}
    \includegraphics[width=0.3\textwidth]{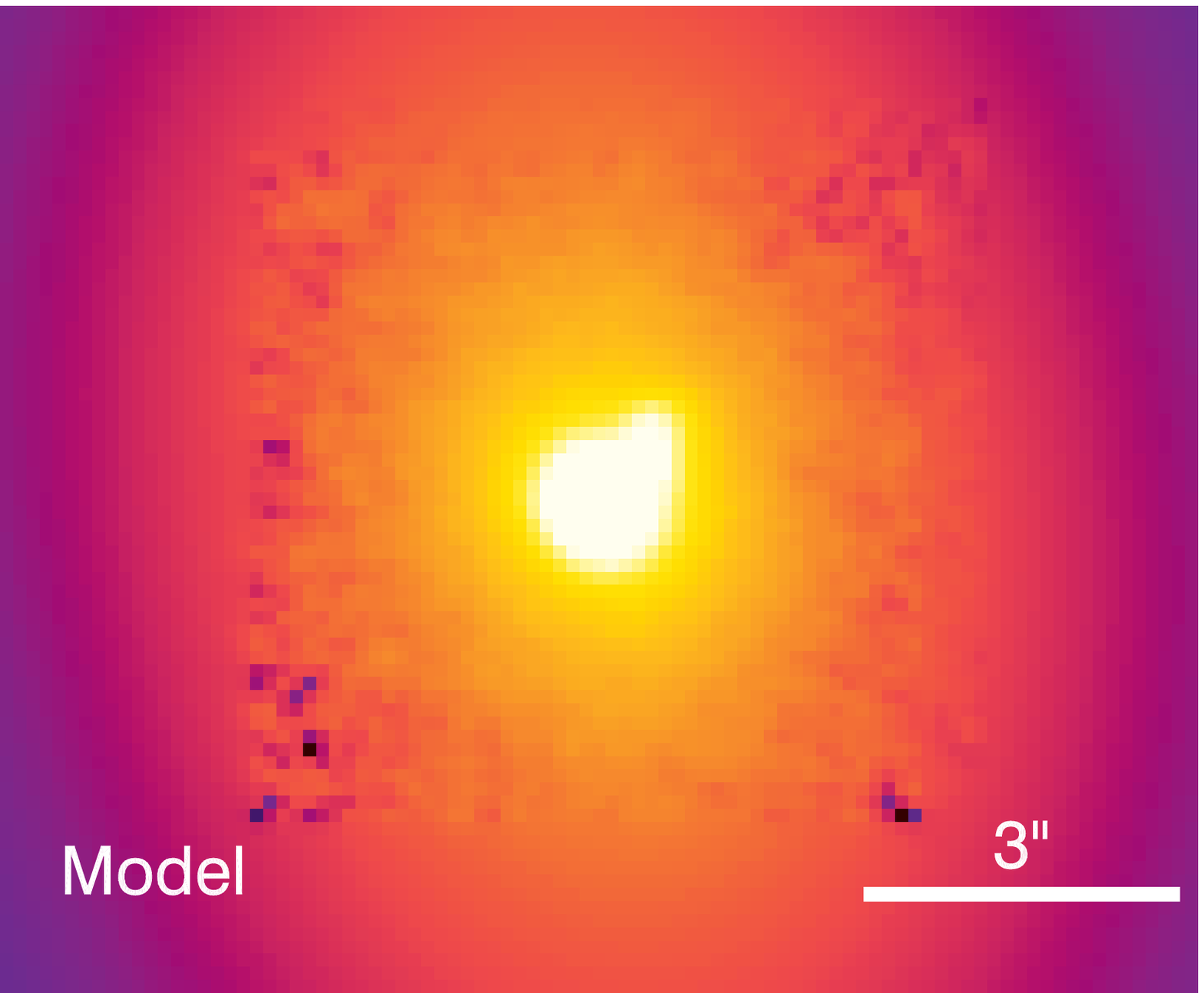}
    \includegraphics[width=0.3\textwidth]{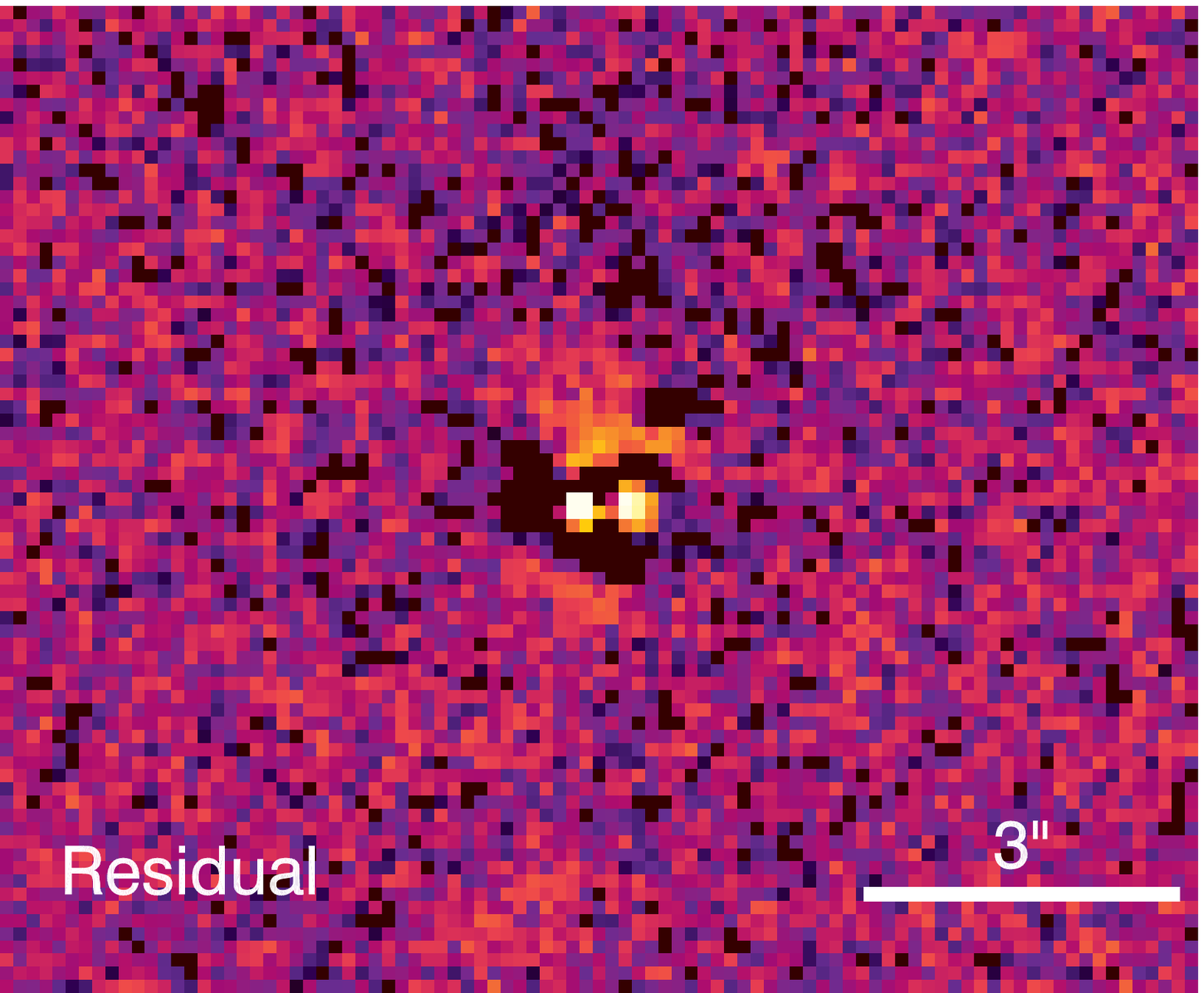}}
    \subfigure{
    \includegraphics[width=0.3\textwidth]{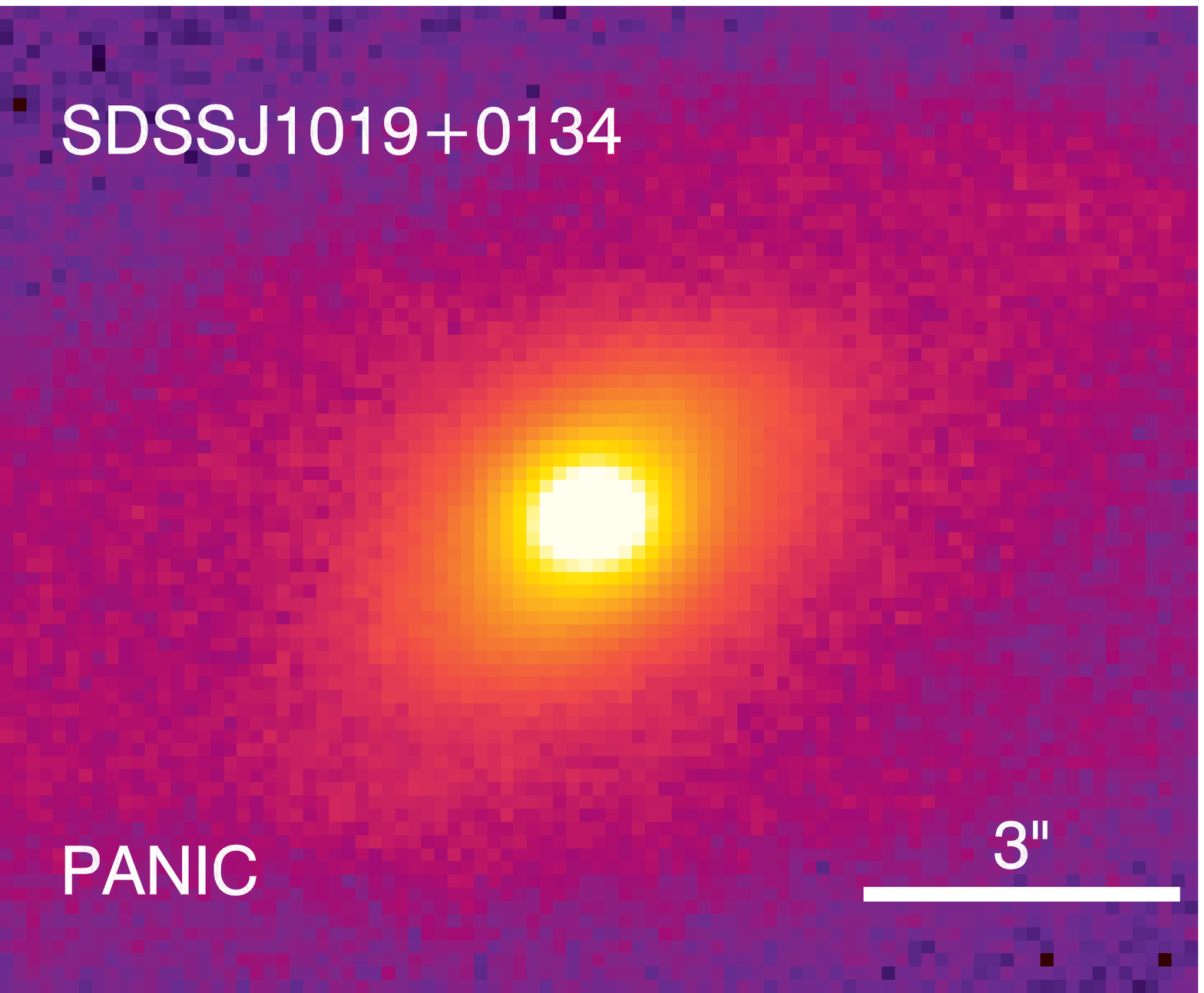}
    \includegraphics[width=0.3\textwidth]{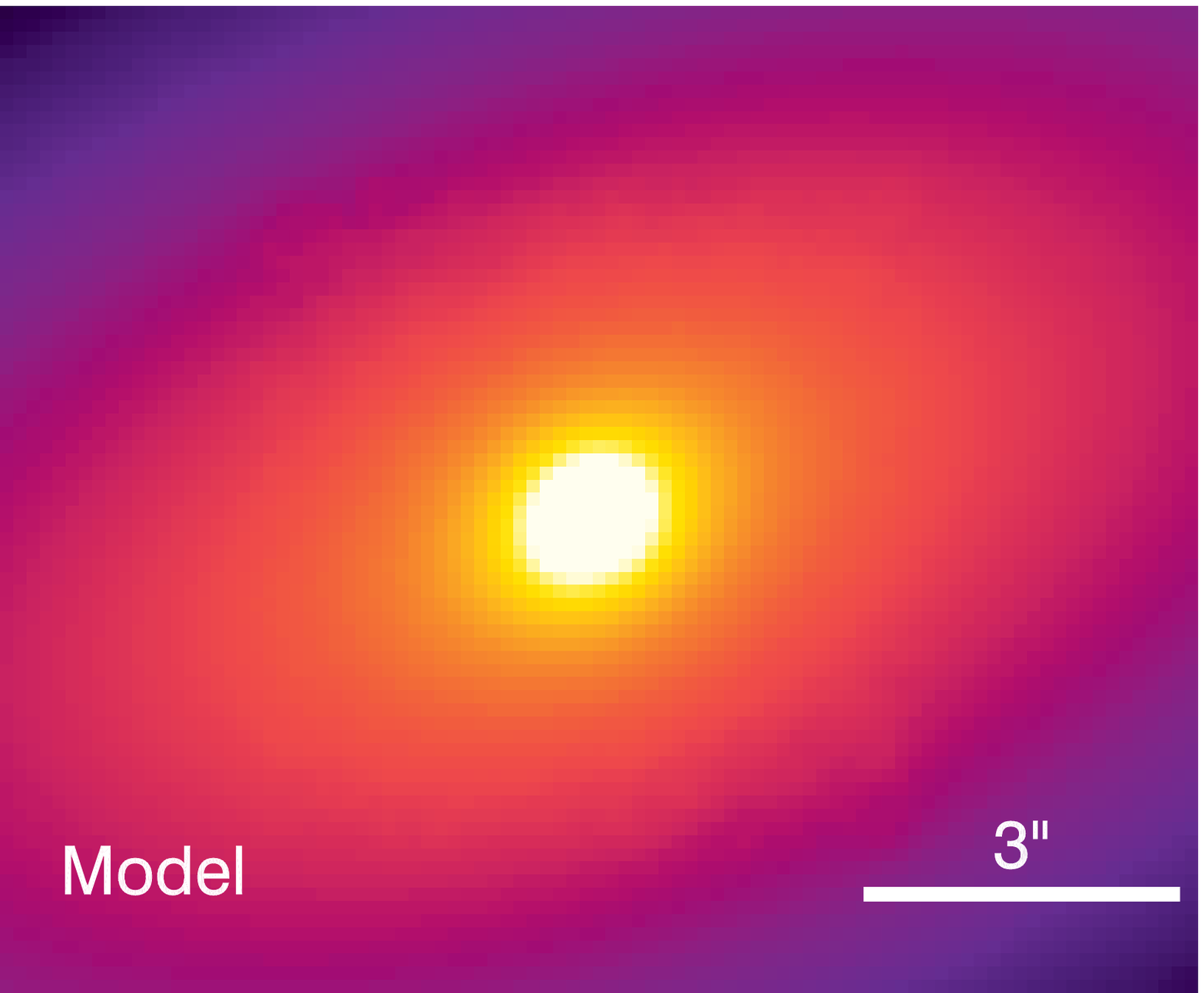}
    \includegraphics[width=0.3\textwidth]{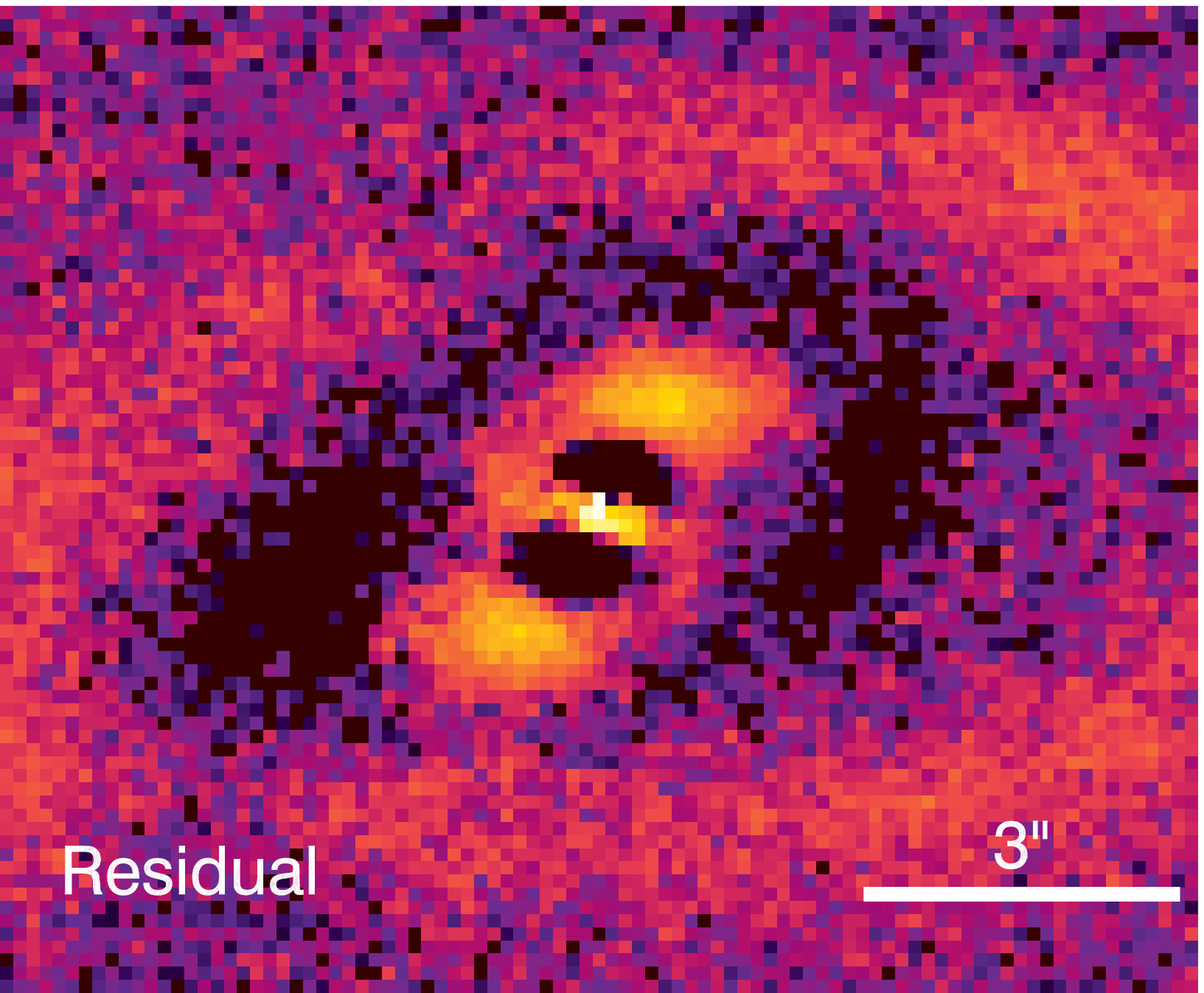}}
    \subfigure{
    \includegraphics[width=0.3\textwidth]{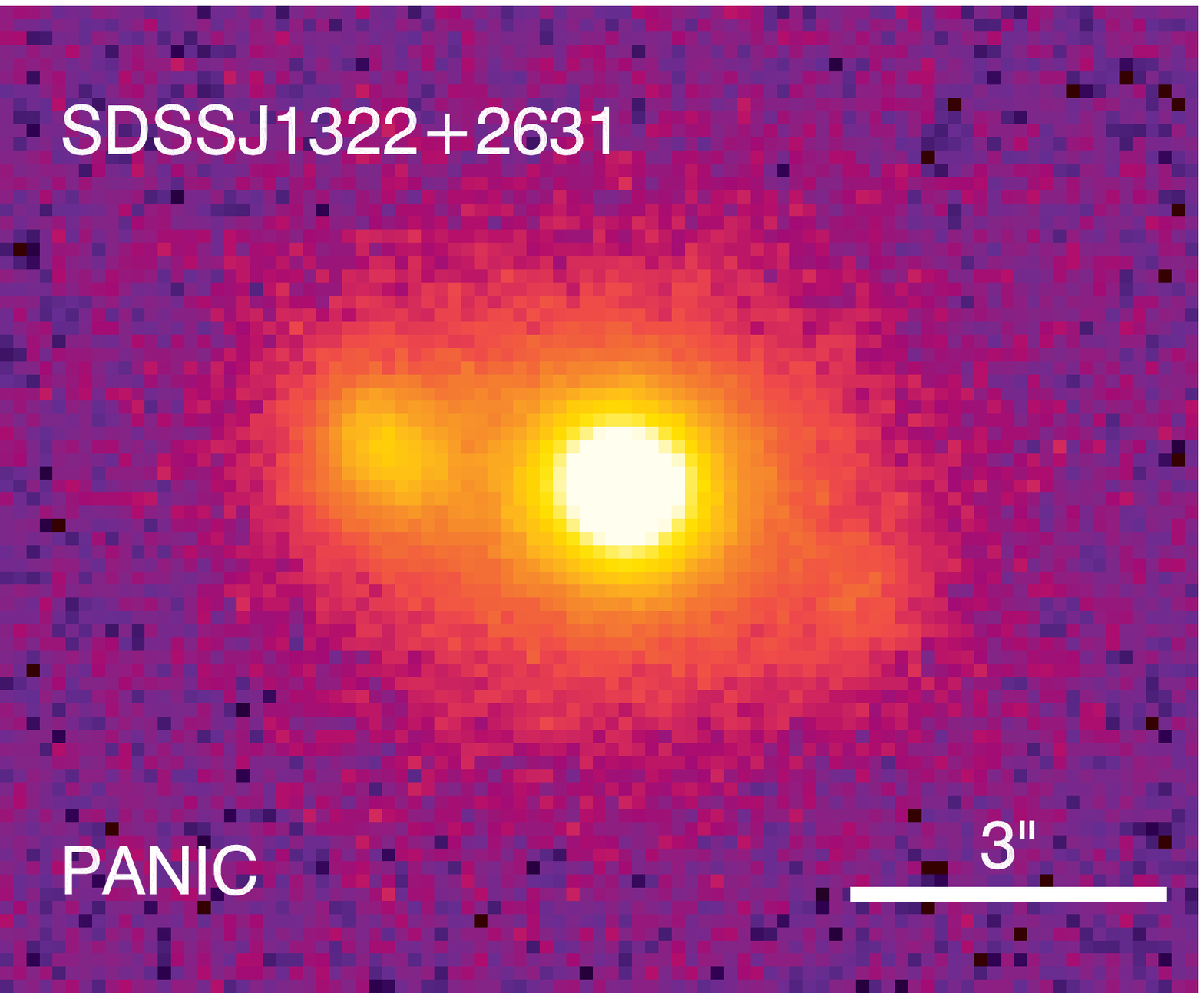}
    \includegraphics[width=0.3\textwidth]{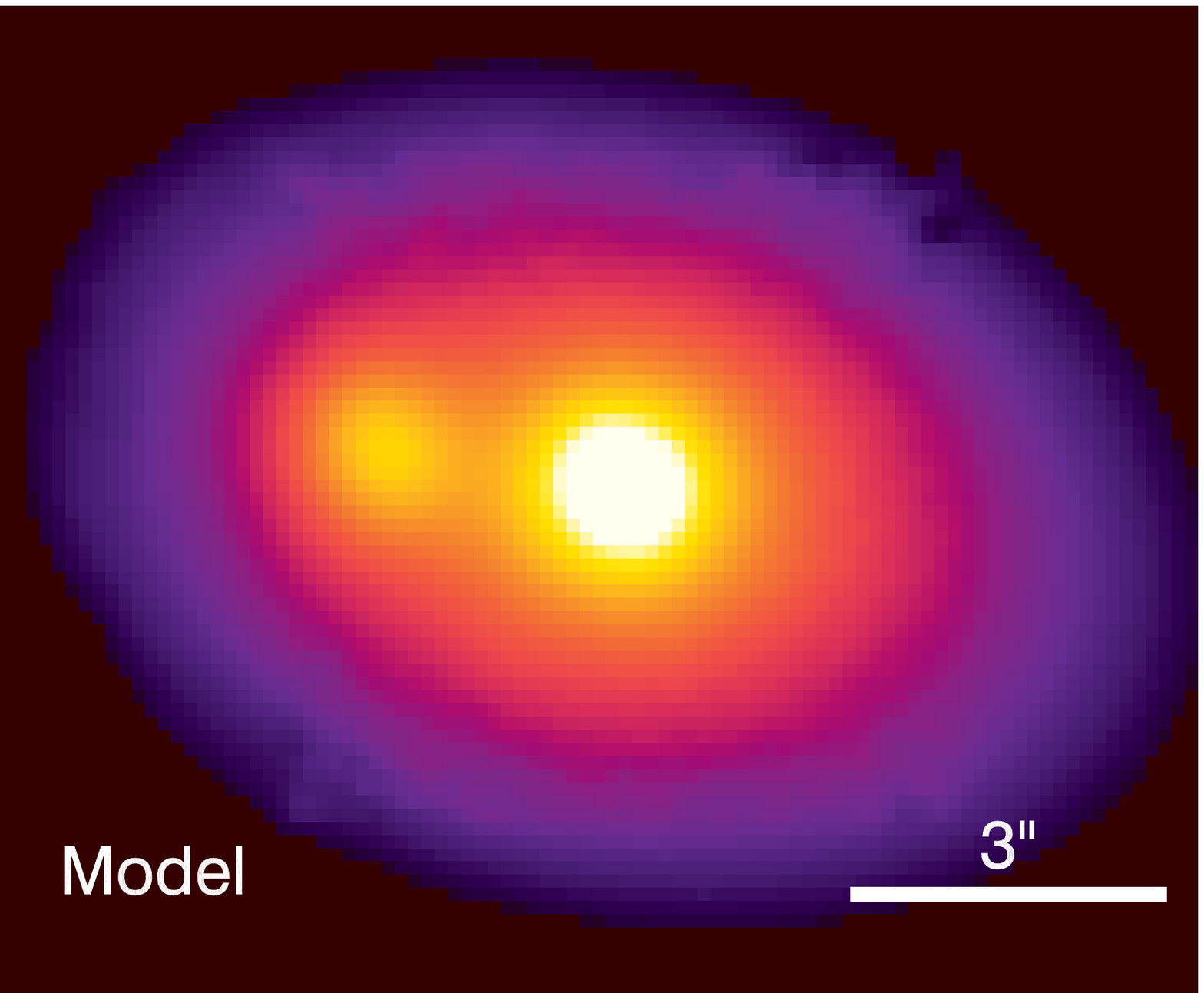}
    \includegraphics[width=0.3\textwidth]{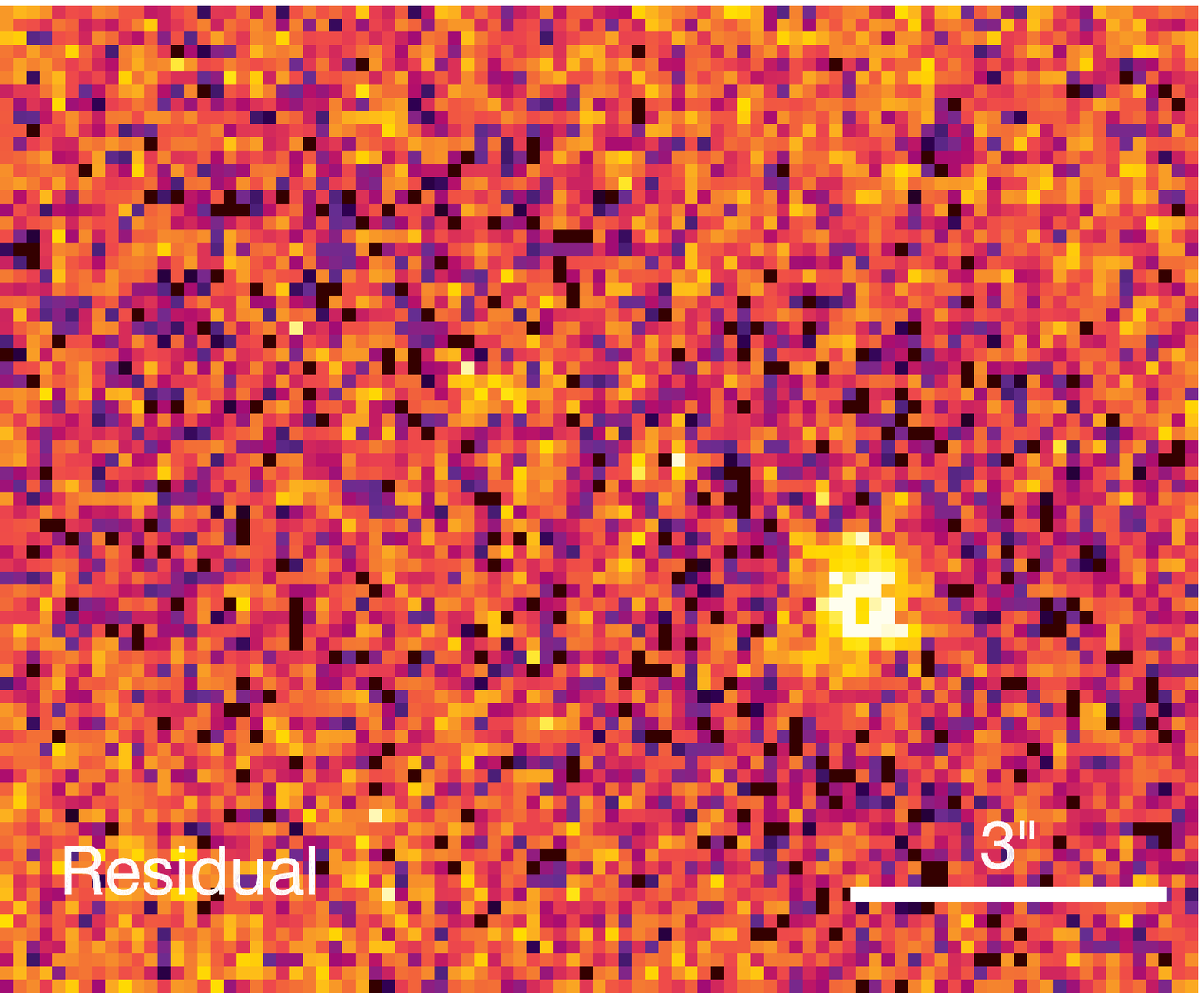}}
    \caption{\footnotesize Model fits of the NIR surface brightness for objects classified as ambiguous cases (\S\ref{sec:amb}) using {\sevenrm
    GALFIT}. The first column shows the data, the second column shows the models and the last column shows the residuals. Note that the residual maps
    have been re-stretched to enhance the contrast and the residuals are not important compared to the observed fluxes (the apparent excess seen in
    some of the residual maps has values less than $5\%$ of the
    observed fluxes at the corresponding locations). }
    \label{fig:amb_galfit}
\end{figure*}

\addtocounter{figure}{-1}
\begin{figure*}
\addtocounter{subfigure}{1}
  \centering
   \subfigure{
    \includegraphics[width=0.25\textwidth]{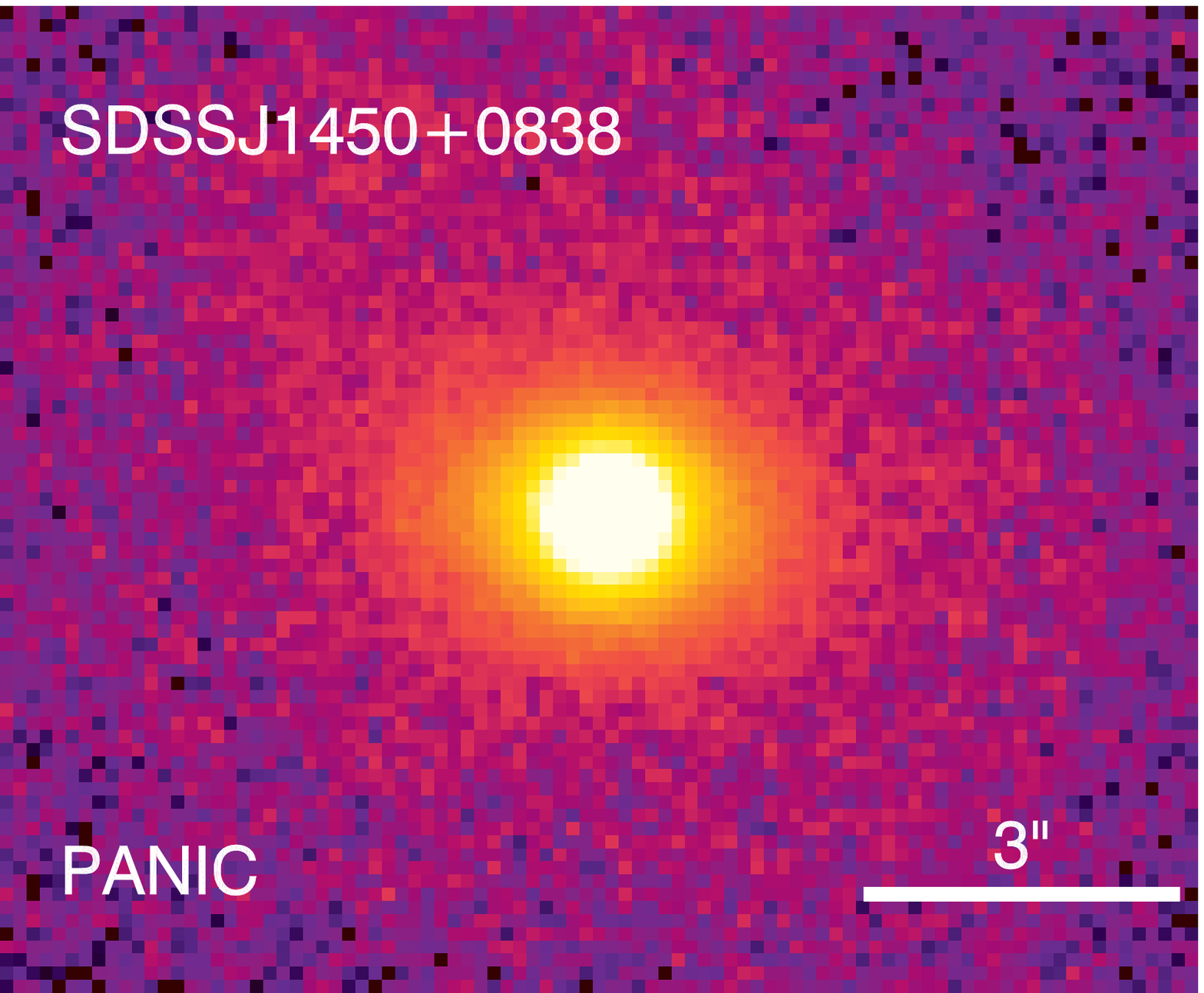}
    \includegraphics[width=0.25\textwidth]{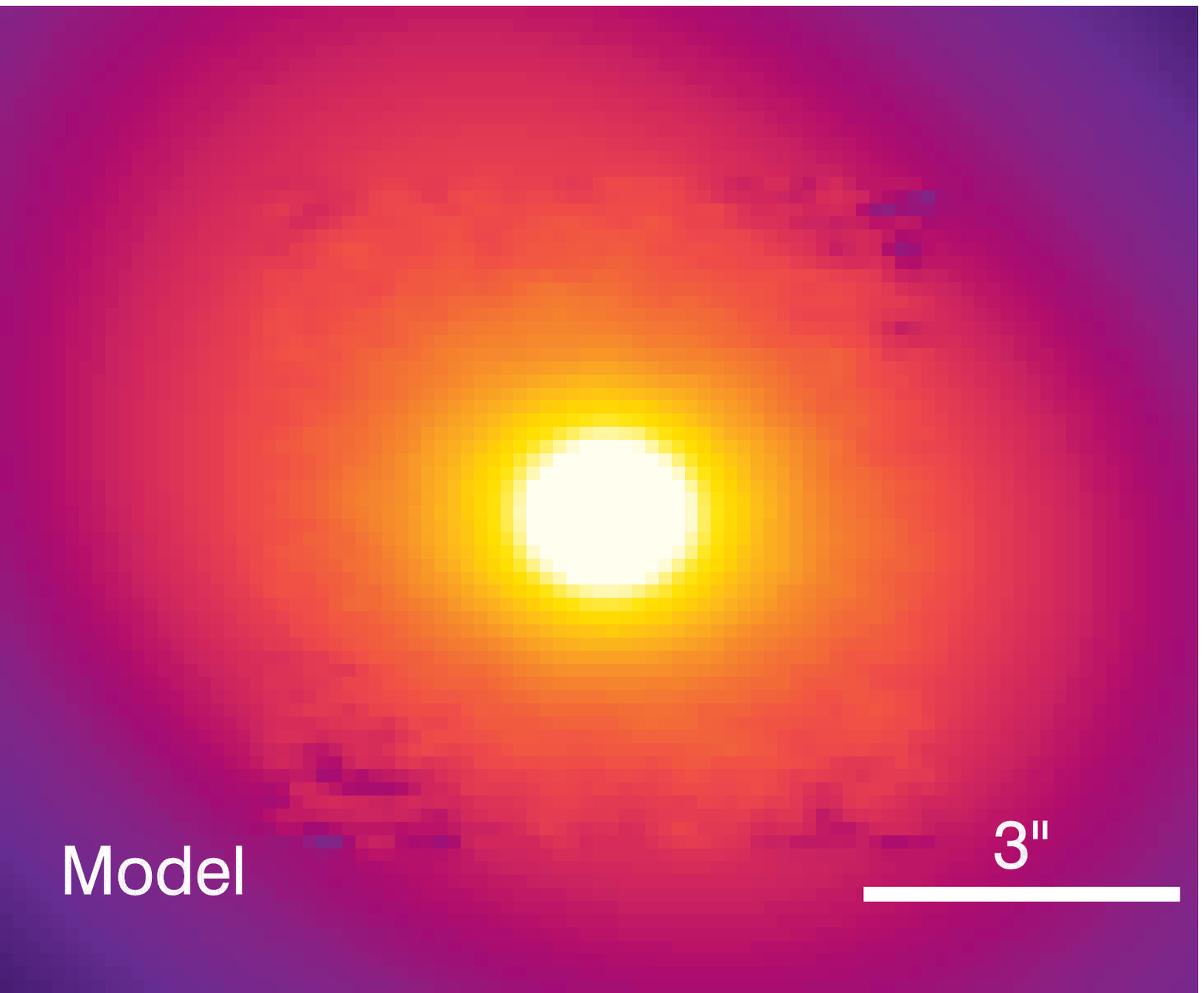}
    \includegraphics[width=0.25\textwidth]{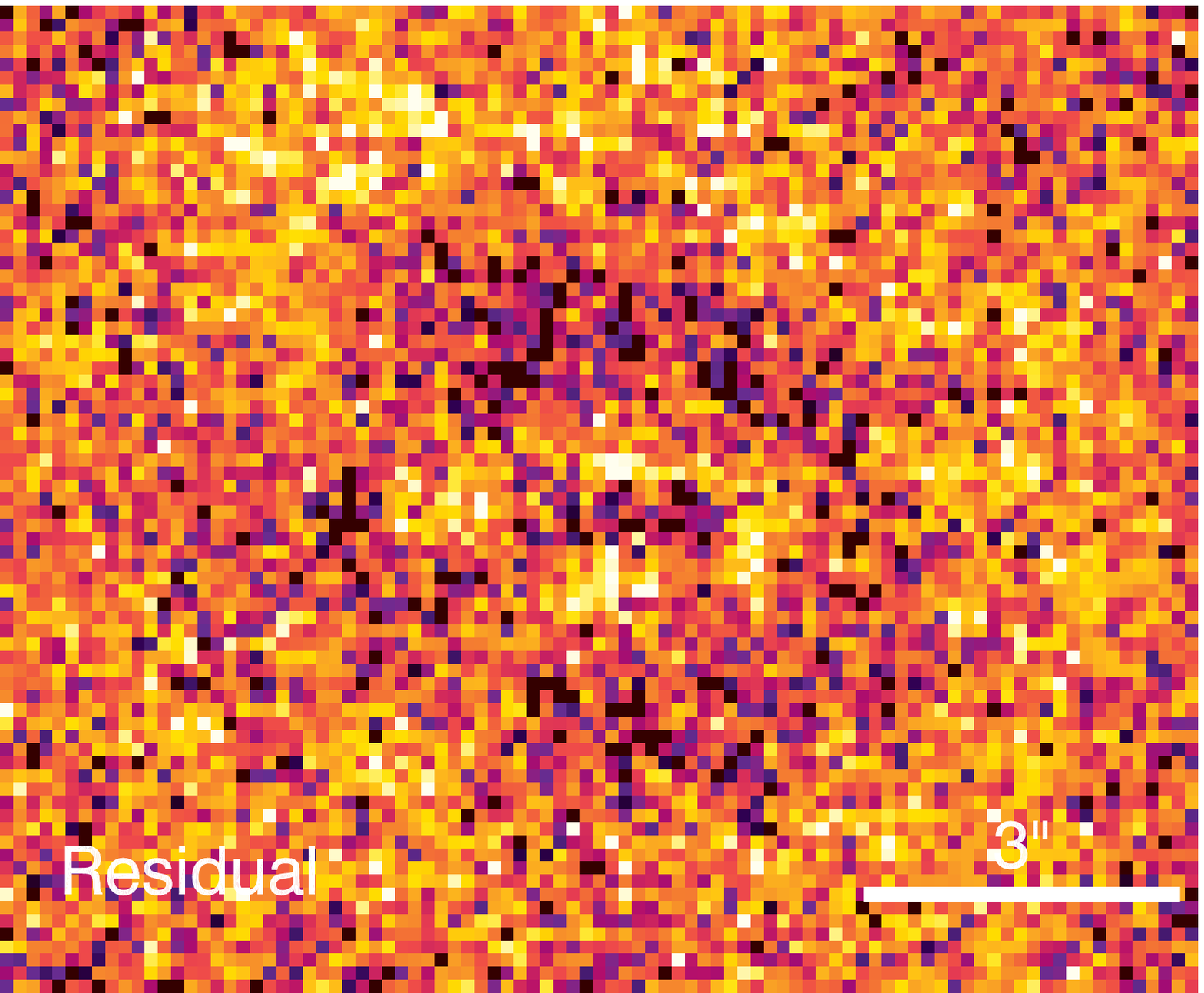}}
    \subfigure{
    \includegraphics[width=0.25\textwidth]{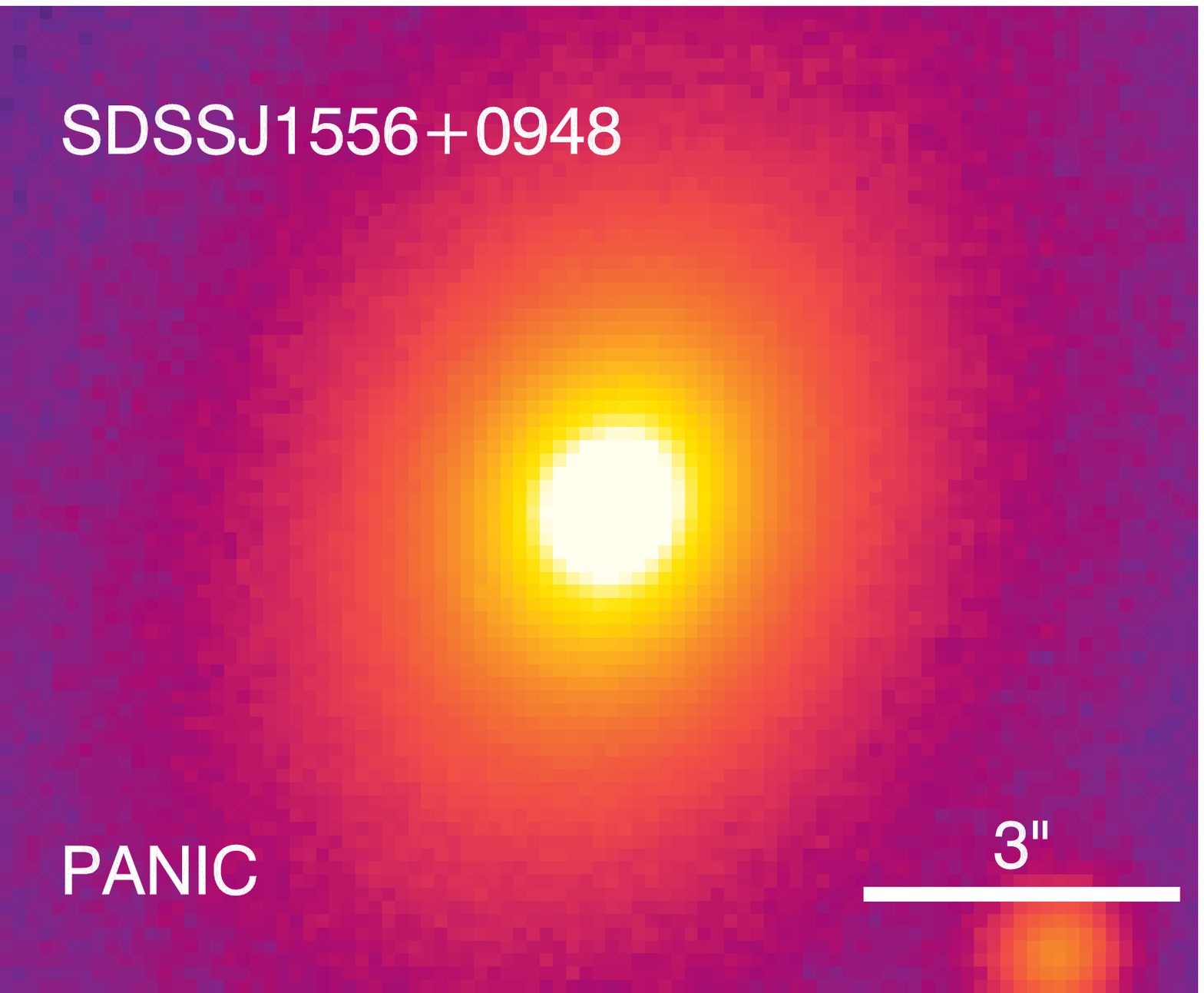}
    \includegraphics[width=0.25\textwidth]{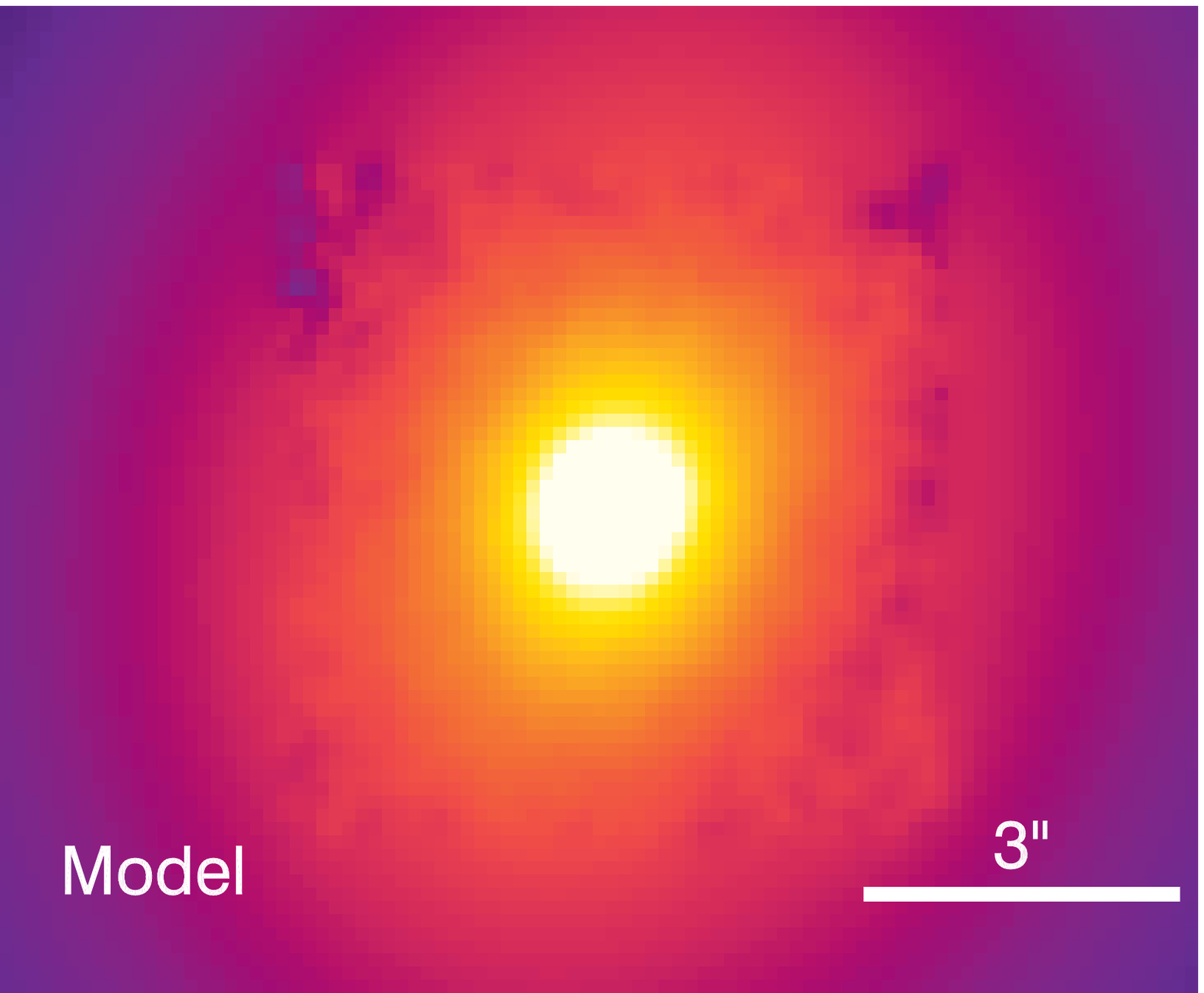}
    \includegraphics[width=0.25\textwidth]{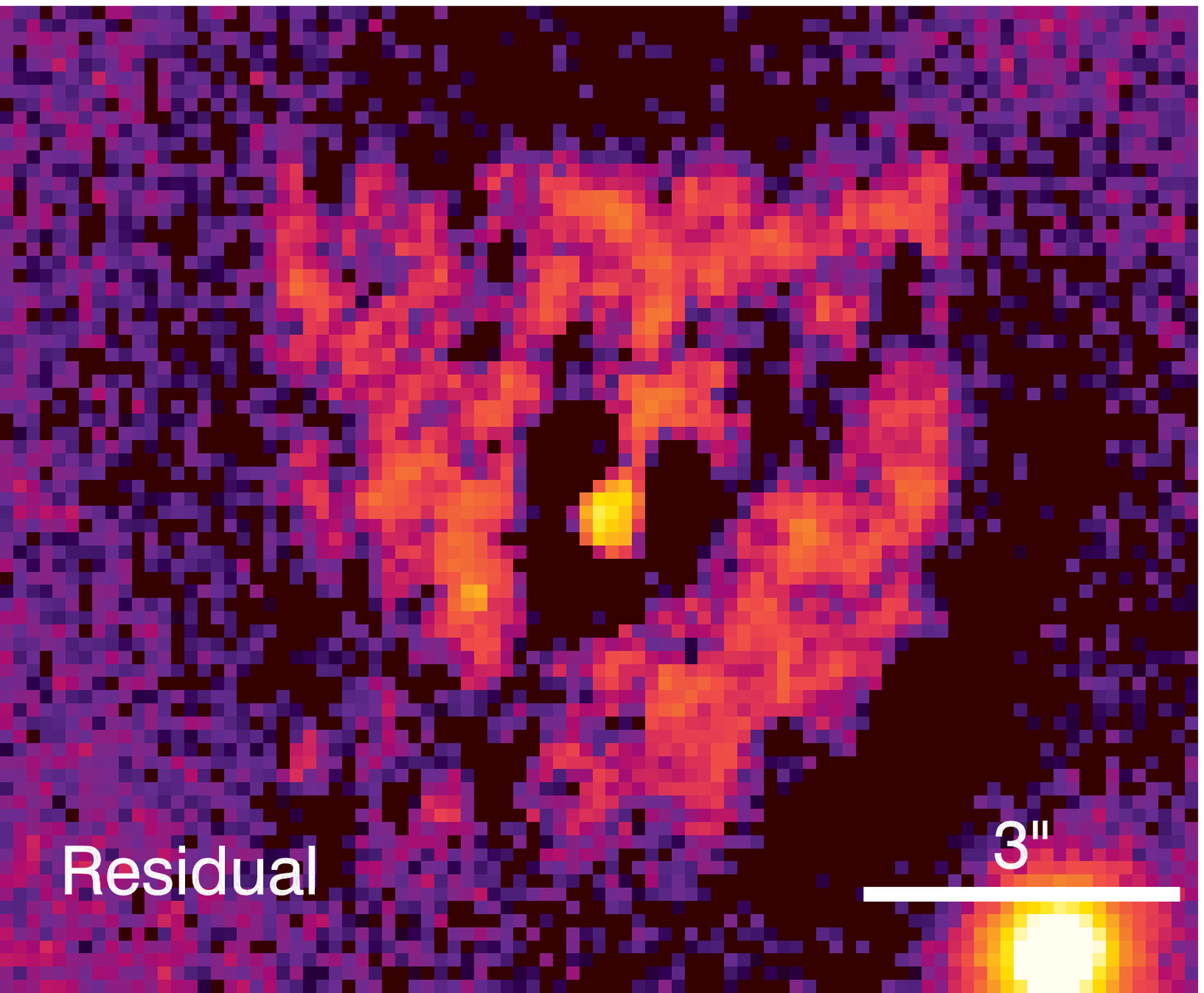}}
    \subfigure{
    \includegraphics[width=0.25\textwidth]{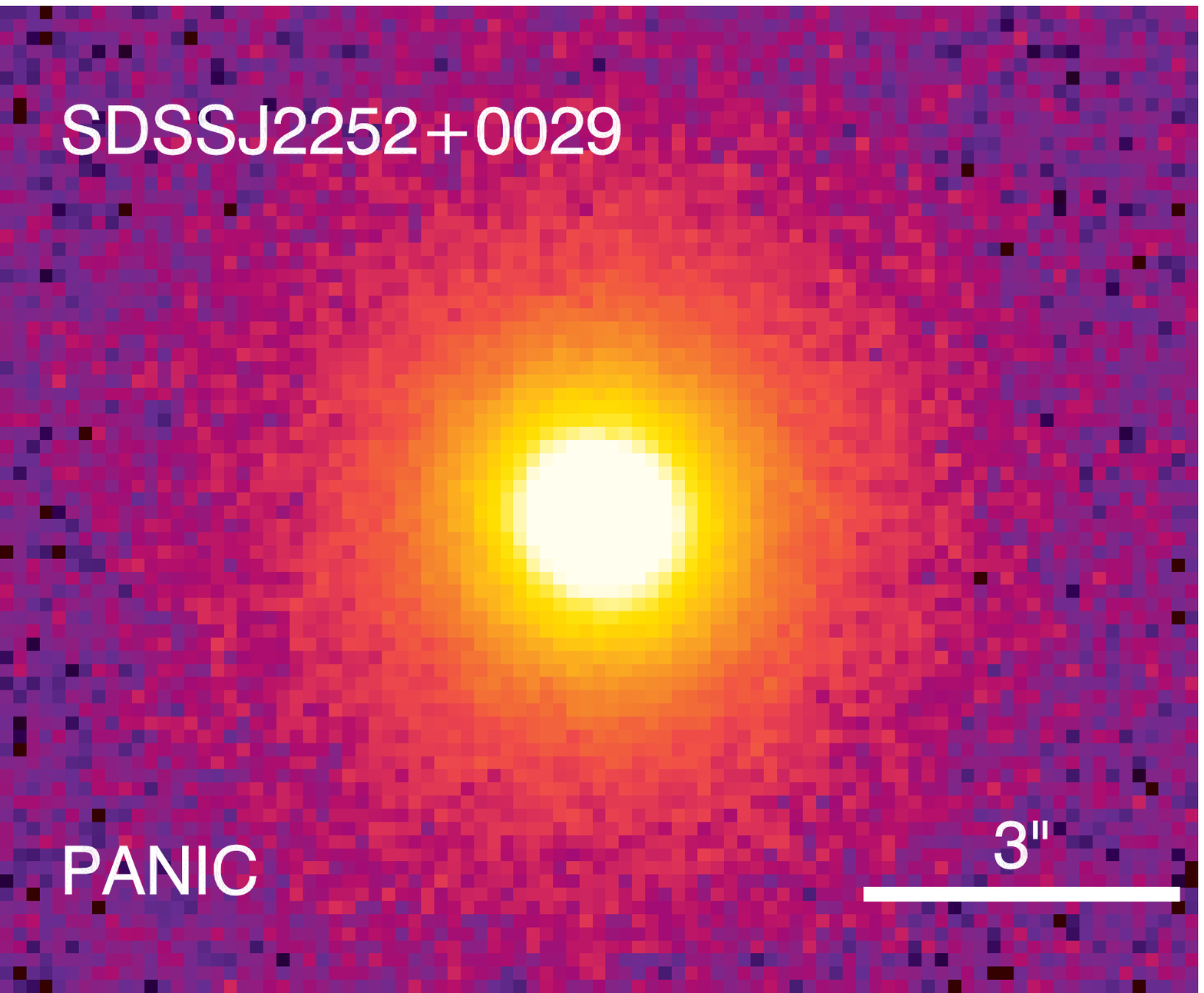}
    \includegraphics[width=0.25\textwidth]{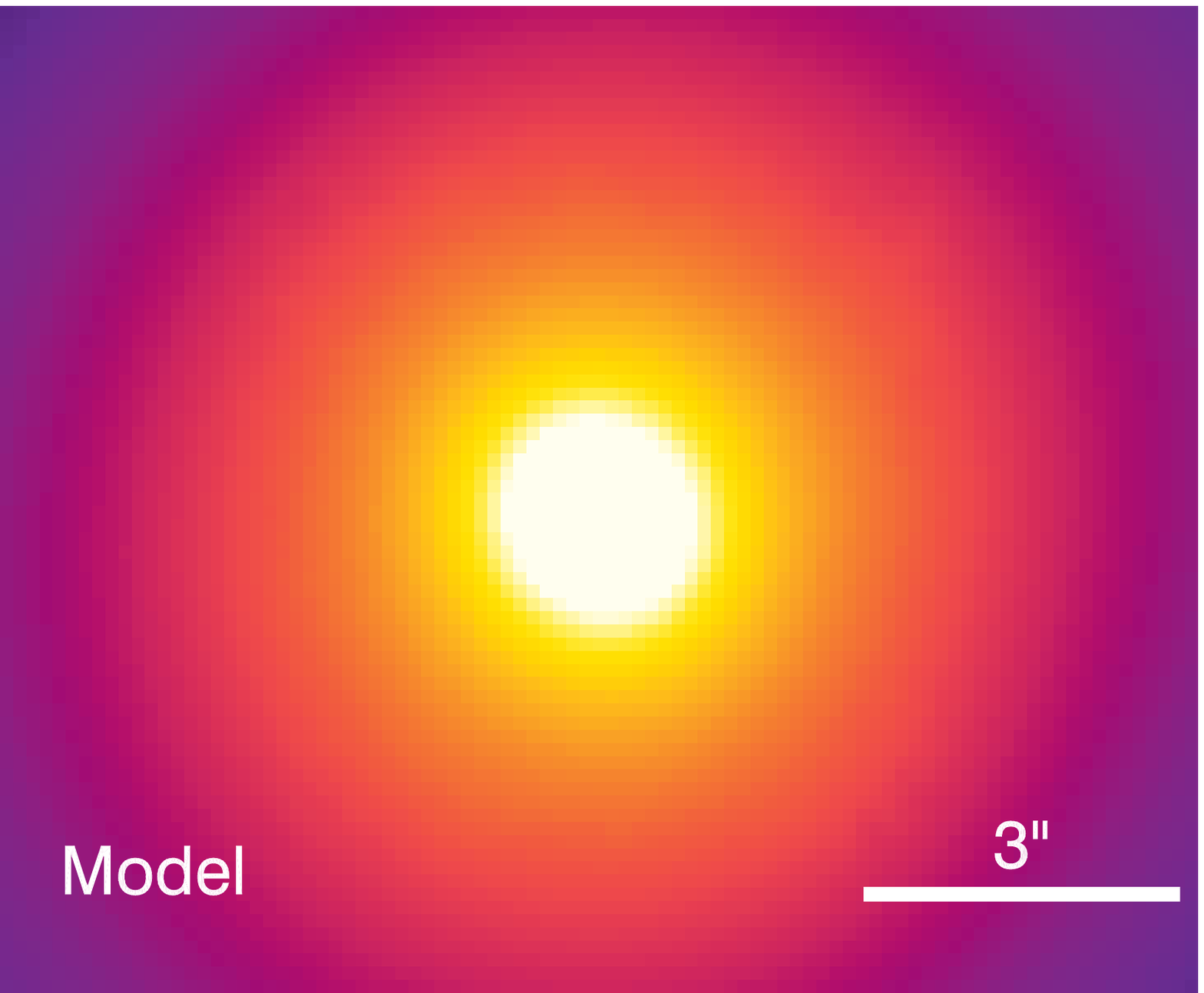}
    \includegraphics[width=0.25\textwidth]{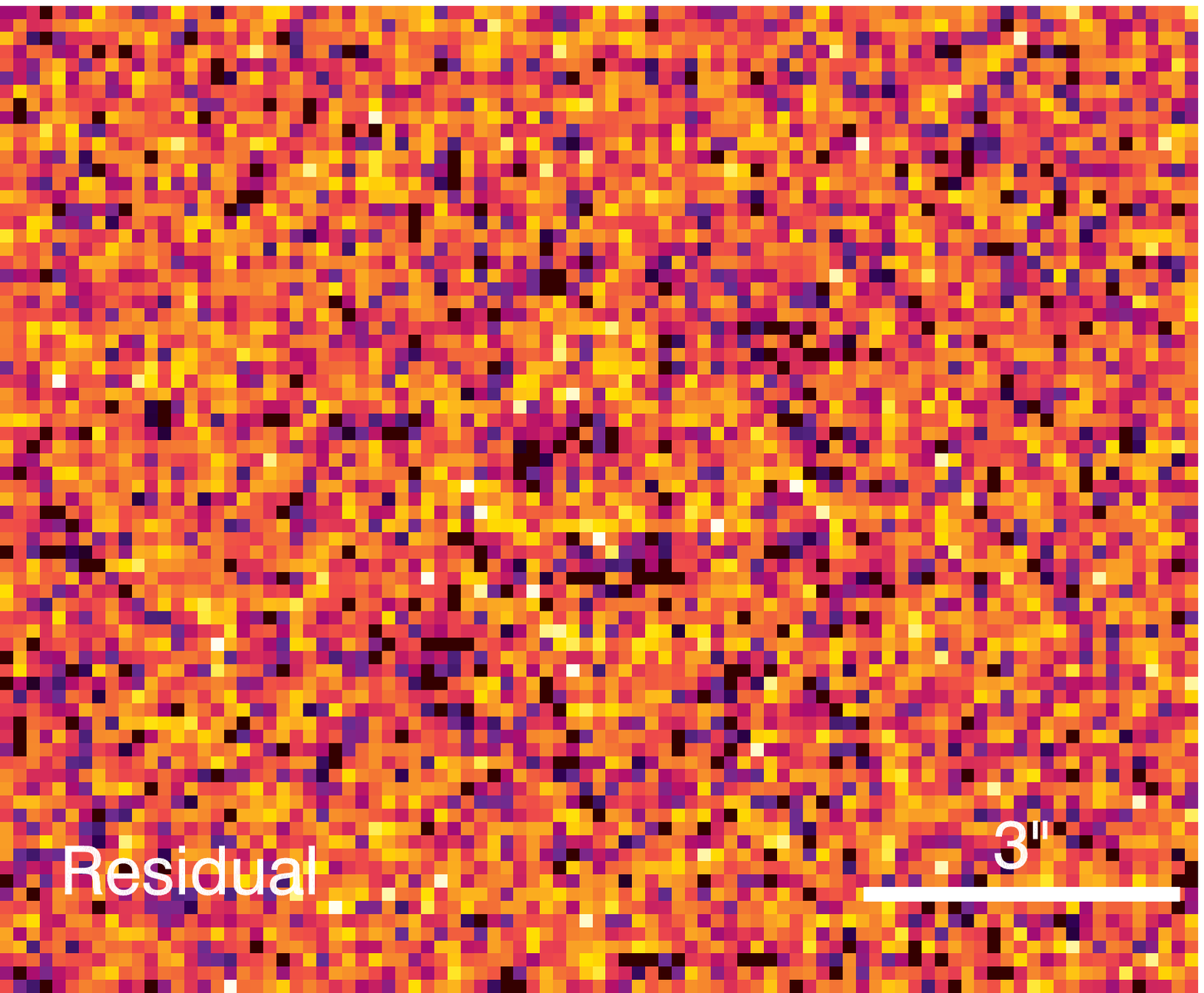}}
    \subfigure{
    \includegraphics[width=0.25\textwidth]{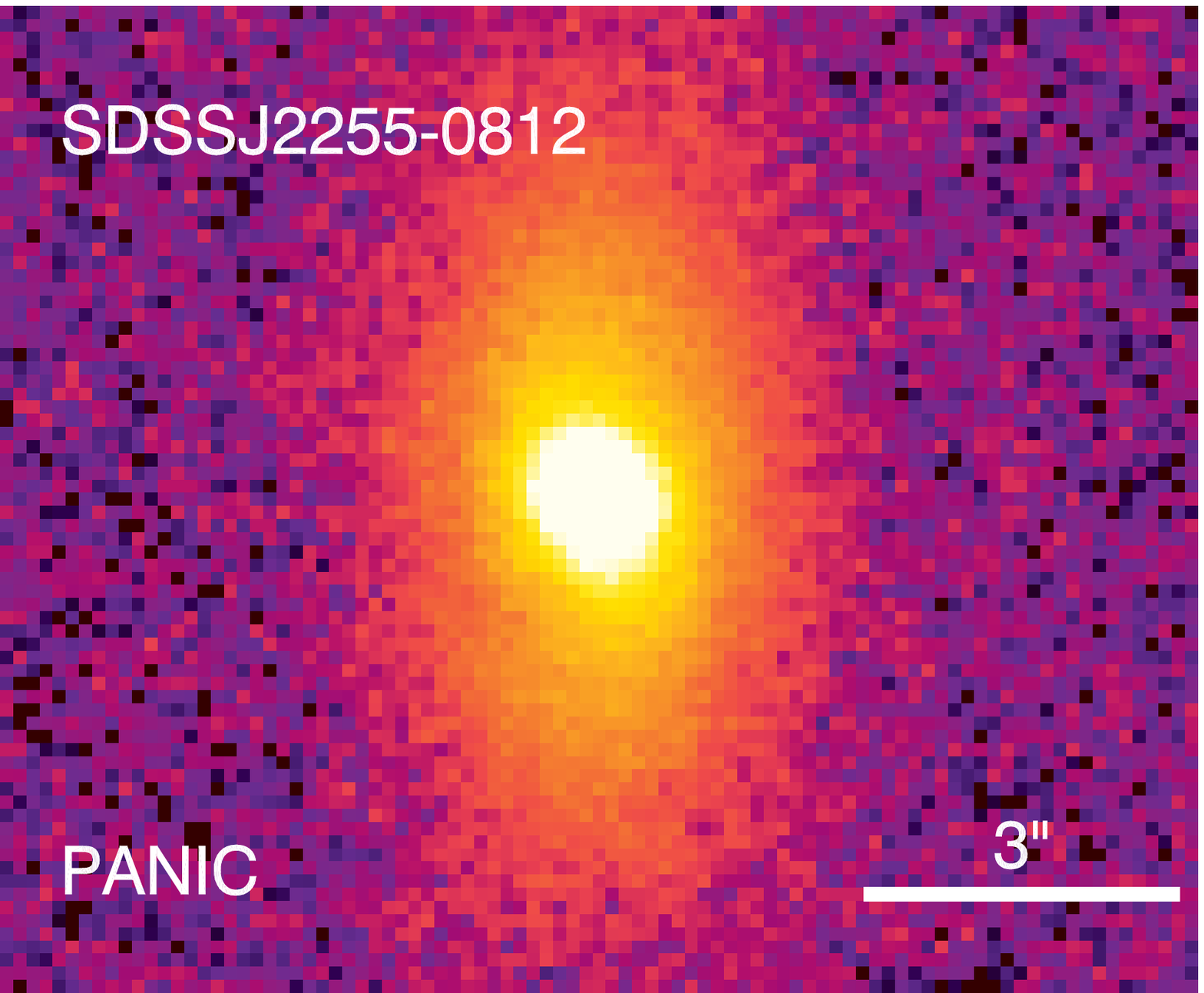}
    \includegraphics[width=0.25\textwidth]{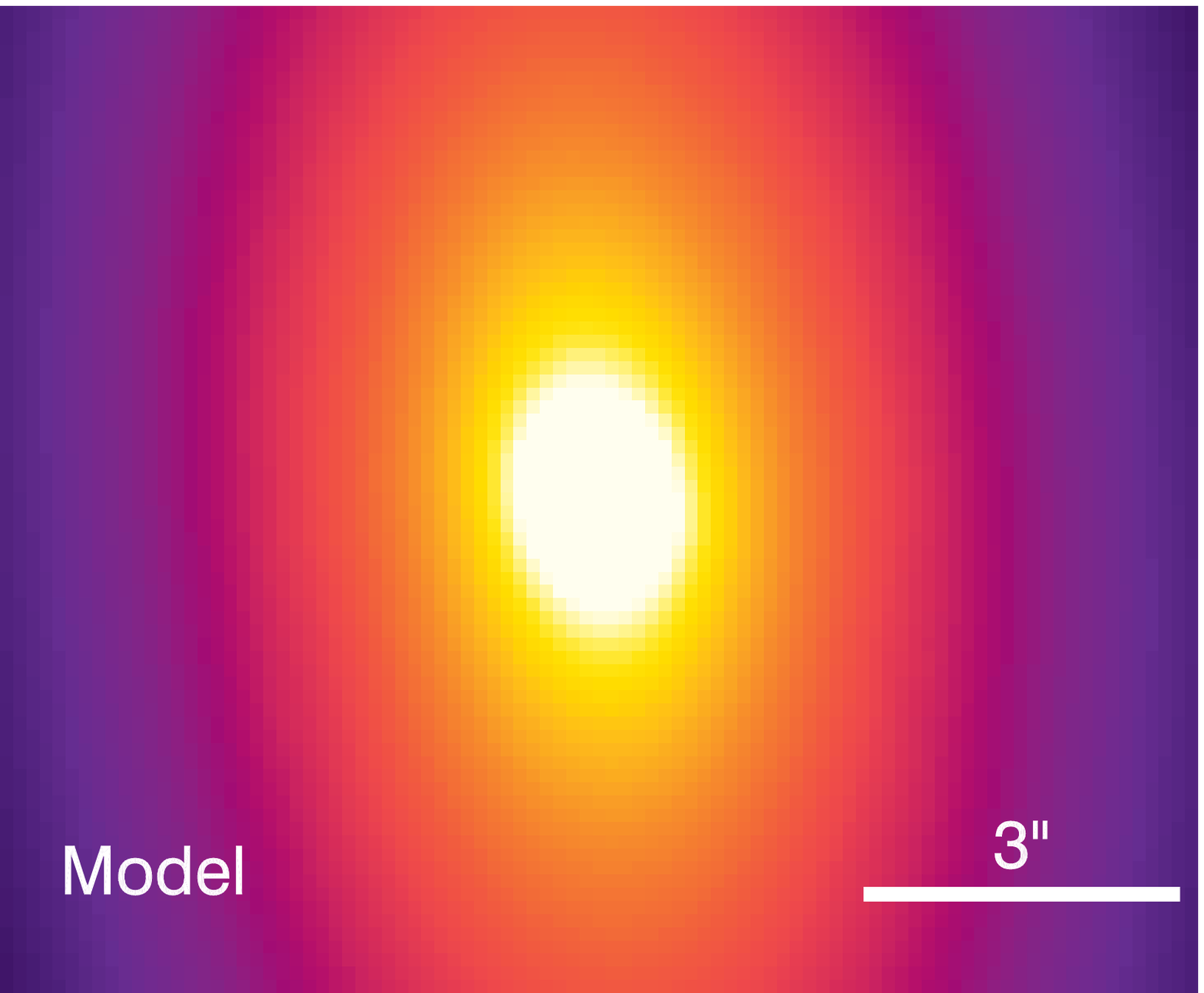}
    \includegraphics[width=0.25\textwidth]{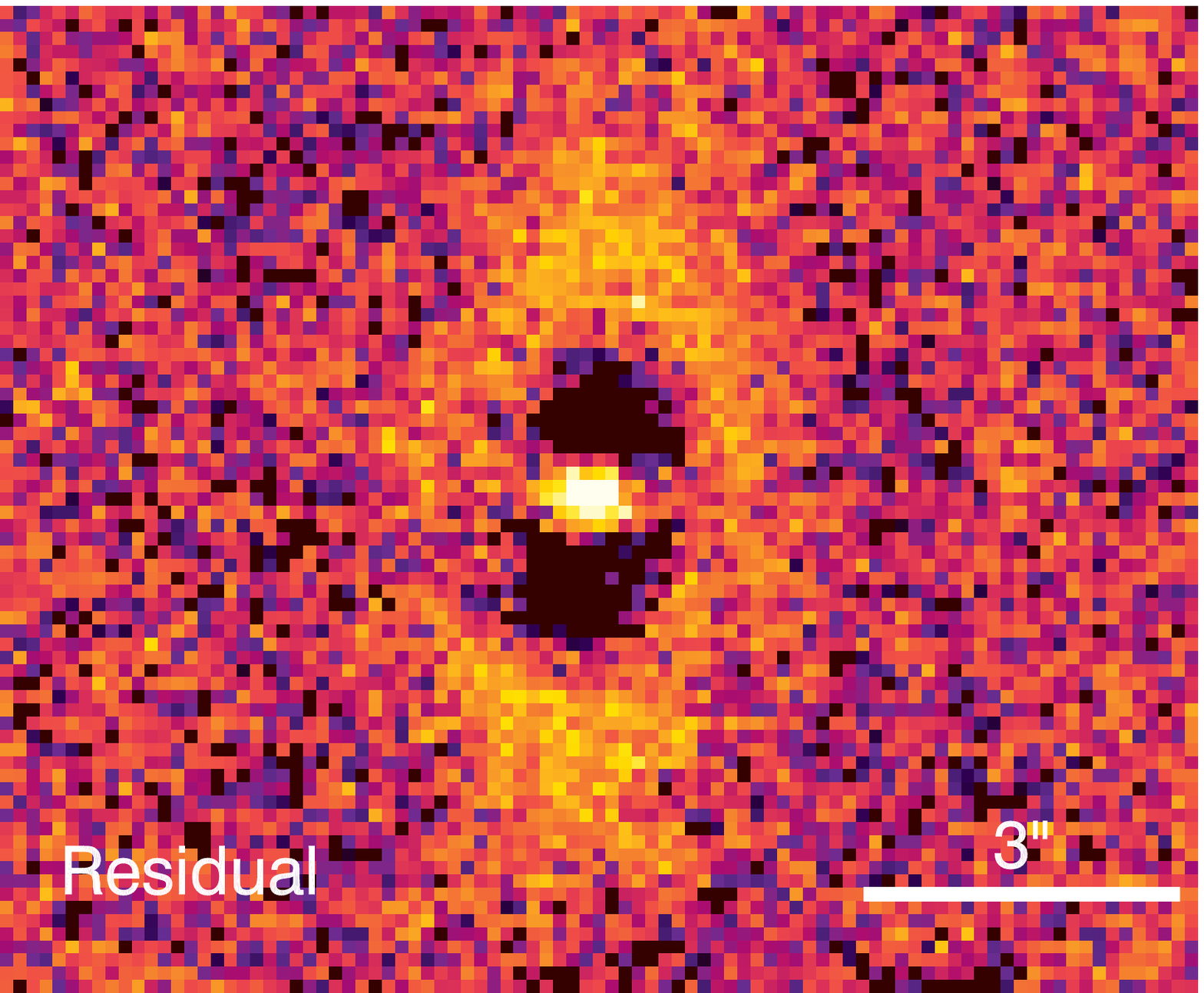}}
    \subfigure{
    \includegraphics[width=0.25\textwidth]{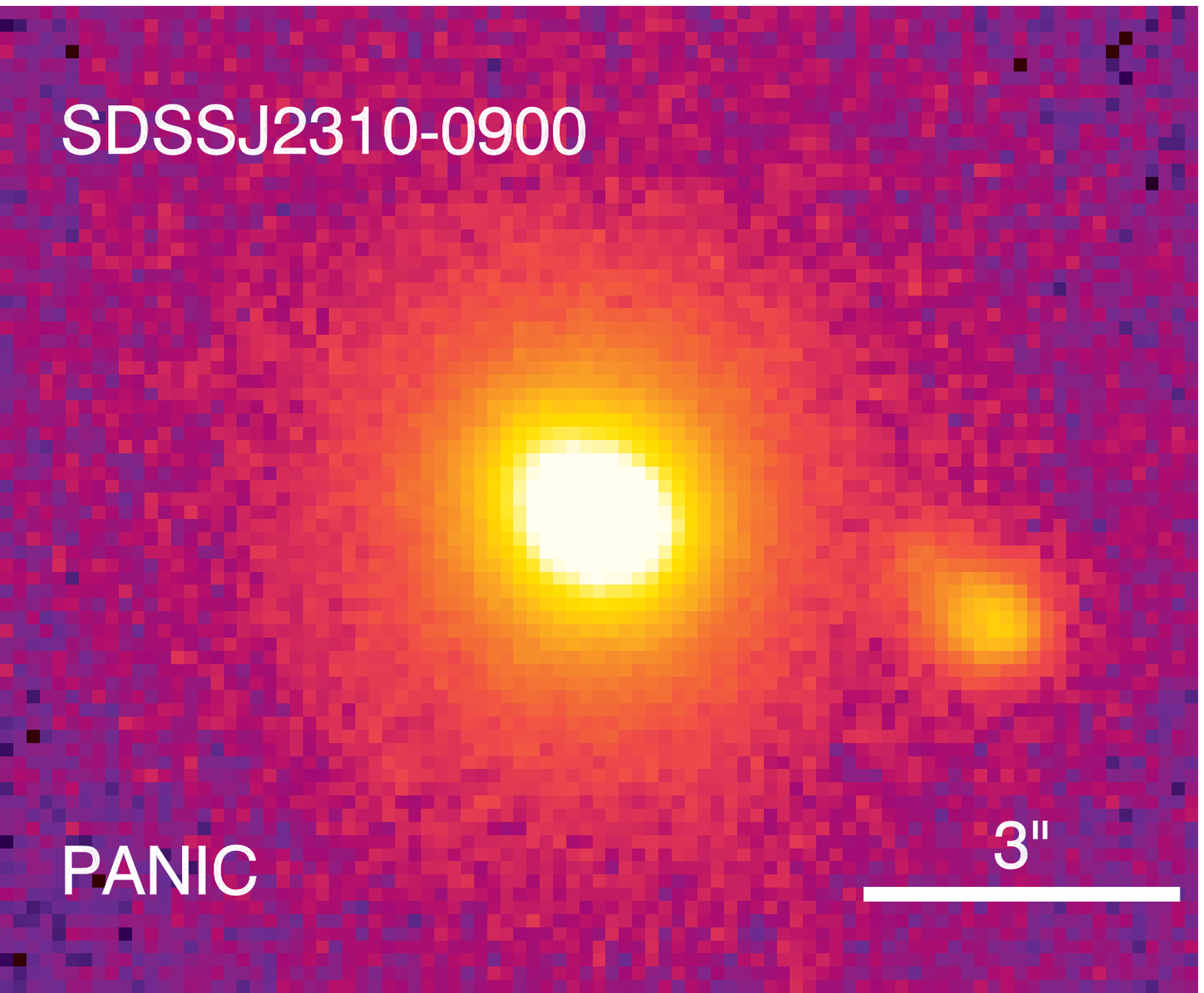}
    \includegraphics[width=0.25\textwidth]{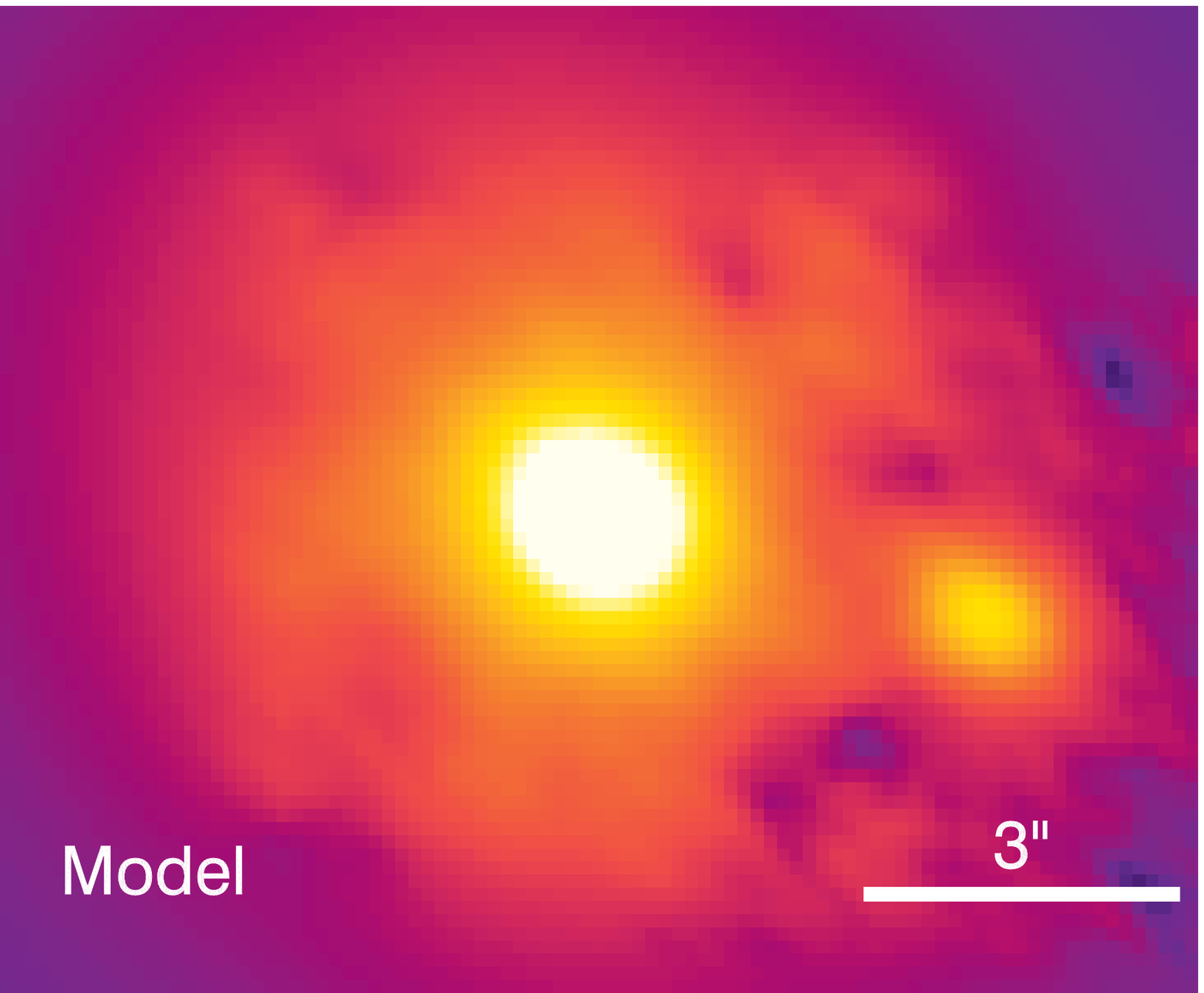}
    \includegraphics[width=0.25\textwidth]{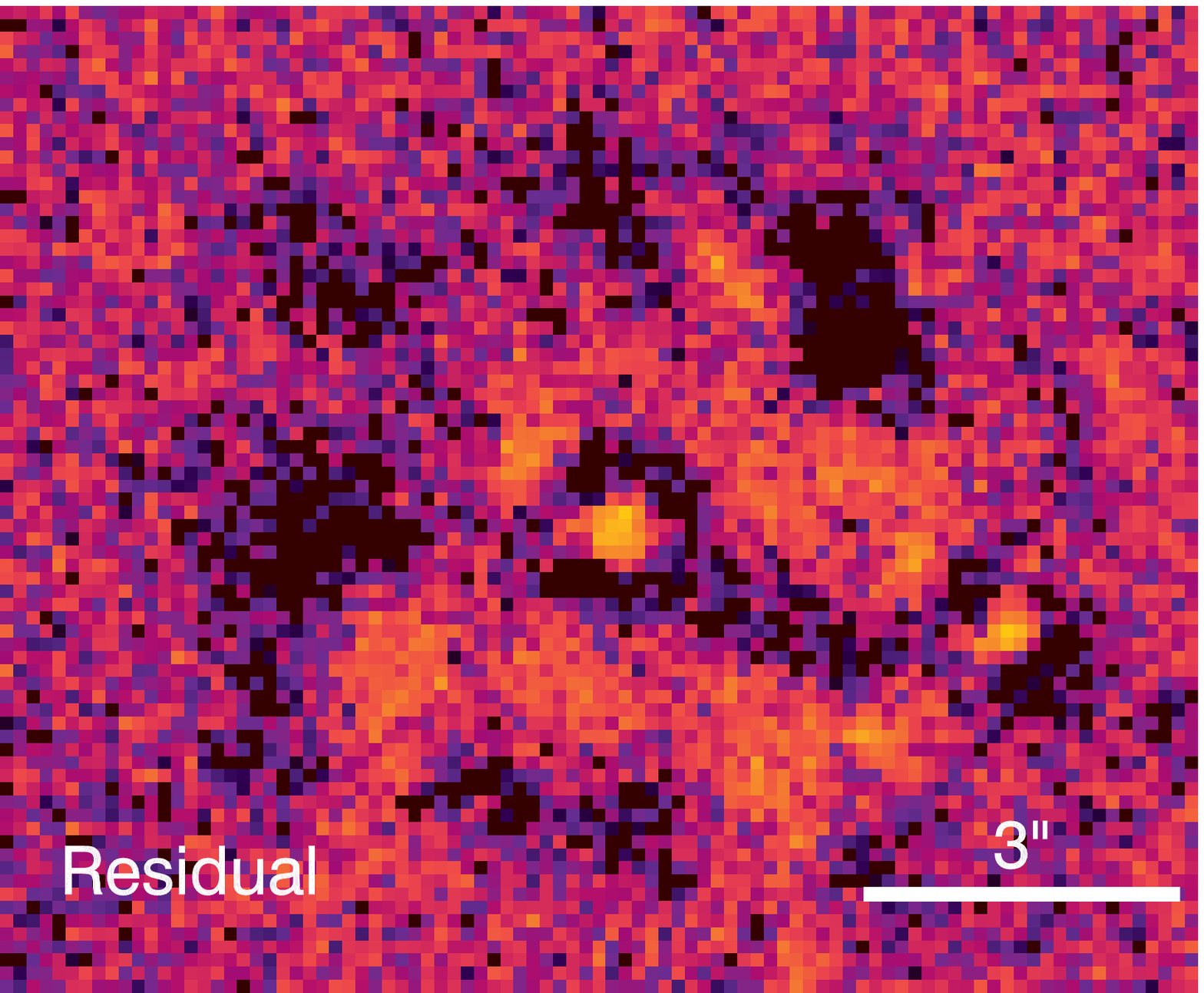}}
    \subfigure{
    \includegraphics[width=0.25\textwidth]{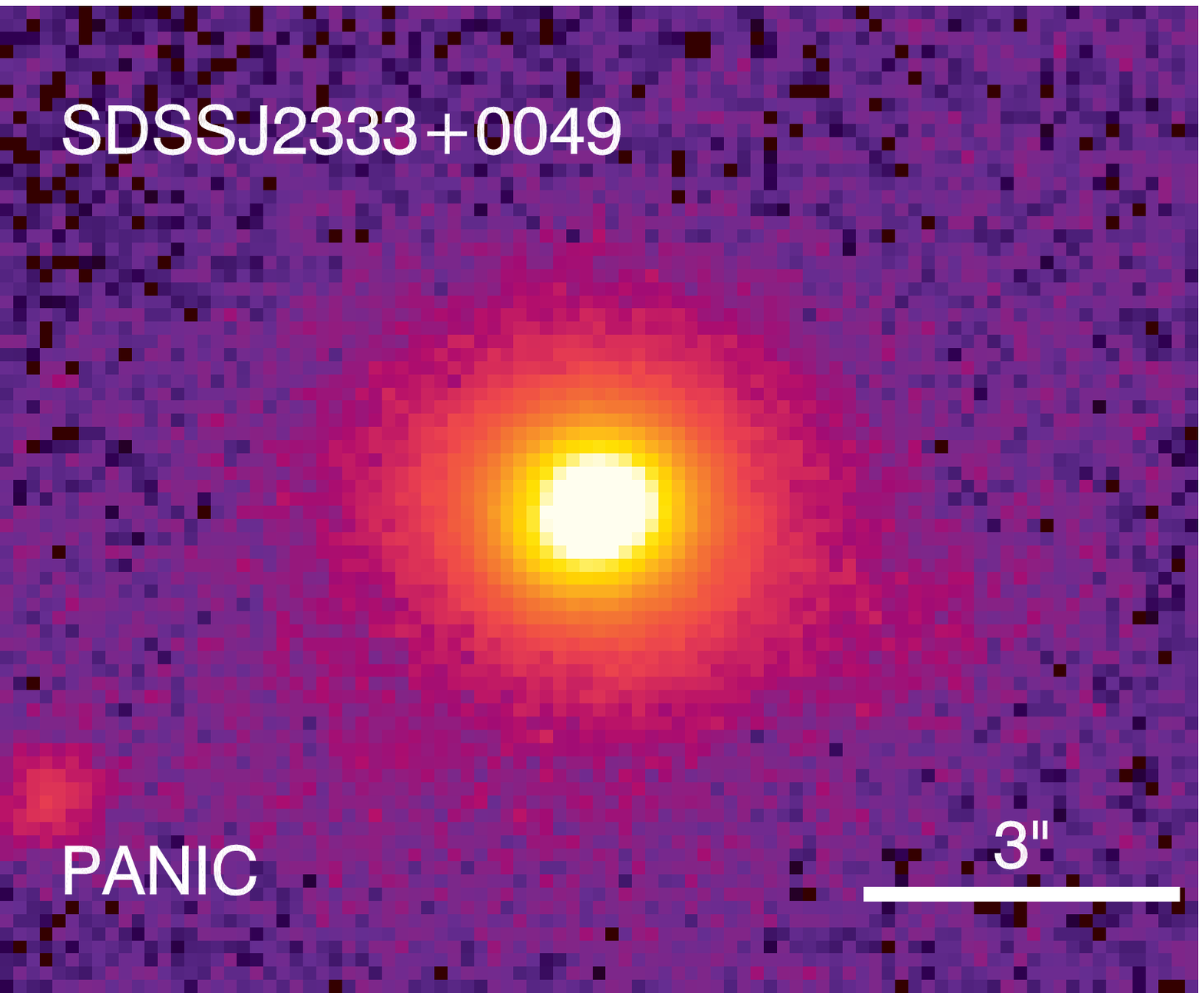}
    \includegraphics[width=0.25\textwidth]{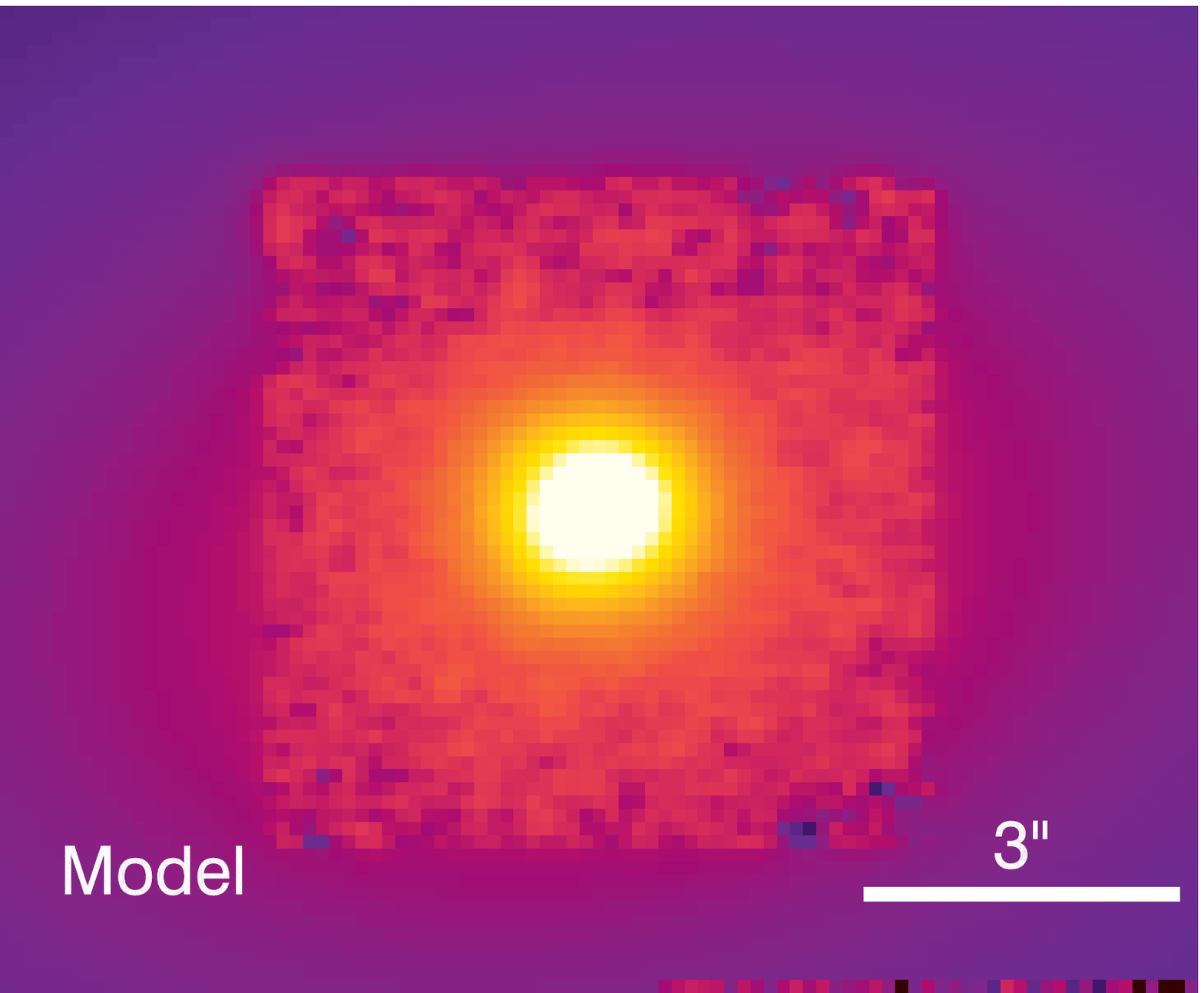}
    \includegraphics[width=0.25\textwidth]{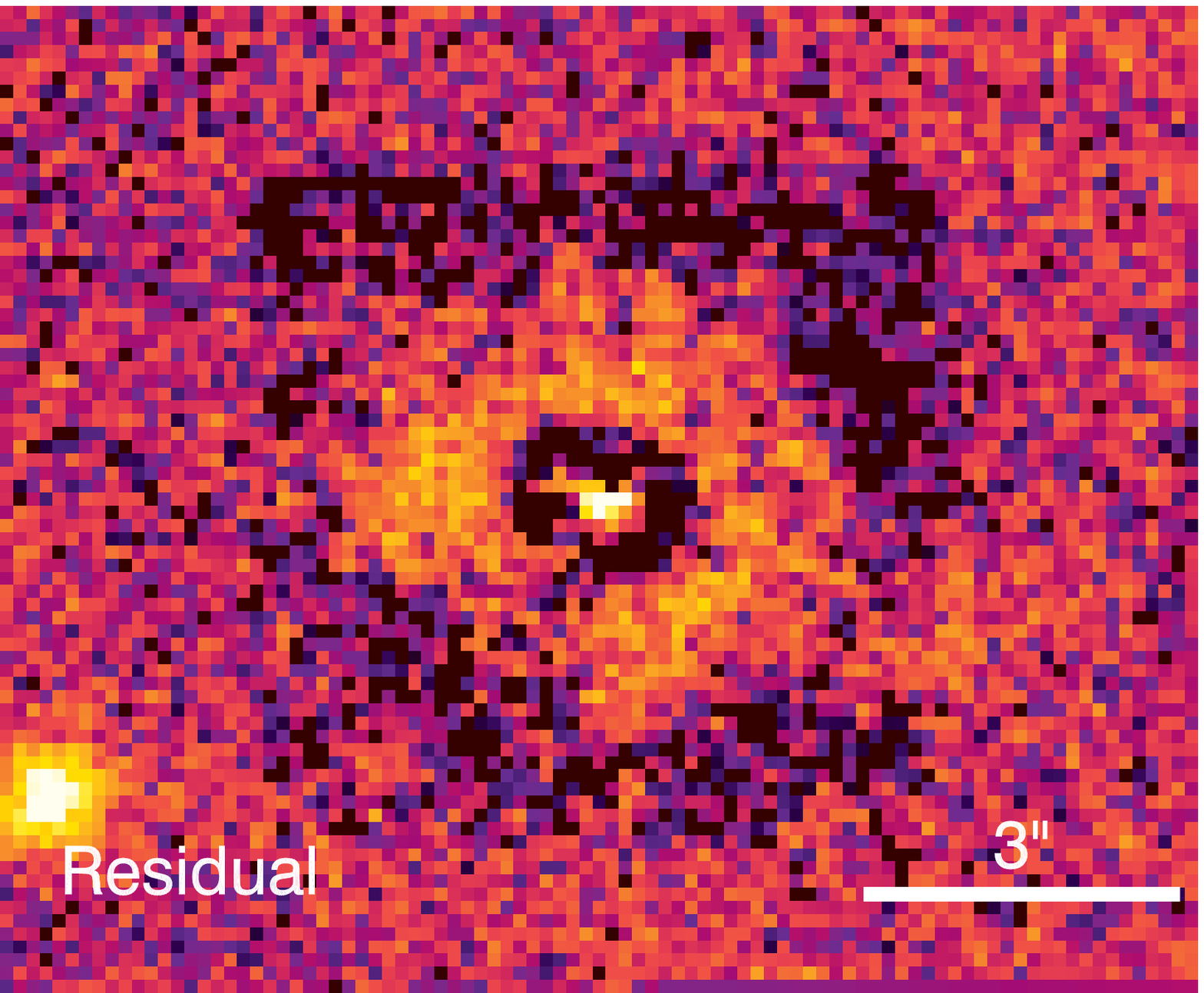}}
\caption{Continued. }
    \label{fig:amb_galfit}
\end{figure*}

\end{document}